\documentclass{like-thesis}
\usepackage[utf8]{inputenc}
\usepackage[english]{babel}

\usepackage{graphicx}
\graphicspath{{\figuredir/}} 
\usepackage{tensor}
\usepackage{siunitx}
\usepackage{float}
\usepackage{import}
\usepackage[numbers]{natbib}

\usepackage[dvipsnames]{xcolor}
\definecolor{Wheat}           {RGB}{96, 87, 70}
\definecolor{Wheat2}          {RGB}{238, 216, 174}
\definecolor{pblue}           {rgb}{0.13, 0.13, 1}
\definecolor{pgreen}          {rgb}{0, 0.5, 0}
\definecolor{pred}            {rgb}{0.9, 0, 0}
\definecolor{pgray}           {RGB}{220,220,220}
\definecolor{IndigoBlue}      {RGB}{9, 131, 146}
\definecolor{RoyalBlueDark}   {RGB}{0, 35, 102}
\definecolor{FullBlue}        {RGB}{0, 0, 255}
\definecolor{NavyBlue}        {RGB}{0, 0, 128}
\definecolor{RoyalBlue}       {RGB}{65, 105, 225}
\definecolor{NavyBlueLight}   {RGB}{25, 116, 210}
\definecolor{AzureBlue}       {RGB}{0, 127, 255}
\definecolor{DodgerBlue}      {RGB}{0, 90, 156}
\definecolor{BlueViolet}      {RGB}{138, 43, 226}
\definecolor{BlueViolet2}     {RGB}{138, 43, 254}
\definecolor{Gold}            {RGB}{255,215,0}
\definecolor{ForestGreen}     {RGB}{34,139,34}
\definecolor{GrassGreen}      {RGB}{126, 211,	33}
\definecolor{CadmiumGreen}    {RGB}{0, 107,	60}
\definecolor{DarkCyanGreen}   {RGB}{0, 120,	60}
\definecolor{LightCyanGreen}  {RGB}{0, 181,	100}
\definecolor{SpanishGreen}    {RGB}{0, 145, 80}
\definecolor{SpringGreen}     {RGB}{0, 178,	88}
\definecolor{SeaGreen}        {RGB}{46, 139, 87}
\definecolor{Pink}            {RGB}{255, 81, 199}
\definecolor{DeepPink}        {RGB}{255, 20, 147}
\definecolor{BrightPurple}    {RGB}{167, 17, 196}
\definecolor{BrightViolet}    {RGB}{127, 0, 255}
\definecolor{FireBrick}       {RGB}{178, 34, 34}
\definecolor{BrightRed}       {RGB}{255, 32, 50}
\definecolor{DarkRed}         {RGB}{208, 2, 27}

\definecolor{DodgerPurple}    {RGB}{144, 19, 254}
\definecolor{DarkYellow}      {RGB}{210, 177, 8}
\definecolor{DarkOrange}      {RGB}{245, 166, 35}
\definecolor{DodgerPurple}    {RGB}{144, 19, 254}
\definecolor{Aqua}            {RGB}{92, 225, 225}
\definecolor{Brownish}        {RGB}{115, 65, 55}

\definecolor{PalePink}        {RGB}{255, 207, 220}
\definecolor{PaleSalmon}        {RGB}{255, 177, 154}

\title{Efficient Algorithms for Optimal Homology Problems \\ and their Applications}
\author{Kostiantyn Lyman}
\thesistype{Dissertation}
\degree{Doctor of Philosophy}
\department{Department of Mathematics and Statistics}
\submitmonth{May}
\submityear{2024}

\newcommand{\phd}{Ph.D.}

\newcommand{\figuredir}{./figs}

\usepackage[T1]{fontenc} 

\usepackage{algorithmic}
\usepackage[linesnumbered, vlined, ruled]{algorithm2e}

\usepackage{amsfonts} 
\usepackage{amsmath}
\usepackage{amssymb}
\usepackage{wasysym}
\usepackage{mathtools}
\usepackage{verbatim}
\usepackage{multirow}
\usepackage[single]{accents}
\usepackage{bm}
\usepackage{xfrac}
\usepackage{dsfont}
\usepackage{fontawesome5}
\usepackage{scalerel}

\usepackage[mathscr]{euscript}

\newcommand*{\dsp}                  {\displaystyle}

\newcommand*{\scale}[1]             {\scalebox{0.6}{$#1$}}

\DeclareMathOperator*{\Img}         {Im}                  
\DeclareMathOperator*{\im}          {Im}                  
\DeclareMathOperator*{\rank}        {rank}                
\DeclareMathOperator*{\supp}        {supp} 
\DeclareMathOperator*{\dist}        {dist} 
 
\DeclareMathOperator*{\cost}        {cost}

\newcommand*{\abs}[1]               {\left| {#1} \right|}                     
\newcommand*{\Abs}[1]               {\left|{#1}\right|}
\newcommand*{\aBs}[1]               {\big|{#1}\big|}

\newcommand*{\norm}[1]              {\left\| {#1} \right\|}                   
\newcommand*{\support}[1]           {\lfloor {#1} \rfloor}                    

\newcommand*{\real}                 {\mathbb{R}}
\newcommand*{\nat}                  {\mathbb{N}}
\newcommand*{\integer}              {\mathbb{Z}}

\newcommand*{\bigO}                 {\mathcal{O}}  
\newcommand*{\integerplus}          {\mathbb{Z}^{\scalebox{0.75}{+}}}
\newcommand*{\realplus}             {\mathbb{R}^{\scalebox{0.75}{+}}}

\newcommand*{\Half}                 {\frac{1}{2}}

\newcommand*{\tld}[1]               {\tilde{#1}}
\newcommand*{\bbar}[1]              {\overline{#1}}

\newcommand*{\iimplies}             {\Longrightarrow}

\newcommand*{\pprime}               {\prime\prime}
\newcommand*{\tr}                   {\scalebox{0.5}{$\mathsf{T}$} }

\newcommand*{\me}                   {\phantom{-}}         

\newcommand*{\splus}                {\scale{+}}                                               
\newcommand*{\sbullet}[1][0.75]     {\mathbin{\vcenter{\hbox{\scalebox{#1}{$\bullet$}}}}}     

\newcommand*{\mc}[1]                {\mathcal{#1}}

\newcommand*{\mbb}[1]               {\mathbb{#1}}

\newcommand*{\embf}[1]              {\textbf{\emph{#1}}}

\newcommand*{\Mobius}               {M{\"o}bius }

\newcommand*{\subto}                {\mathrm{s.t.}}

\newcommand*{\bd}                   {\partial}                            
\newcommand*{\cobd}                 {\delta}                              
\newcommand*{\dart}                 {\uparrow}                            
\newcommand*{\csim}                 {\overset{{c}}{\sim}}                 
\newcommand{\widesim}[2][1.5]{
  \mathrel{\overset{#2}{\scalebox{#1}[1]{$\sim$}}}
}

\newcommand*{\e}                    {\delta}
\newcommand*{\perturb}              {\mathrel{\rightsquigarrow}}
\newcommand*{\perturbe}             {\mathrel{\overset{\e}{\rightsquigarrow}}}
\newcommand*{\perturbeplus}         {\mathrel{\overset{\e^{\scalebox{0.4}{+}}}{\rightsquigarrow}}}

\DeclareRobustCommand{\perturbE}    {{\overset{\e}{\rightsquigarrow}}}
\DeclareRobustCommand{\perturbEplus}         {\mathrel{\overset{\e^{\scalebox{0.4}{+}}}{\rightsquigarrow}}}

\newcommand*{\C}                    {\mathcal{C}}                 
\newcommand*{\Z}                    {\mathcal{Z}}                 
\newcommand*{\B}                    {\mathcal{B}}                 
\renewcommand*{\H}                    {\mathcal{H}}                 

\DeclareFontFamily{U} {MnSymbolA}{}
\DeclareFontShape{U}{MnSymbolA}{m}{n}{
  <-6> MnSymbolA5
  <6-7> MnSymbolA6
  <7-8> MnSymbolA7
  <8-9> MnSymbolA8
  <9-10> MnSymbolA9
  <10-12> MnSymbolA10
  <12-> MnSymbolA12
}{}
\DeclareSymbolFont{MnSyA} {U} {MnSymbolA}{m}{n}

\DeclareFontFamily{U} {MnSymbolC}{}
\DeclareFontShape{U}{MnSymbolC}{m}{n}{
  <-6> MnSymbolC5
  <6-7> MnSymbolC6
  <7-8> MnSymbolC7
  <8-9> MnSymbolC8
  <9-10> MnSymbolC9
  <10-12> MnSymbolC10
  <12-> MnSymbolC12
}{}
\DeclareSymbolFont{MnSyC} {U} {MnSymbolC}{m}{n}

\DeclareFontFamily{U} {MnSymbolD}{}
\DeclareFontShape{U}{MnSymbolD}{m}{n}{
  <-6> MnSymbolD5
  <6-7> MnSymbolD6
  <7-8> MnSymbolD7
  <8-9> MnSymbolD8
  <9-10> MnSymbolD9
  <10-12> MnSymbolD10
  <12-> MnSymbolD12
}{}
\DeclareSymbolFont{MnSyD} {U} {MnSymbolD}{m}{n}

\DeclareMathSymbol{\rcirclearrowright}{\mathrel}{MnSyA}{248}                  
\DeclareMathSymbol{\lcirclearrowright}{\mathrel}{MnSyA}{252}
\DeclareMathSymbol{\lcirclearrowleft}{\mathrel}{MnSyA}{254}                   
\DeclareMathSymbol{\lcurvearrowleft}{\mathrel}{MnSyA}{186}                    
\DeclareMathSymbol{\rcurvearrowright}{\mathrel}{MnSyA}{192}                   

\DeclareMathSymbol{\vdotdot}{\mathbin}{MnSyC}{2}                              

\newcommand*{\coloneq}              {\mathrel{{\vdotdot}{=}}}                 
\newcommand*{\asgn}                 {\coloneq}                                

\DeclareMathSymbol{\sqdoublefrown}{\mathord}{MnSyD}{45}                           
\DeclareMathSymbol{\sqfrowneq}{\mathord}{MnSyD}{51}                           

\newcommand*{\sqfrowneqsmall}     {\mathord{\scalebox{1.07}[0.8]{$\sqfrowneq$}}}

\newcommand*{\hatbar}[1]           {\accentset{\sqfrowneqsmall}{#1}}

\newcommand*{\pvec}[1]           {\accentset{\hookrightarrow}{#1}} 
\newcommand*{\vvec}[1]           {\accentset{\rightarrow}{#1}}   
\newcommand*{\hvec}[1]           {\hat{{\accentset{\hookrightarrow}{#1}}}}   

\usepackage{amsthm}

\theoremstyle{plain}
\newtheorem{theorem}{Theorem}[section]
\newtheorem{lemma}[theorem]{Lemma}
\newtheorem{proposition}[theorem]{Proposition}
\newtheorem{corollary}[theorem]{Corollary}

\theoremstyle{definition}
\newtheorem{definition}[theorem]{Definition}
\newtheorem{assumption}[theorem]{Assumption}

\newtheorem{example}[theorem]{Example}

\allowdisplaybreaks

\usepackage{fullpage}

\begin{document}
\maketitle


\begin{abstract}{\phd}{Bala Krishnamoorthy}

The multiscale simplicial flat norm (MSFN) of a (non-bounding) $d$-cycle $\bm{z}$
is a family of optimal homology problems indexed by a scale parameter $\lambda \geq 0$.
Each instance optimizes a weighted combination of positive integer weights ("volumes")
of the homologous $d$-cycle and the bounded $(d + 1)$-chain,
with the scale factor $\lambda$ applied to either component.
Additionally, one may seek the optimal flat norm decomposition of the input $d$-cycle into corresponding $d$- and $(d + 1)$-components.
Notably, the classical \textbf{optimal homologous cycle problem (OHCP)} represents a special case of mSFN when $\lambda = 0$.

We propose a min-cost flow formulation for solving an instance of MSFN
for the case of $(d + 1)$-dimensional simplicial complexes  embedded in $\real^{d + 1}$.
Furthermore, we establish both weak and strong duality results for mSFN  defined over $\integer$ and $\integerplus$ coefficients.
Additionally, we introduce stronger optimality conditions for directed min-cost flow solutions of OHCP
and offer a topological characterization of these conditions.

  Next, we propose an approach based on the \emph{multiscale flat norm}, a notion of distance between objects defined in the field of geometric measure theory, to compute the distance between a pair of planar geometric networks.
  Using a triangulation of the domain containing the input networks, the flat norm distance between two networks at a given scale can be computed by solving a linear program.
  In addition, this computation automatically identifies the 2D regions (patches) that capture where the two networks are different.
  We demonstrate through 2D examples that the flat norm distance can capture the variations of inputs more accurately than the commonly used Hausdorff distance.
  As a notion of \emph{stability}, we also derive upper bounds on the flat norm distance between a simple 1D curve and its perturbed version as a function of the radius of perturbation for a restricted class of perturbations.
  We demonstrate our approach on a set of actual power networks from a county in the USA.
  Our approach can be extended to validate synthetic networks created for multiple infrastructures such as transportation, communication, water, and gas networks.

\end{abstract}

\tableofcontents
\listoffigures


\begin{mainchapters}

\chapter{Introduction}
\label{sec:intro}

\textbf{Optimal homology problems (OHP)}, the main focus of this paper, 
involve seeking an optimum, typically a minimum, of certain numerical characteristics, 
collectively referred to as \textbf{costs}, 
within an equivalence class of input data $\bm{z}$ defined by the simplicial (co)homology of the input complex $\mc{K}$.
We specifically consider problems in codimension 1, where $\dim  \bm{z} = \dim \mc{K} - 1$, 
with all numerical characteristics defined over the ring of integers $\integer$.
 Throughout this paper, unless explicitly stated otherwise, 
 we maintain fixed dimensions for the input data $\bm{z}$ and simplicial complex $\mc{K}$ at $q$ and $q + 1$, respectively.
  The formal definition of discussed OHPs in Section\ref{subsec:def:opt-homology}. 

The \textbf{optimal homologous cycle problem (OHCP)} is a classical example of OHP. 
OHCP aims to minimize costs related solely to the dimension of the input cycle, i.e., 
costs that are defined as non-negative integer-valued functions on the $q$-simplices of $\mc{K}$.
 For example, this could involve minimizing the length of a $1$-cycle on a $2$-complex.


The \emph{flat norm} is a tool from geometric measure theory to measure distances between shapes~\cite{glaunes2005transport, meyur2023structural}.
The \emph{multiscale flat norm} was introduced by in Morgan and Vixie \cite{MoVi2007}, was used for applications in image analysis and denoising~\cite{vixie2010multiscale}.
Ibrahim, Krishnamoorthy, and Vixie in \cite{ibrahim2011simplicial} introduced the \embf{multiscale Simplicial Flat Norm (MSFN)}, 
which is a discretization of the flat norm to spaces triangulated by a simplicial complex.
MSFN is a family of OHPs indexed by a scale parameter $\lambda \geq 0$ 
that optimizes a weighted combination of costs defined in the data and the input complex's dimensions, 
encompassing both $q$ and $(q + 1)$-simplices. 
If the scale parameter is a positive integer, 
$\lambda \in \integerplus$, then the $(q + 1)$-costs are multiplicatively scaled by $\lambda$, 
resulting in an \textbf{area-scaled SFN} or \textbf{$\lambda$-SFN}.
 Conversely, if the reciprocal scale is a positive integer,
 $\Lambda \asgn \sfrac{1}{\lambda} \in \integerplus$, 
 then the $q$-costs are scaled by $\Lambda$, leading to a \textbf{length-scaled SFN} or \textbf{$\Lambda$-SFN}.
We collectively refer to instances of the multiscale simplicial flat norm with a fixed scale parameter as mSFN.
  Hence, mSFN provides a notion of "distance" at the chosen scale within the homology class of the input cycle $\bm{z}$. 
  A \textit{simplicial flat norm (SFN)} problem is a special case of mSFN defined at the unit scale, $\lambda = 1$. 
  Notably, OHCP corresponds to the special case of mSFN with $\lambda = 0$. 
  Another classical OHP, stemming from the length-scaled SFN with $\Lambda = 0$, 
  where only $(q + 1)$-costs are defined, is known as the \textit{optimal bounding chain problem (OBCP)} \cite{sullivan1990crystalline}. 
  For a broader overview of OHPs and their linear program, see \cite{minRepresentatives2021}.

 In general, mSFN, as well as OHCP, are NP-hard even when homology is defined over integers~\cite{dunfield2011least, chen2011hardness}, 
 however in the case of embedded complexes it can be solved in polynomial time~\cite{ohcp2011, ibrahim2011simplicial}. 
 The standard approach is  to solve optimal homology problems as a linear relaxation of corresponding integer program~\cite{minRepresentatives2021, escolar2016optimal}, 
 which at best has matrix multiplication time {$O(n^{2.38} \log n)$}\cite{van2020deterministic}.
Here we propose a network flow model for solving OHCP and mSFN as a min-cost flow on the dual graph of a $(d + 1)$-simplicial complex embedded in $\real^{d + 1}$. 
The fastest algorithm for solving min-cost problems at the time of writing
has almost linear complexity $m^{1 + o(1)}$ in terms of the number of edges, \cite{chen2022maximum}.
We will present optimality conditions for OHCP  in Section \ref{subsec:ohcp:optimality-chains}, and in Section \ref{subsec:ohcp:optimality-dir} we prove a stronger version of reduced costs optimality condition for directed case.
Then, in Section \ref{sec:fnorm:duality-theorems} we prove dual theorems for mSFN, and the directed case is considered in \ref{subsec:ohcp:optimality-dir}.
As result, we present the Algorithm~\ref{algo:fnorm:flow-network} for construction of a dual flow network and prove that the min-cost flow solution on it corresponds to an optimal solution on the original complex.

We also considered applications of optimal homology problems to practical problems. 
One such problem is the problem of evaluating the quality of reconstructed networks.
The abstract problem can be stated as follows: \emph{compute the similarity between a given pair of embedded planar graphs}.
This is similar to the well-known subgraph isomorphism problem~\cite{eppstein_1995} wherein we look for isomorphic subgraphs in a pair of given graphs.
A major precursor to this problem is that we require a one-to-one mapping between nodes and edges of the two graphs.
While such mappings are well-defined for street networks, the same cannot be inferred for power distribution networks.
Since power network datasets are proprietary, the node and edge labels are redacted from the network before it is shared.
The actual network is obtained as a set of ``drawings'' with associated geographic embeddings. Each drawing can be considered as a collection of line segments termed a \emph{geometry}.
Hence the problem of comparing a set of power distribution networks with geographic embedding can be stated as the following: 
\emph{compute the similarity between a given pair of geometries lying on a geographic plane}.
It is impossible to derive \emph{standard} stability results for this distance measure that imply only small changes in the flat norm metric when the inputs change by a small amount---there is no alternative metric to measure the \emph{small change in the input}.
For instance, our theoretical example (in Fig.~\ref{fig:currents:example-currents-and-neighborhoods}) shows that the commonly used Hausdorff metric cannot be used for this purpose.
Instead, we have derived upper bounds on the flat norm distance between a piecewise linear 1-current and its perturbed version as a function of the radius of perturbation under certain assumptions provided the perturbations are performed carefully (see Section \ref{subsec:FN-BOUND}).
On the other hand, we do get natural stability results for our distance following the properties of the flat norm---small changes in the input geometries lead to only small changes in the flat norm distance between them~\cite{Federer1969,Morgan2016}.

A lack of one-to-one correspondence between edges and nodes in the pair of networks prevents us from performing one-to-one comparison of edges.
Instead we can sample random regions in the area of interest and compare the pair of geometries within each region.
For performing such local comparisons, we define a \emph{normalized flat norm} where we normalize the flat norm distance between the parts of the two geometries by the sum of the lengths of the two parts in the region. 
Such comparison enables us to characterize the quality of the digital duplicate for the sampled region.
Further, such comparisons over a sequence of sampled regions allows us to characterize the suitability of using the entire synthetic network as a duplicate of the actual network.

\section{Summary of results}

In our research, we addressed the problem of efficient solving of the Optimal Homology Cycle Problem  (OHCP) and the Multiscale Simplicial Flat Norm (MSFN)
by formulating them as min-cost flow problems using "flow" complexes.
These complexes have valid dual graphs, which we call dual flow networks.
An important observation is that feasible and optimal solutions on these constructed complexes coincide with solutions on the original complex,
ensuring the effectiveness of our approach.

Our \textbf{main theoretical contributions} regarding optimal homology problems are the following:
We extended Sullivan's duality results for OHCP to complexes with non-trivial homology.
Additionally, we establish stronger optimality conditions for directed min-cost flow solutions of OHCP, and we give a topological characterization of these conditions.
For mSFN, we prove the duality theorems by constructing the "hat"-complex.
This construction preserves the original solutions, and can be extended to a direct hat-complex.
The dual of a direct hat-complex is a dual flow network for a mSFN problem.


Additionally, we investigate the stability of mSFN as a measuring device for comparing pairs of geometries in 2D.
We propose a fairly general metric to compare a pair of network geometries embedded on the same plane.
Unlike standard approaches that map the geometries to points in a possibly simpler space and then measuring distance between those points~\cite{KeBaCaLe2009}, or comparing ``signatures'' for the geometries,  our metric works directly in the input space and hence allows us to capture all details in the input.
The metric uses the multiscale flat norm from geometric measure theory, and can be used in more general settings as long as we can triangulate the region containing the two geometries.
It is impossible to derive \emph{standard} stability results for this distance measure that imply only small changes in the flat norm metric when the inputs change by a small amount---there is no alternative metric to measure the \emph{small change in the input}.
For instance, our theoretical example (in Fig.~\ref{fig:currents:example-currents-and-neighborhoods}) shows that the commonly used Hausdorff metric cannot be used for this purpose.
Instead, we have derived upper bounds on the flat norm distance between a piecewise linear 1-current and its perturbed version as a function of the radius of perturbation under certain assumptions provided the perturbations are performed carefully (see Section \ref{subsec:FN-BOUND}).
On the other hand, we do get natural stability results for our distance following the properties of the flat norm---small changes in the input geometries lead to only small changes in the flat norm distance between them~\cite{Federer1969,Morgan2016}.

A lack of one-to-one correspondence between edges and nodes in the pair of networks prevents us from performing one-to-one comparison of edges.
Instead, we can sample random regions in the area of interest and compare the pair of geometries within each region.
For performing such local comparisons, we define a \emph{normalized flat norm} where we normalize the flat norm distance between the parts of the two geometries by the sum of the lengths of the two parts in the region.
Such comparison enables us to characterize the quality of the digital duplicate for the sampled region.
Further, such comparisons over a sequence of sampled regions allow us to characterize the suitability of using the entire synthetic network as a duplicate of the actual network.

We use the proposed metric to compare a pair of power distribution networks: (i) actual power distribution networks of two locations in a county of USA obtained from a power company and (ii) synthetically generated digital duplicate of the network created for the same geographic location.
The proposed comparison metric is able to perform a global, as well as a local comparison of network geometries for the two locations.
We discuss the effect of different parameters used in the metric on the comparison.
Further, we validate the suitability of using the flat norm metric for such comparisons using computation, as well as theoretical examples.

Our \textbf{main practical contributions} regarding mSFN are the following:
(i) we propose a distance measure for comparing a pair of geometries embedded in the same plane using the flat norm that accounts for deviation in length and lateral displacement between the geometries; and
(ii) we perform a region-based characterization of synthetic networks by sampling random regions and comparing the pair of geometries contained within the sampled region.
The proposed distance allows us to perform global as well as local comparisons between a pair of network geometries.

\section{Related Work} \label{ssec:relwrk}

Our objective is to develop network flow models and algorithms
that offer better efficiency compared to conventional linear programming techniques commonly employed for solving optimal homology problems (OHPs),
for the cases where such methods are applicable.

Chen and Freedman \cite{chen2011hardness} provided fundamental insights into the computational complexity of OHP instances with $\integer_2$-coefficient.
Specifically, they proved the NP-hardness of general instances of the optimal homologous cycle problem (OHCP) when coefficients are drawn from the field $\integer_2$.
Notably, their analysis extended to polynomial cases, particularly when the $(q + 1)$-complex is embedded within $\real^{q + 1}$.

Expanding on this exploration of computational complexity,
 Dunfield and Hirani \cite{dunfield2011least} demonstrated that OHCP instances
 with homology defined over the integers $\integer$ present NP-complete challenges.
  This result encompasses a broader domain, including instances of the optimal bounding chain problem (OBCP)
  when homology is defined over $\integer$.
  Sullivan (1990) provided a noteworthy exception by proposing a polynomial-time algorithm for specific instances of OBCP,
  particularly when the input complex conforms to an $(q + 1)$-manifold in $\real^{(q + 1)}$, leveraging principles of network flow.

Dey, Hirani, and Krishnamoorthy \cite{ohcp2011} contributed to the understanding of polynomial-time solvability in OHPs
 by identifying specific conditions under which OHCP instances with homology defined over $\integer$ can be efficiently solved.
 They established that polynomial-time solutions are attainable when the boundary matrix of the input complex is totally unimodular (TU).
 This property is characterized by the torsion-freeness of certain subcomplexes within the input complex.

Ibrahim, Krishnamoorthy, and Vixie \cite{ibrahim2011simplicial} has extended these results to the multiscale simplicial flat norm (mSFN) problem.
While mSFN problems generally exhibit NP-hardness, even when homology is defined over integers,
their work revealed a notable exception within the domain of embedded complexes.
In these cases, polynomial-time solutions become feasible, offering a glimpse of tractability in otherwise computationally challenging scenarios.

Maxwell, in his PhD thesis \cite{maxwell2021algorithmic},  has broaden the class of simplicial complexes that admit a dual graph structure beyond embeddable complexes.
by extending Mac Lane's palanarity criterion \cite{mac1936combinatorial}:
a graph $G$ is planar if and only if there exists a basis $\mathfrak{B}_{1}$ of $\mc{H}_{1}(G; \integer_2)$  such that every edge of $G$ is contained in at most two cycles of $\mathfrak{B}_{1}$.
The generalization of such a basis, called \textbf{2-basis}, to $(q + 1)$-complexes:
a 2-basis $\mc{H}_{q}(\mc{K}; \integer_2)$ is a homology basis $\mathfrak{B}_{q}$ such that each $q$-simplex is contained in at most two basis elements.
Theorem 5 from \cite{maxwell2021algorithmic} states that $q$-dimensional simplicial complexes with  a 2-basis $\mathfrak{B}_{q}$
have a dual graph that can be constructed in $\bigO(\beta_q^2 \cdot \abs{\mc{K}_{q}})$ time.
Notably, the class of $(q + 1)$-weak pseudomanifolds,
a generalization of embeddable complexes, admits  construction of a dual graph.
This result matches with polynomial-time guarantees for solving OHCP on triangulation of  $(q + 1)$-dimensional compact orientable manifolds\cite{ohcp2011}.
Which suggests a deeper connection between polynomial-time solvabilty of optimal homology problems and existence of a dual graph of a complex.

Several well defined graph structure comparison metrics such as subgraph isomorphism and edit distance have been proposed in the literature along with algorithms to compute them efficiently.
Tantardini et al.~\cite{Tantardini2019} compare graph network structures for the entire graph (global comparison) as well as for small portions of the graph known as motifs (local comparison).
Other researchers have proposed methodologies to identify structural similarities in embedded graphs~\cite{BaDiBiChSuWa2019,graphsim-2020}.
However, all these methods depend on one-to-one correspondence of graph nodes and edges rather than considering the node and edge geometries of the graphs.
The edit distance, i.e., the minimum number of edit operations to transform one network to the other, has been widely used to compare networks having structural properties~\cite{PaGaMiHa2018,riba2020,xu2015}.
Riba et al.~\cite{riba2020} used the Hausdorff distance between nodes in the network to compare network geometries.
Majhi et al.~\cite{MaWe2024} modified the traditional definition of graph edit distance to be applicable in the context of ``geometric graphs'' embedded in a Euclidean space.
Along with the usual insertion and deletion operations, the authors have proposed a cost for translation in computing the geometric edit distance between the graphs.
However, the authors also show that the problem of computing this metric is NP-hard.

Meyur et al.~\cite{rounak_pnas} compared network geometries using the Hausdorff distance after partitioning the geographic region into small rectangular grids and comparing the geometries for each grid.
However, the Hausdorff metric is sensitive to outliers as it focuses only on the maximum possible distance between the pair of geometries.
When the geometries coincide almost entirely except in a few small portions, the Hausdorff metric still records the discrepancy in those small portions without accounting for the similarity over the majority of portions.
The similar approach used by Brovelli et al.~\cite{roads_haus} to compare a pair of road networks in a geographic region suffers from the same drawback.
This necessitates a well-defined distance metric between networks with geographic embedding~\cite{roads2014}.

Several comparison methods have been proposed in the context of planar graphs embedded in a Euclidean space~\cite{streets2006,efficiency2020}.
They include local and global metrics to compare road networks.
The local metrics characterize the networks based on cliques and motifs, while the global metrics involve computing the \emph{efficiency} of constructing the infrastructure network.
The most efficient network is assumed to be the one with only straight line geometries connecting node pairs.
Albeit useful to characterize network structures, these methods are not suitable for a numeric comparison of network geometries.

%


\label{sec:prelims}
\chapter{Preliminaries}
 In this paper, we specifically focus on finite $(q + 1)$-connected \emph{orientable} $(q + 1)$-complexes that can be \emph{embedded} into $\mathbb{R}^{q + 1}$. 
 Such embeddable complexes hold significant interest in computational topology due to their similarity to planar graphs. Like planar graphs, embeddable complexes possess a well-defined dual graph that extends to a dual complex \cite{maxwell2021algorithmic}. Moreover, the structure of their homology groups permits polynomial time solutions for optimal homology problems \cite{ohcp2011}.
 
 Leveraging the guarantees of polynomial time solutions, we harness the dual graph to formulate both the OHCP and mSFN problems as minimum-cost flow problems on corresponding flow networks, which are extensions of graphs with designated source and sink nodes.
 
 The availability of a dual graph not only empowers us to employ the rich and powerful graph-theoretical framework
 to enhance our understanding of high-dimensional geometric and combinatorial problems
 but also enables practitioners to utilize existing, mature,
 and efficient software libraries to solve such problems.
 Practically, the computational examples of OHCP and mSFN solutions presented in the appendix
 were computed using the Python library {\texttt{networkx}} \cite{hagberg2008exploring}.
 
 Furthermore, the practical considerations of solving these problems as network flow problems prompted us to extend our results obtained for (co)chains (integer-valued functions that respect the orientation of simplices) to non-negative integer-valued functions defined on opposite orientations of underlying simplices. These functions correspond to flows in the dual graph.

\section{Simplicial Complexes: orientable and embeddable}

A finite abstract \embf{simplicial complex} $\mc{K}$ over a set $\mc{N}$ of $N$ \emph{vertices} 
is a finite collection of downwardly closed subsets of $\mc{N}$: 
that is for any $\tau \in \mc{K}$ if $\sigma \subset \tau$ then $\sigma \in \mc{K}$. 
The set $\mc{N}$ is called the \emph{vertex set} of $\mc{K}$,
and any subset of $d + 1$ vertices $\sigma \subseteq \mc{N}$ that is included in $\mc{K}$ 
is called a \embf{$d$-simplex}, with $\dim \sigma = d$.
An \embf{oriented} $d$-simplex is defined as an ordered set of $(d + 1)$ vertices, 
denoted as ${\sigma} = [v_0, v_1, \ldots, v_d]$,
while $-\sigma$ denotes the opposite orientation is given by an \emph{odd permutations} of this ordering.
In general, any subset  $\sigma$ of $\tau \in \mc{K}$ is called a \embf{face} of $\tau$,
and, in turn, $\tau$ is called a \embf{coface} of $\sigma$.
However, in this paper, we will apply these terms only in co-dimension 1: $\dim \tau  = \dim \sigma + 1$.
The dimension of $\mc{K}$ is defined as the largest simplicial dimension among all $\mc{K}$'s simplices:
$\dim \mc{K} = \max_{\sigma \in \mc{K}} \dim \sigma$. 
A \embf{$d$-complex} is a $d$-dimensional simplicial complex. 
The set of all $d$-simplices of $\mc{K}$ is called the \embf{$d$-skeleton} of $\mc{K}$, which we denote as $\mc{K}_{d}$,
and $\abs{\mc{K}_{d}}$ denotes its cardinality.
Note that for any $0 \leq d \leq \dim \mc{K}$ a $d$-skeleton is a valid $d$-subcomplex of $\mc{K}$: $\mc{K}_{d} \subset \mc{K}$.
The 0-th skeleton corresponds to the vertex set of a complex.

A $d$-complex $\mc{K}$ is \embf{pure} if every $(d - 1)$-simplex has \emph{at least one} $d$-coface in $\mc{K}$.
A pure $d$-complex $\mc{K}$ is called a \embf{weak $d$-pseudomanifold} 
  if each $(d - 1)$-simplex has \emph{at most two} cofaces that are included in $\mc{K}$.
A $(d - 1)$-simplex of $\mc{K}$ is called a \embf{boundary $(d - 1)$-simplex} if it has \emph{at most one} $d$-coface in $\mc{K}$. 
Two $d$-simplices $\sigma_0$ and $\sigma_k$ are \emph{$d$-connected} on $\mc{K}$ 
if there is a  sequence of $d$-simplices $(\sigma_0, \ldots, \sigma_k)$ such that any two consecutive simplices share a $d-1$-face:
$(\sigma_{j} \cap \sigma_{j  +1} )\in \mc{K}_{d  -1}$. 
$\sigma_0$ and $\sigma_k$ are \emph{$d$-connected} on $\mc{K}$ 
If any two $d$-simplices on $\mc{K}$ are $d$-connected we say that $\mc{K}$ is \embf{$d$-connected}.

\paragraph*{Embedding and cell complex}

We define \embf{embedding} $\support{\sigma}$ of a $d$-simplex $\sigma \in \mc{K}$ for $d \leq q + 1$
as an injection into a subspace of $\real^{q + 1}$, called a \embf{$d$-cell}, 
that is homeomorphic to a $d$-dimensional \textit{closed}  ball $\mathbb{B}^{d}$:
$\support{\sigma} \!: \sigma \mapsto \real^{d} \subseteq \real^{q + 1}$ and $\support{\sigma} \approx \mathbb{B}^{d}$.
We extend this definition to the whole complex $\support{\mc{K}} \!: \mc{K} \mapsto \real^{q + 1}$ 
by additionally  requiring that the intersection of embeddings of any two simplices must be the embedding of their common face:
if $\sigma, \tau \in \mc{K}$ such that $\sigma \cap \tau \in \mc{K}$ then $\support{\sigma} \cap \support{\tau} = \support{\sigma \cap \tau}$.
A $(q + 1)$-complex $\mc{K}$ is \textbf{embeddable} if there exists an embedding $\support{\mc{K}} \subseteq \real^{q + 1}$.
A valid embedding $\support{\mc{K}}$ defines a partition of a connected bounded region in $\real^{q  +1}$ into a collection of downwardly closed $(q + 1)$-cells,
which is called as a \textbf{CW-complex} or \embf{cellular complex}.
Cellular complexes are generalizations of simplicial complexes that maintain similar algebraic properties and, 
for  our purposes, are indistinguishable and both will be referred simply as complexes.
Therefore, when it is clear from the context, we will use $\support{\mc{K}}$ and $\mc{K}$ interchangeably.
  Note that any embeddable complex is a weak pseudomanifold, but not every weak pseudomanifold is embeddable.
  In general, finding a valid embedding is NP-hard even in low dimensions, as for example 2-complexes in $\real^{3}$ \cite{embeddabilityR3}, 
  and even determining whether or not a complex is embeddable in $\real^{q  +1}$~\cite{embeddabilityRd}.

\paragraph*{Orientation}

An \embf{oriented} $d$-simplex  $\vec{\sigma}_{i}  = [v_0, v_1, \ldots, v_d]$ is given by an ordered set of $d+1$ vertices along with the class of even permutations, 
while any \emph{odd permutation} of this ordering, denoted as $\vec{\sigma}_{-i} = -\vec{\sigma}_{i}$, 
is said to be \embf{oppositely oriented} to $\vec{\sigma}_{i}$. 
A $(d - 1)$-simplex $\vec{\tau}_j$ is a \embf{positive face} of $\vec{\sigma}_{i}$, denoted as $\vec{\tau}_j \subset \vec{\sigma}_{i}$,
if there exist an \emph{even} permutation $[v_{(0)}, v_{(1)}, \ldots, v_{(d)}]$ of  $\vec{\sigma}_{i}$ such that $\vec{\tau}_j =  [v_{(1)}, \ldots, v_{(d)}]$.
If instead $\vec{\tau}_j =  [v_{(0)}, \ldots, v_{(d - 1)}]$ then $\vec{\tau}_j$ is a   \embf{negative face} $\vec{\sigma}_i$, i.e. $-\vec{\tau}_j \subset \vec{\sigma}_{i}$.
We assume that all $\sigma_{i} \in \mc{K}$ are oriented and that $\mc{K}$ contains \emph{only one of the two} possible orientations of $\sigma_{i}$: 
$\vec{\sigma}_{i} \in \mc{K}$ and $\vec{\sigma}_{-i} \notin \mc{K}$.
The class of even permutations of $\vec{\sigma}_{i}$ that is included in $\mc{K}$  
is called the \embf{positive} or \embf{natural orientation} of $\sigma_{i}$,  and is denoted as $\vec{\sigma}_{+i}$.
The opposite orientation of $\vec{\sigma}_{+i}$ is called \embf{negative} or \embf{reversed orientation} and is denoted as $\vec{\sigma}_{-i}$\footnote{
      In indexless notation: 
          $\vec{\sigma} = \vec{\sigma}_{i}$ then $\vec{\sigma}_{-}  = -\vec{\sigma}$;
          $\vec{\sigma} \in \mc{K}$ then $\vec{\sigma}_{+} = \vec{\sigma}$ and $\vec{\sigma}_{-} = -\vec{\sigma}$.
}.
Consequently, we define \embf{positive} and \embf{negative $d$-skeletons} as the sets of all positive and, respectively, all negative orientations of every $d$-simplex in $\mc{K}$:
$\vec{\mc{K}}^{+}_{d}  = \left\{ \vec{\sigma}_{+i} \mid \sigma_i \in \mc{K}_{d} \right\}$ and 
$\vec{\mc{K}}^{-}_{d}  = \left\{ \vec{\sigma}_{-i} \mid \sigma_i \in \mc{K}_{d} \right\}$.
We call the union of two $\vec{\mc{K}}_{d} = \vec{\mc{K}}^{+}_{d} \cup \vec{\mc{K}}^{-}_{d}$ a \embf{directed $d$-skeleton}.
A simplicial complex $\vec{\mc{K}}$ constructed from $\mc{K}$ by replacing its $d$-skeleton with $\vec{\mc{K}}_{d}$
we call a \emph{$d$-directed simplicial complex} or simply \embf{$d$-directed complex}.

We use the vector-accent notation $\vec{\sigma}$ when we want to emphasise that we are dealing with the two classes of orientations.
In most instances, we use  $\sigma_{i} \in \mc{K}_{d}$ to denote the natural orientation of a $d$-simplex (the one that is present in $\mc{K}$),
and $\sigma_{-i} = -\sigma_{i}$ to denote the opposite orientation.
When the orientation class of a $d$-simplex is unknown we use $\vec{\sigma}_{i}$.
We then define the \emph{orientation} function 
$\Omega_{d}\!: \vec{\mc{K}}_{d} \mapsto \{-1, +1\}$  as 
$\Omega_{d}\!: \vec{\sigma}_{-i} \mapsto -1$ and  $\Omega_{d}\!: \vec{\sigma}_{+i} \mapsto +1$.
In particular, for any  $\sigma \in \mc{K}_{d}$ of $\mc{K}$ we always have $\Omega_{d}\!: \sigma \mapsto +1$.

\section{Simplicial Homology}
  The following overview of simplicial homology is and definitions are applicable to a wide class of finite abstract complexes
  

\subsection{Simplicial complexes and homology}

Let $\mc{K}$ be a finite (pure) abstract $(q  +1)$-complex.
Then for any $0 \leq d \leq q + 1$ we define following components of its simplicial homology and cohomology.

\paragraph*{Chains \& Cochains}
A \embf{$d$-chain} $\bm{x}$ on $\mc{K}$ is 
a formal sum of $d$-simplices of $\mc{K}$ with coefficients over a scalar group $\mathbb{K}$ 
defined for any $0 \leq d \leq q + 1$.
A $d$-cochain $\bm{f}\!: \mc{K}_{d} \mapsto \mathbb{K}$ is  a $\mathbb{K}$-valued homomorphism (analog of a linear function) defined on $d$-simplices.
Let 
$x_i, f_i \in \mathbb{K}$ for any $\sigma_{i} \in K_{d}$, then:
\begin{align}
\label{eq:def:chain-cochain}
    \bm{x} = \sum_{\sigma_{i} \in K_{d}} \bm{x} (\sigma_i)
           = \sum_{\sigma_{i} \in K_{d}} x_i \sigma_i
    &&
        \bm{f} = \sum_{\sigma_{i} \in K_{d}}  \bm{f}(\sigma_i)
         = \sum_{\sigma_{i} \in K_{d}}  f_i
\end{align}

In practice, common choices for $\mathbb{K}$ include binary and real fields, $\integer_2$ and $\real$,  
\emph{commutative ring} of integers $\integer$, 
 \emph{commutative cancellative semirings} $\integerplus$ and $\realplus$ of non-negative integers and reals, 
(note $0 \in \integerplus$ and $0 \in \realplus$).
Both $d$-chains and $d$-cochains are closed under addition on $\mathbb{K}$, 
thus forming \emph{additive free abelian} chain and cochain groups 
$\mc{C}_{d}(\mc{K}; \mathbb{K})$ and $\mc{C}^{d}(\mc{K}; \mathbb{K})$, respectively.
If $\mathbb{K}$ is a field (such as $\integer_2$ and $\real$), 
then $\mc{C}_{d}(\mc{K}; \mathbb{K})$ and $\mc{C}^{d}(\mc{K}; \mathbb{K})$ are vector spaces.
Note that in the case of commutative semirings, the (co)chain groups are abelian $\mathbb{K}$-semimodules, not groups. 
Despite this distinction and  differences in the definition of boundary and coboundary operators, 
we do not distinguish the two and maintain the terminology of (co)chain \emph{groups} for simplicity and consistency.
More details on homology over semirings can be found in \cite{mendez2023directed, patchkoria1977cohomology}.

In this paper we consider only $\integer$  and $\integerplus$ coefficients.
In this paper $\integer$ is our default choice for chains on oriented complexes, 
since we can use the sign of chain's value to determine the orientation of a $d$-simplex,
and we use $\integerplus$ on directed $d$-complexes .
to model positive and negative parts of $\integer$-(co)chain:
$\bm{x} = \bm{x}^{+} - \bm{x}^{-}$ 
where $\bm{x} \in \mc{C}(\mc{K}; \integer)$
and $\bm{x}^{\pm} \in \mc{C}(\vec{\mc{K}}; \integerplus)$.
If it is clear from the context which group of coefficients we are working with, $\integer$ or $\integerplus$,
we will be dropping it and simply write $\mc{C}_{d}(\mc{K})$.

Note that the $d$-th chain group is isomorphic to $\abs{\mc{K}_{d}}$-dimensional lattice $\integer^{\abs{\mc{K}_{d}}}$ (analog of a vector space),
$\mc{C}_{d}(\mc{K}; \integer) \cong \integer^{\abs{\mc{K}_{d}}}$,
and $d$-chains are isomorphic to integer tuples (analog of vectors).
Note that $d$-th chain group $\mc{C}_{d}(\mc{K})$ is generated by individual $d$-simplices, 
hence $\mc{K}_{d}$ is a \emph{basis} of $\mc{C}_{d}(\mc{K})$, which we denote as $\left[ \mc{C}_{d}(\mc{K}) \right]$. 
The  group of $d$-cochains, on the other hand, is generated by elementary $d$-cochains $\bm{\sigma}$ that map all $d$-simplices, except one, to zero: 
$\bm{\sigma}\!: \sigma \mapsto 1, \, \tau \mapsto 0$ for $\tau \neq \sigma$.
This means that the groups of $d$-chains and $d$-cochains are isomorphic to each other $\mc{C}_{d}(\mc{K}) \cong \mc{C}^{d}(\mc{K})$.
Because of this isomorphism, while keeping the formal distinction between the two, we will treat $d$-chains and $d$-cochains either as integer vectors or functions depending on the context.

  A $d$-(co)chain $\bm{p}\!: {\mc{K}}_{d} \mapsto \integer$
  is  \embf{simple} if it assigns no more than a unit value to $d$-simplices:
  $\bm{p} \!: \sigma \mapsto \{-1, 0, +1 \}$  for all $\sigma \in \mc{K}_{d}$.
  A \emph{positive simple} $\bm{p}$ assigns binary values to $d$-simplices $\bm{p} \!: \sigma \mapsto \{ 0, +1 \}$.
  A \emph{directed} $d$-(co)chain $\vec{\bm{p}}\!: \vec{\mc{K}}_{d} \mapsto \integerplus$ 
  is a \emph{positive simple} $d$-(co)chain defined on the directed $d$-skeleton:
  $\vec{\bm{p}}( \vec{\sigma}_i) \in \{0, 1\}$ for all $\sigma \in \vec{\mc{K}}_{d}$.

\paragraph*{Boundaries}
The \embf{$d$-boundary operator}  is a homomorphism   $\bd_{d}\!: \mc{C}_{d}(\mc{K}) \mapsto \mc{C}_{d - 1}(\mc{K})$
defined for every basis $d$-chain $\bm{\tau} \in \mc{C}_{d}(\mc{K}) \!: \sigma \mapsto 1$ 
as a an alternating sum over $(d - 1)$-faces $\sigma \subset \tau$ of corresponding $d$-simplex $\tau = [v_0, v_1, \ldots, v_d]$: 
\begin{align}
\label{eq:def:boundary}
   \bd_d \tau = \sum_{\sigma \subset \tau} \Omega_{\bd \tau}(\sigma) \cdot \sigma
                &= \sum_{j = 0}^{d} (-1)^j [v_0, v_1, \ldots, \hat{v}_{{j}}, \ldots, v_d]
\end{align}
where $\hat{v}_{{j}}$ means that the vertex $v_j$ was omitted,
and $\Omega_{\bd \sigma}(\tau)\!: \tau \mapsto \in \{-1, +1\}$ is the orientation of a face $\tau$ on the boundary of $\sigma$:
$\Omega_{\bd \sigma}(\tau) = \bd_d \sigma (\tau)$.
Hence, $\bd_{d}$ respects orientation of oriented $d$-simplices and their $(d - 1)$-faces: 
$\bd_{d} \vec{\tau}_{-}  = -\bd_{d} \vec{\tau}_{+}$ and $\bd_{d} \tau (\vec{\sigma}_{-}) = -\bd_{d} \tau (\vec{\sigma}_{+})$.
The \textit{fundamental property of homology} is:
$\bd_{d-1} \circ \bd_{d} = 0$. 
The \embf{chain complex} of $\mc{K}$ is be the following sequence of chain groups connected by the boundary operators:
  \begin{align}
  \label{eq:def:chain-complex}
    \ldots \overset{\bd_{d + 2}}{\longrightarrow}\mc{C}_{d + 1}
           \overset{\bd_{d + 1}}{\longrightarrow}\mc{C}_{d}
           \overset{\bd_{d}}    {\longrightarrow}\mc{C}_{d - 1}
           \overset{\bd_{d - 1}}{\longrightarrow} \ldots
  \end{align}

  The image of a $\bd_{d + 1}$ operator
  forms the \textbf{group of $d$-boundaries}
  \begin{align*}
    \mc{B}_{d }(\mc{K})  \asgn \Img \bd_{d + 1 } = \left\lbrace \bm{w} \in \mc{C}_{d}(\mc{K}) \mid \exists \bm{\pi} \in \mc{C}_{d + 1}(\mc{K}) \!: \bm{w} = \bd_{d + 1}  \bm{\pi} \right\rbrace 
  \end{align*}
  where $d$-chains $\bm{w} \in \mc{B}_{d}(\mc{K})$ are called \textbf{$d$-boundaries} or \embf{null-homologous} $d$-cycles.
  Note that generators of $\mc{B}_{d}$ are in 1-to-1 correspondence with $(d  +1)$-simplices:
  $\ring{\bm{w}_j} \in \left[ \mc{B}_{d }(\mc{K}) \right] \iff \ring{\bm{w}}_j =  \bd_{d  +1} \tau_j $
  for $\tau_j \in \mc{K}_{d + 1}$.

  The kernel of the $d$-boundary operator, called the \embf{group of $d$-cycles},
   consist of $d$-chains without a boundary:
  \begin{align*}
    \mc{Z}_d(\mc{K}) \asgn \ker \bd_d = \left\lbrace \bm{z} \in \mc{C}_{d}(\mc{K}) \mid \bd_d \bm{z} = 0 \right\rbrace 
  \end{align*}
  The elements of $\mc{Z}_d(\mc{K})$ are called \embf{$d$-cycles}.
  Any $d$-boundary  is a $d$-cycle, 
  but not every  $d$-cycle is  a $d$-boundary.
  Therefore we have the inclusion $\mc{B}_d(\mc{K}) \subseteq \mc{Z}_d(\mc{K}) \equiv  \Img \bd_{d + 1 } \subseteq \ker \bd_{d}$.
  The $d$-cycles such that $\bm{z} \in \mc{Z}_d(\mc{K}) \setminus \mc{B}_d(\mc{K})$ 
  we call \embf{non-bounding} or \emph{non-trivial} $d$-cycles.


    As a homomorphism between free abelian groups $\bd_d$ has a unique matrix representation,
    which we denote $[\bd_d]$, 
    with respect to a choice of basis.
    With respect to $(d - 1)$-th and $d$-skeletons $\mc{K}_{d - 1}$ and  $\mc{K}_{d}$, 
    that  are bases of  $\mc{C}_{d - 1}(\mc{K})$ and $\mc{C}_{d}(\mc{K})$,
     $[\bd_d]$ is a $\abs{\mc{K}_{d - 1}} \times \abs{\mc{K}_{d }}$ matrix:
    \begin{align*}
      [\bd_d]_{ij} = \begin{cases}
                        \pm 1, &\text{ if } \sigma_i \subset \tau_j \\
                        0, & \text{ otherwise}
                     \end{cases}
       && \text{ where } \sigma_i \in \mc{K}_{d - 1} \text{ and } \tau_j \in \mc{K}_{d}
    \end{align*}

\paragraph*{Coboundaries}
The \embf{$d$-coboundary operator} $\cobd_{d}: \mc{C}^{d - 1}(\mc{K}) \to \mc{C}^{d}(\mc{K})$ 
is the dual homomorphism of $\bd_{d}$
that is defined for each basis $(d - 1)$-cochain ${\bm{\sigma}} \in \mc{C}^{d - 1}(\mc{K})$ 
as a formal sum of all of its $d$-cofaces.
It can be shown that $\cobd_{d} \bm{f}$ is equivalent to the composition $\bm{f} \circ \bd_{d}$,
i.e.  $\cobd_{d} \bm{f} (\tau) = \bm{f} (\bd_{d} \tau)$ for any $\tau \in \mc{K}_{d}$.
This implies that $d$-cochains can be seen as integer functions defined on $d$-cycles $\bm{f}\!: \Z(\mc{K}) \mapsto \integer$.
The matrix of a coboundary operator is transpose of corresponding boundary matrix: $[\cobd_{d}] = [\bd_{d}]^{\tr}$.
  \begin{align*}
    \cobd_d \sigma 
          &= \sum_{\tau \supset \sigma } \Omega_{\bd \tau}(\sigma) \cdot \tau
           = \sum_{{\tau} \supset \sigma } (-1)^{j} [v_0, v_1, \ldots, \check{u}_{j}, \ldots, v_{d -1}]  
  \end{align*}
  and $\check{u}_{j}$ means that the vertex $u_j$ was inserted on $j$-th position, 
  $\tau = [v_0, v_1, \ldots, \check{u}_{j}, \ldots, v_{d -1}] \in \mc{K}_{d}$ 
  is a $d$-coface of a $(d - 1)$-simplex $\sigma = [v_0, v_1, \ldots,  v_{d -1}]$.
  Hence, $\cobd_{d}$ respects the orientation of oriented $(d - 1)$-simplices and their $d$-cofaces (if they exist): 
  $\cobd_{d} \vec{\sigma}_{-}  = -\cobd_{d} \vec{\sigma}_{+}$ and $\cobd_{d} \sigma (\vec{\tau}_{-}) = -\cobd_{d} \sigma (\vec{\tau}_{+})$.
  The \emph{fundamental property of cohomology} is: $\cobd_{d + 1} \circ \cobd_{d} = 0$.
  The \embf{cochain complex} of $\mc{K}$ is be the following sequence of cochain groups connected by the coboundary operators:
    \begin{align}
    \label{eq:def:cochain-complex}
      \ldots \overset{\cobd^{d + 2}}{\longleftarrow}\mc{C}^{d + 1}
             \overset{\cobd^{d + 1}}{\longleftarrow}\mc{C}^{d}
             \overset{\cobd^{d}}    {\longleftarrow}\mc{C}^{d - 1}
             \overset{\cobd^{d - 1}}{\longleftarrow} \ldots
    \end{align}

  The image of a $d$-coboundary operator
    forms the \textbf{group of $d$-coboundaries} 
    \begin{align*}
        \mc{B}^{d }(\mc{K}) \asgn \Img \cobd_{d } = \left\lbrace \bm{w} \in \mc{C}^{d}(\mc{K}) \mid \exists \bm{\rho} \in \mc{C}^{d - 1}(\mc{K}) \!: \bm{w} = \cobd_{d }  \bm{\rho} \right\rbrace   
        \text{ for some } \bm{\rho} \in \mc{C}_{d - 1}(\mc{K})
    \end{align*}
    with $d$-cochains $\bm{w} \in \mc{B}^{d}(\mc{K})$ referenced as \textbf{$d$-coboundaries}, 
    or if $\bm{w}$ is simple,  as \embf{$d$-cuts} (cuts on graphs are 1-coboundaries).
      Let $\bm{\rho} \in \mc{C}_{d - 1}$ be a simple positive $(d - 1)$-chain, 
      then its $d$-coboundary, $\bm{w} = \cobd_{d} \bm{\rho}$, \emph{cuts} through all (simple) $d$-chains 
      that have at least two $(d - 1)$-boundary components mapped to different values by $\bm{\rho}$.

  Similarly, the kernel of $\cobd_{d + 1}$ defines the group of \embf{$d$-cocycles} 
  -- $d$-cochains with an empty coboundary:  
  \begin{align*}
      \mc{Z}^d(\mc{K}) \asgn \ker \cobd_{d + 1} = \left\lbrace \bm{f} \in \mc{C}^{d}(\mc{K}) \mid \cobd_{d + 1} \bm{f} = 0 \right\rbrace 
  \end{align*}
    Again, any $d$-coboundary  is a $d$-cocycle, 
    but not every  $d$-cocycle  is  a $d$-coboundary.
    Therefore we have the inclusion $\mc{B}^d(\mc{K}) \subseteq \mc{Z}^d(\mc{K}) \equiv  \Img \cobd_{d  } \subseteq \ker \cobd_{d + 1}$.
    
    In literature, $d$-cycles also are referred as \emph{flows}, and the equation $\bd_d \bm{z} = 0$, equivalent to $\bm{z}(\cobd_{d} \sigma) = 0$ for all $\sigma \in \mc{K}_{d - 1}$, 
      as the \emph{flow conservation} condition: in-flow is equal to out-flow.
    In the context of flow models considered in this paper, we will refer to    $d$-cocycles $\bm{f} \in \mc{Z}^d(\mc{K}) $ 
    as \embf{$d$-coflows},
    and to the condition $\cobd_{d + 1} \bm{f} = 0 $, equivalent to $\bm{f}(\bd_{q + 1} \tau) = 0$ for all $\tau \in \mc{K}_{d + 1 }$,
    as the  \embf{circulation conservation} condition.
      Then any $d$-coflow $\bm{f} \in \mc{Z}^{d}(\mc{K})$ 
       can be seen as a function defined on $d$-cycles:
      $\bm{f}\!: \mc{Z}_{d}(\mc{K}) \mapsto \integer$ such that
      $\bm{f}\!: \mc{B}_{d}(\mc{K}) \mapsto 0$,
      i.e.  $d$-boundaries  are mapped to 0.
        \begin{align*}
            \bm{f} \in \mc{Z}^{d}(\mc{K})\!: \ 
            &\cobd_{d + 1} \bm{f}(\tau) = \bm{f}(\bd_{d + 1} \tau)= 0      
                  && \text{for all }  \tau \in  \mc{K}_{d + 1}
       \\
          & \bm{f}({\bm{z}}) = \gamma \in \integer
                  && \text{for some }  {\bm{z}} \in  \mc{Z}_{d}(\mc{K}) \setminus \mc{B}_{d}(\mc{K})
        \end{align*}
    
      We refer to the value $\bm{f}(\bm{z})$ of a $q$-coflow  $\bm{f} \in \Z^{q}(\mc{K})$  on $\bm{z} \in \Z_{q}(\mc{K})$ 
      as \embf{coflux of $\bm{f}$ through $\bm{z}$}. 
       By the Hodge decomposition Eq.~\eqref{eq:hodge-decomposition}, $q$-coboundaries always have 0 coflux through any $\bm{z} \in \Z_{q}(\mc{K})$,
        \begin{align*}
            \bm{g} \in \mc{B}^{q}(\mc{K})\!: \ 
          & \bm{g}(\bm{z}) = 0
                  && \text{for any }  \bm{z} \in   \Z_{q}(\mc{K}) 
        \end{align*}
    
    \begin{definition}
        A \textbf{$d$-copath}  $\bm{p}  = \bm{p}(\eta, \tau)$ between two $(d + 1)$-simplices
        $\eta, \tau \in {\mc{K}}_{d + 1}$, 
        is a simple $d$-cochain $\bm{p} \in \mc{C}^{d}(\mc{K})$
        with a non-empty coboundary 
        ${\cobd}_{d + 1} \bm{p} = \tau - \eta$.
        We refer to the $(d + 1)$-facets $\eta$ and $\tau$ as the \textbf{endpoints} of $\bm{p}$.
        
        Alternatively, a \textbf{$d$-copath}  $\bm{p}  = \bm{p}(\bm{w}^{\prime}, \bm{w}^{\pprime})$ between two simple  $d$-cycles
        $\bm{w}^{\prime}, \bm{w}^{\pprime} \in \mc{Z}_{q}(\mc{K})$
        is a simple $d$-cochain $\bm{p} \in \mc{C}^{d}(\mc{K})$
        such that $\bm{p}(\bm{w}^{\prime}) = -1$ and $\bm{p}(\bm{w}^{\pprime}) = +1$.
        We use the combinatorial up-Laplacian to denote the end-cycles of such copath:
         $\mc{L}^{q} \bm{p} \asgn \bm{w}^{\pprime} - \bm{w}^{\prime}$.

    \end{definition}

  Note that if the end-cycles are boundaries on $\mc{K}$: $\bm{w}^{\prime}, \bm{w}^{\pprime} \in \B_{d}(\mc{K})$, 
  then the coboundary of $\bm{p}$ is well defined: $\cobd_{d  +1} \bm{p} = \tau^{\pprime} - \tau^{\prime}$ such that 
  $ \bd_{d + 1} \tau^{\pprime} = \bm{w}^{\prime}$ and $\bd_{d + 1} \tau^{\pprime} = \bm{w}^{\pprime}$.
  If the end-cycles are non-bounding $\bm{w}^{\prime}, \bm{w}^{\pprime} \in \Z_{d}(\mc{K}) \setminus \B_{d}(\mc{K})$, 
  then such $d$-copath is a $d$-cocycle, 
  and we call it the \textbf{augmenting $d$-copath} and denote it as $\bar{\bm{p}} \in \Z^{d}(\mc{K})$.

  \begin{definition}
    A \textbf{directed $q$-copath} 
    $\vec{\bm{p}} \in \mc{C}^{q}(\vec{\mc{K}}; \integerplus) $
    given as 
    $\vec{\bm{p}} = \vec{\bm{p}}(\eta, \tau) = \vec{\bm{p}}[\eta \to \tau]$
    for $\eta, \tau \notin \mc{K}_{q + 1}$
    or
    $\vec{\bm{p}} = \vec{\bm{p}}(\bm{w}^{\prime}, \bm{w}^{\pprime}) = \vec{\bm{p}}[\bm{w}^{\prime} \to \bm{w}^{\pprime}]$
    for  $\bm{w}^{\prime}, \bm{w}^{\pprime} \in \mc{Z}_{q}(\vec{\mc{K}})$: 
    is a simple directed $q$-cochain that 
    has a non-empty coboundary on the acyclization $\bar{\mc{K}}$: 
        $\bar{\cobd}_{q + 1} \vec{\bm{p}} = \tau - \eta$ or $\mc{L}^{q} \bm{p} = \bm{w}^{\pprime} - \bm{w}^{\prime}$
        
     
    %
  \end{definition}

  If $\mc{K}$ is a pure $(q  +1)$-complex then the cochain complex in Eq.~\eqref{eq:def:cochain-complex} terminates at the $(q  +1)$-th cochain group defined on the facets of $\mc{K}$, 
  that have no cofaces: $\cobd_{q + 2}\!: \mc{C}^{q + 1} \mapsto 0$.
    It implies that the $(q + 1)$-facets of $\mc{K}$ are $(q + 1)$-cocycles, i.e. $\mc{C}^{q + 1}(\mc{K})  = \mc{Z}^{q + 1}(\mc{K}) $, 
    and since all $q$-simplices have at least one $(q + 1)$-coface on $\mc{K}$,
  by the fundamental property of cohomology,
  $\mc{C}^{q + 1}(\mc{K}) = \Img \cobd_{q + 1} = \mc{B}_{q + 1}(\mc{K})$.
  On the other hand, the chain complex in Eq.~\eqref{eq:def:chain-complex} extends all the way down to  $\mc{C}_{0}(\mc{K})$ 
       -- the chain group of vertices of $\mc{K}$ such that $\bd_{0}\!: \mc{C}_{0}(\mc{K}) \mapsto 0$.
      It implies that the vertices are $0$-cycles, i.e. $\mc{C}_{0}(\mc{K})  = \mc{Z}_{0}(\mc{K}) $, 
      and since all vertices belong to some $(q + 1)$-simplex, by the fundamental property $\bd$-operator,
      $\mc{C}_{0}(\mc{K}) = \Img \bd_1 = \mc{B}_{0}(\mc{K})$.

\paragraph*{Homology \& Cohomology} 
Let $G$ be an abelian group and $H \subseteq G$ its subgroup.
A \embf{coset} $a+H$ of an abelian group $G$ relative to $H$
is defined as $a+H = \left\{ a+h \mid h \in H , a \in G \right\} \subset G$.
The \embf{quotient group} $G / H$ is the set of all cosets of $H$ in $G$,
i.e. $G / H = \big\{ (a + H) \mid a \in G \big\}$. 
$G / H$ is a group with the group operation $\bm{+}_{G/H}$ defined as: 
$(a + H) \bm{+}_{G/H} (b + H) = (a + b) + H \in G/H$, for $(a+H), (b+H) \in G/H$.
    
A subgroup of an abelian additive group $G$ that is \embf{generated} by a single element $x_i \in G$ is denoted as $[x_i] = \{ a_i x_i \mid a_i \in \integer \}$.
An abelian additive group $G$ is called \embf{finitely generated} 
if $\exists x_1, \ldots, x_n \in G$, $n < \infty$ such that $\forall x \in G : x = a_1 x_1 +  \ldots + a_n x_n$,
which we denote $G = [x_1] \oplus \ldots \oplus [x_n] = \bigoplus_{i  =1}^{n} [x_i]$. 
We refer to the set $\{ x_1, \ldots, x_n \}$ as  a \embf{generating set} of $G$ or as a \embf{basis} of $G$,
and denote it as $[G] = \{ x_1, \ldots, x_n \}$. 
By the \textit{fundamental theorem of finitely generated abelian groups}
any such group $G$ can be written as a direct sum (cartesian product) of two groups $G = F \oplus T$,
where  $F \cong (\integer \oplus \ldots \oplus \integer) = \integer^k$ for some $k \in \nat$, 
and $T \cong (\integer/t_1 \oplus \ldots \oplus \integer/t_u)$ 
such that $t_i > 1$ and $t_i$ divides $t_{i + 1}$, $t_i \leq t_{i + 1}$. 
The subgroup $T$ is called the {\embf{torsion}} subgroup of $G$,
and if $T$ is trivial, $T \equiv \varnothing$, we say that $G$ is \embf{torsion-free}.
The \embf{rank} of $G$ is given by the rank of its torsion free subgroup: $\rank G = k$ where $F = G/T \cong  \integer^k$.

Let $\mc{R} \subseteq \mc{K}$ be a pure $(q + 1)$-dimensional subcomplex of a $(q  +1)$-complex  $\mc{K}$, 
and $\mc{R} _0 \subseteq \mc{R} $ be a pure $q$-dimensional subcomplex.
We can divide $\C_q(\mc{R} )$ into cosets of $q$-chains that are different in $q$-simplices from $\mc{R}_0$,
  but not from $\mc{R}  \setminus \mc{R}_0$. 
  In other words we treat $\mc{R} _0$ as something ``trivial'' or ``reduced to emptiness''.
  It forms the quotient group $\C_q(\mc{R} , \mc{R} _0) = \C_q(\mc{R} ) / \C_{q}(\mc{R} _0)$, 
  which we call the \embf{group of relative chains of $\mc{R} $ modulo $\mc{R}_0$}. 
  We can consider the restriction of the boundary operator $\bd_{q + 1}: \C_{q + 1}(\mc{R}) \to \C_{q}(\mc{R})$
  to $\mc{R}_0$ and an induced homomorphism $\bd_{q + 1}^{(\mc{R}, \mc{R}_0)}: \C_{q + 1}(\mc{R}, \mc{R}_0) \to \C_{q}(\mc{R}, \mc{R}_0)$.
Then the matrix $[\bd_{q + 1}^{(\mc{R}, \mc{R}_0)}]$ of the relative boundary operator 
$\bd_{q + 1}^{(\mc{R}, \mc{R}_0)}: \C_{q + 1}(\mc{R}, \mc{R}_0) \to \C_q(\mc{R}, \mc{R}_0)$ 
is obtained from $[\bd_{q + 1}]$ 
by leaving columns corresponding to $(q + 1)$-simplices from $\mc{R}$,
and then removing all zero rows as well as rows corresponding to $q$-simplices from $\mc{R}_0$.

 The $d$-th \embf{homology group} is the $d$-th cycle group modulo the $d$-th boundary group,
  $\mc{H}_{d}(\mc{K}) = \mc{Z}_{d}(\mc{K}) /  \mc{B}_{d}(\mc{K})$. 
  In other words, we partition the set of all $d$-cycles into classes of equivalence 
  relative to the set of $d$-boundaries of all of the $(d + 1)$-chains:
  two $d$-cycles $\bm{z}$ and $\bm{x}$ belong to the same homology class,
  and we say they are \embf{homologous}, $\bm{z} \sim \bm{x}$, 
  if there exists a  $(d + 1)$-chain $\bm{\pi} \in \mc{C}_{d + 1}(\mc{K})$
  such that $\bd_{d + 1} \bm{\pi} = \bm{z} - \bm{x}$.
  Alternatively,  $\bm{z} \sim \bm{x}$ if and only if $\bm{z} - \bm{x} \in \mc{B}_{d}(\mc{K})$.
  All  $d$-boundaries belong to the same homology class, given by $0$-element of $\mc{H}_{d}(\mc{K})$:
   $\bm{w} \in \mc{B}_{d}(\mc{K})$ then we write  $\bm{w} \sim 0$.

  The $d$-th \embf{cohomology group} of $\mc{K}$ is defined in the same fashion as the quotient 
   $\mc{H}^{d}(\mc{K}) = \mc{Z}^{d}(\mc{K}) /  \mc{B}^{d}(\mc{K})$. 
   We say that two $d$-coflows $\bm{f}$ and $\bm{g}$ belong to the same cohomology class,
     i.e. they are \embf{cohomologous}, $\bm{f} \csim \bm{g}$, 
     if there exists a  $(d - 1)$-cochain $\bm{\rho} \in \mc{C}^{d - 1}(\mc{K})$
     such that $\bm{g} = \bm{f} + \cobd_{d } \bm{\rho}$.

  There is an isomorphism $\mc{H}_{q}(\mc{K}) \cong \mc{H}^{q}(\mc{K})$, 
  and hence $\rank \mc{H}_{q}(\mc{K}) = \rank \mc{H}^{q}(\mc{K}) = \beta_{d} $,
  where $\beta_{q}$ denotes the $q$-th \embf{Betti number}  that corresponds to $(q + 1)$-dimensional ``holes'' in $\mc{K}$.


\paragraph*{Hodge decomposition} 

The \textbf{Hodge decomposition} theorem is a very important result that we utilize throughout the paper.
For more details on Hodge decomposition and Combintorial Laplacian see \cite{lim2020hodge} and \cite{goldberg2002combinatorial}
Recall $\cobd_{q}\!: \C^{q - 1}\to {\C^{q}}$ \ and \  
$\bd_{q+1}\!: \C_{q + 1}\to {\C_{q}}$.
Given a $(q + 1)$-complex $\mc{K}$ we have the following decomposition of its $d$-(co)chain group for all $d \leq q$:
  \begin{align}
  \label{eq:hodge-decomposition}
      \C_d(\mc{K}) \cong \Img (\cobd_d) \oplus \H_d(\mc{K}) \oplus \Img(\bd_{d + 1})
                       = \B^{d}(\mc{K}) \oplus \H_d(\mc{K}) \oplus \B_{d}(\mc{K})
  \end{align}
  where by leveraging the isomorphism $\H_{d}(\mc{K}) \cong \H^{d}(\mc{K})$,
  we have $\Z_{d}(\mc{K}) = \H_{d}(\mc{K}) \oplus \B_{d}(\mc{K})$
  and $\Z^{d}(\mc{K}) = \H^{d}(\mc{K}) \oplus \B^{d}(\mc{K})$.

\begin{corollary}[Hodge Theorem]
 The {Hodge decomposition} implies following important properties of $d$-cycles and $d$-cocycles:
  \begin{itemize}
        \item 
              $d$-cocycles are orthogonal to $d$-boundaries:
              $\Z^{d}(\mc{K}) \perp \B_{d}(\mc{K})$:
              
              $\bm{f} \circ \bd_{d + 1}  = \bm{f}( \bd_{d + 1} \tau) = \cobd_{d + 1} \bm{f}( \tau)= 0 $
              where $\bm{f} \in \Z^{d}(\mc{K})$ and $\tau \in \mc{K}_{d  +1}$
              
        \item 
              $d$-cycles are orthogonal to $d$-coboundaries: 
              $\B^{d}(\mc{K}) \perp \Z_{d}(\mc{K})$:

               $\cobd_{d} \circ \bm{z} = \cobd_{d} {\rho} ( \bm{z} ) = \bd_{d} \bm{z}({\rho} ) = 0$
              where $\bm{z} \in \Z_{d}(\mc{K})$ and $\rho \in \mc{K}_{d - 1}$

        \item 
              non-bounding $d$-cycles and $d$-coflows are orthogonal to $d$-boundaries and $d$-coboundaries:
              $\H_{d}(\mc{K}) \perp \B_{d}(\mc{K}) \oplus \B^{d}(\mc{K})$.
  \end{itemize}
\end{corollary}

The linear operator (homomorphism) 
$\cobd_{q} \bd_{q} + \bd_{q + 1} \cobd_{q + 1}: \mc{C}_{q}(\mc{K}) \mapsto \mc{C}_{q}(\mc{K})$
is called \textbf{combinatorial $q$-Laplacian} of $\mc{K}$
its kernel contains generators of  $\mc{H}_q(\mc{K})$ and $\mc{H}^q(\mc{K})$:
\begin{align*}
  \mc{H}_q(\mc{K}) \cong \ker(\cobd_{q} \bd_{q} + \bd_{q + 1} \cobd_{q + 1})
         \equiv \ker(\bd_{q}) \cap \ker(\cobd_{q + 1})
         = \mc{Z}_q(\mc{K}) \cap \mc{Z}^q(\mc{K})
\end{align*} 
     
We define the \textbf{up} and \textbf{down combinatorial Laplacians} of $\mc{K}$ as   the following linear operators:
\begin{align*}
    \mc{L}^q= \mc{L}^{UP}_q &= \bd_{q + 1} \cobd_{q + 1}: \C^{q}(\mc{K}) \to \C^{q + 1}(\mc{K}) \to \C_{q}(\mc{K})
    \\
    \mc{L}_q = \mc{L}^{DN}_{q}&= \cobd_{q } \bd_{q}: \C_{q}(\mc{K}) \to \C_{q - 1}(\mc{K}) \to \C^{q}(\mc{K})
\end{align*}

  This decomposition appears not only in the context of simplicial complexes, 
  but in many others areas of mathematics too.
   Alternative names for the components of the decomposition borrowed from vector calculus:
   \begin{itemize}
      \item $\cobd_{q + 1}$ -- \textsl{curl} operator, 
            and $\mc{B}_q(\mc{K}) = \Img(\bd_{q + 1})$ -- the \textit{curl} subspace, curl is the circulation on infinitesimal boundaries; 
      \item $\cobd_{q}$ -- \textsl{grad}, gradient opearator, 
                        $\bm{f} = \cobd_{q} \bm{\rho}$ -- $\bm{f}$ flows from maxima of $\bm{\rho}$ to its minima, $\bm{\rho}$ - potential function
                        -- the \textit{gradient} subspace
      \item $\bd_{q}$ -- \textsl{div}, divergence operator,
                          $\ker \bd_{q} = \mc{Z}_{q}(\mc{K})$  -- divergence free flows, flow conservation.
   \end{itemize}

\subsection{Boundary matrix and totally unimodularity}
         
   An $n \times m$ integer matrix $A$ with linearly independent rows 
   is called \embf{totally unimodular} (TU) if every square submatrix of $A$ has $\det(A) \in \{ -1, 0, 1 \}$.
 
   \begin{theorem}[\textbf{\textit{Integrality of TU, Theorem 2.1}} \cite{ohcp2011}]
     Let $A$ be an $n \times m$ totally unimodular matrix and $\bm{b} \in \integer^n$. 
     Then the polyhedron $\mc{P} = \left\{ \bm{x} \in \real^m \mid A \bm{x} = \bm{b}, \bm{x} \geq \bm{0}_m \right\}$
     \textbf{is integral}, 
     which means that the vertices of $\mc{P}$ are $m$-dimensional integer vectors 
     embedded in the $m$-dimensional real space, $\integer^m \subset \real^m$.
   \end{theorem}
   
   The condition for polynomial time solution of OHCP on $(q  +1)$-complex  $\mc{K}$.
   Hence, if the boundary matrix $[\bd_{q + 1}]$ of $\mc{K}$ is \textbf{totally unimodular},
       then the OHCP given in Eq.~\ref{eq:def:ohcp} can be solved in polynomial time (\textbf{\textit{Theorem 3.6}} \cite{ohcp2011}).
   The topological characteristic of $\mc{K}$ when the boundary matrix $[\bd_{q + 1}]$ is TU are given by the following theorem.
 
   \begin{theorem}[\textbf{\textit{Theorem 5.2}} \cite{ohcp2011}]
   \label{th:ohcp:tu-bd-mattrix}
     Let $\mc{R} \subseteq \mc{K}$ be a pure subcomplex of dimension $q + 1$ 
     and $\mc{R}_0 \subset \mc{R}$ be a pure subcomplex of dimension $q$. 
     Then $[\bd_{q + 1}]$ is TU if and only if $\H_q(\mc{R}, \mc{R}_0)$ is torsion-free
      \textbf{for all} such pure subcomplexes $\mc{R}_0$ and $\mc{R}$ of $\mc{K}$.
   \end{theorem}

   Two special topological cases of simplicial $(q + 1)$-complexes that adhere to the conditions of Th.\ref{th:ohcp:tu-bd-mattrix}
   are given by complexes \emph{embedded} in $\real^{q + 1}$ and by complexes that
   \emph{triangulate a compact orientable $(q + 1)$-manifold} (with or without boundary).

   \begin{theorem}[\textbf{\textit{Theorem 4.1}} \cite{ohcp2011}]
     Let $\mc{K}$ be a finite simplicial complex \textbf{triangulating a compact orientable $(q + 1)$-manifold} (with or without boundary).
     Then $[\bd_{q + 1}]$ is TU irrespective of the orientations of the simplices.
   \end{theorem}
   
   \begin{theorem}[\textbf{\textit{Theorem 5.7}} \cite{ohcp2011}]
     Let $\mc{K}$ be a finite simplicial complex \textbf{embedded} in $\real^{q + 1}$.
     Then, $\H_{q}(\mc{R}, \mc{R}_0)$ is torsion-free for all pure subcomplexes $\mc{R}_0$ and $\mc{R}$
     of dimensions $q$ and $q + 1$ respectively, such that $\mc{R}_0 \subset \mc{R}$.
     Or in other words, $[\bd_{q + 1}]$ is TU.
   \end{theorem}

\subsection{Directed Homology over \texorpdfstring{$\integerplus$}{Z+}}

When we consider $d$-(co)chains $\bm{f} \in \mc{C}_{d}(\mc{K}; \integer)$ over abelian groups, like ring of integers,
 the underlying simplices may not be endowed with some natural sense of orientation, 
since we always can define some simple non-zero $d$-chain $\Omega_d\! : \sigma \mapsto \{-1, +1\}$ 
which we choose to represent the natural orientation of $\vec{\sigma}_{+} \in \mc{K}_{d}$.
Then $d$-cochains are given as odd function relative to chosen orientation: 
$f(\vec{\sigma}) = f(\Omega_d(\vec{\sigma})) = \Omega_d(\vec{\sigma}) f(\sigma)$.
For example, integer-valued function on undirected edges  assigns direction and magnitude of flow.
Same function could be defined on the opposite $d$-simplices of directed $d$-complex  
$\accentset{\leftrightarrows}{\bm{f}} \in \mc{C}_{d}(\vec{\mc{K}}; \integer)$
as an odd  function
$\accentset{\leftrightarrows}{\bm{f}} (\vec{\sigma}_{-i}) = -\accentset{\leftrightarrows}{\bm{f}} (\vec{\sigma}_{+i})$
 and then associate the value on  ``undirected'' simplex $f(\sigma)$,
 for example, with  $\accentset{\leftrightarrows}{\bm{f}} (\vec{\sigma}_{+i}$).
This is a popular approach for graph-theoretical analysis of optimal homology problems \cite{chambers2012homology, generalizedMaxflow2021}
since all symmetries of (co)chains are preserved.

Our goal was to formulate the flow models and prove the correctness of solutions for OHCP and mSFN
 in a way amenable to most network-flow software libraries.
Since many classical implementations assume non-negative flow on di-arcs,
and we aim to compute the original function on  ``undirected'' $d$-simplex $f(\sigma)$,
we model the value of $f$ by defining  a \emph{directed non-negative function} 
$\vec{f} \in \mc{C}_{d}(\vec{\mc{K}}; \integerplus): \vec{\mc{K}}^{+}_{d} \cup  \vec{\mc{K}}^{-}_{d} \mapsto \integerplus$
such that 
$f(\sigma)  = \vec{f}(\vec{\sigma}_{+}) - \vec{f}(\vec{\sigma}_{-})$ and
$\abs{f(\sigma)}  = \vec{f}(\vec{\sigma}_{+}) + \vec{f}(\vec{\sigma}_{-})$,
i.e. $\vec{f}: \vec{\mc{K}}^{+}_{d}  \mapsto \integerplus$ computes the positive part 
and $\vec{f}: \vec{\mc{K}}^{-}_{d}  \mapsto \integerplus$ the negative part of  $f(\sigma)$.
But in this case $\mc{C}_{d}(\vec{\mc{K}}; \integerplus)$ is not a group but a semimodule,
and thus the previous construction of a (co)chain complex of groups does not apply here.

Here, we present fundamental definitions and results, developed  by Patchkoria \cite{patchkoria1977cohomology}
 and further extended by Mendez and Sanchez-Garcia~\cite{mendez2023directed}, 
 that are essential for establishing directed homology over the semiring $\integerplus$.
Introducing (co)chain complexes in the context of (co)chain semimodules presents an immediate challenge: 
defining alternating sums becomes problematic due to the absence of inverses for elements in a semimodule. 
To address this challenge, 
authors introduce  two maps, 
one for the positive components and another for the negative components of the boundary operators.


\begin{definition}
    A semiring $(\mathbb{K}, +, \cdot)$ is a set together with two operations such that
    \begin{itemize}
        \item $(\mathbb{K}, +)$ is an abelian monoid whose identity element we denote $0_{\mathbb{K}}$,
        \item $(\mathbb{K}, \cdot)$ is a monoid whose identity element we denote $1_{\mathbb{K}}$,
        \item $\cdot$ is distributive with respect to $+$ from either side,
        \item $0 \cdot \kappa = \kappa \cdot 0 = 0$ for all $\kappa \in \mathbb{K}$.
    \end{itemize}

\end{definition}

A semiring is commutative if $(\mathbb{K}, \cdot)$ is a commutative monoid,
 and cancellative if $(\mathbb{K}, +)$ is a cancellative monoid, 
 that is, $\kappa + \kappa' = \kappa + \kappa''$ 
 implies $\kappa' = \kappa''$ for all $\kappa, \kappa', \kappa'' \in \mathbb{K}$.
  A semiring is a semifield if every $0 \neq \kappa \in \mathbb{K}$ 
  has a multiplicative inverse.
   A semiring is zerosumfree if no element other than $0$ has an additive inverse.

\begin{definition}
    Let $\mathbb{K}$ be a semiring. A (left) $\mathbb{K}$-semimodule is an abelian monoid $(A, +)$ 
    with identity element $0_A$ together with a map $\mathbb{K} \times A \rightarrow A$ 
    which we denote $(\kappa, a) \mapsto \kappa a$ and such that for all $\kappa, \kappa' \in \mathbb{K}$ and $a,a' \in A$:
    \begin{itemize}
        \item $(\kappa \kappa')a = \kappa(\kappa' a)$
        \item $\kappa(a + a') = \kappa a + \kappa a'$
        \item $(\kappa + \kappa')a = \kappa a + \kappa' a$
        \item $1_{\mathbb{K}}a = a$
        \item $\kappa 0_A = 0_A = 0_{\mathbb{K}} \kappa$
    \end{itemize}
\end{definition}

A $\mathbb{K}$-semimodule $A$ is cancellative if $a + a' = a + a''$ implies $a' = a''$, 
and zerosumfree if $a + a' = 0$ implies $a = a' = 0$, for all $a, a', a'' \in A$.

The Cartesian product of $\mathbb{K}$-semimodules forms a $\mathbb{K}$-semimodule as well. 
The concept of a direct sum of modules extends naturally to $\mathbb{K}$-semimodules, 
where a finite direct sum of $\mathbb{K}$-semimodules is isomorphic to the Cartesian product of the same $\mathbb{K}$-semimodules. 
Quotient $\mathbb{K}$-semimodules can be defined using congruence relations.

\begin{definition}
    Let $A$ be a left $\mathbb{K}$-semimodule. 
    An equivalence relation $\rho$ on $A$ is a $\mathbb{K}$-congruence if, 
    for all $a, a' \in A$ and all $\kappa \in \mathbb{K}$, the following conditions hold:
    \begin{itemize}
        \item
              if $a \sim_{\rho} a'$ and $b \sim_{\rho} b'$, then $(a + b) \sim_{\rho} (a' + b')$.
        \item
              if $a \sim_{\rho} a'$, then $\kappa a \sim_{\rho} \kappa a'$.
    \end{itemize}
\end{definition}

Let $\vec{\mc{K}}$ be a $(q  +1)$-complex directed in all dimensions: 
$\vec{\mc{K}}_{d} = \vec{\mc{K}}^{+}_{d} \cup \vec{\mc{K}}^{-}_{d}$ for all $0 \leq d \leq q  +1$.
Let $\mathbb{K} = \integerplus$ be a semiring of coefficient non-negative integers, 
and consider a sequence of $\mathbb{K}$-semimodules  
$\vec{\mc{C}}_{d} = {\mc{C}}_{d}(\vec{\mc{K}}; \integerplus)$  and homomorphisms indexed by $0 \leq d \leq q  +1$:
\begin{align*}
    \ldots \mathrel{\mathop{\rightrightarrows}^{\bd^{+}_{d + 2}}_{\bd^{-}_{d + 2}}}  \vec{\mc{C}}_{d + 1}
           \mathrel{\mathop{\rightrightarrows}^{\bd^{+}_{d + 1}}_{\bd^{-}_{d + 1}}} \vec{\mc{C}}_{d}
           \mathrel{\mathop{\rightrightarrows}^{\bd^{+}_{d }}_{\bd^{-}_{d }}} \vec{\mc{C}}_{d - 1}
           \mathrel{\mathop{\rightrightarrows}^{\bd^{+}_{d - 1}}_{\bd^{-}_{d - 1}}} \ldots
\end{align*}

 We denote this sequence as 
 ${\mc{C}}(\vec{\mc{K}}; \integerplus) = \{\vec{\mc{C}}_d, \bd_{d}^{+}, \bd_{d}^{-}\}$, 
 which constitutes a chain complex of $\integerplus$-semimodules if the following condition is satisfied:
\begin{align}
    \bd_{d}^{+} \bd_{d + 1}^{+} + \bd_{d}^{-} \bd_{d + 1}^{-} 
    = \bd_{d}^{+} \bd_{d+1}^{-} + \bd_{d}^{-} \bd_{d+1}^{+}
\end{align}

The $\integerplus$-semimodule of $d$-cycles of ${\mc{C}}(\vec{\mc{K}}; \integerplus)$ is defined as:
\begin{align*}
     {\mc{Z}_{d}}(\vec{\mc{K}}; \integerplus)
     = \left\{ \vec{\bm{z}} \in  {\mc{C}}_{d}(\vec{\mc{K}}; \integerplus) \mid \bd_{d}^{+}(\vec{\bm{z}}) = \bd_{d}^{-}(\vec{\bm{z}}) \right\}
\end{align*}

The $d$-th homology of $\vec{\mc{K}}$ is then the quotient $\integerplus$-semimodule 
\begin{align*}
\mc{H}_{d}(\vec{\mc{K}}; \integerplus) = \mc{Z}_{d}(\vec{\mc{K}}; \integerplus) / \rho_{d}(\vec{\mc{K}}; \integerplus)
\end{align*}
where $\vec{\rho}_d = \rho_d(\vec{\mc{K}}; \integerplus)$ is the following $\integerplus$-congruence relation on $\mc{Z}_{d}(\vec{\mc{K}}; \integerplus)$: 
\begin{align*}
    \vec{\bm{x}} \widesim[2]{\vec{\rho}_d}    \vec{\bm{z}}
    \iff &\exists \bm{\xi}, \bm{\psi} \in  {\mc{C}}_{d + 1}(\vec{\mc{K}}; \integerplus) \text{ such that }
    \\
     &\vec{\bm{x}} + \bd^{+}_{d+1} \bm{\xi}  + \bd^{-}_{d+1}\bm{\psi}
    = \vec{\bm{z}} + \bd^{+}_{d+1} \bm{\psi} + \bd^{-}_{d+1}\bm{\xi}
\end{align*}


For the boundary relation, note that we may need two different $(d + 1)$-chains $\bm{\xi}$ and $\bm{\psi}$ 
in order to establish two classes as homologous. 
Intuitively, these two classes are the ``positive'' and ``negative'' part of $\bm{\pi}$, 
where $\bm{x} = \bm{z} - \bd_{d + 1} \bm{\pi}$ in the classical setting.


\section{Optimal homology problems}
\label{subsec:def:opt-homology}

In this section we give formal statements of optimal homology problems encountered in this paper. 
We begin with the OHCP, and then continue with mSFN which is a generalization of the former. 
These problems are stated for a general case of a finite $(q + 1)$-dimensional simplicial complex $\mc{K}$ with homology defined over $\integer$,
and $\integerplus$-weighted $q$ and $(q + 1)$-skeletons.
However, the general case is {NP-hard} not only over $\integer$~\cite{dunfield2011least}, but over $\integer_2$ and $\real$ as well \cite{chen2011hardness}.

Let $\mc{K}$  be a finite $(q + 1)$-dimensional simplicial complexes (not necessary embeddable),
with $\abs{\mc{K}_{q}} = E$ and $\abs{\mc{K}_{q  +1}} = F$,
and let $\bm{z} \in \Z_{q}({\mc{K}}; \integer)$ be the \emph{input}  $q$-cycle.
We assume that $q$ and $(q + 1)$-simplices of $\mc{K}$ are weighted by non-negative  functions ${c}: \mc{K}_q \to \integerplus$  and ${a}: \mc{K}_{q + 1} \to \integerplus$.
We  refer to these weights as $q$ and $(q + 1)$-\embf{volumes},  or  $q$ and $(q + 1)$-\embf{costs},  or  \embf{length-costs} and \embf{area-costs}, respectively.
 We denote $\bm{c} = \big( c({e}_1), \ldots, c({e}_E) \big) = (c_1, \ldots, c_E)$
and $\bm{a} = \big( a({\tau}_1), \ldots, a({\tau}_F) \big) = (a_1, \ldots, a_F)$.
The costs of a $q$-chain $\bm{x} \in \mc{C}_{q}(\mc{K}; \integer)$ 
and a $(q + 1)$-chain $\bm{\pi} \in C_{q + 1}(\mc{K}; \integer)$ are given by their total $q$-volume and total $(q + 1)$-volume:
given:
\begin{align*}
  \abs{{c}(\bm{x})}
  = \sum_{e \in \bar{K}_{q}} {c}(e_i) \cdot \abs{\bm{x}(e_i)}  \geq 0
  && &&
  \abs{{a}(\bm{\pi})}
    = \sum_{\tau_j \in \mc{K}_{q + 1}} {a}(\tau_j) \cdot \abs{\bm{\pi}(\tau_j)}  \geq 0
\end{align*}

We extend the area-volume function to \emph{any}  extension $\tld{\mc{K}} \supset \mc{K}$ of $\mc{K}$ by defining 
$a(\nu) = +\infty$ (or some big number) for all $\nu \notin \mc{K}_{q + 1}$.

\subsubsection{Optimal homologous cycle problem}
\label{subsec:def:ohcp}

\begin{definition}[OHCP]
    Given a finite simplicial $(q + 1)$-complexes $\mc{K}$ and a  $q$-cycle $\bm{z} \in \mc{Z}_{q}(\mc{K})$,
    the \emph{optimal homologous cycle  problem} (OHCP) asks to find 
    a $q$-cycle $\bm{x} \sim \bm{z}$ of minimal cost $\abs{c(\bm{x})}$:
    \begin{align}
      \label{eq:def:ohcp}
        \min\limits_{\bm{x}, \bm{\pi}}\ & 
        \abs{c({\bm{x}}) }  \geq 0        \notag \\
        \subto\ & \bd_{q + 1} \bm{\pi}   = \bm{z} -\bm{x}                  \\
              & \bm{x} \in \mc{Z}_{q}({\mc{K}}), \ \bm{\pi} \in \mc{C}_{q + 1}({\mc{K}})  
              \notag
      \end{align}
      
\end{definition}

We want to guarantee that the optimal solution $(\bm{x}_{\star}, \bm{\pi}_{\star})$ found on \emph{any}  extension $\tld{\mc{K}}$ of $\mc{K}$
is feasible and optimal on the original complex  $\mc{K}$ as well. 

\begin{align}
\label{eq:def:ohcp-on-K}
  \begin{array}{rrcll}\dsp
        \min\limits_{\bm{x}, \bm{\pi}}\ 
              &\dsp  \abs{{c}({\bm{x}}) }  \geq 0 
              \\
        \subto&\dsp \bd_{q + 1} \bm{\pi} = \bm{z} - \bm{x}              \\
              & \bm{\pi} \in \mc{C}_{q + 1}(\tld{\mc{K}})\!: 
              &\dsp \bm{\pi}(\tld{\nu})  = 0 
               &\text{for all } \tld{\nu} \notin \mc{K}_{q  +1}      
        \\
             & \bm{x} \in \mc{Z}_{q}(\tld{\mc{K}})\!: 
              &\bm{x}(\tld{e})  = 0 
               & \text{for all } \tld{e} \notin \mc{K}_{q}      
        \\
  \end{array}
\end{align}

The linear program for OHCP,
and its  dual problem is the problem of the  \embf{maximum $q$-coflux} through $\bm{z}$:
\begin{align}
\label{eq:def:ohcp-max-flux-lp}
      \begin{array}{ll}\dsp
        \min\limits_{\bm{x}, \bm{\pi}}\ 
        &\dsp {c}({\bm{x}^{+}}) + {c}(\bm{x}^{-})         \geq 0 
        \\
        \subto&\dsp  [\bd_{q + 1}] (\bm{\pi}^{+} - \bm{\pi}^{-}) +\bm{x}^{+} - \bm{x}^{-} = \bm{z}    
        \\
              &\dsp \bm{x}^{\pm} \geq \bm{0}_{E}, \ \bm{\pi}^{\pm} \geq \bm{0}_{F}
      \end{array}
      &&\overset{dual}{\Longrightarrow}&&
      \begin{array}{lllll}\dsp
            \max\limits_{\bm{f}}\ &\dsp \abs{{\bm{f}}({\bm{z}})}   \geq 0       \\
            \subto\ &\dsp  \cobd_{q + 1} \bm{f} = \bm{0}_{F}  
            \\
                    &\dsp  -\bm{c} \leq  \bm{f} \leq \bm{c}          
        \end{array}
\end{align}
where $\cobd_{q + 1} \bm{f} = \bm{0}_{F} \equiv \bm{f}(\bd_{q+1} {\tau}_j) = 0  $
implies that  $\bm{f}$ is a \embf{$q$-coflow} $\bm{f} \in \mc{Z}^{q}({\mc{K}})$.

A $q$-coflow is \embf{feasible} if:
\begin{align}
\label{eq:def:feasible-coflow}
    -\bm{c} \leq  \bm{f} \leq \bm{c}:
    &&\dsp  -{c_i} \leq  \bm{f}(e_i) \leq {c}_i    &&
    &\forall e_i \in  \mc{K}_{q}    
\end{align}

\subsubsection{Multiscale Simplicial Flat Norm}
\label{subsec:def:msfn}
The notion of the \emph{flat norm} comes from geometric measure theory where it is used to measure distances between shapes given by directed \textit{currents} (generalizations of oriented surfaces/manifolds) \cite{glaunes2005transport}.
The \emph{multiscale flat norm}, \cite{MoVi2007}, was used for applications in image analysis and denoising, \cite{vixie2010multiscale}.
The \embf{Simplicial Flat Norm (SFN)} with multiple scales, 
introduced in \cite{ibrahim2011simplicial}, 
is a discretization of the flat norm to spaces triangulated by a simplicial complex.
SFN is closely related to the OHCP, and  provides a sense of ``distance'' within some homology class.

%

\begin{definition}[SFM]
    The \embf{simplicial flat norm} of $\bm{z} \in \mc{Z}_{q}(\mc{K}; \integer)$ is defined as 
    \begin{align}
    \label{eq:def:simplex-flat-norm}
        {\mbb{F}}(\bm{z}) 
         = \min\limits_{\bm{x} \sim \bm{z}} {\mbb{F}}(\bm{z}; \bm{x}) 
         = \min\limits_{\bm{x} \sim \bm{z}} 
            \big\{ \abs{{{c}}({\bm{x}})} + \abs{ {{a}}({\bm{\pi}})} \ \big\vert \  \bd_{q + 1} \bm{\pi} = \bm{z} - \bm{x} \big\}
    \end{align}
    where $\bm{x} \in \mc{Z}_{q}(\mc{K}; \integer)$ and $\bm{\pi} \in \mc{C}_{q + 1}(\mc{K}; \integer)$.
    The pair $(\bm{x}_{\star}, \bm{\pi}_{\star})$ that minimizes ${\mbb{F}}(\bm{z}; \bm{x})$
    is called the \embf{(optimal) flat norm decomposition (SFN-decomposition)} of $\bm{z}$.
\end{definition}

\begin{definition}[mSFM]
    Given a scale parameter $\lambda \in \integerplus$ or $\Lambda = \sfrac{1}{\lambda} \in \integerplus$,
    the \embf{multiscale simplicial flat norm} of $\bm{z} \in \mc{Z}_{q}(\mc{K}; \integer)$ is defined as 
    \begin{align}
    \label{eq:prelim:def:scaled-flat-norm}
        &&{\mbb{F}}_{\lambda}(\bm{z}) 
        &= \min\limits_{\bm{x} \sim \bm{z}} {\mbb{F}}_{\lambda}(\bm{z}; \bm{x}) 
         = \min\limits_{\bm{x} \sim \bm{z}} 
           \big\{  
                   \abs{{{c}}({\bm{x}})} + \lambda \abs{ {{a}}({\bm{\pi}})} 
                   \ \big\vert \  \bd_{q + 1} \bm{\pi} = \bm{z} - \bm{x} 
           \big\}&
    \\[3pt]
         \text{or} &&&
    \notag \\[3pt]
       & &{\mbb{F}}_{\Lambda}(\bm{z}) 
        &= \min\limits_{\bm{x} \sim \bm{z}} {\mbb{F}}_{\Lambda}(\bm{z}; \bm{x}) 
         = \min\limits_{\bm{x} \sim \bm{z}} 
            \big\{  
                    \Lambda \abs{{{c}}({\bm{x}})} + \abs{ {{a}}({\bm{\pi}})} 
                    \ \big\vert \  \bd_{q + 1} \bm{\pi} = \bm{z} - \bm{x} 
            \big\}&
    \end{align}
    where $\bm{x} \in \mc{Z}_{q}(\mc{K}; \integer)$ and $\bm{\pi} \in \mc{C}_{q + 1}(\mc{K}; \integer)$.
    The pair $(\bm{x}_{\star}, \bm{\pi}_{\star})$ that minimizes ${\mbb{F}}_{\lambda}(\bm{z}; \bm{x})$
    is called the \embf{(optimal) scaled flat norm decomposition ($\lambda$-FN  or $\Lambda$-FN decomposition)} of $\bm{z}$.
\end{definition}

The default simplicial flat norm of $\bm{z}$ corresponds to the case of $\lambda = \Lambda =  1$,
the OHCP corresponds to $\lambda = 0$, and the OBCP to $\Lambda = 0$. 

Consider the following integer linear program with an objective function that computes ${\mbb{F}}(\bm{z})$ in Eq.~\eqref{eq:def:simplex-flat-norm} and its dual program:
\begin{align}
\label{eq:prelim:flat-norm-lp}
        \begin{array}{rl}\dsp
            \min\limits_{\bm{x}, \bm{\pi}}\ 
            &\dsp  
                    {c}(\bm{x}^{+}) + {c}(\bm{x}^{-}) + {a}(\bm{\pi}^{+}) + {a}(\bm{\pi}^{-})
            \\[2pt]
            \text{s.t.} &\dsp 
                    \bd_{q + 1} (\bm{\pi}^{+} - \bm{\pi}^{-}) + \bm{x}^{+} - \bm{x}^{-} = \bm{z}         
            \\ 
            &\dsp 
                    \bm{x}^{\pm}  \geq \bm{0}_{E}, \ \bm{\pi}^{\pm}  \geq \bm{0}_{F}
        \end{array}
    &&\overset{dual}{\Longrightarrow}&&
        \begin{array}{lll}\dsp
            \max\limits_{ \tilde{\bm{f}} }\ 
            &\dsp 
                    \abs{\tilde{\bm{f}}(\bm{z})} \geq 0
            \\[2pt]
            \text{s.t.} &\dsp  
                    -\bm{a}_{F} \leq \cobd_{q + 1} \tilde{\bm{f}} \leq \bm{a}_{F}
            \\  
            &\dsp  
                    -\bm{c}_{E} \leq \tilde{\bm{f}} \leq \bm{c}_{E}
        \end{array}
\end{align}
where the dual variable is a $q$-cochain $\tilde{\bm{f}} \in  \mc{C}^{q}(\mc{K}; \integer)$, 
which we refer to as \embf{$q$-pseudocoflow},
and $\bm{\pi} = \bm{\pi}^{+} - \bm{\pi}^{-} \in \mc{C}_{q + 1}(\mc{K}; \integer)$ together with
$\bm{x} = \bm{x}^{+} - \bm{x}^{-} \in \mc{Z}_{q}(\mc{K}; \integer)$ constitute  the FN-decomposition of $\bm{z}$.

A $q$-pseudocoflow is \embf{feasible} if:
\begin{align}
\label{eq:def:feasible-pseudocoflow}
    -\bm{c} \leq  \bm{f} \leq \bm{c}:
    &&\dsp  -{c_i} \leq  \bm{f}(e_i) \leq {c}_i    &&
    &\forall e_i \in  \mc{K}_{q}    
    \\
    -\bm{a} \leq  \tld{\bm{f}} \leq \bm{a}:
    &&\dsp  -{a_j} \leq  \tld{\bm{f}}(\bd_{q + 1} \tau_j) \leq {a}_i    &&
    &\forall \tau_j \in  \mc{K}_{q + 1}    
\end{align}

Recall that the embedding assumptions guarantee that the boundary matrix $[\bd_{q + 1}]$ is \emph{totally unimodular} \cite{ohcp2011}, 
and since all weights $\bm{c}$ and $\bm{a}$ are integral, we can find an optimal integral solution in polynomial time for both programs in Eq.~\eqref{eq:prelim:flat-norm-lp}, \cite{veinott1968integral, ibrahim2011simplicial}.

%
%
%

\chapter{Network flow models for optimal homology}


\section{Co-homology of embedded complex and Dual complex}

Embeddable complexes are of great interest in computational topology because, like planar graphs, 
In this paper, we focus on finite $(q + 1)$-connected \emph{orientable} $(q + 1)$-complexes with an \emph{embedding} into $\real^{q + 1}$.
they  have a well-defined dual graph that extends to a dual complex \cite{maxwell2021algorithmic},
and the structure of their homology groups admits polynomial time solutions for optimal homology problems \cite{ohcp2011}.

Let $\mc{K}$ be a $(q + 1)$-connected \emph{pure} orientable $(q + 1)$-dimensional simplicial complex embedded in $\real^{q + 1}$.
The connected components of 
$\real^{q + 1} \setminus \mc{K} = \real^{q + 1} \setminus \support{\mc{K}}$ are well defined,
and by the Alexander duality theorem can be partitioned into a set of $\beta_{q} + 1$ open $(q + 1)$-cells,
where $\beta_{q}$, called the \embf{$q$-th Betti number}, is the rank of $q$-th (co)homology group of $\mc{K}$: 
$\beta_{q} = \rank \mc{H}_{q}(\mc{K})$.
Let  $\mc{V} = \left\{ \nu_{0}, \nu_{1}, \ldots, \nu_{\beta_{q}} \right\}$ be the closures of connected components of $\real^{q + 1} \setminus {\mc{K}}$,
so that $\mc{K} \cup \mc{V} = \real^{q  +1}$.
Given a valid embedding $\support{\mc{K}}$, such partition can be computed in polynomial time, \cite{dey2020computing} ($n \log n$).
The (topological) boundary of a void $\nu_k$ is given by $\support{\bd \nu_k} = \support{\nu_k} \cap \support{\mc{K}}$,
then the $q$-surface 
$\support{\bd \mc{K}} = \support{\mc{K}} \cap \support{\mc{V}} = \support{\mc{K}} \cap \support{ \nu_0 } \cap \ldots \cap \support{ \nu_{\beta_q} }$ 
is the (topological) boundary of the complex.
Only one of the voids cannot be bounded by a $(q + 1)$-ball of large enough radius, we call it the \embf{outer void} and use $\nu_{0}$ to reference it. 
The rest of the (bounded) voids we call \embf{inner voids} and denote $\mc{V}_{I} = \mc{V} \setminus \{\nu_0\}$.
Note that the image of any $q$-simplex under the embedding $\support{\mc{K}}$ lies on the boundary of at most two $(q + 1)$-cells in $\mc{K} \cup \mc{V}$,
which means that it can be dualized as a dual edge 
and that the embedding of any $(q + 1)$-simplex can be identified as a subspace of $\real^{q+1}$ bounded by the $q$-cells of its faces.

\paragraph*{Fixing orientation on embedded complex}
The embeddability, among others, also guarantees that that $\mc{K}$ is \embf{orientable}, 
which means there is no relative torsion in $q$-th dimension and there is no \Mobius subcomplexes of dimension $d \leq q$, \cite{ohcp2011},
and hence a consistent assignment of orientations in all dimensions $d \leq q + 1$ is possible.

We \emph{fix the positive orientation} on $\mc{K}$ by requiring that the $(q + 1)$-simplices included $\mc{K}$ to be \embf{consistently oriented}:
if two $(q+ 1)$-simplices $\eta, \tau \in \mc{K}_{q + 1}$ share a $q$-face $e = \eta \cap \tau$
then $e$ must be a positive face of one of them, and a negative of the other.
The embedding assumptions guarantee that it is always possible to choose the natural orientation of $(q + 1)$-simplices to be included in $\mc{K}$, 
since any $q$-simplex is embedded on the boundaries of at most two $(q + 1)$-cells.
Without loss of generality, let's assume that $e \subset \tau$ and $-e \subset \eta$,
which is equivalent to $\bd \tau (e) = +1$ and $\bd \eta (e) = -1$.
Then we denote the natural orientation of $e \in \mc{K}$ as $e = (\eta \dart \tau)$ 
and define its coboundary as $\cobd e = \tau - \eta$.
We call $\eta$ and $\tau$ the \embf{left} and \embf{right coface} of $e$ respectively,
which we denote as $\eta = L(e)$ and $\tau = R(e)$. 
Hence, by \textit{the left-hand rule}, the \emph{positive} faces of $\tau \in \mc{K}_{q + 1}$ 
define the \emph{clockwise} direction of curling along its boundary $\bd \tau$, 
and negative -- \emph{counter-clockwise}. 
Since the $q$-boundaries cancel out on common faces, $\bd \eta (e) + \bd \tau (e) = 0$,
the boundary of the whole complex $\bd \mc{K} = \sum_{\mc{K}_{q + 1}} \bd \tau $ is oriented clockwise.
It implies that 
the outer void is the left-coface of all $q$-simplices $e$ on the its boundary, 
$\support{e} \subset  \support{\bd \nu_{0}} \iimplies \nu_0 = L(e)$, 
and all inner voids are right cofaces of $q$-simplices on their boundaries,
$\support{e} \subset  \support{\bd \nu_{k}} \iimplies \nu_k = R(e)$ for $1 \leq k \leq \beta_{q}$.
We fix the orientation of $q$-simplices in these cases as $e = (\nu_0 \dart \tau)$ and $e = (\eta \dart \nu_k)$, respectively, 
for some $\eta, \tau \in \mc{K}_{q + 1}$\footnote{This choice is no more than a convenience, and does not affect our proves and constructions.}.
We also assume that the input $q$-cycle is directed clockwise.
\begin{align}
    &e = (\eta \dart \tau) \in \mc{K}_{q}:&\\
    &\cobd_{q + 1} e = \tau - \eta&\iff&\left\lbrace\begin{array}{ccccc}
        \me e \subset  \tau &\text{ and }  & \bd_{q + 1} \tau (e) = +1 &\text{ hence }  &\tau = R(e)  \\    
            -e \subset \eta &\text{ and }  & \bd_{q + 1} \eta (e) = -1 &\text{ hence }  &\eta = L(e)
    \end{array}\right.
\end{align}

   To emphasize  the analogy of $\mc{K}$ with planar graphs
   we refer to the top two dimensional simplices of $\mc{K}$, 
   $e \in \mc{K}_{q}$ and $\tau \in \mc{K}_{q + 1}$,
   as \embf{$q$-edges} and \embf{$(q + 1)$-facets}, or \embf{edge-simplices} and \embf{facet-simplices}, respectively.
   The cardinalities of corresponding skeletons are denoted as $E = \abs{\mc{K}_{q}}$ and $F = \abs{\mc{K}_{q  +1}}$.
   We assume that each $q$- and $(q + 1)$-simplex has a non-negative integer weight assigned to it, 
   which we collectively refer to as its $q$-th or $(q + 1)$-th dimensional \emph{volumes}.
   A function  ${c}: \mc{K}_{q} \to \integerplus$ defined on the  $q$-skeleton assigns the \embf{length-volumes} or \embf{$q$-length} to the $q$-edges, and is also called \emph{capacity} function.
   The \embf{area-volumes} or \embf{$(q + 1)$-areas} of the facet-cells are defined by a  function ${a}: \mc{K}_{q + 1} \to \integerplus$.

\begin{figure}
  \centering
  \includegraphics[width=0.9\linewidth]{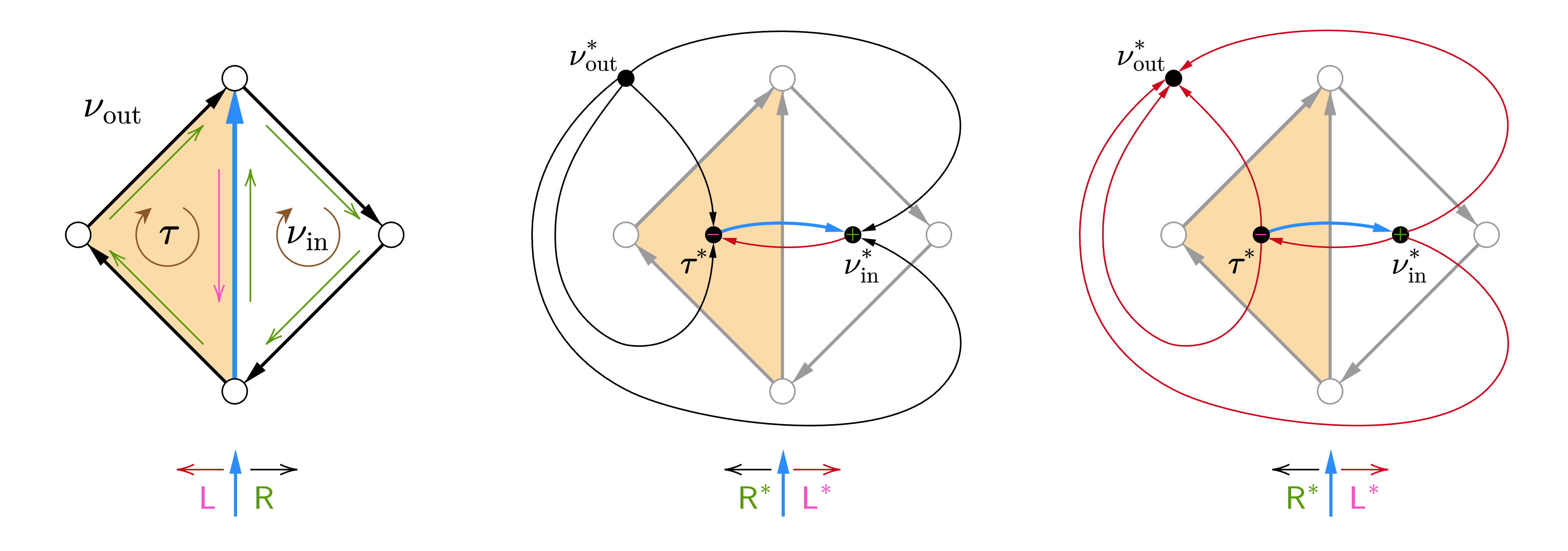}
  \caption[Natural orientation and duality.]{
        Natural orientation and duality.
        \textbf{Left:} A facet $\tau$ is on the left of the blue edge $e$, and an inner void $\nu_{{in}}$ is on the right.
        \textbf{Middle:}  The positive direction is clockwise, and hence $e$ is negeative on the boundary of $\tau$ and positive on the boundary of $\nu_{{in}}$.
        Therefore, the positive dual di-arc $(\tau^{*} \to \nu_{{in}})$ (black in the middle) is oriented from left to right.
        \textbf{Right:} Negative dual di-arcs are shown in red.
    }
  \label{fig:ortanddual-clockwise-02}
\end{figure}

\paragraph*{Acyclization of embedded complex}
The Alexander duality theorem implies a bijection between generators of $\mc{H}_{q}(\mc{K})$ 
and inner voids (or any subset of $\beta_q$ voids) of $\mc{K}$'s complement.
Moreover, the inner voids produced by the embedding  of the $q$-th skeleton, $\left( \real^{q + 1}\setminus {\mc{K}_{q}} \right)$, 
are in bijection with generators of $q$-th cycle group $\mc{Z}_{q}(\mc{K})$ of $\mc{K}$.
The key idea behind the construction of a dual graph 
is to put the generators of $\mc{Z}_{q}(\mc{K})$ into one-to-one correspondence with dual nodes.
Adapting the definition from \cite{duval2015cuts}:
\begin{definition}
  A \textbf{$q$-acyclization} of a $(q + 1)$-dimensional simplicial complex $\mc{K}$ embedded in $\real^{q +1}$
  is a $(q + 1)$-dimensional cellular complex $\bar{\mc{K}}$ with same embedding and $\mc{K}$ as its subcomplex,
  but acyclic in $q$-th dimension:
  \begin{align}
  \label{eq:def:acyclization}
      \bar{\mc{K}}_{q^{\prime}} = {\mc{K}}_{q^{\prime}} 
      &&
      {\mc{K}}_{q + 1} \subseteq \bar{\mc{K}}_{q + 1} 
      &&
      \mc{H}_{q}(\bar{\mc{K}}; \integer) =  0
  \end{align}
  for all $0 \leq q^{\prime} \leq q$.

\end{definition}



We construct the \embf{acyclization} $\bar{\mc{K}}$ of $\mc{K}$
 as a $(q + 1)$-complex \emph{without} a boundary that includes \emph{all the voids} as new basis generators to the $(q+1)$-th chain group: 
$\bar{\mc{K}} = \mc{K} \cup \mc{V} \cong \bar{\real}^{q + 1} \approx \mathbb{S}^{q + 1}$,
where $\bar{\real}^{d} = \real^{d} \cup \{ \infty \}$ and  $\mathbb{S}^{d}$ is the $d$-sphere.
Note that $\mc{H}_{q + 1}(\bar{\mc{K}}) = 1$, and hence  $\bar{\mc{K}}$ wraps around a $(q + 2)$-void.
To construct a chain complex of $\bar{\mc{K}}$, we need to define the boundary homomorphism for new elements. 

Since we assume that all $(q + 1)$-simplices of $\mc{K}$ are consistently oriented, 
the boundary of $\mc{K}$ is given by summation of all basis element in 
$\mc{B}_{q}(\mc{K}) = \im \bd_{q +1}$:
$\bd \mc{K} = \bd_{q + 1} \mc{K} = \sum_{\mc{K}_{q  +1}} \bd_{q + 1} \tau_j$.
Then simple $q$-cycles $\ring{\bm{w}}_{k}$ defined for each void $\nu_k \in \mc{V}$ as $\ring{\bm{w}}_{k} = \sum_{\support{e_i} \subset \support{\bd \nu_k} } -h_i\cdot {e}_i$ 
constitute the independent components of $\mc{K}$'s boundary:
$\bd \mc{K} = -\sum_{k = 0}^{\beta_{q}} \ring{\bm{w}}_{k}$,
where $h_i =  \bd{\mc{K}} ({e}_i) \in \{-1, +1\}$ is the orientation of ${e}_i$ on the boundary of $\mc{K}$.
Recall that we fixed the orientation of the boundary $q$-simplices so that $h_i = +1$ if $\support{e_i} \subset \support{\bd \nu_0}$,
and $h_i = -1$ if $\support{e_i} \subset \support{\bd \nu_k}$ for $k \neq 0$.
We set $\bd_{q + 1}\nu_k \asgn \ring{\bm{w}}_{k}$, and then $\bd_{q + 1}: \C_{q + 1}(\bar{\mc{K}}) \mapsto \C_{q}(\bar{\mc{K}})$.
We also update the coboundary operator for the boundary $q$-simplices by setting:
$\cobd_{q + 1 } e_i (\nu_0) = -1$ for $\support{e_i} \subset \support{\bd \nu_0}$,
and 
$\cobd_{q + 1 } e_i (\nu_k) = +1$ for $\support{e_i} \subset \support{\bd \nu_k}$.
Since $\mc{K}_{q} = \bar{\mc{K}}_{q}$ we will not differentiate the $(q + 1)$-boundary operators on $\mc{K}$ and $\bar{\mc{K}}$, and will use $\bd_{q  +1}$ in  both cases.
But when it is necessary we will distinguish the original coboundary operator $\cobd_{q  +1} : \mc{C}^{q}(\mc{K}) \mapsto  \mc{C}^{q +1}(\mc{K})$ 
and the coboundary operator on $\bar{\mc{K}}$: $\bar{\cobd}_{q  +1} : \mc{C}^{q}(\bar{\mc{K}}) \mapsto  \mc{C}^{q +1 }(\bar{\mc{K}}) \neq \mc{C}^{q +1}(\mc{K})$.
We  extend the upper Lagrangian on $\mc{K}$, $\mc{L}^q \!:  \C^{q}(\mc{K}) \mapsto \B_{q}(\mc{K})$, by including the non-bounding $q$-cycles in its image:
$\mc{L}^{q} \asgn \bar{\mc{L}}^q=   \bd_{q + 1} \bar{\cobd}_{q + 1}: \C^{q}(\bar{\mc{K}}) \mapsto  \B_{q}(\bar{\mc{K}}) \equiv \C^{q}({\mc{K}}) \mapsto  \Z_{q}({\mc{K}})$.


Any $\beta_{q}$ of the $q$-cycles $\{ \ring{\bm{w}}_0, \ring{\bm{w}}_1, \ldots, \ring{\bm{w}}_{\beta_q} \}$ form a basis of $\mc{H}_{q}(\mc{K})$,
yet, by default, we choose boundary components corresponding to the inner voids
$\left[\H_{q}(\mc{K}) \right] = \bigoplus_{\nu_k \in \mc{V}_{I}} [ \ring{\bm{w}}_{k} ] = \bigoplus_{k \neq 0} [ \ring{\bm{w}}_{k} ]$.
Note that if $\beta_{q} = 0$ then $\mc{K}$ either does not have a boundary ($\bd \mc{K} = \varnothing$), 
or it has only one boundary component  associated with the outer void ($\bd \mc{K} = -\bd \nu_0 = -\ring{\bm{w}}_{0}$);
and either way $\Z_{q}(\mc{K}) = \B_{q}(\mc{K})$, i.e. all $q$-cycles are null-homologous.
Hence, $\ring{\bm{w}}_{0} \sim 0$ and we can choose it as 0 of $\H_{q}(\mc{K})$, 
and then just write $\left[\H_{q}(\mc{K}) \right] = \bigoplus_{k =0}^{\beta_k} [ \ring{\bm{w}}_{k} ] $.
 Any non-bounding $q$-cycle $\bm{z} \in \mc{H}_{q}(\mc{K})$ is homologous to some linear combination of the void boundaries:
  $\bm{z}~\sim~\sum_{k = 0}^{\beta_q} \zeta_k \cdot \bd_{q + 1} \nu_k = \sum_{k = 0}^{\beta_q} \zeta_k \cdot \ring{\bm{w}}_{\beta_k}$.
The integer vector of coefficients of such combination  $[\bm{z}] = [\zeta_0, \zeta_1, \ldots, \zeta_{\beta_{q}}] $ 
defines the \embf{homology signature} of a $q$-cycle $\bm{z}$:
$\bm{x} \sim \bm{z}$ if and only if $[\bm{x}]  = [\bm{z}] $.
 Note that if $\beta_q \geq 1$ then $\zeta_0  = 0$ for any $\bm{z}$.

 A basis of $\H^{q}(\mc{K})$ should consist of $\beta_{q}$ linearly independent integer functions defined on the generators of $\H_{q}(\mc{K})$. /
 Let  $\{  \ring{\bm{p}}_1, \ldots, \ring{\bm{p}}_{\beta_q} \}$ be a collection of $q$-copath on $\mc{K}$ of the form 
 $\ring{\bm{p}}_{k} = \bm{p}(\ring{\bm{w}}_0, \ring{\bm{w}}_k)$, 
 i.e. $\ring{\bm{p}}_{k}(\ring{\bm{w}}_0) = -1$ and $\ring{\bm{p}}_{k}(\ring{\bm{w}}_k) = +1$,
 or $\mc{L}^{q} \ring{\bm{p}}_{k} = \ring{\bm{w}}_k -  \ring{\bm{w}}_0$,
 or $\bar{\cobd}_{q  +1}  \ring{\bm{p}}_{k} = \nu_k -  \nu_0$ .
 This collection forms, what we call, the $\ring{\bm{w}}_0$-rooted or \embf{$\nu_0$-rooted basis} of the $q$-th cohomology group: 
 $\H^{q}(\mc{K}) \cong \bigoplus_{k = 1}^{\beta_q} [\ring{\bm{p}}_{k}]$.
 We can see each $q$-copath as a ``branch'' of a simple $q$-cochain $\bm{p}_{0}$ with 
 $\bar{\cobd}_{q  +1} \bm{p}_{0} = \sum_{k = 1}^{\beta_q} \nu_k - \nu_0 = \mc{V}_{I} - \nu_0$,
 which we call a  \emph{$\nu_0$-rooted \textbf{$q$-cotree}} and denote $\bm{p}_{0} = \C^{q}(\nu_0 \to \mc{V}_{I})$.
 In general, let $\mc{S}, \mc{T} \subset \mc{V}$ such that $\mc{S} \cap \mc{T} = \varnothing$,
 then a \embf{$\mc{ST}$-cotree} $\bm{p}_{\mc{ST}} =   \C^{q}(\mc{S} \to \mc{T} )$
 is a simple $q$-cochain such that  $\bar{\cobd}_{q  +1} \bm{p}_{\mc{ST}} = \sum_{\mc{T}} \nu_t - \sum_{\mc{S}} \nu_s$.

In the case of the acyclization $\bar{\mc{K}}$, as name suggests, all of the $\beta_q$ homology classes of non-trivial $q$-cycles are destroyed 
by including their generators $\support{\bd \nu_k}$ as the boundaries of newly added $(q + 1)$-cells.
As a result, we have 
$\mc{H}_{q}(\bar{\mc{K}})  = \varnothing$ and 
$\mc{Z}_{q}(\bar{\mc{K}}) = \mc{B}_{q}(\bar{\mc{K}})$,
while $\mc{Z}_{q}({\mc{K}}) = \mc{Z}_{q}(\bar{\mc{K}})$.
So, the acyclization has not changed   the group of  $q$-cycles as a whole, but some cycles have changed their allegiance:
 non-bounding $q$-cycles turned  into $q$-boundaries $\mc{H}_{q}(\mc{K}) \mapsto \mc{B}_{q}(\bar{\mc{K}})$.

Similarly,
by the isomorphism between co-homology groups, the $q$-th cohomology group of $\bar{\mc{K}}$ is trivial $\H^{q}(\bar{\mc{K}})  = \varnothing$, 
and therefore as previously $\Z^{q}(\bar{\mc{K}}) = \B^{q}(\bar{\mc{K}})$.
But there is no more equality between $q$-th cocycle groups: $\Z^{q}({\mc{K}}) \neq \Z^{q}(\bar{\mc{K}})$.
The inclusion of $\nu_k$'s as new basis generators of $\C_{q +1}(\bar{\mc{K}})$ 
defined coboundaries for non-trivial cocycles(coflows) of $\mc{K}$, 
$\bar{\cobd}_{q+1} \ring{\bm{p}}_{k}(\nu_{k}) = +1 $,
and turned them into bounded $q$-cochains: 
$\H^{q}(\mc{K}) \mapsto \C^{q}(\bar{\mc{K}}) \setminus  \Z^{q}(\bar{\mc{K}})$.
Hence, unlike homology, the acyclization preserves \emph{only} the group of $q$-coboundaries $\B^{q}(\mc{K}) = \B^{q}(\bar{\mc{K}})$,
see Lemma~\ref{lm:barK:coboundaries}.




\begin{lemma}
\label{lm:barK:cycles}
    If a simple $q$-chain $\bm{w} \in \C_{q}(\mc{K})$ has an empty boundary on $\bar{\mc{K}}$,
    i.e. $\bar{\bd}_{q} \bm{w} = 0$,
    then it is a $q$-cycle on $\mc{K}$:  $\bm{w} \in \Z_{q}({\mc{K}})$.
\end{lemma}

\begin{lemma}
\label{lm:barK:coboundaries}
    If a simple $q$-cochain $\bm{p} \in \mc{C}^{q}(\mc{K})$ has an empty coboundary on $\bar{\mc{K}}$,
    i.e. $\bar{\cobd}_{q + 1} \bm{p} = 0$,
    then it is a $q$-coboundary on $\mc{K}$:  $\bm{p} \in \mc{B}^{q}({\mc{K}})$.
\end{lemma}
\begin{proof}

By the isomorphism between co-homology groups, the $q$-th cohomology group of $\bar{\mc{K}}$ is trivial $\H^{q}(\bar{\mc{K}})  = \varnothing$, 
and therefore $\Z^{q}(\bar{\mc{K}}) = \B^{q}(\bar{\mc{K}})$.
But there is no  equality between $q$-th cocycle groups: $\Z^{q}({\mc{K}}) \neq \Z^{q}(\bar{\mc{K}})$.
The inclusion of $\nu_k$'s as new basis generators of $\C_{q +1}(\bar{\mc{K}})$ 
defined the coboundaries for every basis $q$-cocycle $\ring{\bm{p}}_{k} \in \mc{H}^{q}(\mc{K})$: 
and turned them into bounded $q$-cochains: 
$\H^{q}(\mc{K}) \mapsto \C^{q}(\bar{\mc{K}}) \setminus  \Z^{q}(\bar{\mc{K}})$.
\begin{align*}
  \ring{\bm{p}}_{k} = {\bm{p}}(\ring{\bm{w}}_{0}, \ring{\bm{w}}_{k}) \in \H^{q}(\mc{K}) \!: 
   \cobd_{q + 1} \ring{\bm{p}}_{k} = 0
  && \mapsto
  && \bar{\cobd}_{q + 1} \ring{\bm{p}}_{k} = \nu_k - \nu_0 
  && \implies \ring{\bm{p}}_{k}  \notin \Z^{q}(\bar{\mc{K}})
\end{align*}

So, if $\bar{\cobd}_{q + 1} \bm{p} = 0$ then  $\bm{p} \in \Z^{q}(\bar{\mc{K}})$, and hence $\bm{p} \notin \H^{q}({\mc{K}})$.
But $\bm{p}$ also has empty coboundary on $\mc{K}$: ${\cobd}_{q + 1} \bm{p} = 0$,
which means it is a $q$-cocycle. 
Thus,  $\bm{p}$ must be a $q$-coboundary $\bm{p} \in \Z^{q}(\mc{K}) \setminus \H^{q}(\mc{K}) = \B^{q}(\mc{K})$.

Hence, unlike homology, the acyclization preserves \emph{only} the group of $q$-coboundaries $\B^{q}(\mc{K}) = \B^{q}(\bar{\mc{K}})$,

\end{proof}

The summary of co-homology on $\bar{\mc{K}}$:
\begin{align*}
  && \B_{q}(\mc{K}) \subset \B_{q}(\bar{K})
  &&& \H_{q}(\mc{K}) \subset \B_{q}(\bar{K})
  && \Z_{q}(\mc{K}) = \Z_{q}(\bar{K}) = \B_{q}(\bar{K})
  \\
  && \B^{q}(\mc{K}) = \B^{q}(\bar{K})
  && &\H^{q}(\mc{K}) \subset \C^{q}(\bar{K}) \setminus \Z^{q}(\bar{\mc{K}})
  && \Z^{q}(\mc{K}) \neq \Z^{q}(\bar{K}) = \B^{q}(\bar{K})
\end{align*}

\subsection{Dual graph and dual complex}

The dual graph of $\mc{K}$ is defined in the following manner. 
The \embf{dual graph} $\bar{\mc{K}}^{*} = (\bar{\mc{F}}^{*}, {\mc{E}}^{*})$ 
has a node for each $(q + 1)$-simplex and each void, including the outer void,
and a dual edge for each $q$-simplex.
A dual edge $e^{*} = (\eta^{*}, \tau^{*}) = \eta^{*} \tau^{*}$ is added between nodes $\eta^{*}$ and $\tau^{*}$ if and only if 
the embedding of the primal $q$-simplex $e \in \mc{K}_{q}$ 
is given by the intersection of the two corresponding $(q + 1)$-cells in 
${\mc{K}} \cup \mc{V}$:  $\support{e} = \support{\eta} \cap \support{\tau}$.
Therefore, $\bar{\mc{F}}^{*} = \mc{F}^{*} \cup \mc{V}^{*}  \cong \mc{K}_{q + 1} \cup \mc{V}$
and $\mc{E}^{*} \cong \mc{K}_{q}$. 
One of our core assumptions is  that the underlying complex is orientable, 
which means that $\bar{\mc{K}}^{*}$ is directed and the dual 1-skeleton ${\mc{E}}^{*}$ consists of \emph{directed arcs} (di-arcs). 
For each oriented $q$-simplex 
$e = \eta \cap \tau  \in \mc{K}_{q}$ there are two alternative dual di-arcs to choose as its dual:
$(\eta^{*} \to \tau^{*})$ and $(\tau^{*} \to \eta^{*})$.

\begin{figure}[!t]
 \centering
 \includegraphics[width=0.3\linewidth]{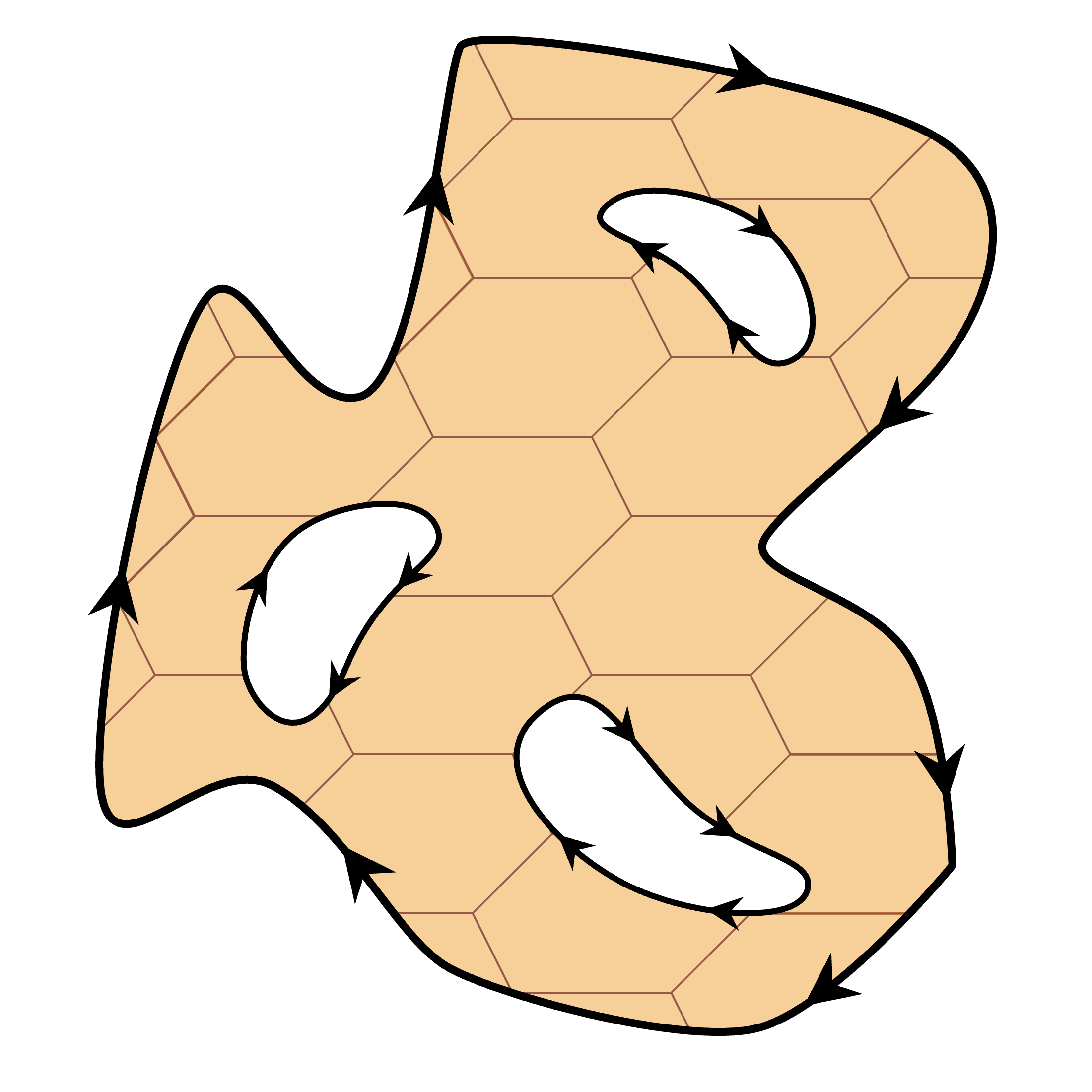}
 \includegraphics[width=0.3\linewidth]{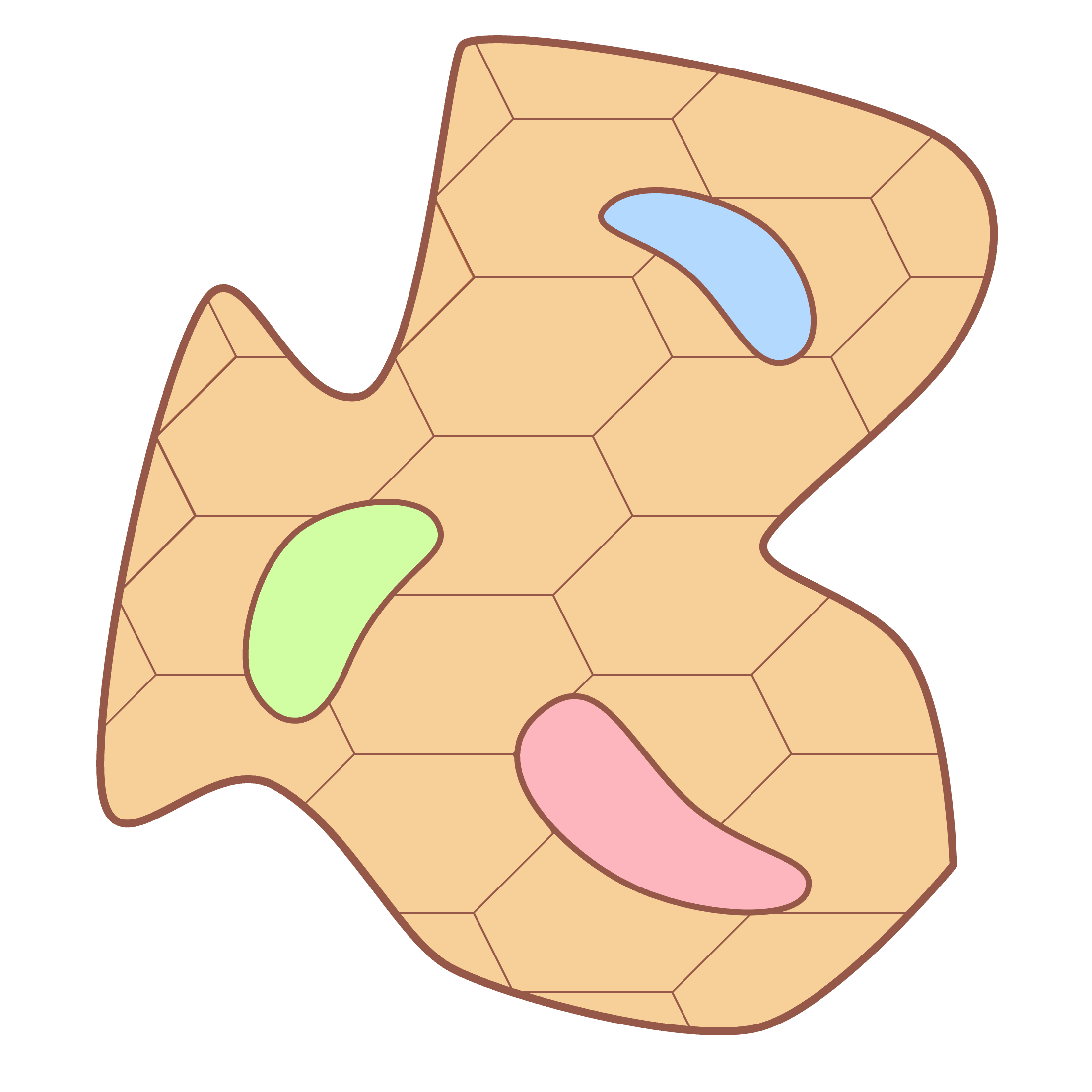}
 \includegraphics[width=0.3\linewidth]{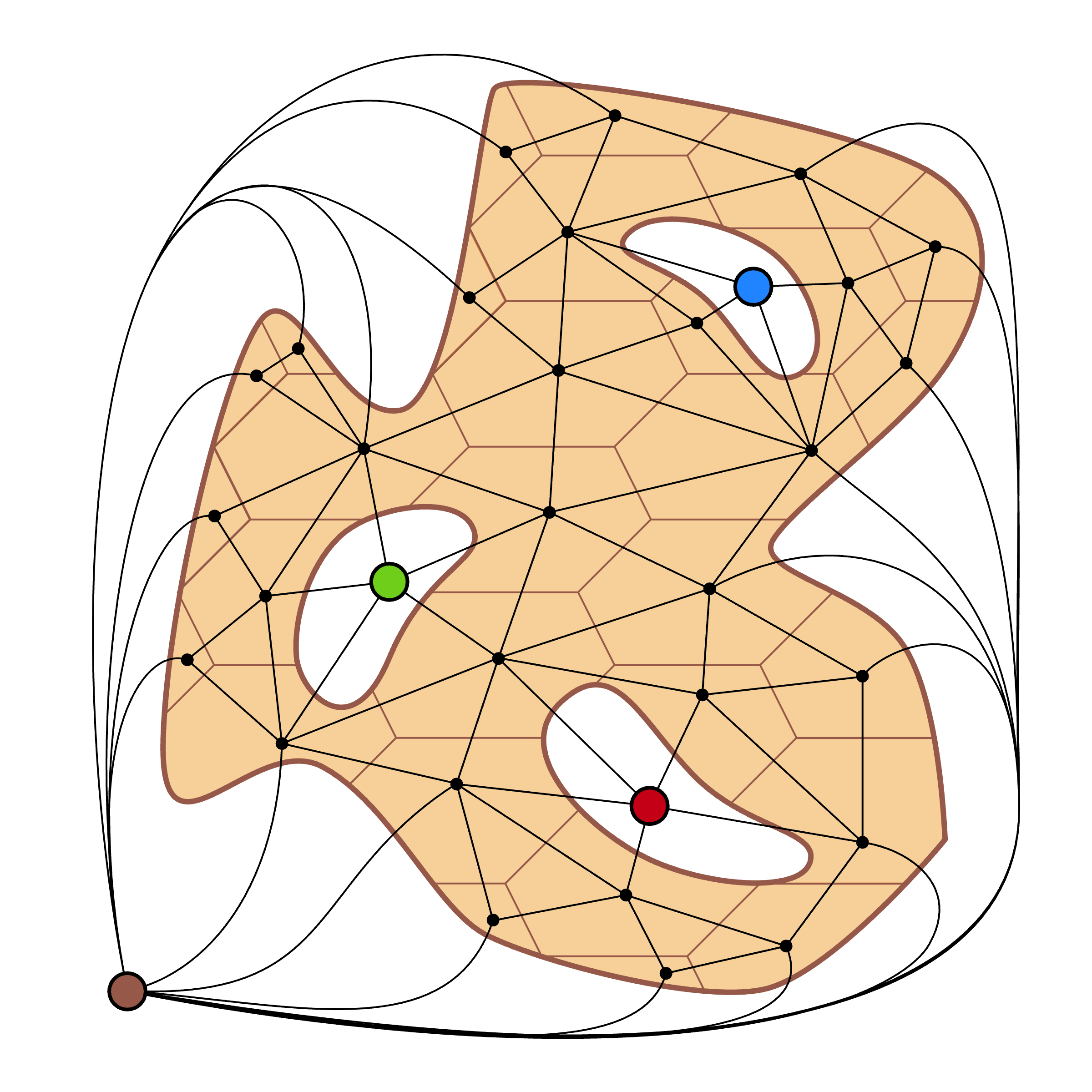}
 
 \caption[Example of a $(q + 1)$-complex and its dual graph.]{
     \textbf{Left:} Example of a $(q + 1)$-complex $\mc{K}$  with three inner voids (referred to as blue, green, and red).
     \textbf{Middle:} Its acyclization $\bar{\mc{K}}$ with inner voids plugged by corresponding $(q + 1)$-cells. The cell corresponding to the outer void $\nu_0$ occupies the whole space outside of the complex.
     \textbf{Middle:} The dual graph  $\mc{K}^{*}$ with a node for every $(q + 1)$-cell of $\bar{\mc{K}}$ (the brown node corresponds to $\nu_0$). 
     Every undirected edge in this figure represents a pair of opposite dual arcs, see Fig.~\ref{fig:ortanddual-clockwise-02} for details.
 }
 \label{fig:complex-and-dual-graph}
\end{figure}

The natural dual di-arc of $e = (\eta \dart \tau) \in \mc{K}_{q}$ is  given as $e^{*} = (\eta^{*} \to \tau^{*}) \in \mc{E}^{*}$.
In other words, we imply a duality between incoming arcs of a node $\tau^{*}$ and positive $q$-faces of its primal $(q+1)$-cell $\tau$,
or more broadly between \emph{inward direction} and \emph{clockwise orientation} on $\mc{K}$.
The boundary $q$-simplices 
$(\nu_0 \dart \tau)$ and $(\eta \dart \nu_k)$
are dualized as tail and head arrows respectively:
$ (\nu^{*}_0 \to \sbullet)$ and $e = (\sbullet \to \nu^{*}_k)$,

\paragraph*{Dual complex $\mc{K}^{*}$}
As was mentioned above, the set of voids defined by the embedding of the $q$-skeleton, $ \left( \real^{q + 1}\setminus {\mc{K}_{q}} \right)$, 
contains all the $(q + 1)$-cells and voids defined by $\support{\mc{K}}$, 
in fact  $\abs{ \left( \real^{q + 1}\setminus {\mc{K}_{q}} \right) } = \abs{\mc{K}_{q}} + \abs{\mc{V}}$. 
This implies that the dual graph constructed from $\support{\mc{K}}$ is equivalent to 
the dual graph constructed from the embedding of the $q$-skeleton $\support{\mc{K}_{d}}$.

The dual graph $\bar{\mc{K}}^{*} = (\bar{\mc{F}}^{*}, {\mc{E}}^{*})$, as defined above,  is a $1$-skeleton subcomplex of a dual complex $\bar{\mc{K}}^{*}$.
$\bar{\mc{K}}^{*}$ is given by the isomorphism of chain groups $\mc{C}_{d}(\bar{\mc{K}}^{*}) \cong \mc{C}_{q - d + 1}(\bar{\mc{K}})$,
that are linked by dual homomorphisms $\cobd^{*}_{d}$ and $\bd^{*}_{d}$ that commute with primal boundary and coboundary  operators $\bd_{q - d + 2}$ and $\cobd_{q - d + 2}$.  
The sole $(q + 1)$-cycle  that generates $\mc{H}_{q + 1}(\bar{\mc{K}})$ is dual to the single connected component in $\mc{H}_{0}(\bar{\mc{K}}^{*})$,
and $\rank \mc{H}_{q}(\bar{\mc{K}}) = 0$ implies $\rank \mc{H}_{1}(\bar{\mc{K}}^{*}) = 0$.


 The dual complex $\mc{K}^{*}$ of the original complex $\mc{K}$, 
 that preserves the structure of the $q$-th cycle group under the duality isomorphism, can be defined algebraically. 
 While the  duality 
  between $(q + 1)$-simplices and dual nodes remains unchanged, $\mc{C}_{q+1}(\mc{K}_{q}) \cong \mc{C}_{0}(\mc{K}^{*})$
  adjustments are needed to accommodate boundary $q$-simplices. 
  The duality between primal boundaries and dual coboundaries also has not changed: $\im \cobd^{*}_{1} \cong \im \bd_{q + 1}$.

We introduce a new duality 
 between $q$-chains and dual 1-chains by establishing the duality between primal coboundaries and dual boundaries:
  $\im \bd^{*}_{1} \cong \im \cobd_{q + 1}$.
  This means that a boundary $q$-simplex $\sigma$ with only one coface on $\mc{K}$, 
  $\support{\sigma} \subset \support{\bd \mc{K}} \iff \sigma = \eta \cap \tau \implies \eta = \tau$,
  has a dual $\sigma^{*}$ that can found on the coboundary of only one  node,
   the one that is  dual to its sole coface $\tau^{*}$.
  We call such elements \embf{arrows}.
  The dual boundary of an arrow $\sigma^{*} \in \mc{K}^{*}_{1}$  contains only element:
   $\bd^{*}_{1} \sigma^{*} = \tau^{*}$  or  $\bd^{*}_{1} \sigma^{*} = -\tau^{*}$, 
   depending on the orientation of $\sigma$ and $\tau$.
  The dual boundary of an arrow $\sigma^{*}$ contains a single element: 
  $\bd^{*}_{1} \sigma^{*} = \tau^{*}$ or $\bd^{*}_{1} \sigma^{*} = -\tau^{*}$, 
  depending on the orientation of $\sigma$ and $\tau$. 
  In the former case, $\sigma^{*}$ we referred to as the \embf{arrow-head}, 
  while in the latter case, we call it the \embf{arrow-tail}.
  

Of course, since the construction of dual complex  of $\mc{K}$ is purely algebraic,
the 1-skeleton of $\mc{K}^{*}$ is not a graph. 
Hence we use the 1-skeleton of $\bar{\mc{K}}^{*}$ as the dual graph of $\mc{K}$.

\subsection{Coflows and augmenting copaths}


Let $\dist^{*}_{\bm{z}} = \dist^{*}({\bm{z}}; \nu^{*}_{0}) \in \mc{C}^{0}(\bar{\mc{K}}^{*}; \integer) \leftarrow 
    \mathtt{ShortestPathDist} \big( \mathtt{From} = \nu^{*}_{0}, \mathtt{To} = \bar{\mc{K}}_{0}^{*}, \mathtt{Length} = \bm{z}^{*} \big)$
be a dual $0$-cochain on the nodes of the dual graph $\bar{\mc{K}}^{*}$ 
whose values are given by the shortest path distances from $\nu_{0}^{*}$, -- the node dual to the outer void,
to all other nodes. 
Note that because $\bm{z}$ is a bounding $q$-cycle on $\bar{\mc{K}}$, for any $\tau^{*} \in \bar{\mc{F}}^{*}$ the values $\dist^{*}_{\bm{z}}(\tau^{*})$ do not depend on the actual path taken from $\nu^{*}_{0}$ to $\tau^{*}$ -- any path will yield same distance. 
For any two nodes $\eta^{*}, \tau^{*} \in \bar{\mc{F}}^{*}$,
the dual distance cochain satisfies $\dist^{*}_{\bm{z}}(\tau^{*}) - \dist^{*}_{\bm{z}}(\eta^{*}) = \bm{z}^{*}(\eta^{*} \to \tau^{*})$, which implies that the primal $(q + 1)$-chain isomorphic to $\dist^{*}_{\bm{z}}$
is bounded by $\bm{z}$ on $\bar{\mc{K}}$, i.e. $\bd_{q + 1} \dist_{\bm{z}} = \bm{z}$ where  $\dist_{\bm{z}} \in \mc{C}_{q + 1}(\bar{\mc{K}}; \integer)$.
The positive values of $\dist_{\bm{z}}$ are assigned to the facets of $\bar{\mc{K}}$ around which $\bm{z}$ circles in the \emph{clockwise} direction, negative values to those that are circled in the \emph{counter-clockwise} direction, and zero values to the facets that are ``outside'' of $\bm{z}$. 
The magnitude of $\dist_{\bm{z}}$ tells how many times a facet is encircled by $\bm{z}$. 

The LPs on $\bar{\mc{K}}$ dual to shortest path LPs on the dual graph:
\begin{align}
\label{eq:barK:dist-lp}
      \begin{array}{llll}\dsp
        \max\limits_{\bm{d}}\ 
        &\dsp \sum_{\mc{V}_{I}} \abs{\bm{d}(\nu_t) - \bm{d}(\nu_0)}         \geq 0 
        \\
        \subto&\dsp  \bm{d}(\tau) \leq \bm{z}(\eta \dart \tau) + \bm{d}(\eta) &\forall e_i \in \bar{\mc{K}}_{q}
        \\
                    &\dsp \bm{d}(\nu_0) = 0
        \\
              &\dsp \bm{d} \in \C^{q + 1}(\bar{\mc{K}})
      \end{array}
      &&\overset{dual}{\Longrightarrow}&&
      \begin{array}{lllll}\dsp
            \min\limits_{\bm{p}}\ &\dsp \abs{{\bm{p}}({\bm{z}})}   \geq 0       \\
            \subto\ &  \bar{\cobd}_{q + 1} \bm{p} = \sum_{\mc{V}_{I}} \nu_t  - \beta_{q} \nu_0 
            \\
                    &\dsp  \bm{p} \in \C^{q}(\nu_0 \to \mc{V}_{I})
        \end{array}
\end{align}
where $\dist_{\bm{z}} \asgn \bm{d}$ such that  $\bar{\bd}_{q + 1} \bm{d} = \bm{z}$ is the max-bounding chain.
This result corresponds to the maximum $\gamma$-flow from \cite{generalizedMaxflow2021}.
Note that $\bm{p} \in \H^{q}(\mc{K})$.

When $\dist_{\bm{z}}$ is restricted to the void-facets $\dist_{\bm{z}}$ defines the homology signature of the input $q$-cycle: 
$[\bm{z}] = [\zeta_0, \zeta_1, \ldots, \zeta_{\beta_{q}}] 
          = [\dist_{\bm{z}}(\nu_0), \dist_{\bm{z}}(\nu_1), \ldots, \dist_{\bm{z}}(\nu_{\beta_q})] \in \integer^{{\beta}_q + 1} $ 
such that $\bm{z}$ is homologous to the corresponding linear combination of void boundaries:
$\bm{z}~\sim~\sum_{k = 0}^{\beta_q} \zeta_k\cdot\bd_{q + 1} \nu_k$.
Note that if $\beta_q \geq 1$ then $\zeta_0 = \dist_{\bm{z}}(\nu_0) = 0$ for any $\bm{z}$.

The cost of sending $\gamma$ units of flow along a copath $\bm{p} = \bm{p}(\eta, \tau)$ is given by the value of $\dist_{{z}}$ on its endpoints:
\begin{align}
  \cost(\bm{p}; \gamma) = \gamma \cdot \bm{p}(\bm{z}) 
                        = \gamma \cdot  \dist\nolimits_{\bm{z}}(\cobd_{q + 1} \bm{p})
                        = \gamma \cdot  (\dist\nolimits_{\bm{z}}(\tau) - \dist\nolimits_{\bm{z}}(\eta))
\end{align}

\begin{figure}
  \centering
  \includegraphics[width=0.45\linewidth]{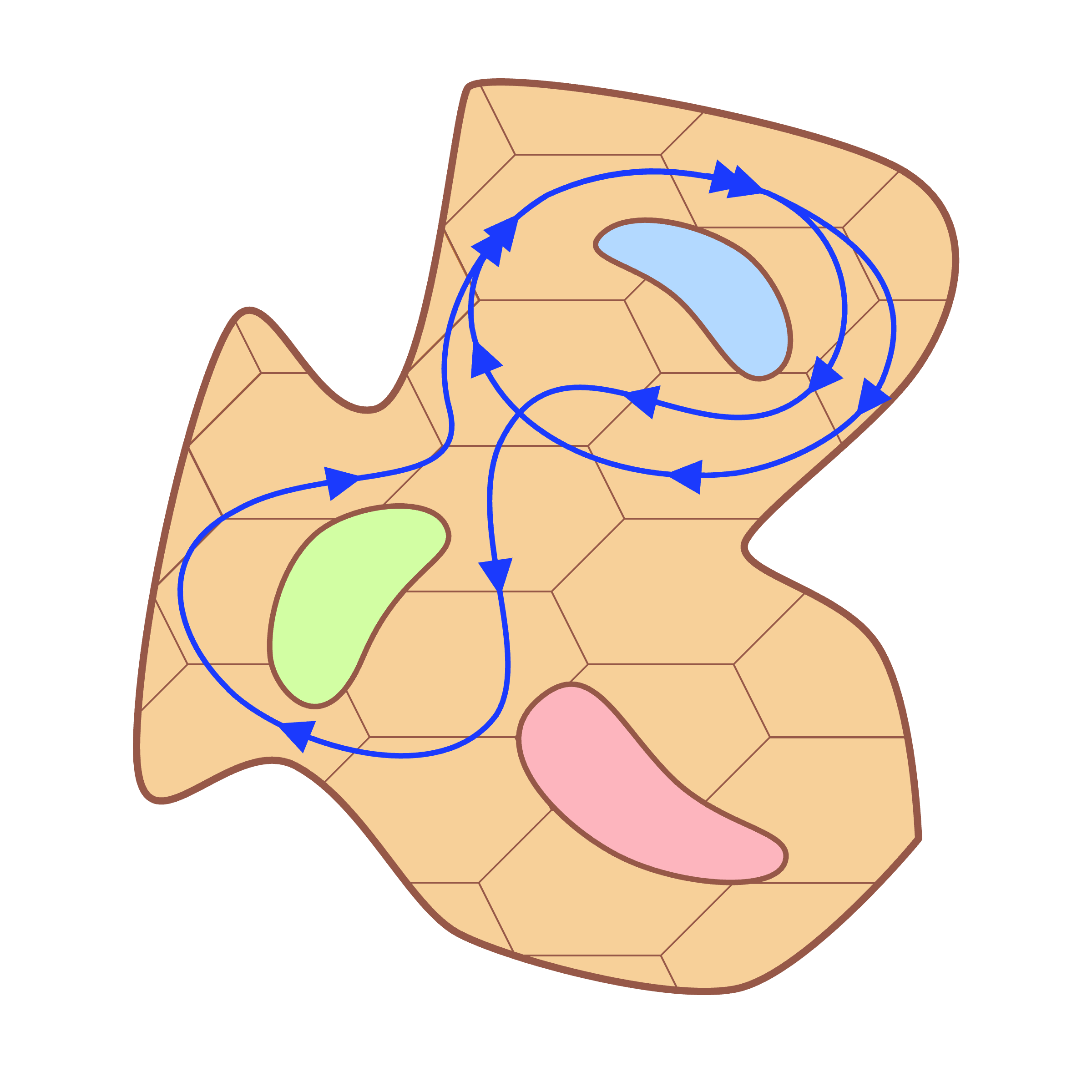}   \  
  \includegraphics[width=0.45\linewidth]{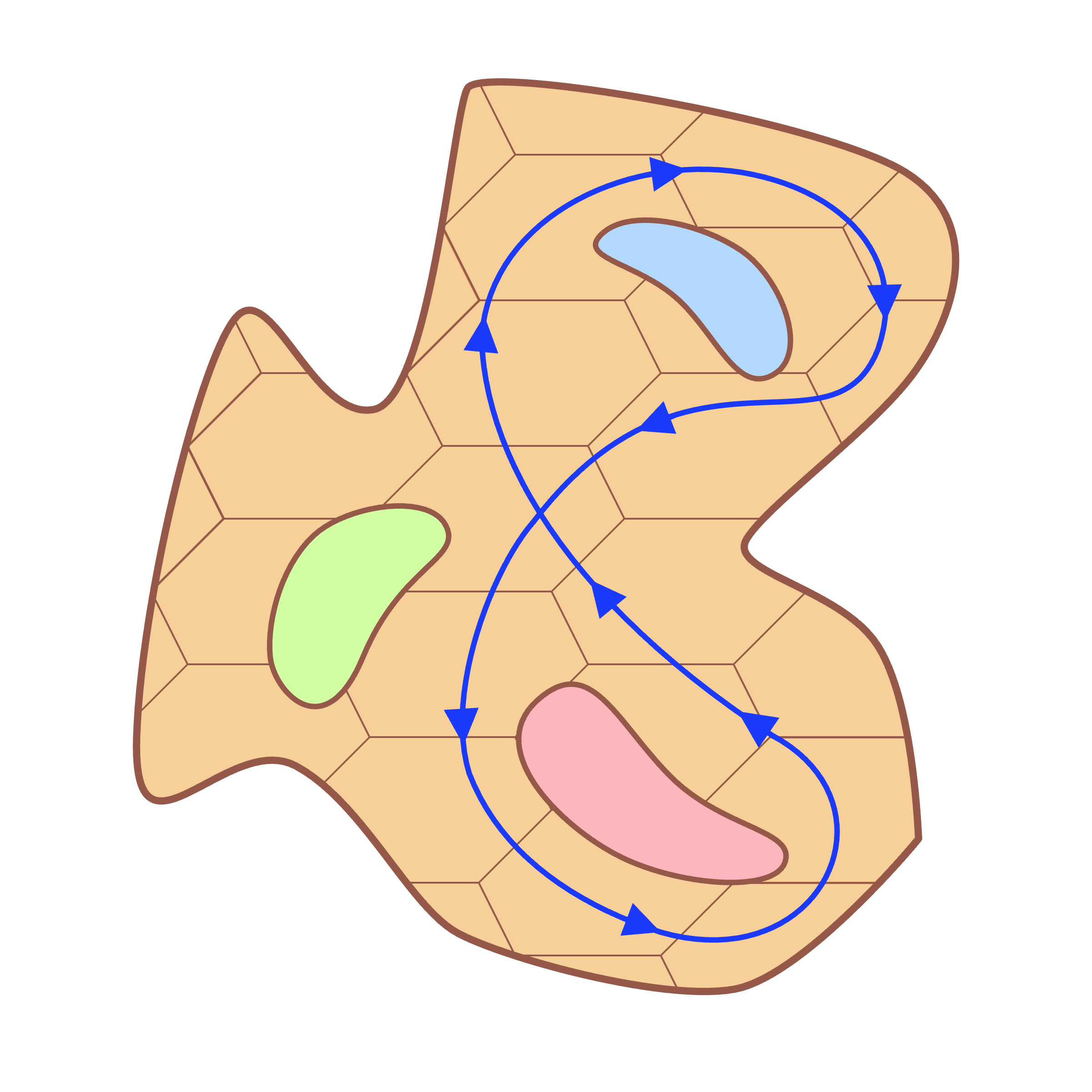}   
  \caption[Two types of $q$-cycles.]{
        The two kinds of $q$-cycles $\bm{z}$: 
        those that curl in a single direction (\textbf{Left}), and those that curl in  both directions  -- \textit{mixed} curling (\textbf{Right}).
        Double arrows mean that the corresponding $q$-edges appear with coefficient $2$ in the input $q$-cycle.
    }
  \label{fig:types-of-cycles}
\end{figure}

\subsection{Recyclization of \texorpdfstring{$\bar{\mc{K}}$}{K-bar}}

The key observation, that is fundamental for our approach to solving OHPs presented in Section~\ref{subsec:def:opt-homology},
is that in order to optimize the flux of coflow through $\bm{z}$ it is sufficient that the $q$-coflow $\bm{f}$ is non-zero only on the non-zero homology classes of $\bm{z}$.
Let $\mc{S} \subset \mc{V} = \{ \nu_s \mid \dist_{\bm{z}}(\nu_s) > 0 \}$,
 $\mc{Q} \subset \mc{V} = \{ \nu_k \mid \dist_{\bm{z}}(\nu_k) = 0 \}$,
 and 
 $\mc{T} \subset \mc{V} = \{ \nu_t \mid \dist_{\bm{z}}(\nu_k) < 0 \}$.
 Note that $\mc{Q}$ is non-empty, since it always contains  $\nu_0$. 
 There are three possibilities:
 \begin{align}
   \mc{S} = \varnothing \text{ and } \mc{T} \neq \varnothing &&
   \mc{S} \neq \varnothing \text{ and } \mc{T} = \varnothing &&
   \mc{S} \neq \varnothing \text{ and } \mc{T} \neq \varnothing 
   \label{eq:recyclization-01}
 \\
   \bm{z} \text{ is curling in CCW}
   &&
   \bm{z} \text{ is curling in CW}
   &&
   \bm{z} \text{ has CW and CCW parts}
   \notag
 \\
  \mc{S} \asgn \mc{Q} &&
  \mc{T} \asgn \mc{Q} &&
  \label{eq:recyclization-02}
 \end{align}

There are 2 choices for the direction of a coflow $\bm{f}$:
from $\mc{S}$ to $\mc{T}$ -- $\bm{f} \in \C^{q}(\mc{S} \to \mc{T})$, 
and from $\mc{T}$ to $\mc{S}$ -- $\bm{f} \in \C^{q}(\mc{T} \to \mc{S})$. 
Let $\bar{\bm{p}}_{s t} = \bar{\bm{p}}(\nu_s, \nu_t) \in \mc{H}^{q}(\mc{K})$ be a { $q$-copath} between a pair of voids $\nu_s \in \mc{S}$ and $\nu_t \in \mc{T}$
 since $\dist_{\bm{z}}(\nu_s) \geq \dist_{\bm{z}}(\nu_t)$, the cost of sending a unit flow is negative:
 $\bar{\bm{p}}_{s t}(\bm{z}) = \dist\nolimits_{\bm{z}}(\nu_t) - \dist\nolimits_{\bm{z}}(\nu_s) \leq 0$. 
The cost of sending a unit of flow in the opposite direction along a $q$-copath $\bar{\bm{p}}_{t s} = \bar{\bm{p}}(\nu_t, \nu_s) \in \mc{H}^{q}(\mc{K})$ 
is, respectively, positive:
 $\bar{\bm{p}}_{t s}(\bm{z}) = \dist\nolimits_{\bm{z}}(\nu_s) - \dist\nolimits_{\bm{z}}(\nu_t) \geq 0$. 
 Note that the min-cost flow along $\bar{\bm{p}}_{t s}$ is same as in the $\dist$-LP in Eq.~\eqref{eq:barK:dist-lp},
 and the minimizing flow cost along $\bar{\bm{p}}_{s t}$'s is the equivalent to maximizing the flux through $\bm{z}$.

\begin{figure}
  \centering
  \includegraphics[width=0.45\linewidth]{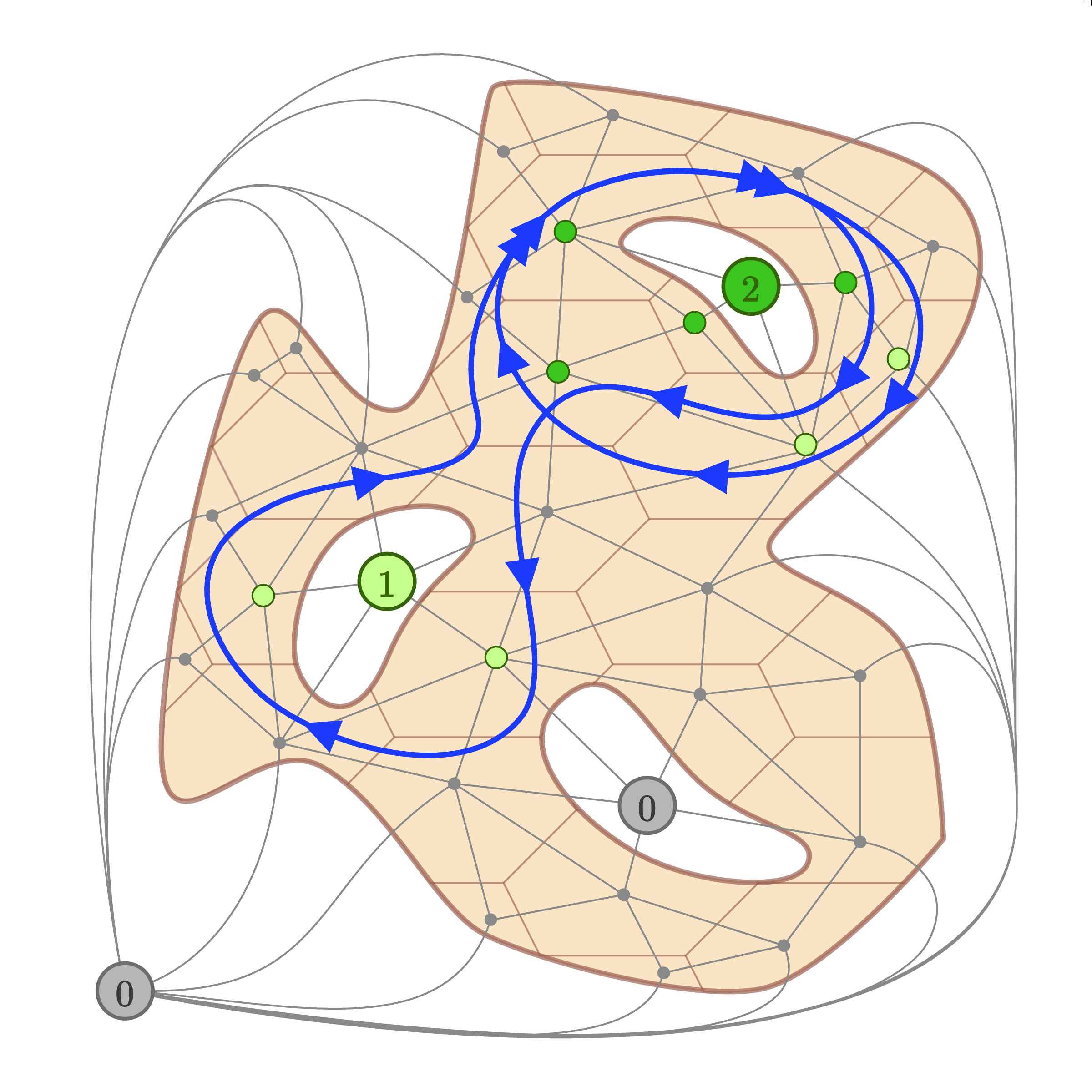} \
  \includegraphics[width=0.45\linewidth]{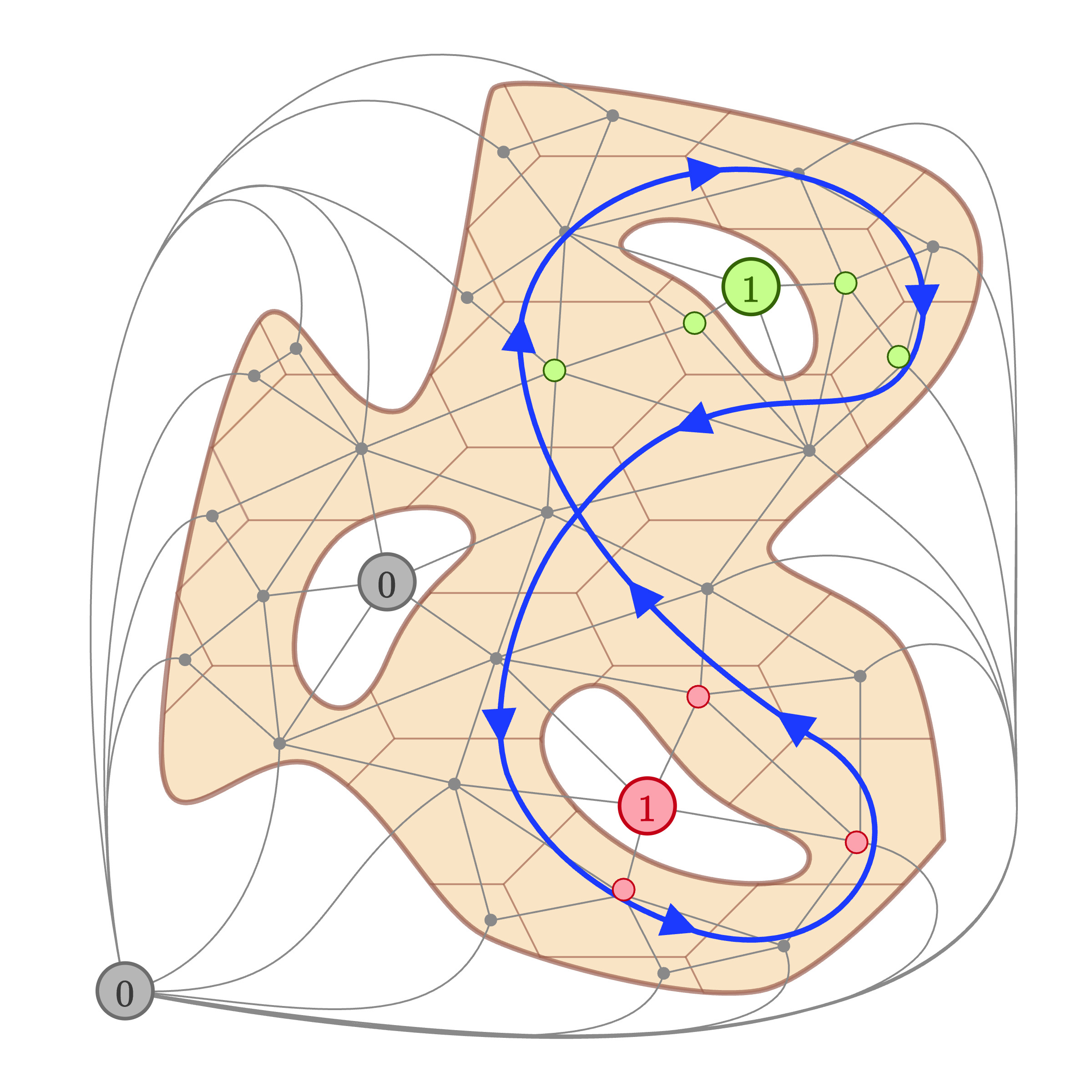}
  \caption[Shortest path distances $\dist_{\bm{z}}$.]{
        The dual cochain $\dist^{*}_{\bm{z}}$ that is constructed for the two examples from Fig.~\ref{fig:types-of-cycles}.
        The node's value is defined by 
        its shortest path distance from $\nu_0^{*}$, with the length of dual arcs given by the value of the input $q$-cycle $\bm{z}$.
        The node's color represents the sign of $\dist^{*}_{\bm{z}}$, and hence the direction of $\bm{z}$ curling around the corresponding $(q + 1)$-cell: green is positive (CW curl), red is negative (CCW curl), 
        and grey is for nodes that are outside of $\bm{z}$, i.e. $\dist_{\bm{z}}$ is zero on such nodes.
    }
  \label{fig:sh-path-dist}
\end{figure}


Let $\bm{f} \in \mc{H}^{q}(\mc{K})$ be some feasible $q$-coflow, including the zero coflow $\bm{f}_0\!: \mc{K}_{q} \mapsto 0$, such that $\bm{f}(\bm{z}) \leq 0$.
We say that an augmenting $q$-copath $\bar{\bm{p}}\in \mc{H}^{q}(\mc{K})$ \embf{follows} coflow $\bm{f}$ 
if all $q$-edges of $\bar{\bm{p}}$ have same orientation on $\bm{f}$ or are 0:
$\bar{\bm{p}}(e_i) \cdot \bm{f}(e_i) \geq 0$.
We \embf{augment} $q$-coflow  $\bm{f}$ for $\gamma > 0$ units of flow \embf{along} $\bm{p}$
as follows: $\bm{f} \asgn \bm{f} + \gamma \bar{\bm{p}}$.

We define the \embf{augmenting capacity} of $\bm{p}$ as follows:
\begin{align}
  c_{f}(\bar{\bm{p}}) = r_{f}(\bar{\bm{p}}) 
  = \min\limits_{e_i \!: \bar{\bm{p}}(e_i) \neq 0} \big\{ c(e_i) - \bar{\bm{p}}(e_i) \cdot \bm{f}(e_i) \big\} \geq 0
\end{align}

\begin{algorithm}[!ht]
    
    \SetKwData{From}{from} \SetKwData{To}{to} \SetKwData{Length}{arc length}
    \SetKwFunction{Capacity}{capacity} \SetKwFunction{Cost}{cost}
    \SetKwFunction{ShortestPathDist}{ShortestPathDist}
    \SetKwInOut{Input}{input}\SetKwInOut{Output}{output}
    
    \Input{ 
            $\bar{\mc{K}}^{*} = (\mc{F}^{*} \cup \mc{V}^{*}, \vec{\mc{E}}^{*} = \vec{\mc{E}}^{*}_{+} \cup \vec{\mc{E}}^{*}_{-} )$, 
            $\bm{z}^{*} \in \mc{Z}^{1}(\mc{K}^{*}; \integer) $,
            $\bm{c} \in \integer^{E}$, $\bm{a} \in \integer^{F}$, $\lambda \in \integerplus$ 
    }
    \Output{ 
            $\hat{\mc{K}}^{*}_{\vvec{\textsc{st}}} 
                  = (\hat{\mc{F}}^{*}_{\vvec{\textsc{st}}}, \hat{\mc{E}}^{*}_{\vvec{\textsc{st}}})$ -- flow network for OHCP,
    }
    \BlankLine
    
    \tcp*[h]{DIST LABELS of VOIDS $\equiv$ HOMOLOGY CLASS of $\bm{z}$}
    
    $\dist^{*}_{\bm{z}} \in \mc{C}^{0}(\mc{K}^{*}; \integer) \leftarrow$ 
    \ShortestPathDist {\From = $\nu^{*}_{0}$, \To = $\bar{\mc{F}}^{*}$, \Length = $\bm{z}^{*}$}\;
    \BlankLine

    \tcp*[h]{VOIDS = SOURCES + TRANSITS + SINKS}
    
    $\mc{S} \leftarrow \big\{ \nu^{*}_{s} \in \mc{V}^{*} \mid \dist^{*}_{\bm{z}}(\nu^{*}_{s}) > 0   \big\}$,  
    $\mc{T} \leftarrow \big\{ \nu^{*}_{t} \in \mc{V}^{*} \mid \dist^{*}_{\bm{z}}(\nu^{*}_{t}) < 0   \big\}$,
    $\mc{Q} \leftarrow \emptyset$\;
    \lIf { $\mc{S} \neq \emptyset$ and $\mc{T} = \emptyset$}{
            $\mc{T} \leftarrow \big\{ \nu^{*}_{t} \in \mc{V}^{*} \mid \dist^{*}_{\bm{z}}(\nu^{*}_{t}) = 0 \big\}$ 
    }
    \lElseIf{ $\mc{S} = \emptyset$ and $\mc{T} \neq \emptyset$}{
            $\mc{S} \leftarrow \big\{ \nu^{*}_{s} \in \mc{V}^{*} \mid \dist^{*}_{\bm{z}}(\nu^{*}_{s}) = 0 \big\}$ 
    }
    \lElse{
            $\mc{Q} \leftarrow \big\{ \nu^{*}_{k} \in \mc{V}^{*} \mid \dist^{*}_{\bm{z}}(\nu^{*}_{k}) = 0   \big\}$
    }
    \BlankLine

    \tcp*[h]{CONSTRUCT FLOW NETWORK}

    $\hat{\mc{F}}^{*}_{\vvec{\textsc{fn}}} \leftarrow \mc{F}^{*} \cup \mc{V}^{*} \cup \{ S^{*}, T^{*}, U^{*}, D^{*} \}$  
    \tcp*[f]{{\small facets*, voids*, \{source, sink, up, down\}}}
    
    $\vec{\mc{E}}^{*}_{\textsc{st}} \leftarrow 
            \vec{\mc{E}}^{*}
            \cup \{ (S^{*} \to \nu^{*}_k) \}_{\mc{S} \cup \mc{Q}}
            \cup \{ ( \nu^{*}_k \to T^{*}) \}_{\mc{T} \cup \mc{Q}}
            \cup \{ ( T^{*} \to S^{*}) \}
    $
    \tcp*[f]{OHCP network}



    \lForEach{$\vec{e}^{*}_{+i} \in \vec{\mc{E}}^{*}_{+}$}{
            \Cost {$\vec{e}^{*}_{+i}$} $\leftarrow  \bm{z}^{*}(\vec{e}^{*}_{+i}) \equiv \phantom{-} \bm{z}({e}_{i})$,
            \Capacity {$\vec{e}^{*}_{+i}$} $\leftarrow  c_i$,  
    }
    
    \lForEach{$\vec{e}^{*}_{-i} \in \vec{\mc{E}}^{*}_{-}$}{
            \Cost {$\vec{e}^{*}_{-i}$} $\leftarrow  \bm{z}^{*}(\vec{e}^{*}_{-i}) \equiv -\bm{z}({e}_{i})$,
            \Capacity {$\vec{e}^{*}_{-i}$} $\leftarrow  c_i$,  
    }

    \lForEach{$\vec{e}^{*} \in \{ \cobd^{*}_{1} S^{*} \} \cup \{ \cobd^{*}_{1} T^{*} \} $}{
            \Cost {$\vec{e}^{*}$} $\leftarrow  0$,
            \Capacity {$\vec{e}^{*}$} $\leftarrow  \infty$,  
    }

     \label{algo:ohcp:flow-network}
     \caption{Flow Network Setup.}
\end{algorithm}

\begin{figure}[hb!]
  \centering
  \includegraphics[width=0.45\linewidth]{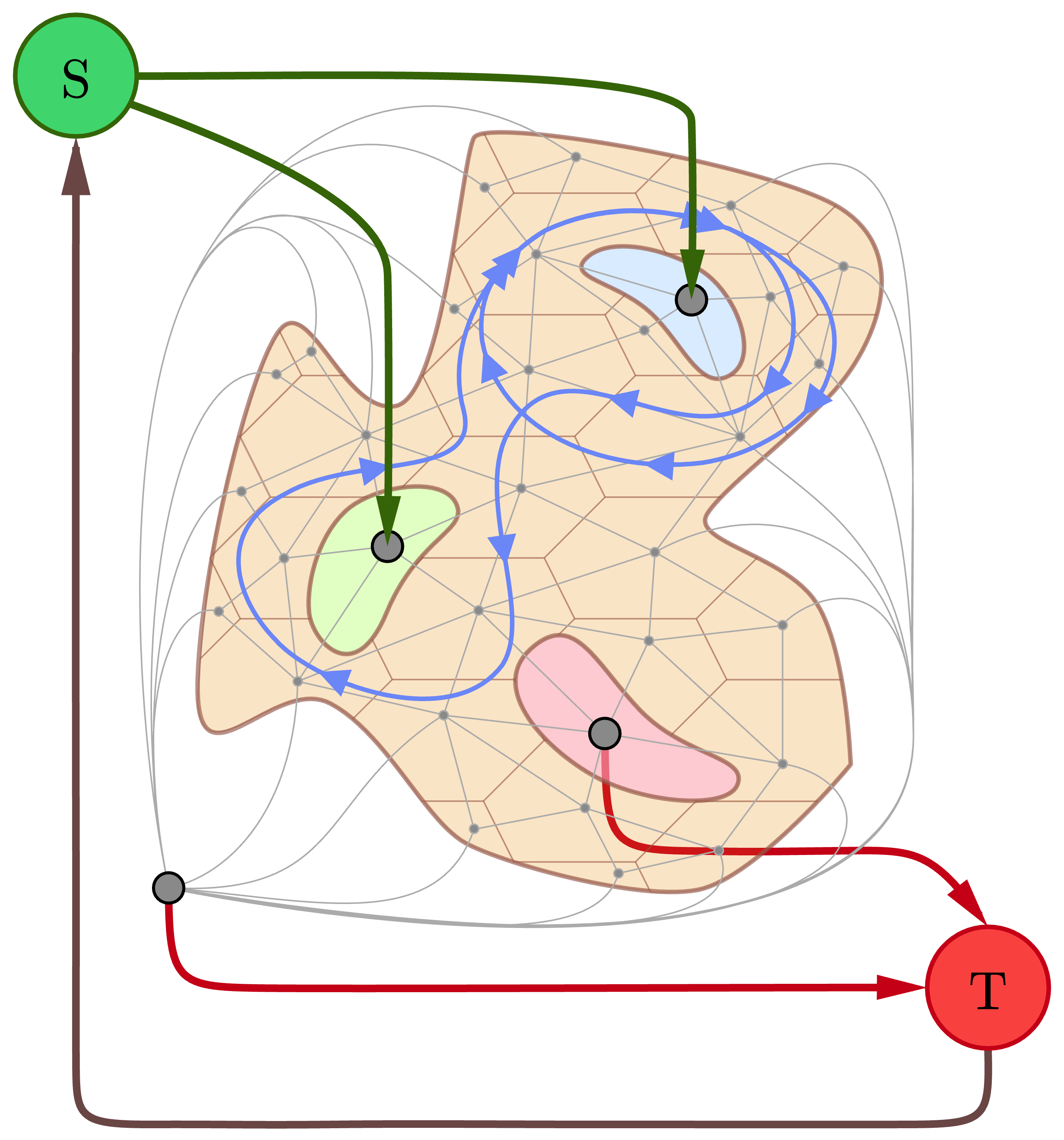}   \  
  \includegraphics[width=0.45\linewidth]{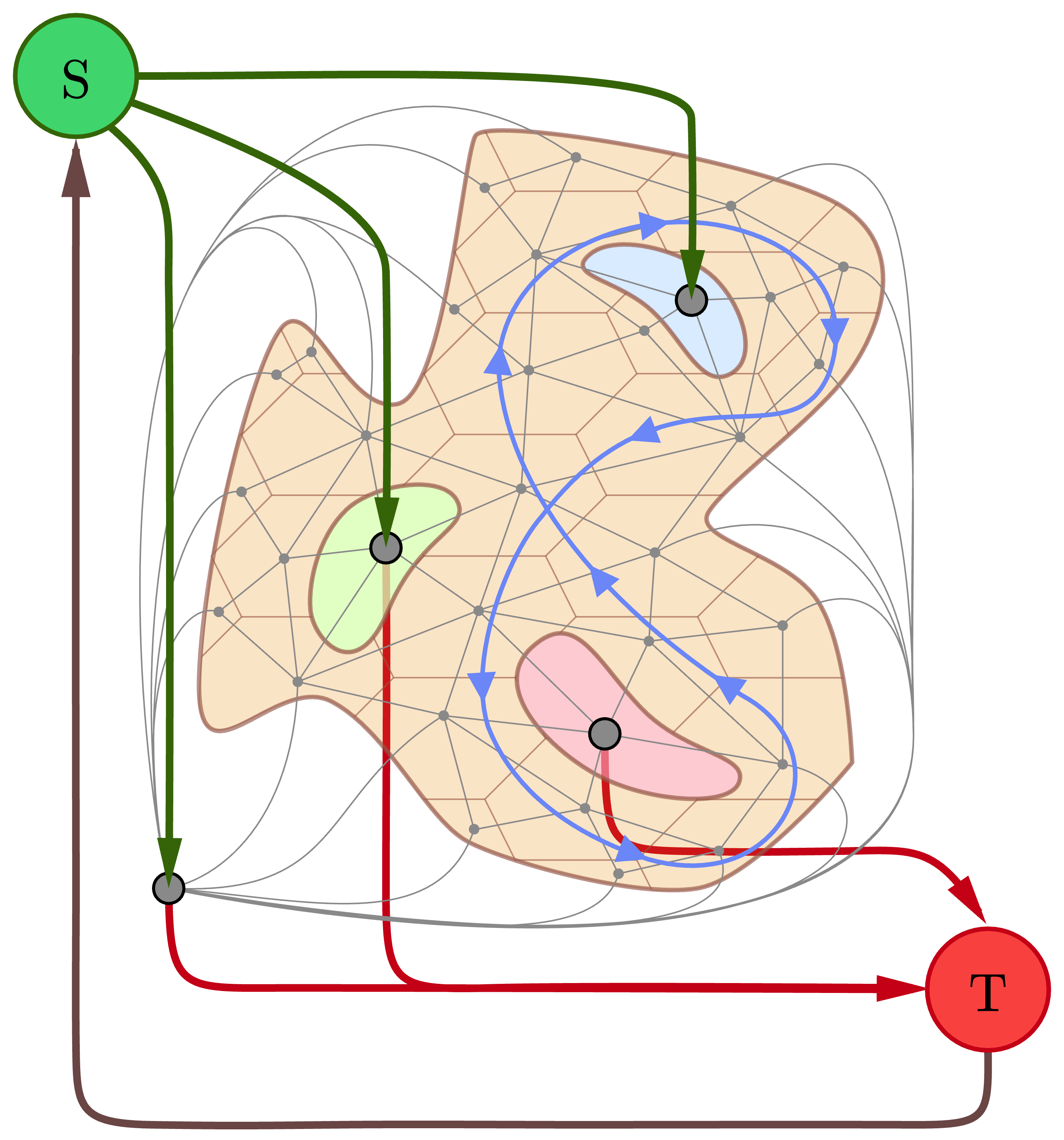}   
  \caption[Flow networks for OHCP.]{
        %
        The dual flow networks produced by Algorithm~\ref{algo:ohcp:flow-network} for the two examples in Fig.~\ref{fig:types-of-cycles}.
        \textbf{Left}: In the case of a  $q$-cycle curling in a single direction, as a consequence of the partition in Eq.~\eqref{eq:recyclization-01} and \eqref{eq:recyclization-02},
        the void-nodes are connected either to the source node $S$, or to the sink node $T$, but never to both. 
        \textbf{Right}:
        On the other hand, the homology class of a mixed curling $q$-cycle always contains generators of both CW and CCW directions that, consequently, are connected to the $S$ and $T$ respectively. 
        The voids that are ``outside'' of $\bm{z}$, i.e. those with $\dist_{\bm{z}} = 0$,
        become transit nodes -- they are connected to both the source and sink nodes of the flow network.

        A number of nodes in an OHCP-network is $F+ \beta_q + 1 + 2 = F + \beta_q + 3$,
        and the number of di-arcs is at least  $2E  + \beta_q  + 1 + 1  = 2E +  \beta_q  + 2$.        
    }
  \label{fig:ohcp:flow-network}
\end{figure}

\label{subsec:residual-complex-}
\subsection{Residual complex}
Let $\bm{f} \in \mc{Z}^{q}(\bar{\mc{K}}; \integer)$ be some $q$-coflow:   
$\cobd_{q + 1} \bm{f} = \sum_{\mc{T}}\gamma_t \nu_{t} - \sum_{\mc{S}} \gamma_{s} \nu_{s}$,
and $\vec{\bm{f}} \in \mc{Z}^{q}(\vec{\mc{K}}; \integer_{+})$ the corresponding non-negative directed coflow. 
The \textbf{residual capacity} of ${\bm{f}}$ is defined as follows:
\begin{align}
  {c}^{+}_{f}(e_i) =  {c}(e_i) - {\bm{f}}(e_i) &&
  {c}^{-}_{f}(e_i) =  {c}(e_i) + {\bm{f}}(e_i) &&
  {c}_{f}(e_i) = \min \left\{ {c}^{+}_{f}(e_i); \, {c}^{-}_{f}(e_i) \right\}
\end{align}
and for any $q$-chain $\bm{w} \in \mc{C}_{q}(\mc{K}; \integer)$:
\begin{align*}
c_{f}(\bm{w}) = c(\bm{w}) - \bm{f}(\bm{w}) 
= \sum_{\bm{w}(e_i) \neq 0} c(e_i) \cdot \abs{\bm{w}(e_i)} - \sum_{\bm{w}(e_i) \neq 0} \bm{f}(e_i) \cdot \bm{w}(e_i)
\end{align*}

The \textbf{(directed) residual capacity} of $\vec{\bm{f}}$ is defined as follows:
\begin{align}
\label{eq:def:residual-capacity}
  \vec{{c}}_{f}(\eta \dart \tau) &=     
      {c}(e_i) - \vec{\bm{f}}(\eta \dart \tau) + \vec{\bm{f}}(\tau \dart \eta) 
      \equiv {c}(e_i) - {\bm{f}}(e_i) 
  \\ 
\label{eq:def:residual-capacity-rev}
  \vec{{c}}_{f}(\tau \dart \eta) &=     
      {c}(e_i) + \vec{\bm{f}}(\eta \dart \tau) - \vec{\bm{f}}(\tau \dart \eta) 
      \equiv {c}(e_i) + {\bm{f}}(e_i) 
  \\[4pt] 
\label{eq:def:residual-capacity-general}
  \vec{{c}}_{f}(\vec{e}_{i}) &= c(e_i) - \vec{\bm{f}}(\vec{e}_{i}) + \vec{\bm{f}}(\vec{e}_{-i})
\end{align}

\begin{lemma}
  If $\vec{{c}}_{f}(\eta \dart \tau) \geq 0$ or ${{c}}_{f}(e_i) \geq 0$ for all $(\eta \dart \tau) \in \mc{K}_{q}$
  then $\bm{f}$ and $\vec{\bm{f}}$ are \textbf{feasible}.
\end{lemma}
\begin{proof}
  \begin{align*}
      \vec{{c}}_{f}(\eta \dart \tau) \geq 0 
      && \iff && 
      {c}^{+}_{f}(e_i) &=  {c}(e_i) - {\bm{f}}(e_i)  \geq 0
      && \iff && 
      {\bm{f}}(e_i) \leq {c}(e_i) 
  \\
      \vec{{c}}_{f}(\tau \dart \eta) \geq 0 
      && \iff && 
      {c}^{-}_{f}(e_i) &= {c}(e_i) + {\bm{f}}(e_i)  \geq 0
      && \iff && 
      {\bm{f}}(e_i) \geq - {c}(e_i) 
  \end{align*}
  which also implies that ${c}_{f}(e_i) \geq 0$.
  Hence, $\bm{f}$ is feasible.
  
  $\vec{\bm{f}}$ is non-negative and such that $\bm{f} = \vec{\bm{f}}_{+} - \vec{\bm{f}}_{-}$, and $ \vec{\bm{f}}(\tau \dart \eta) = 0$ or $ \vec{\bm{f}}(\eta \dart \tau) = 0$.
  If ${\bm{f}}$ is feasible then $\vec{\bm{f}}_{+}$ and $\vec{\bm{f}}_{-}$ are non-negative and such that 
  $-\bm{c} \leq \vec{\bm{f}}_{+} - \vec{\bm{f}}_{-} \leq \bm{c}$. 
  Which implies that $0 \leq \vec{\bm{f}}_{+} \leq \bm{c}$ and  $0 \leq \vec{\bm{f}}_{-} \leq \bm{c}$.
\end{proof}

\begin{theorem}[Negative $q$-cycles]
  Two feasible $q$-coflows  $\bm{f}$ and $\bm{\phi}$ are cohomologous, $\bm{f} \csim \bm{\phi}$,
  if any $q$-cycle ${\bm{w}} \in \mc{Z}_{q}(\mc{K}; \integer) $ has same residual capacity:
  \begin{align*}
      {{c}}_{f}({\bm{w}}) =  {{c}}_{\phi}({\bm{w}})  \geq 0
  \end{align*}
  where $c_{f}(\bm{w}) = c(\bm{w}) - \bm{f}(\bm{w}) = \sum_{\bm{w}(e_i) \neq 0} c(e_i) \cdot \abs{\bm{w}(e_i)} - \sum_{\bm{w}(e_i) \neq 0} \bm{f}(e_i) \cdot \bm{w}(e_i)$,
\end{theorem}
\begin{proof}
  For $\bm{f}, \bm{\phi} \in \mc{Z}^{q}(\mc{K}; \integer)$, 
  $\bm{f} \csim \bm{\phi}$ means that there exists  $\bm{\rho} \in \mc{C}^{q - 1}({\mc{K}}; \integer)$ such that:
  \begin{align*}
      {\bm{\phi}} &= {\bm{f}} + \cobd_{q}\bm{\rho}
  \end{align*}

  Then, since $\mc{B}^{q} \perp \mc{Z}_{q}$:
  \begin{align*}
      {\bm{\phi}}({\bm{w}}) 
      &= {\bm{f}}({\bm{w}}) + \cobd_{q}\bm{\rho}({\bm{w}})
      = {\bm{f}}({\bm{w}}) 
  \\[2pt]
    {{c}}_{\phi}({\bm{w}})  
    &= {{c}}({\bm{w}}) - {\bm{\phi}}({\bm{w}})
    = {{c}}({\bm{w}}) - {\bm{f}}({\bm{w}})
    = {{c}}_{f}({\bm{w}})  
  \end{align*}
  
  So, if ${\bm{f}} \in \mc{Z}^{q}(\mc{K}; \integer)$ is feasible than any cohomologous $q$-coflow $\bm{\phi} \csim \bm{f}$ is also feasible. 
\end{proof}

\begin{definition}($q$-Coflow Network)
\label{def:coflow-network}
  A simplicial \textbf{$q$-coflow network ($q$-CFN)}
  is a tuple $(\mc{K}, \bm{z}, c)$ 
  where a \textbf{$q$-coflow complex} $\mc{K}$ is an oriented $(q + 1)$-dimensional simplicial complex embedded in $\real^{q + 1}$, 
  $c \!: \mc{K}_{q} \mapsto \integerplus$ is the non-negative  \emph{capacity function}, 
  and $\bm{z} \in \mc{Z}_{q}(\mc{K}; \integer)$ is a $q$-cycle on $\mc{K}$.
\end{definition}

\begin{definition}(Directed $q$-Coflow Network)
\label{def:coflow-network-dir}
  Given a $q$-coflow network $(\mc{K}, \bm{z}, c)$,
  a \textbf{directed coflow network (dCFN)} is defined as a tuple $(\vec{\mc{K}}, \vec{\bm{z}}, \vec{c}) \equiv (\vec{\mc{K}}, {\bm{z}}, {c})$ 
  where the \textbf{directed coflow complex} $\vec{\mc{K}}$ 
  is constructed from $\mc{K}$ by including the opposite orientation of $q$-simplices as basis elements of $q$-th chain groups:
  \begin{align}
  \label{eq:def:coflow-network-dir}
    \vec{\mc{K}}_{q + 1} = {\mc{K}}_{q + 1}
    &&
    \vec{\mc{K}}_{q} = \vec{\mc{K}}^{+}_{q} \cup \vec{\mc{K}}^{-}_{q}
    &&
    \vec{\mc{K}}_{q^{\prime}} = {\mc{K}}_{q^{\prime}} \quad \text{ for } q^{\prime} < q
  \end{align}
  
  The input $q$-cycle is replaced by $\vec{\bm{z}} \in \mc{Z}_{q}(\vec{\mc{K}}; \integer)$ such that 
  $\vec{\bm{z}}(\eta \dart \tau) = - \vec{\bm{z}}(\tau \dart \eta) = \bm{z}(e_i)$.
  
  The new capacity function 
  $\vec{c}: \vec{\mc{K}}_{q} \mapsto \integerplus$ is extension of $c$ on $\vec{\mc{K}}_{q}$:
  $\vec{c}(\tau \dart \eta) = \vec{c}(\eta \dart \tau) = {c}(e_i)$.
\end{definition}


\begin{definition}(Residual Complex)
  Let $(\mc{K}, \bm{z}, c)$ be a simplicial coflow network and $\bm{f} \in \mc{Z}^{q}(\mc{K}; \integer)$
  be a feasible $q$-coflow such that $\bm{f}(\bm{z}) \leq 0$.
  Then the \textbf{residual complex} is defined as a directed simplicial coflow network 
  $\vec{\mc{K}}_{f} = (\vec{\mc{K}}, \vec{\bm{z}}, \vec{c}_{f})$
  with the residual capacities $\vec{\bm{c}}_{f}$ instead of the original capacities.

\end{definition}

%

\section{Optimality results for OHCP}
\label{subsec:ohcp:optimality-chains}
\label{subsec:ohcp:undir}
\subsection{Oriented complex: homology over $\integer$}
\label{subsec:ohcp:weak-duality}
\subsubsection{Weak Duality (OHCP)}
\begin{theorem}[Weak duality]
\label{th:ohcp:weak-duality}
  Let $(\bm{x}, \bm{\pi})$ 
  be a feasible solution for the OHCP problem in Eq.~\eqref{eq:def:ohcp}
  such that 
  $\bm{x} \in \mc{Z}_{q}(\bar{\mc{K}}; \integer)\!:   \bm{x} \sim \bm{z}$,
  $\bm{\pi} \in \mc{C}_{q + 1}(\bar{\mc{K}}; \integer)\!: \bd_{q + 1} \bm{\pi} = \bm{z} - \bm{x}$,
  and $\bm{\pi}(\nu_k) = 0$ for all $\nu_k \in \mc{V}$.

  
  Then for any \textbf{feasible} $q$-coflow  
  $\bm{f} \in \mc{C}^{q}(\bar{\mc{K}}; \integer) \!:\, \cobd_{q + 1} \bm{f} = \sum_{\mc{T}} \gamma_{t} \nu_{t} - \sum_{\mc{S}} \gamma_{s} \nu_{s} 
  \equiv \cobd_{q + 1} \bm{f} = \gamma \mc{T} - \gamma \mc{S} $:
  \begin{align}
  \label{eq:ohcp:weak-duality}
      {\bm{f} (-\bm{z}) = } \ 
      \abs{\bm{f} (\bm{z})} \leq  \abs{\bm{c}(\bm{x})} \ 
  \end{align}
\end{theorem}
\begin{proof}
    By Hodge decomposition theorem: Coflows/Cocycles are orthogonal to boundaries. 
    In particular:
    \begin{align*}
      \bm{f} (\bd_{q + 1} \bm{\pi}) 
        = \bm{\pi} (\cobd_{q + 1} \bm{f})
        = \bm{\pi} (\gamma T - \gamma S)
        = \gamma \pi_T - \gamma \pi_S
        = \gamma \cdot 0 - \gamma \cdot 0
        = 0
    \end{align*}
    
    \begin{align*}
      \bm{f} (\bd_{q + 1} \bm{\pi}) 
        = \bm{\pi} (\cobd_{q + 1} \bm{f})
        &= \bm{\pi} \big( {\textstyle \sum_{\mc{T}} \gamma_{t} \nu_{t} - \sum_{\mc{S}} \gamma_{s} \nu_{s}} \big) 
        \\
        &= {\textstyle \sum_{\mc{T}} \gamma_{t} \pi_{t} - \sum_{\mc{S}} \gamma_{s} \pi_{s} +  \sum_{\mc{F}} 0 \cdot \pi_{j} } 
        \\
        &=
         {\textstyle \sum_{\mc{T}} \gamma_{t} \cdot 0 - \sum_{\mc{S}} \gamma_{s} \cdot 0 + 0} 
        &= 0
    \end{align*}
    
    Then 
    \begin{align*}
      0 = \bm{f} (\bd_{q + 1} \bm{\pi}) 
        = \bm{f} (\bm{z} - \bm{x}) 
        = \bm{f} (\bm{z}) - \bm{f}(\bm{x}) 
        &&\iimplies
        \bm{f} (\bm{z}) = \bm{f}(\bm{x}) 
    \end{align*}
    
    Since $\bm{f}$ is feasible, it obeys the capacity constraints $-\bm{c} \leq  \bm{f} \leq \bm{c}$,
    and hence:
    \begin{align*}
    -\abs{\bm{c}(\bm{x})} \leq \bm{f}(\bm{x})  \leq \abs{\bm{c}(\bm{x})}
    && \iff&&
    \abs{\bm{f}(\bm{x}) } \leq \abs{\bm{c}(\bm{x})}
    && \iimplies &&
    \abs{\bm{f}(\bm{z}) } \leq \abs{\bm{c}(\bm{x})}
    \end{align*}
    
    \begin{itemize}
      \item Recall that $\bm{f} (\bm{z}) \leq 0$, therefore $\bm{f}(\bm{x}) \leq 0$ for any $\bm{x} \sim \bm{z}$, and thus
            \begin{align*}
            -\abs{\bm{c}(\bm{x})} \leq \bm{f}(\bm{x})  \leq 0
            && \iff &&
            0 \leq \bm{f}(-\bm{x})  \leq \abs{\bm{c}(\bm{x})}
            && \iimplies&&
            0 \leq \bm{f}(-\bm{z}) \leq \abs{\bm{c}(\bm{x})}
            \end{align*}
    \end{itemize}

\end{proof}

\begin{corollary}
Let a $q$-cycle $\bm{x}_{\star} \sim \bm{z}$ and $q$-coflow $\bm{f}_{\star} \!: \cobd_{q + 1} \bm{f}_{\star} = \gamma (T - S)$
be optimal solutions to the min-homologous cycle Eq.~\eqref{eq:def:ohcp-on-K} and the max-coflux problems Eq.~\eqref{eq:def:ohcp-max-flux-lp}.
Then
  \begin{align}
  \label{eq:ohcp:weak-duality-opt}
     \bm{f}_{\star}(-\bm{z}) = \abs{\bm{f}_{\star}(\bm{z})} = 
     \max\limits_{\bm{f} : feasible} \abs{\bm{f}(\bm{z}) } 
     \leq \min\limits_{\bm{x} \sim \bm{z}} \abs{\bm{c}(\bm{x})} = \abs{\bm{c}(\bm{x}_{\star})}
  \end{align}
\end{corollary}

\begin{corollary}
\label{th:property_9_2}
Let  $\bm{x}_{\circ} \sim \bm{z}_{\circ}$ be some homologous $q$-cycle 
and $\bm{\pi} \in C_{q + 1}(\bar{K}; \integer)$ such that $ \bd_{q + 1} \bm{\pi} = \bm{z}_{\circ} - \bm{x}_{\circ}$
\begin{itemize}
\item[a)]
    For any $q$-copath $\bm{p} \in C^{q}(\bar{K}; \integer)$ between two $(q + 1)$-faces, $\cobd_{q + 1} \bm{p} = \tau - \eta$:
    \begin{align*}
        \bm{p} (\bd_{q + 1} \bm{\pi}) 
            &= \bm{\pi} (\cobd_{q + 1} \bm{p})
            = \bm{\pi} (\tau - \eta)
            = \pi_{\tau} - \pi_{\eta}
    \\
        \bm{p}(\bm{x}_{\circ}) 
            &= \bm{p}(\bm{z}_{\circ}) - \bm{p} (\bd_{q + 1} \bm{\pi}) 
             = \bm{p}(\bm{z}_{\circ}) - \pi_{\tau} + \pi_{\eta}
    \end{align*}
    
\item[b)]
    For any $q$-cocycle $\bm{w} \in \mc{Z}^q(\bar{\mc{K}}; \integer) : \cobd_{q + 1} \bm{w} = 0$:
    \begin{align*}
        \bm{w} (\bd_{q + 1} \bm{\pi}) 
            &= \bm{\pi} (\cobd_{q + 1} \bm{w})
            = \bm{\pi} (0)
            = 0
    \\
        \bm{w}(\bm{x}_{\circ}) 
            &= \bm{w}(\bm{z}_{\circ}) - \bm{w} (\bd_{q + 1} \bm{\pi}) 
             = \bm{w}(\bm{z}_{\circ}) 
    \end{align*}
    
\end{itemize}
    
\end{corollary}

\label{subsec:ohcp:strong-duality}
\subsubsection{Complementary Slackness (OHCP)}
Sullivan \cite{sullivan1990crystalline} in his PhD dissertation has shown that the following  \emph{strong duality}\emph{complementary slackness} conditions hold for optimal solutions to the minimal $q$-chain problem stated on an acyclic $(q+1)$-complex -- homologous $q$-chains need to share a $(q - 1)$-boundary. 
Below we extend these results to the optimal homologous $q$-cycle problem on the complexes with non-trivial $q$-th (co)homology group.

\begin{theorem}[Complementary Slackness]
\label{th:ohcp:strong-duality}
  Equality in Eq.~\eqref{eq:ohcp:weak-duality-opt}
  is achieved whenever the Complementary Slackness conditions (orthogonality conditions) hold:
  \begin{align}
    \begin{array}{ccll}
        \bm{x}_{\star}(e_i) > 0 &\iff& \bm{f}_{\star}(e_i) = - \bm{c}(e_i) \\
        \bm{x}_{\star}(e_i) < 0 &\iff& \bm{f}_{\star}(e_i) = \me\bm{c}(e_i) 
    \end{array}
  \end{align}
\end{theorem}
\begin{proof}
  The optimal $q$-cycle $\bm{x}_{\star}$ has smallest capacity out of all $q$-cycles homologous to $\bm{z}$, 
  $\abs{\bm{c}(\bm{x}_{\star})} \leq \abs{\bm{c}(\bm{z})}$.
  Meanwhile the value of any feasible $q$-coflow on $\bm{z}$ and $\bm{x}_{\star}$ is the same: 
  $\bm{f}(\bm{x}_{\star}) = \bm{f}(\bm{z}) \leq 0$.
  Out of all feasible $q$-coflows $\bm{f}_{\star}$ has the largest absolute value on $\bm{x}_{\star}$,
  which, if equality in Eq.~\eqref{eq:ohcp:weak-duality}/\eqref{eq:ohcp:weak-duality-opt} is reached,
  is equal to the capacity of $\bm{x}_{\star}$: 
  $\bm{f}_{\star}(-\bm{x}_{\star})  = \abs{\bm{f}_{\star}(\bm{x}_{\star})} = \abs{\bm{c}(\bm{x}_{\star})}$.
  It can happen only if $\bm{f}_{\star}$ saturates all edges of $-\bm{x}_{\star}$ to their full capacity. 
  \begin{align*}
      \bm{f}_{\star}(-\bm{x}_{\star}(e_i)) 
      = \bm{c}(\abs{\bm{x}_{\star}(e_i)}) 
      &&\iff&&
        -\bm{x}_{\star}(e_i) \cdot \bm{f}_{\star}(e_i)
      =  \abs{\bm{x}_{\star}(e_i)} \cdot \bm{c}(e_i)
  \\[3pt]
      &&\iff&&
      \begin{cases}
          \bm{f}_{\star}(e_i) = - \bm{c}(e_i) ,& \text{if } \bm{x}_{\star}(e_i) > 0 \\
          \bm{f}_{\star}(e_i) = \me \bm{c}(e_i) ,& \text{if } \bm{x}_{\star}(e_i) < 0 \\
      \end{cases}
  \end{align*}
\end{proof}

\subsection{Directed complex: homology over $\integerplus$}
\label{subsec:ohcp:optimality-dir}
Let $\vec{\bm{f}}= [\vec{\bm{f}}_{+} \ \vec{\bm{f}}_{-}] \in \mc{C}^{q}(\vvec{\mc{K}}; \integer_{+})$ be a non-negative directed $q$-coflow,
$\vec{\bm{f}} \!: \vec{\mc{K}}_{q} = \vec{\mc{K}}^{+}_{q} \cup \vec{\mc{K}}^{-}_{q} \to \integer_{+}$,
where $\vec{\mc{K}}_{q}^{+} = {\mc{K}}_{q}$ and $\vec{\mc{K}}_{q}^{-} = -{\mc{K}}_{q}$ are \emph{natural} and \emph{opposite} orientations of edge-simplices. 
A $q$-coflow $\bm{f} \in \mc{C}^{q}({\mc{K}}; \integer)$ is given as $\bm{f}(e_i) = \vec{\bm{f}}(\eta \dart \tau) - \vec{\bm{f}}(\tau \dart \eta) \equiv \vec{\bm{f}}_{+} - \vec{\bm{f}}_{-}$.


Note that $\vec{\bm{z}}(\tau \dart \eta) = - \vec{\bm{z}}(\eta \dart \tau) = - {\bm{z}}(e_i)$.
and ${{c}}(\tau \dart \eta) = {{c}}(\eta \dart \tau) = {{c}}(e_i)$.
Thus $\bm{f}(\bm{z}) = \vec{\bm{f}}_{+}(\bm{z}) - \vec{\bm{f}}_{-}(\bm{z}) 
                     = [\bm{z} \ -\bm{z}] \cdot [\vec{\bm{f}}_{+} \ \vec{\bm{f}}_{-}]^{\tr} 
                     = [\vec{\bm{z}}_{+} \ \vec{\bm{z}}_{-}] \cdot [\vec{\bm{f}}_{+} \ \vec{\bm{f}}_{-}]^{\tr} 
                     = \vec{\bm{f}}(\vec{\bm{z}})$ 

Consider the directed max-coflux problem: 
\begin{align*}
  \begin{array}{lll}\dsp
      \max\limits_{\vec{\bm{f}}}\ &\dsp \vec{\bm{f}} (-\vec{\bm{z}})   \geq 0       \\
      \subto\ &\dsp  \cobd_{q + 1} \vec{\bm{f}} = {\textstyle \sum_{\mc{T}} \gamma_t \nu_t - \sum_{\mc{S}} \gamma_s \nu_s}            \\
              &\dsp  \bm{0}_{2E} \leq  \vec{\bm{f}} \leq \vec{\bm{c}}_{2E}          
  \end{array}
\end{align*}

or in more detail:
\begin{align}
\label{eq:ohcp:dir-max-flux}
  \begin{array}{rllrrr}\dsp
      \max\limits_{\vec{\bm{f}}}\ &\dsp - \sum_{(\eta \dart \tau) \in \vec{\mc{E}}} \vec{\bm{f}}(\eta \dart \tau) \cdot \vec{\bm{z}}(\eta \dart \tau)   \geq 0
\\
      \subto\ 
          &\dsp  \cobd_{q + 1} \vec{\bm{f}}(\tau) = \vec{\bm{f}}(\bd_{q + 1} \tau) 
                                                  = \sum_{ e_i \in K_q } \bigg( \vec{\bm{f}}(\eta \dart \tau) - \vec{\bm{f}}(\tau \dart \eta) \bigg) = 0        
          &&\forall \tau \in  \mc{K}_{q + 1}    
  \\[2pt]
          &\dsp  \cobd_{q + 1} \vec{\bm{f}}(\nu_{s})    = \vec{\bm{f}}(\bd_{q + 1} \nu_{s}) = \sum_{e_s \in \mc{K}_q } -\vec{\bm{f}}(\nu_{s} \dart \tau)   = - \gamma_s            
  \\[2pt]
          &\dsp  \cobd_{q + 1} \vec{\bm{f}}(\nu_{t})    = \vec{\bm{f}}(\bd_{q + 1} \nu_{t}) = \sum_{e_t \in \mc{K}_q }  \vec{\bm{f}}(\tau \dart \nu_{t}) = + \gamma_t             
  \\[2pt]
          &\dsp 0 \leq \vec{\bm{f}}(\eta \dart \tau) \leq {c}(\eta \dart \tau)
          &&\forall \vec{e}_{i}  \in \vec{\mc{K}}_{q} 
  \end{array}
\end{align}

The dual of this LP has a variable $-\bm{\pi}(\tau)$ for each $(q + 1)$-face $\tau$ and 
a non-negative variable $\vec{\bm{x}}(\eta \dart \tau) \geq 0$ for each oriented $q$-simplex $(\eta \dart \tau) \in \vec{\mc{K}}_{q}$.
\begin{align}
\label{eq:ohcp:dir-prob}
  \begin{array}{rllrrr}\dsp
      \min\limits_{\vec{\bm{x}}, \bm{\pi} }\ 
          &\dsp \sum_{(\eta \dart \tau) \in \vec{\mc{K}}_{q}} \vec{\bm{x}}(\eta \dart \tau) \cdot {{c}}(\eta \dart \tau)  
              + \sum_{\nu_{t} \in \mc{T}} \gamma_t \pi_t - \sum_{\nu_{s} \in \mc{S}} \gamma_s \pi_s
  \\[2em]
      \subto\ 
          &\dsp  \bm{\pi}(\eta) - \bm{\pi}(\tau) + \vec{\bm{x}}(\eta \dart \tau) \geq -\vec{\bm{z}}(\eta \dart \tau) = -    \bm{z}(e_i)           &&\forall \vec{e}_{+i} \in \vec{\mc{K}}^{+}_{q}    \\[1em]
          &\dsp  \bm{\pi}(\tau) - \bm{\pi}(\eta) + \vec{\bm{x}}(\tau \dart \eta) \geq -\vec{\bm{z}}(\tau \dart \eta) = \me {\bm{z}}(e_i)          &&\forall \vec{e}_{-i} \in \vec{\mc{K}}^{-}_{q}     \\[1em]
          &\dsp \vec{\bm{x}}(\eta \dart \tau) \geq 0
          &&\forall \vec{e}_{i}  \in \vec{\mc{K}_{q}} 
          \\[2pt]
          &\dsp {\bm{\pi}}(\nu_s) = 0; \  {\bm{\pi}}(\nu_t) = 0
          &&\forall \nu_k \in {\mc{V}} 
  \end{array}
\end{align}
where $\pi_t = 0$ and $\pi_s = 0$  $\iff \bm{\pi} \in \mc{C}_{q + 1}(\mc{K}; \integer)$.

\begin{theorem}
\label{th:ohcp:dir-prob-solution}
Let $\bm{\pi}_{\star}$ and $\vec{\bm{x}}_{\star}$ be the optimal solution for the dual problem above, Eq.~\eqref{eq:ohcp:dir-prob}. 
Then 
\begin{itemize}
  \item[(a)]
      $\pi^{\star}_{t} = \bm{\pi}_{\star}(\nu_t) = 0$ and $\pi^{\star}_{s} = \bm{\pi}_{\star}(\nu_s) = 0$.
  \item[(b)]
      A $q$-chain $\bm{x}_{\star} \in C_q(K; \integer)$ defined as 
      $\bm{x}_{\star}(e_i) = \vec{\bm{x}}_{\star}(\tau \dart \eta) - \vec{\bm{x}}_{\star}(\eta \dart \tau) \equiv \bm{x}_{\star}^{-} - \bm{x}_{\star}^{+}$
      is a minimal $q$-cycle homologous to $\bm{z}$ on $K$ such that $\bd_{q + 1} \bm{\pi}_{\star} = \bm{z} - \bm{x}_{\star}$. 
\end{itemize}
\end{theorem}
\begin{proof}[Proof: (b)]
  Because every primal capacity ${c}(\eta\dart \tau)$ is non-negative, 
  the dual variables $\vec{\bm{x}}_{\star}(\eta \dart \tau )$ are \textit{individually as small as possible without violating any constraints}:
  \begin{align*}
      \vec{\bm{x}}_{\star}&(\eta \dart \tau ) \geq -\vec{\bm{z}}(\eta \dart \tau) - \bm{\pi}_{\star}(\eta) + \bm{\pi}_{\star}(\tau) 
      \qquad\text{ or }\qquad
      \vec{\bm{x}}_{\star}(\eta \dart \tau ) \geq 0
      \iimplies \\
      &\iimplies 
          \vec{\bm{x}}_{\star}(\eta \dart \tau ) 
          =\max \left\lbrace  0;\,  -{\bm{z}}(\eta \dart \tau) - \bm{\pi}_{\star}(\eta) + \bm{\pi}_{\star}(\tau)  \right\rbrace
          =\max \left\lbrace  0;\,  -{\bm{z}}(e_i) + \bd_{q + 1}\bm{\pi}_{\star} (e_i)  \right\rbrace
  \\[4pt]
      \vec{\bm{x}}_{\star}&(\tau \dart \eta ) \geq -\vec{\bm{z}}(\tau \dart \eta) - \bm{\pi}_{\star}(\tau) + \bm{\pi}_{\star}(\eta) 
      \qquad\text{ or }\qquad
      \vec{\bm{x}}_{\star}(\tau \dart \eta ) \geq 0
      \iimplies \\
      &\iimplies 
          \vec{\bm{x}}_{\star}(\tau \dart \eta ) 
          =\max \left\lbrace  0;\,  -\vec{\bm{z}}(\tau \dart \eta) - \bm{\pi}_{\star}(\tau) + \bm{\pi}_{\star}(\eta)  \right\rbrace
          =\max \left\lbrace  0;\,  {\bm{z}}(e_i) - \bd_{q + 1}\bm{\pi}_{\star} (e_i)  \right\rbrace
  \end{align*}

  Which implies
  \begin{align*}
      \bm{x}_{\star}(e_i) 
      &= \vec{\bm{x}}_{\star}(\tau \dart \eta) - \vec{\bm{x}}_{\star}(\eta \dart \tau) 
       = {\bm{z}}(e_i) - \bd_{q + 1}\bm{\pi}_{\star} (e_i) 
  \end{align*}
  for all $ e_i \in K_q$. 
  And hence $\bm{x}_{\star} \sim \bm{z}$ on $\mc{K}$.
\end{proof}

\begin{proof}[Proof: (a)]
  If $\pi^{\star}_{t} = \bm{\pi}_{\star}(\nu_t) = 0$ or $\pi^{\star}_s = \bm{\pi}_{\star}(\nu_s) = 0$ then $\bm{\pi}_{\star} \in C_{q + 1}(\mc{K}; \integer)$
  and, from \textbf{(b)}, $\bm{x}_{\star} \sim \bm{z}$ on $\mc{K}$.
  If $\bm{x}_{\star} \sim \bm{z}$ on $\mc{K}$ then it must be $\bm{\pi}_{\star} \in C_{q + 1}(\mc{K}; \integer)$ for any $\bm{\pi}_{\star}$ such that $\bm{x}_{\star} = {\bm{z}} + \bd_{q + 1}\bm{\pi}_{\star}$,
  and hence
  $\pi^{\star}_{t} = \bm{\pi}_{\star}(\nu_t) = 0$ and $\pi^{\star}_{s} = \bm{\pi}_{\star}(\nu_s) = 0$.
  
  If $\pi^{\star}_{t} = \bm{\pi}_{\star}(\nu_t) \neq 0$ or $\pi^{\star}_s = \bm{\pi}_{\star}(\nu_s) \neq 0$
  then $\bm{x}_{\star} \sim \bm{z}$ on $\bar{\mc{K}}$ but not on $\mc{K}$.

\end{proof}

\subsubsection{Non-positive \texorpdfstring{$q$}{q}-cocycles}

Here we restate classical optimality conditions in the context of max-flux/ohcp problems in Eq.~\eqref{eq:ohcp:dir-max-flux} and Eq.~\eqref{eq:ohcp:dir-prob}.

\begin{theorem}[Non-positive $q$-cocycle optimality condition]
\label{th:ohcp:dir-optimality:neg-cocycles}
    Let $\vec{\bm{f}} \in \mc{C}^{q}(\vec{\mc{K}}; \integerplus)$ be a feasible solution for the max-coflux/min-cost coflow problem. 
    Then $\vec{\bm{f}}$ is \textbf{optimal} if and only if the residual complex $\vec{\mc{K}}_{f}$ 
    \textbf{has no augmenting $q$-copath $\vec{\bm{p}}$} with 
    \textbf{non-positive flux through $\vec{\bm{z}}$}:
    $\vec{\bm{p}}(\vec{\bm{z}}) < 0$ 
    and $\vec{\bm{p}}$ is augmenting on  residual complex $\vec{\mc{K}}_{f}$:
    \begin{align*}
        \vec{c}_{f}(\vec{\bm{p}}) = \min_{\vec{e}_{i} \in \vec{\bm{p}}} \vec{c}_{f}(\vec{e}_{i}) \geq 0
        &&\text{ or }&&
        {c}_{f}(\vec{\bm{p}}) = \min_{\vec{\bm{p}}(e_i) \neq 0} \big\{ {c}({e}_{i}) - \vec{\bm{p}}(e_i) \cdot \bm{f}(e_i) \big\} \geq 0
    \end{align*}

\begin{figure}
  \centering
  \includegraphics[width=0.45\linewidth]{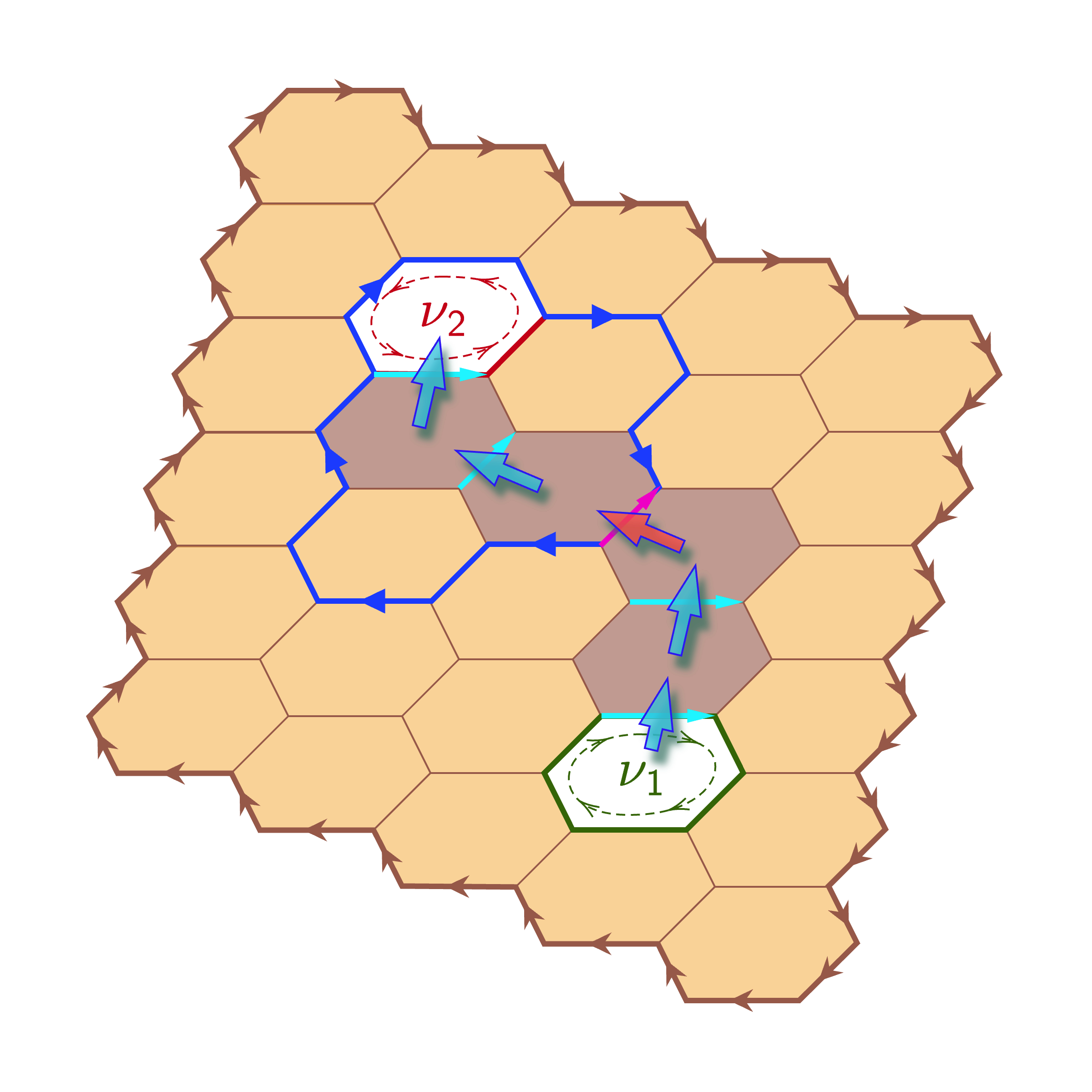} \ 
  \includegraphics[width=0.45\linewidth]{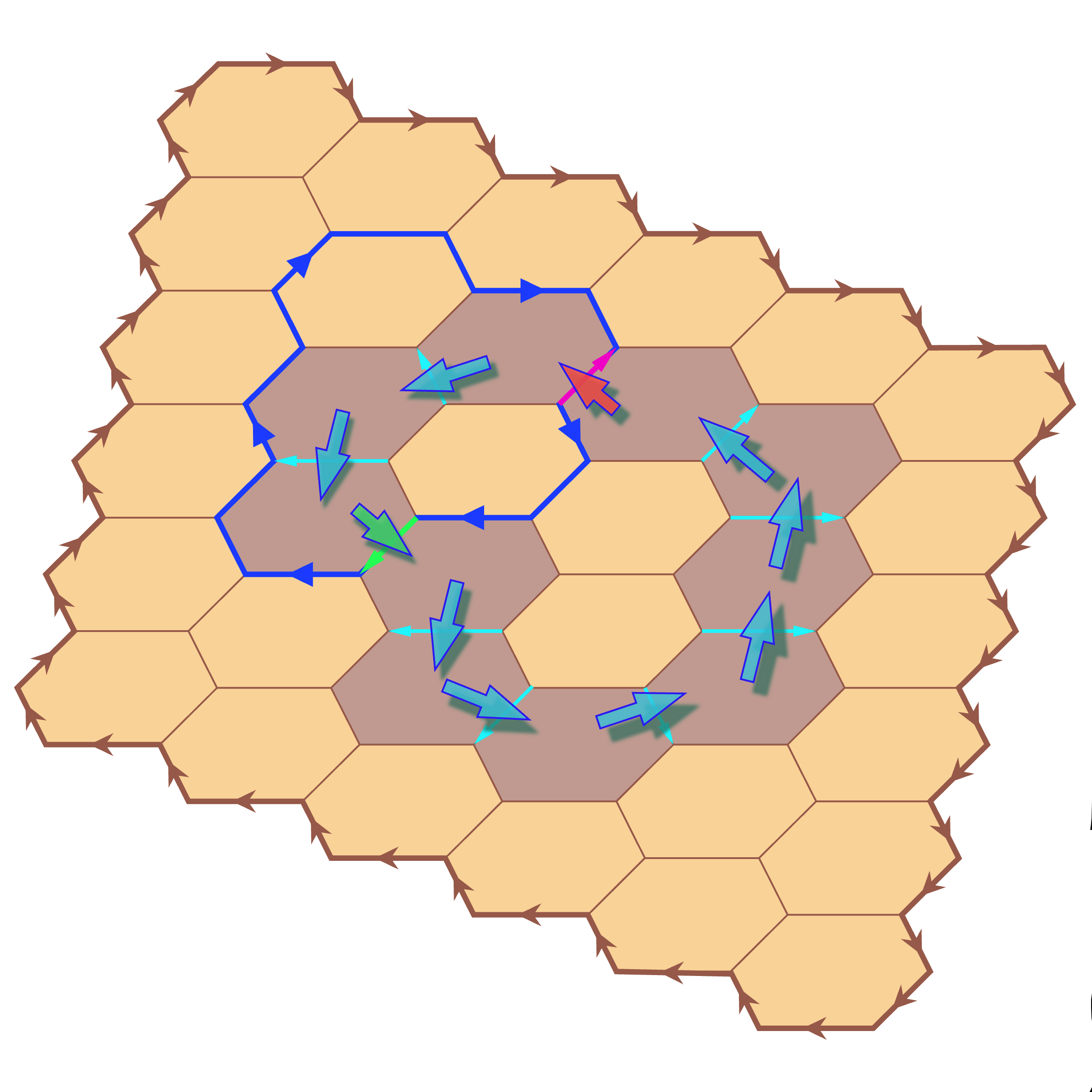}
  \caption[Non-positive augmenting $q$-cocycles.]{
      The illustration to the Theorem~\ref{th:ohcp:dir-optimality:neg-cocycles}. 
      \textbf{Left:} A $q$-cocyle \textcolor{Aqua}{$\vec{\bm{p}}(\bd \nu_1, \bd \nu_2)$} crossing the input $q$-cycle \textcolor{RoyalBlue}{$\bm{z}$} has strictly negative cost of coflow if $\dist_{\bm{z}}(\nu_2) < \dist_{\bm{z}}(\nu_1)$. 
      \textbf{Right:} 
      The cost of sending coflow along a simple coboundary \textcolor{Aqua}{$\bm{w}$} that crosses \textcolor{RoyalBlue}{$\bm{z}$} is always 0: 
      since for each $q$-edge \textcolor{Pink}{$e_i$} that is oriented against $\bm{z}$ in $\bm{w}$ there is a $q$-edge \textcolor{GrassGreen}{$e_j$} that is oriented along $\bm{z}$.
    }
  \label{fig:ohcp:neg-cocycles}
\end{figure}

\end{theorem}
\begin{proof}
  If there exist a directed cocycle $\vec{\bm{w}}$ in $\vec{\mc{K}}_{f}$  with negative cost $\vec{\bm{w}}({\bm{z}}) = \vec{\bm{w}}(-\vec{\bm{z}}) < 0$ 
  then the coflow $\vec{\bm{f}}$ is \textbf{not optimal}, since we can improve the objective function by sending a unit of coflow along it. 

              If a $q$-cocycle $\bm{w}$  on  the residual complex $\vec{\mc{K}}_{f}$ 
              has 0 flux through $\bm{z}$ then it is either 0 everywhere or $\bm{w}$ is a $q$-coboundary.
              Note that  augmenting along a $q$-coboundary does not change $\bm{f}(\bm{z})$ and, equivalently, $\bm{f}^{\prime} \csim \bm{f}$,
              where $\bm{f}^{\prime}$ is the augmented coflow:
              $\bm{f}^{\prime} = \bm{f} + \bm{w}$.

              Despite that the cost $\bm{w}(\bm{z})$ of augmenting coflow along a $q$-coboundary $\bm{w}$ is 0, 
              when  the coflow $\bm{f}$ is optimal there is no augmenting  $q$-coboundary $\bm{w}$ that crosses $\bm{z}$.
              Let's assume that such a coboundary exists. 
              Then we can break it into 
              a pair of copathes $\bm{p}_1$ and $\bm{p}_2$ 
                  such that $\bm{w} = \bm{p}_1 + \bm{p}_2$ and  $\bm{p}_1(\bm{z}) = -\bm{p}_2(\bm{z}))$. 
              This means that sending one unit of coflow along one of them strictly improves the objective function, which means that $\bm{f}$ is not optimal.

        Note that we can augment along a $q$-coboundary $\bm{w}$ only if all $(q + 1)$-facets visited by it have the same value of $\dist_{\bm{z}}$.
              It means that an optimal coflow  $\bm{f}_{\star}$ can saturate some $q$-edges that are not part of $\bm{x}_{\star} \sim \bm{z}$, hence the ``non-positive'' (co)cycles optimality conditions instead of the classical ``negative'' (co)cycles.

\end{proof}

\subsubsection{Reduced costs are homologous cycles}

\begin{theorem}
\label{th:ohcp:dir-optimality:reduced-costs}
  A feasible directed $q$-coflow $\vec{\bm{f}} \!: \vec{\mc{K}}_{q}  \to \integerplus$
  is an optimal solution of the max-flux/min-cost-flow problem in Eq.~\eqref{eq:ohcp:dir-max-flux}
  if and only if there is some $(q + 1)$-chain $\bm{\pi} \in \mc{C}_{q + 1}(\vec{\mc{K}}; \integer)$ 
  that satisfies the following \embf{reduced cost optimality conditions} on the residual complex $\vec{\mc{K}}_{f}$:
  \begin{align}
  \label{eq:ohcp:dir-optimality:reduced-costs-residual-complex}
    \vec{\bm{x}}^{\pi} (\eta \dart \tau) &= \vec{\bm{z}}(\eta \dart \tau) - \bm{\pi} (\tau) + \bm{\pi}(\eta) \geq 0
  \end{align}
  for any $\vec{e}_i = (\eta \dart \tau) \in \vec{\mc{K}}_{f}$ such that  $\vec{c}_{f}(\vec{e}_{i}) > 0$.
  
  Or equivalently on the directed coflow complex $\vec{\mc{K}}$:
  \begin{align}
  \label{eq:ohcp:dir-optimality:reduced-costs-dir-complex}
    \vec{\bm{x}}^{\pi} (\eta \dart \tau) &= \vec{\bm{z}}(\eta \dart \tau) - \bm{\pi} (\tau) + \bm{\pi}(\eta) \geq 0
    && 
      \iff &&&  0 < &\vec{\bm{f}}(\vec{e}_{i}) < c(e_i)
  \\[4pt]
  \label{eq:ohcp:dir-optimality:reduced-costs-dir-complex:saturated}
    \vec{\bm{x}}^{\pi} (\eta \dart \tau) &= \vec{\bm{z}}(\eta \dart \tau) - \bm{\pi} (\tau) + \bm{\pi}(\eta) \leq 0
    && 
      \iff &&&  \vec{\bm{f}}(\vec{e}_{i}) = 0 &\text{ or } \vec{\bm{f}}(\vec{e}_{i}) = c(e_i)
  \end{align}
  for all  $\vec{e}_i = (\eta \dart \tau) \in \vec{\mc{K}}$. 
\end{theorem}
\begin{proof}
  The reduced cost opt-conditions above are equivalent to the {negative-cocycle optimality condition}, Theorem~\ref{th:ohcp:dir-optimality:neg-cocycles}.
  
  $\bm{\Longrightarrow}$
  If the directed $q$-coflow $\vec{\bm{f}}$ satisfies the reduced cost optimality conditions Eq.~\eqref{eq:ohcp:dir-optimality:reduced-costs-residual-complex} or Eq.~\eqref{eq:ohcp:dir-optimality:reduced-costs-dir-complex},
  then for any directed $q$-cocycle/copath $\vec{\bm{w}} \in \mc{Z}^{q}(\vec{\mc{K}}_{f}; \integerplus)$:
  $\vec{\bm{w}} (\vec{\bm{x}}^{\pi}) = \vec{\bm{w}} (\vec{\bm{z}}) \geq 0$ 
  which implies that $\vec{\mc{K}}_{f}$ contains no negative $q$-cocycles/copaths.
  
  $\bm{\Longleftarrow}$
  Now let's assume that $\vec{\mc{K}}_{f}$ contains no negative $q$-cocycles.
  
  Let $\dist_{f} = \dist_{f}^{\bm{z}}  = \dist_{f} (\bm{z}) \in \mc{C}_{q + 1}(\vec{\mc{K}}_{f}; \integer) \leftarrow 
      \mathtt{ShortestCoPathDist} \big( \mathtt{From} = \nu_{0}, \mathtt{To} = \vec{\mc{K}}_{f}, \mathtt{Length} = \vec{\bm{z}} \big)$
  be a $q + 1$-chain dual to shortest path distances in dual residual network $\vec{\mc{K}}^{*}_{f}$  from the node dual to $\nu_0$
  to all other nodes with dual arc-length given by $\vec{\bm{z}}^{*}$.
  It satisfies the distance conditions:
  \begin{align*}
      \dist\nolimits_{f}(\tau) \leq \dist\nolimits_{f}(\eta) + \vec{\bm{z}}(\eta \dart \tau)
      && \text{for all } \vec{e}_i = (\eta \dart \tau) \in \vec{\mc{K}}_{f} 
              \text{ such that }  \vec{c}_{f}(\vec{e}_{i}) > 0
  \end{align*}
  We rewrite these inequalities as:
  \begin{align*}
    \vec{\bm{z}}(\eta \dart \tau)  + \dist\nolimits_{f}(\eta) - \dist\nolimits_{f}(\tau) 
    = \vec{\bm{z}}(\eta \dart \tau)  - \bd_{q + 1} \dist\nolimits_{f}(\eta \dart \tau) 
    \geq 0
  \end{align*}
  or $\vec{\bm{x}}^{\dist_{f}} \geq 0$.
  
\end{proof}

\subsubsection{Complementary Slackness }

\begin{theorem}
\label{th:ohcp:dir-optimality:complementary-slackness}
  A feasible directed $q$-coflow $\vec{\bm{f}}_{\star} \!: \vec{\mc{K}}_{q}  \to \integerplus$
  is an optimal solution of the max-flux/min-cost-flow problem in Eq.~\eqref{eq:ohcp:dir-max-flux}
  if and only if for some $(q + 1)$-chain $\bm{\pi} \in \mc{C}_{q + 1}(\vec{\mc{K}}; \integer)$ 
  the reduced costs $\vec{\bm{x}}^{\pi}$ and coflow values satisfy the following \embf{complementary slackness optimality conditions} for all 
  $\vec{e}_{i} = (\eta \dart \tau)$ on the directed coflow complex $\vec{\mc{K}}$:
  \begin{align}
  \label{eq:ohcp:dir-optimality:complementary-slackness}
      \vec{\bm{x}}^{\pi} (\eta \dart \tau) > 0 &&\iimplies&& \vec{\bm{f}}_{\star}(\eta \dart \tau) &= 0       \\[2pt]
      \vec{\bm{x}}^{\pi} (\eta \dart \tau) < 0 &&\iimplies&& \vec{\bm{f}}_{\star}(\eta \dart \tau) &= c(e_i)  \\[4pt]
      0 < \vec{\bm{f}}_{\star}(\eta \dart \tau) < c(e_i) &&\iimplies&& \vec{\bm{x}}^{\pi} (\eta \dart \tau) &= 0
  \end{align}
  And 
    \begin{align*}
       (\vec{\bm{f}}_{\star}(\eta \dart \tau) = 0  \text{ or } 
       \vec{\bm{f}}_{\star}(\eta \dart \tau) = c(e_i))
       \text{ and } \vec{\bm{x}}^{\pi} (\eta \dart \tau) = 0
       &&\iimplies&& {\bm{\pi}} (\eta) =  {\bm{\pi}}  (\tau) 
    \end{align*}
  
\end{theorem}
\begin{proof}
  Recall that $\vec{\bm{z}}(\eta \dart \tau) = -\vec{\bm{z}}(\tau \dart \eta)$, then
  \begin{align*}
    \vec{\bm{x}}^{\pi} (\tau \dart \eta) 
        &= \vec{\bm{z}}(\tau \dart \eta) - \bm{\pi} (\eta) + \bm{\pi}(\tau)
         = -\vec{\bm{z}}(\eta \dart \tau) - \bm{\pi} (\eta) + \bm{\pi}(\tau)
         = -\vec{\bm{x}}^{\pi} (\eta \dart \tau) 
  \end{align*}
  or $\vec{\bm{x}}^{\pi}(\vec{e}_{i}) = -\vec{\bm{x}}^{\pi}(\vec{e}_{-i})$
  for all $\vec{e}_{i} \in \vec{\mc{K}}_{q}$.

  \begin{itemize}
      \item[Case 1: ] 
            If $\vec{\bm{x}}^{\pi}(\vec{e}_{i}) = \vec{\bm{x}}^{\pi} (\eta \dart \tau) > 0$
            then the reversed $q$-edge $\vec{e}_{-i} = (\tau \dart \eta)$ has negative reduced cost $\vec{\bm{x}}^{\pi}(\vec{e}_{-i}) < 0$,
            and hence $\vec{c}_{f}(\vec{e}_{i}) > 0$ and $\vec{c}_{f}(\vec{e}_{-i}) \leq 0$
            by Eq.~\eqref{eq:ohcp:dir-optimality:reduced-costs-residual-complex}.
            From the definition of residual capacities in Eq.~\eqref{eq:def:residual-capacity} and  Eq.~\eqref{eq:def:residual-capacity-rev},
            and since $\vec{\bm{f}}_{\star}$ is non-negative $\vec{\bm{f}}_{\star} \geq 0$
            since :
            \begin{align*}
                \vec{c}_{f}(\vec{e}_{-i})  \leq 0 
                &&\iff&&
                {c}(e_i) + \vec{\bm{f}}(\vec{e}_{i}) - \vec{\bm{f}}(\vec{e}_{-i}) \leq 0 
                &&\iimplies&&
                \vec{\bm{f}}_{\star}(\vec{e}_{i}) = 0
                \text{ and }
                \vec{\bm{f}}_{\star}(\vec{e}_{-i}) \geq {c}(e_i) 
            \end{align*}
            Therefore,
            since $\vec{\bm{f}}_{\star}$ is feasible and obeys capacity constraints,
            $\vec{\bm{f}}_{\star}(\vec{e}_{i}) = 0$ and $\vec{\bm{f}}_{\star}(\vec{e}_{-i}) = {c}(e_i) = c_i$.
            
            If $\vec{\bm{f}}_{\star}(\vec{e}_{+i}) = 0$ and $\vec{\bm{f}}_{\star}(\vec{e}_{-i}) = {c}(e_i) = c_i$
            then  $\vec{c}_{f_{\star}}(\vec{e}_{+i})>0$ and $\vec{c}_{f_{\star}}(\vec{e}_{-i})=0$,
            and $\bm{f}_{\star}(e_i) = \vec{\bm{f}}_{\star}(\vec{e}_{+i}) - \vec{\bm{f}}_{\star}(\vec{e}_{-i})) = 0 - c_i = -c_i$.
            Then from Eq.~\eqref{eq:ohcp:dir-optimality:reduced-costs-residual-complex} 
            the corresponding residual costs are:
            $\vec{\bm{x}}^{\pi}(\vec{e}_{+i}) \geq 0$ and $\vec{\bm{x}}^{\pi}(\vec{e}_{-i}) \leq 0$.
            
            $\vec{e}_{+i}$ is non-negative: $\vec{\bm{x}}^{\pi}(\vec{e}_{+i}) \geq 0$
            
            
      \item[Case 2: ] 
            If $\vec{\bm{x}}^{\pi}(\vec{e}_{i}) = \vec{\bm{x}}^{\pi} (\eta \dart \tau) < 0$
            then
            $\vec{c}_{f}(\vec{e}_{i}) \leq 0$ and $\vec{c}_{f}(\vec{e}_{-i}) \leq 0$
            by Eq.~\eqref{eq:ohcp:dir-optimality:reduced-costs-residual-complex}.
%
            From the definition of residual capacities in Eq.~\eqref{eq:def:residual-capacity} and  Eq.~\eqref{eq:def:residual-capacity-rev},
            and since $\vec{\bm{f}}$ is feasible = obeys capacity constrains:
            \begin{align*}
                \vec{c}_{f}(\vec{e}_{i})  \leq 0 
                &&\iff&&
                {c}(e_i) - \vec{\bm{f}}(e_{i}) + \vec{\bm{f}}(e_{-i}) \leq 0 \\
                &&\iimplies&&
                \vec{\bm{f}}(e_{i}) \geq {c}(e_i) = c_i
                \text{ and }
                \vec{\bm{f}}(e_{-i}) = 0
            \end{align*}
            Therefore $\vec{\bm{f}}(e_{i}) = {c}(e_i) = c_i$ and $\vec{\bm{f}}(e_{-i}) = 0$.
            
%
            

      \item[Case 3: ] 
            If $0 < \vec{\bm{f}}_{\star}(\eta \dart \tau) < c(e_i)$ 
            the the residual complex contains both $(\eta \dart \tau)$ and $(\tau \dart \eta)$  
            $\iimplies$
            $0 < \vec{\bm{f}}_{\star}(\tau \dart \eta) < c(e_i)$ 
            $\iimplies$
            $\vec{\bm{x}}^{\pi} (\eta \dart \tau) \geq 0$ and 
            $\vec{\bm{x}}^{\pi} (\tau \dart \eta) \geq 0$,
            which implies\textit{}
            $\vec{\bm{x}}^{\pi} (\eta \dart \tau) = \vec{\bm{x}}^{\pi} (\tau \dart \eta) = 0$
            
  \end{itemize}
\end{proof}

\section{Conclusions}
We extended Sullivan's duality results for OHCP to complexes with non-trivial homology in Theorem~\ref{th:ohcp:strong-duality}.
Additionally, we established stronger optimality conditions for directed min-cost flow solutions of OHCP in Theorem~\ref{th:ohcp:dir-optimality:complementary-slackness}. 
The negative $q$-cocycle optimality condition \ref{th:ohcp:dir-optimality:neg-cocycles}
provides a topological characterization of these conditions.

  \chapter{Network flow models for multiscale simplicial flat norm}

\section{Cohomology flows and MSFN}
\label{sec:coflows-msfn}

Consider the following integer linear program with an objective function that computes ${\mbb{F}}(\bm{z})$ in Eq.~\eqref{eq:def:simplex-flat-norm} and its dual program:
\begin{align}
\label{eq:fnorm:flat-norm-lp}
        \begin{array}{rl}\dsp
            \min\limits_{\bm{x}, \bm{\pi}}\ 
            &\dsp  
                    {c}(\bm{x}^{+}) + {c}(\bm{x}^{-}) + {a}(\bm{\pi}^{+}) + {a}(\bm{\pi}^{-})
            \\[2pt]
            \text{s.t.} &\dsp 
                    \bd_{q + 1} (\bm{\pi}^{+} - \bm{\pi}^{-}) + \bm{x}^{+} - \bm{x}^{-} = \bm{z}         
            \\ 
            &\dsp 
                    \bm{x}^{\pm}  \geq \bm{0}_{E}, \ \bm{\pi}^{\pm}  \geq \bm{0}_{F}
        \end{array}
    &&\overset{dual}{\Longrightarrow}&&
        \begin{array}{lll}\dsp
            \max\limits_{ \tilde{\bm{f}} }\ 
            &\dsp 
                    \tilde{\bm{f}}(-\bm{z}) \geq 0
            \\[2pt]
            \text{s.t.} &\dsp  
                    -\bm{a}_{F} \leq \cobd_{q + 1} \tilde{\bm{f}} \leq \bm{a}_{F}
            \\  
            &\dsp  
                    -\bm{c}_{E} \leq \tilde{\bm{f}} \leq \bm{c}_{E}
        \end{array}
\end{align}
where the dual variable is a $q$-cochain $\tilde{\bm{f}} \in  \mc{C}^{q}(\mc{K}; \integer)$, 
which we refer to as \embf{$q$-pseudocoflow},
and $\bm{\pi} = \bm{\pi}^{+} - \bm{\pi}^{-} \in \mc{C}_{q + 1}(\mc{K}; \integer)$ together with
$\bm{x} = \bm{x}^{+} - \bm{x}^{-} \in \mc{Z}_{q}(\mc{K}; \integer)$ constitute  the FN-decomposition of $\bm{z}$.
Recall that the embedding assumptions guarantee that the boundary matrix $[\bd_{q + 1}]$ is \emph{totally unimodular} \cite{ohcp2011}, 
and since all weights $\bm{c}$ and $\bm{a}^{\lambda}$ are integral, we can find an optimal integral solution in polynomial time for both programs in Eq.~\eqref{eq:fnorm:flat-norm-lp}, \cite{veinott1968integral, ibrahim2011simplicial}.

Its extension to $\bar{\mc{K}}$:
\begin{align}
\label{eq:fnorm:flat-norm-lp-barK}
        \begin{array}{rll}\dsp
            \min\limits_{\bm{x}, \bm{\pi}}\ 
            &\dsp  
                    \abs{\bm{c}(\bm{x})} + \abs{\bm{a}(\bm{\pi})}
            \\[4pt]
            \text{s.t.} &\dsp 
                    \bd_{q + 1} \bm{\pi} + \bm{x} = \bm{z}         
            \\ 
            \forall \nu_k \in \mc{V}      
                    &\dsp  \bm{\pi}(\nu_k)  = 0 
            \\[4pt]
            &\dsp 
                    \bm{x} \in \mc{Z}_{q}(\bar{\mc{K}}; \integer), \ \bm{\pi} \in \mc{C}_{q + 1}(\bar{\mc{K}}; \integer)
        \end{array}
    &&\overset{dual}{\Longrightarrow}&&
        \begin{array}{lll}\dsp
            \max\limits_{ \tilde{\bm{f}} }\ 
            &\dsp 
                    \tilde{\bm{f}}(-\bm{z}) \geq 0
            \\[4pt]
            \text{s.t.} &\dsp  
                    -\bm{a}_{\bar{F}} \leq \cobd_{q + 1} \tilde{\bm{f}} \leq \bm{a}_{\bar{F}}:
            \\[2pt]  
            &\dsp \quad
                    \cobd_{q + 1} \tilde{\bm{f}}(\nu_t) = \gamma_{t} > 0 \\ 
            &\dsp \quad
                    \cobd_{q + 1} \tilde{\bm{f}}(\nu_s) = -\gamma_{s} <0 \\
            &\dsp \quad
                  -a_j \leq  \cobd_{q + 1} \bm{f}(\tau_j)  \leq a_j 
            \\[4pt]  
            &\dsp  
                    -\bm{c}_{E} \leq \tilde{\bm{f}} \leq \bm{c}_{E}
        \end{array}
\end{align}

Now let's try to fix the $q$-pseudocoflow $\tilde{\bm{f}}$ from Eq.~\eqref{eq:fnorm:flat-norm-lp}:
it obeys capacity constraints on edge-simplices of $\mc{K}$, and only the circulation conservation conditions are broken.
Let's say  $\tilde{\bm{f}}$ assigns a non-zero value to a boundary circulation of some facet-simplex from $\mc{K}$:
$\tilde{\bm{f}}(\bd_{q + 1} \tau_j) = \gamma_j$ such that $-a_j \leq \gamma_j \leq a_j \iff \abs{\gamma_j} \leq a_j$ by the circulation imbalance bounds.
Let $\gamma^{U}_{j} \geq 0$ and  $\gamma^{D}_{j} \geq 0$ 
such that $\gamma_{j} = \gamma^{D}_{j} - \gamma^{U}_{j}$ and $\abs{\gamma_{j}} =\gamma^{D}_{j} + \gamma^{U}_{j}$,
therefore only one of them is non-zero and both of them are upper-bounded by the area-volumes of $\tau_j$: 
$\gamma^{D}_{j} \leq {a}_j$ and $\gamma^{U}_{j} \leq {a}_j$.
Let's define the following non-negative $(q + 1)$-cochains: 
$\bm{g}^{U}\!: \tau_j \to \gamma^{U}_j \geq 0 $ and $\bm{g}^{D}\!: \tau_j \to \gamma^{D}_j \geq 0 $ -- UP-inflow and  DOWN-drain. 
Their composition $\bm{g} = \bm{g}^{U} - \bm{g}^{D}$ must obey the  circulation imbalance conditions: 
$-\bm{a}_{F} \leq \bm{g} \leq \bm{a}_{F}$.

Let's denote the fixed pseudocoflow $\bm{\phi} \in \mc{H}^{q}(\mc{K}; \integer)$
such that it originates and terminates on the same voids as $\tilde{\bm{f}}$, 
obeys capacity constraints $-\bm{c}_{E} \leq {\bm{\phi}} \leq \bm{c}_{E}$, 
and it satisfies the \textit{circulation conservation} constraints: 
$\cobd_{d + 1} \bm{\phi} = \cobd_{q + 1} \tilde{\bm{f}}  + \bm{g} = 0$.

Let's now rewrite the fixed max-flux problem from Eq.~\eqref{eq:fnorm:flat-norm-lp} and its dual:
\begin{align}
\label{eq:fnorm:flat-norm-fixed}
        \begin{array}{lll}\dsp
            \max\limits_{ \tilde{\bm{f}}, \bm{g} }\ 
            &\dsp 
                    \tilde{\bm{f}}(-\bm{z}) + 0 \cdot \bm{g}
            \\[4pt]
            \text{s.t.} &\dsp  
                     \tilde{\bm{f}}(\bd_{d + 1} \tau_j) +  \bm{g}(\tau_j) = 0
            \\[4pt]
            &\dsp  
                   \tilde{\bm{f}}\in \mc{C}^{q}(\mc{K}; \integer)\! : \:  
                   -\bm{c}_{E} \leq \tilde{\bm{f}} \leq \bm{c}_{E}
            \\[2pt]
            &\dsp  
                   {\bm{g}}\in \mc{C}^{q + 1}(\mc{K}; \integer)\! : \:  
                    -\bm{a}_{F} \leq \bm{g}  \leq \bm{a}_{F}
            \\[2pt]
        \end{array}
    &&\overset{dual}{\Longrightarrow}&&
        \begin{array}{rl}\dsp
            \min\limits_{\bm{x}, \bm{\pi}, \bm{\psi}}\ 
            &\dsp  
                    0 \cdot \bm{\pi} + 
                    {a}(\bm{\psi}^{U}) + {a}(\bm{\psi}^{D}) + 
                    {c}(\abs{\bm{x}})  
            \\[2pt]
            \text{s.t.} &\dsp 
                    \bd_{d + 1} \bm{\pi} + \bm{x}^{+} - \bm{x}^{-} = \bm{z}         
            \\[2pt] 
            &\dsp 
                    \bm{\pi} + \bm{\psi}^{U} - \bm{\psi}^{D} = \bm{0}_{F}         
            \\[2pt] 
            &\dsp 
                    \bm{\pi} \in \integer^{F}; \ 
                    \bm{\psi}^{U / D} \geq \bm{0}_{F}; \
                    \bm{x}^{\pm}  \geq \bm{0}_{E} 
        \end{array}
\end{align}
where we have $\bm{g} = \bm{g}^{U} - \bm{g}^{D} = -\cobd_{q + 1} \tilde{\bm{f}}$ and
 $\bm{\pi} = \bm{\psi}^{D} - \bm{\psi}^{U}$.

\section{The extended complex \texorpdfstring{$\hat{\mc{K}}$}{K-hat}}

Let's consider a simplicial complex $\hat{\mc{K}}$ that is constructed from $\mc{K}$ 
by adding a pair of new $q$- and $(q + 1)$-simplices to corresponding skeletons for each existing facet-simplex in $\mc{K}_{q + 1}$,
and then extending/updating the related chain groups and (co)boundary operator.
Let 
$\mc{U} = \mc{U}_{q} = \big\{ u_j \mid \dim u_j = q \text{ and } \tau_j \in \mc{K}_{q + 1} \big\}$
and
$\mc{M} = \mc{M}_{q + 1} = \big\{ \mu_{j} \mid \dim \mu_j = q + 1 \text{ and } \tau_j \in \mc{K}_{q + 1} \big\}$ 
be the sets of new $q$- and $(q + 1)$-simplices defined for each $\tau_j \in \mc{K}_{q + 1}$, 
to which we refer as \embf{$u$-edges} and \embf{$\mu$-facets} or \embf{upper-edges} and \embf{upper-facets}, respectively.
Note that $\abs{\mc{U}_{q}} = \abs{\mc{M}_{q + 1}} = \abs{\mc{K}_{q + 1}} = F$.
We emphasize that these simplices and related constructions are \textit{purely algebraical} 
(``we are only concerned with the algebraic properties of the construction and do not actually need to modify the simplicial complex''\cite{generalizedMaxflow2021}).
Then we formally define $\hat{\mc{K}}$, its chain complexes, and extended capacity function as follows:
\begin{align}
\label{eq:up-complex}
  && &\underline{ \hat{\mc{K}} = \mc{K} \cup \mc{U}_{q} \cup \mc{M}_{q + 1} } & &&
\\[4pt]
  \label{eq:up-complex-skeletons}
    &\hat{\mc{K}}_{q} \asgn  \mc{K}_{q} \cup \mc{U}_{q} &
    &\hat{\mc{K}}_{q + 1} \asgn \mc{K}_{q + 1} \cup \mc{M}_{q + 1} &
    &\hat{\mc{K}}_{q^{\prime}} = \mc{K}_{q^{\prime}} \text{ for } q^{\prime} < q
  \\[1pt]
  \label{eq:up-complex-chains}
    &\mc{C}_{q}(\hat{\mc{K}}) \asgn \mc{C}_{q}({\mc{K}}) \bigoplus\limits_{j = 1}^{F} [u_j] &
    &\mc{C}_{q + 1}(\hat{\mc{K}}) \asgn \mc{C}_{q + 1}({\mc{K}}) \bigoplus\limits_{j = 1}^{F} [\mu_j] &
    &\mc{C}_{q^{\prime}}(\hat{\mc{K}}) = \mc{C}_{q^{\prime}}({\mc{K}})
\\ &\rule{100pt}{0.5pt}& &\rule{100pt}{0.5pt}& &\rule{100pt}{0.5pt}& \notag \\
  \label{eq:up-complex-cobd-up}
    &\hat{\cobd}_{q + 1} u_j = \tau_j - \mu_j & 
    &\hat{\bd}_{q + 1} \mu_j = - u_j & 
    &\forall u_j \in \mc{U}_{q}, \ \forall \mu_j \in \mc{M}_{q + 1}
  \\[1pt]
  \label{eq:up-complex-cobd}
    &\hat{\cobd}_{q + 1} e_i = \cobd_{q + 1} e_i &
    &\hat{\bd}_{q + 1} \tau_j = \bd_{q + 1} \tau_j + u_j& 
    &\forall e_i \in \mc{K}_{q}, \ \forall \tau_j \in \mc{K}_{q + 1}
\\ &\rule{100pt}{0.5pt}& &\rule{100pt}{0.5pt}& &\rule{100pt}{0.5pt}& \notag \\
  \label{eq:up-complex-bd-q}
    &\hat{\bd}_{q} u_j = 0 & 
    &\hat{\bd}_{q} e_i = \bd_{q} e_i & 
    &\forall u_j \in \mc{U}_{q}, \ \forall e_i \in \mc{K}_{q }
\\ &\rule{100pt}{0.5pt}& &\rule{100pt}{0.5pt}& &\rule{100pt}{0.5pt}& \notag \\
  \label{eq:up-complex-capacity}
    &\hat{c}: \hat{\mc{K}}_{q} \to \integer_{+}&
    &\hat{c}(e_i) = c(e_i) = c_i &
    &\hat{c}(u_j) = a(\tau_j) = a_j &
\end{align}
where $\cobd_{q + 1}: \mc{C}_{q}(\mc{K}) \to \mc{C}_{q + 1}(\mc{K})$ 
and $\bd_{q + 1}: \mc{C}_{q + 1}(\mc{K}) \to \mc{C}_{q}(\mc{K})$ are the original coboundary and boundary operators defined on $\mc{K}$.
We denote the sizes of the extended skeletons as follows:
$\hat{E} \asgn \abs{\hat{\mc{K}}_{q}} = E + F$ and $\hat{F} \asgn \abs{\hat{\mc{K}}_{q + 1}} = 2F $.

We also extend the voidless complex 
$\bar{\mc{K}} = \mc{K} \oplus \mc{V}$ 
in a similar fashion:
$\hatbar{\mc{K}} \asgn \bar{\mc{K}} \oplus \bar{\mc{U}}_{q} \oplus \bar{\mc{M}}_{q + 1}$,
where 
$\bar{\mc{M}}_{q + 1} = \big\{ \mu_{k} \mid \dim \mu_k = q + 1 \text{ and } \sigma_k \in {\mc{K}}_{q + 1} \cup \mc{V} \big\}$
and
$\bar{\mc{U}}_{q} = \big\{ u_k \mid \dim u_k = q \text{ and } \hat{\cobd}_{q + 1} u_k = \sigma_k - \mu_k \big\}$.
The extended capacity function $\hatbar{c}: \hatbar{\mc{K}}_{q} \to \integer_{+}$ is defined as follows:
$\hatbar{c}(e_i) = c(e_i) = c_i$ for all $e_i \in \mc{K}_{q}$, 
$\hatbar{c}(u_j) = a(\tau_j) = a_j $ for all $\tau_j \in \mc{K}_{q + 1}$, 
and $\hatbar{c}(u_k) = \infty $ for all $\nu_k \in \mc{V}_{q + 1}$.

\subsection{Pserudocoflows on extended complexes}

\paragraph*{Pseudocoflows on $\bar{\mc{K}}$:}

Let $\tld{\bm{f}} \in \mc{C}^{q}(\bar{\mc{K}}; \integer)\!: \sum_{\mc{S}} \nu_s \to \sum_{\mc{T}} \nu_t$ 
be a {$q$-pseudocoflow} on $\bar{\mc{K}}$ \emph{directed from inner to outer voids of $\bm{z}$}, 
and hence \embf{curling against $\bm{z}$}:
\begin{align}
\label{eq:pseudocoflow-barK}
  \tld{\bm{f}}(\bm{z}) \leq 0 
  \iff 
  \cobd_{q + 1} \tld{\bm{f}} = \sum\limits_{\mc{T}} \gamma_t \nu_t - \sum_{\mc{S}} \gamma_s \nu_s 
                             + \sum_{\mc{F}} \gamma_j \tau_j
\end{align}
where $\mc{V} = \mc{S} \cup \mc{T}$, and $\mc{F} = \mc{K}_{q + 1}$.
Note that the coboundary of $\tld{\bm{f}}$ on the original complex $\mc{K}$ 
has only the last sum: 
$\cobd_{q + 1} \tld{\bm{f}} \rvert_{\mc{K}} =  \cobd_{q + 1}^{\mc{K}} \tld{\bm{f}} =  \cobd_{q + 1}^{F} \tld{\bm{f}} = \big[ \cobd_{q + 1} \tld{\bm{f}} \big]_{\mc{K}} 
= \sum_{\mc{F}} \gamma_j \tau_j$.
The {pseudocoflow} $\tld{\bm{f}}$ is \embf{feasible} if:
\begin{align}
\label{eq:fnorm:feasible-pseudocoflow}
    &-{\bm{c}}_{E} \leq  \tld{\bm{f}} \leq \bm{c}_{E}:&
    &\dsp  -{c_i} \leq  \tld{\bm{f}}(e_i) \leq {c}_i    &
    &\forall e_i \in \bar{\mc{E}} = \bar{\mc{K}}_{q} = {\mc{K}}_{q}
\\[4pt]\dsp
    &-{\bm{a}}_{\bar{F}} \leq  \tld{\bm{f}} \leq \bm{a}_{\bar{F}}:&
    &\dsp  -{a}_j \leq  \tld{\bm{f}}(\bd_{q + 1} \tau_j) \leq {a}_j    &
    &\forall \tau_j \in {\mc{F}} = {\mc{K}_{q + 1}} 
  \\
    &&&\dsp  -\infty < \tld{\bm{f}}(\bd_{q + 1} \nu_k) < \infty    &
    &\forall \nu_k \in {\mc{V}} 
\end{align}

Previously, we have patched the pseudocoflow  in Eq.~\eqref{eq:fnorm:flat-norm-fixed} to a $q$-coflow $\bm{\phi} \in \mc{H}^{q}(\mc{K}; \integer)$
by adding a $(q + 1)$-cochain 
$\bm{g} \in \mc{C}^{q + 1}(\mc{K}; \integer)\!: \bm{g} = \bm{g}^{U} - \bm{g}^{D} = - \sum_{\mc{F}} \gamma_j \tau_j $ 
to the coboundary of  $\tld{\bm{f}}$: 
$\cobd_{d + 1} \bm{\phi} = \cobd_{q + 1} \tilde{\bm{f}}  + \bm{g} = 0$.
Note that on $\bar{\mc{K}}$ it entails 
$\cobd_{d + 1} \bm{\phi} = \cobd_{q + 1} \tilde{\bm{f}}  + \bm{g} = \sum_{\mc{T}} \gamma_t \nu_t - \sum_{\mc{S}} \gamma_s \nu_s$,
and therefore ${\bm{\phi}} \in \mc{C}^{q}(\bar{\mc{K}}; \integer)\!: \sum_{\mc{S}} \nu_s \to \sum_{\mc{T}} \nu_t$
and $\bm{\phi}(\bm{z}) \leq 0$. 
From the Weak Duality theorem~\ref{th:fnorm:weak-duality-SFN}:
\begin{align*}
  \bm{\phi}(\bm{z}) &= \tilde{\bm{f}}(\bm{z}) + \bm{g}(\bm{0}_{F}) = \tilde{\bm{f}}(\bm{z}) \leq 0
\\[2pt]
{\bm{\phi}}(\bm{z})  
        &= {\bm{\phi}}\big( \bm{x} + \bd_{d + 1} \bm{\pi} \big) 
        = \tilde{\bm{f}}(\bm{x}) + \tilde{\bm{f}}(\bd_{d + 1} \bm{\pi}) 
        = \tilde{\bm{f}}(\bm{x}) - \bm{g}(\bm{\pi})
        \leq 0
  \\
      &\iimplies
      {\abs{\bm{a}(\bm{\pi})} \geq } \
      \bm{g}(\bm{\pi}) \geq \tilde{\bm{f}}(-\bm{x}) \geq 0
\end{align*}

Since our goal is to minimize ${\bm{\phi}}(\bm{z})$, 
and the co-flux through $\bm{x}$ is bounded from below $\tilde{\bm{f}}(\bm{x}) \geq -\abs{\bm{c}(\bm{x})}$,
we need to maximize the value $\bm{g}(\bm{\pi})$ as much as possible without violating other constrains:
\begin{align*}
{\bm{\phi}}(\bm{z}) \leq 0 \to \min 
&&\iff&& 
  0 \leq \tld{\bm{f}}(-\bm{x}) \leq \abs{\bm{c}(\bm{x})} \to \max
\\
 &&&& 0 \leq \bm{g}(\bm{\pi}) \leq \abs{\bm{a}(\bm{\pi})} \to \max
\end{align*}

\paragraph*{Pseudocoflows on $\hat{\mc{K}}$:}
Now let's consider the max-flux formulation of a SFN problem on the extended complex $\hat{\mc{K}}$.
Let's denote the subgroups of $q$-th and $(q + 1)$-th chain groups generated by new elements from ${\mc{U}}_{q}$ and ${\mc{M}}_{q+ 1}$
as $\mc{C}({\mc{U}}_{q}; \integer) = \bigoplus{\mc{F}} [u_j]$ and $\mc{C}({\mc{M}}_{q + 1}; \integer) = \bigoplus_{\mc{F}} [\mu_j]$, respectively.
Then 
$\mc{C}_{q}(\hat{\mc{K}}; \integer) = \mc{C}_{q}({\mc{K}}; \integer) \oplus \mc{C}({\mc{U}}_{q}; \integer)$ 
and
$\mc{C}_{q + 1}(\hat{\mc{K}}; \integer) = \mc{C}_{q + 1}({\mc{K}}; \integer) \oplus  \mc{C}({\mc{M}}_{q + 1}; \integer)$.
For example, $\tld{\bm{f}} \in \mc{C}^{q}({\mc{K}}; \integer) \subset \mc{C}^{q }(\hat{\mc{K}}; \integer) \ \iff \ \tld{\bm{f}}(u_j) = 0$ for all $u_j \in \mc{U}_{q}$.

Let $\tld{\bm{g}} \in \mc{C}^{q }(\hat{\mc{K}}; \integer)\!: \tld{\bm{g}}(e_i) = 0$
for all $e_i \in \hat{\mc{K}}_{q} \setminus \mc{U}_q = \mc{K}_{q}$,
be a $q$-pseudocoflow defined on new $q$-cells such that 
\begin{align}
\label{eq:fnorm:hatK:auxcoflow}             %
  \tld{\bm{g}} 
      &= \sum_{u_j \in {\mc{U}}_{q}} -\gamma_{j} u_j 
       = \sum_{u_j \in {\mc{U}}_{q}} (\gamma^{U}_{j} - \gamma^{D}_{j}) u_j 
\\[2pt]
\label{eq:fnorm:hatK:auxcoflow:cobd}             %
  \hat{\cobd}_{q + 1} \tld{\bm{g}} 
      &= \bm{g} - \bm{g}_{\mc{M}} 
       = - \sum_{u_j \in \mc{U}_{q}} \gamma_j (\tau_j - \mu_j)
\\[2pt]
\label{eq:fnorm:hatK:auxcoflow:cobd-tau}         %
  \hat{\cobd}_{q + 1} \tld{\bm{g}}(\tau_j) 
      &= \bm{g}(\tau_j) = -\gamma_j 
       = - \cobd_{q + 1} \tld{\bm{f}}(\tau_j) = - \hat{\cobd}_{q + 1} \tld{\bm{f}}(\tau_j)
\end{align}
where $\bm{g}_{\mc{M}} \!: {\mc{M}}_{q + 1} \mapsto \integer$ is defined as $\bm{g}_{\mc{M}}(\mu_j) = -\gamma_j = \bm{g}(\tau_j) $.
Then let's define the fixed $q$-coflow 
$\hat{\bm{\phi}} \in \mc{C}^{q}(\hat{\mc{K}}; \integer)\!: \sum_{\mc{S}} \nu_s \to \sum_{\mc{T}} \nu_t$
as follows
\begin{align}
\label{eq:fnorm:hatK:fixed-coflow}
  \hat{\bm{\phi}} \asgn \tld{\bm{f}}  + \tld{\bm{g}}:
  && 
  \hat{\cobd}_{d + 1}  \hat{\bm{\phi}}(\tau_j) 
      = \hat{\cobd}_{q + 1} \tilde{\bm{f}}(\tau_j)  + \hat{\cobd}_{q + 1} \tilde{\bm{g}}(\tau_j) 
      = 0
  &&
  \text{for all } \tau_j \in \mc{K}_{q + 1}
  \\
\label{eq:fnorm:hatK:fixed-coflow-mu}
  &&
  \hat{\cobd}_{d + 1}  \hat{\bm{\phi}}(\mu_j) 
      = \hat{\cobd}_{q + 1} \tilde{\bm{f}}(\mu_j)  + \hat{\cobd}_{q + 1} \tilde{\bm{g}}(\mu_j)  
      = \gamma_j
  &&
  \text{for all } \mu_j \in \mc{M}_{q + 1}
  \\[4pt]
\label{eq:fnorm:hatK:fixed-coflow-Ecap}
  -{\bm{c}}_{E} \leq  \hat{\bm{\phi}}_{E} \leq \bm{c}_{E}:
  &&  -{c}_i \leq  \hat{\bm{\phi}}(e_i) = \tld{\bm{f}}(e_i)\leq {c}_i 
  && \forall e_i \in {\mc{K}_{q}}
  \\ 
\label{eq:fnorm:hatK:fixed-coflow-Ucap}
  -{\bm{a}}_{{F}} \leq  \hat{\bm{\phi}}_{U} \leq \bm{a}_{{F}}:
  &&  -{a}_j \leq  \hat{\bm{\phi}}(u_j) = \tld{\bm{g}}(u_i)\leq {a}_j 
  && \forall u_j \in {\mc{U}_{q}}
\end{align}
Although, $\hat{\bm{\phi}}$ is a pseudocoflow on the extended complex 
$\hat{\mc{K}}$: $\hat{\cobd}_{q + 1} \hat{\bm{\phi}} = \sum_{\mc{M}} \gamma_j \mu_j$,
it is a feasible coflow on the original complex $\mc{K}$ because $\hat{\cobd}_{d + 1}  \bm{\phi}(\tau_j)  = 0$,
thus we write $\hat{\bm{\phi}} \in \mc{Z}^{q}(\hat{\mc{K}}; \integer)$.
Note that $ \hat{\cobd}_{q + 1} \tilde{\bm{f}}(\mu_j) = 0$, since $\tld{\bm{f}}(u_j) = 0$, 
which implies $\hat{\cobd}_{q + 1} \tilde{\bm{f}} \equiv {\cobd}_{q + 1} \tilde{\bm{f}}$. 
Hence we have the following lemma.

\begin{lemma}
\label{lm:feasible-coflow-hatK}
Let $\hat{\bm{\phi}} = (\tilde{\bm{f}}, \tilde{\bm{g}})$ be a feasible solution for the Max-Flux problem on $\hat{\mc{K}}$ in Eq.~\eqref{eq:fnorm:flat-norm-fixed-lp-hatK}
    such that $\hat{\bm{\phi}}(\bm{z}) = \tilde{\bm{f}}(\bm{z}) \leq 0$.
    
Then $(\tilde{\bm{f}}, [\hat{\cobd}_{q + 1}\tilde{\bm{g}}]_{\mc{K}_{q + 1}} ) = (\tilde{\bm{f}}, {\bm{g}})$
    is a feasible solution for the Max-Flux problem on ${\mc{K}}$ in Eq.~\eqref{eq:fnorm:flat-norm-fixed}.
    
\end{lemma}

\subsection{Homologous cycles on \texorpdfstring{$\hat{\mc{K}}$}{K-hat}}
Let ${\bm{\tau}} = \sum_{\mc{F}} \tau_j \in \mc{C}^{q + 1}(\hat{\mc{K}}; \integer)$ 
and ${\bm{\mu}} = \sum_{\mc{M}} \mu_j \in \mc{C}^{q + 1}(\hat{\mc{K}}; \integer)$.
Let $\vec{\bm{u}} = \sum_{\mc{U}} u_j \in \mc{C}^{q}(\hat{\mc{K}}; \integer)$ be a $q$-copath defined on $\mc{U}_{q}$,
oriented towards the original $(q + 1)$-skeleton, 
$\vec{\bm{u}}\!:~(\mc{M}_{q + 1} \to \mc{K}_{q + 1})$ and
$\hat{\cobd}_{q + 1} \vec{\bm{u}} = \bm{\tau} - \bm{\mu}$.
Let $\hat{\bd}^{u}_{q + 1}  \bm{\pi} = \hat{\cobd}_{q + 1} \vec{\bm{u}} (\bm{\pi})$ 
for some  $\bm{\pi} \in \mc{C}_{q + 1}(\hat{\mc{K}}; \integer)$ such that  $\bm{\pi}(\mu_j) = 0$ for all $\mu_j \in \mc{M}_{q + 1}$.

Let $(\bm{x}, \bm{\pi})$ 
be a feasible solution for the SFN problem in Eq.~\eqref{eq:fnorm:flat-norm-lp}
such that 
\begin{itemize}
  \item $\bm{x} \in \mc{Z}_{q}(\hat{\mc{K}}; \integer)\!:   \bm{x} \sim \bm{z}$ on $\mc{K}$, and $\bm{x}(u_j) = \bm{z}(u_j) = 0$ for all $u_j \in \mc{U}_{q}$;
  \item $\bm{\pi} \in \mc{C}_{q + 1}(\hat{\mc{K}}; \integer)\!: \bd_{q + 1} \bm{\pi} = \bm{z} - \bm{x}$ and $\bm{\pi}(\mu_j) = 0$ for all $\mu_j \in \mc{M}_{q + 1}$.
\end{itemize}

Let $\hat{\bm{x}} \in \mc{C}_{q}(\hat{\mc{K}}; \integer)$ be a \emph{$q$-chain on $\hat{\mc{K}}$ homologous to $\bm{z}$ relative to $\hat{\bd}_{q + 1}$}:
\begin{align}
  \hat{\bm{x}} 
      &= \bm{z} - \hat{\bd}_{q + 1} \bm{\pi}
     \ = \bm{z} - {\bd}_{q + 1} \bm{\pi} - \hat{\cobd}_{q + 1} \vec{\bm{u}} (\bm{\pi})
       = \bm{x} + \bm{y}
\end{align}
where $\bm{y} \in \mc{C}_{q}(\hat{\mc{K}}; \integer)$ such that $\bm{y}(e_i) = 0$ for all $e_i \in \hat{\mc{K}}_{q} \setminus \mc{U}_{q}$,
whose value is defined by the boundary of $-\bm{\pi}$ on the new $q$-cells $u_j \in \mc{U}_{q}$,
i.e. $\bm{y}(u_j) = \big[ \hat{\bd}_{q + 1} (-\bm{\pi}) \big](u_j)$, see Eq.~\eqref{eq:up-complex-cobd-up}: 
\begin{align}
  \bm{y} &\asgn \vec{\bm{u}} \big( \hat{\bd}_{q + 1} (-\bm{\pi}) \big)
         = \hat{\cobd}_{q + 1} \vec{\bm{u}} (-\bm{\pi}) 
         = \sum_{j = 1}^{F} -\pi_j u_j
  \\
  \hat{\cobd}_{q + 1} \bm{y} 
        &= \sum_{j = 1}^{F} -\pi_j \cdot \hat{\cobd}_{q + 1} u_j
         = \sum_{j = 1}^{F}  \pi_j (\mu_j - \tau_j)
         = \bm{\pi}_{\mc{M}} - \bm{\pi}
\end{align}
where $\bm{\pi}_{\mc{M}}\!: \mc{M}_{q + 1} \mapsto \integer^{F} \in \mc{C}_{q + 1}(\hat{\mc{K}}; \integer)$
defined as $\bm{\pi}_{\mc{M}}(\mu_j) = \pi_j$ and $\bm{\pi}_{\mc{M}}(\tau_j) = 0$ 
for all $\mu_j \in \mc{M}_{q + 1}$ and $\tau_j \in \hat{\mc{K}}_{q + 1} \setminus \mc{M}_{q + 1}$.

Then, from the relevant  definitions of $\hat{\mc{K}}$ in Eq.~\eqref{eq:up-complex-bd-q} and \eqref{eq:up-complex-capacity}, 
the cost-capacity of $\hat{\bm{x}}$ and its $q$-boundary are:
\begin{align}
  \hat{{c}}({\hat{\bm{x}}} )
      &= \hat{{c}}(\bm{x}) + \hat{{c}}(\bm{y})
       = {{c}}(\bm{x}) + {{a}}(\bm{y})
       = {{c}}(\bm{x}) + {{a}}(-\bm{\pi})
  \\
  \hat{\bd}_{q}{\hat{\bm{x}}} 
      &= \hat{\bd}_{q} \bm{x} + \hat{\bd}_{q} \bm{y}
       = {\bd}_{q} \bm{x} + {\textstyle \sum_{\mc{U}_{q}} -\pi_j \cdot \hat{\bd}_{q} u_j}
       = 0 + {\textstyle \sum_{\mc{U}_{q}} -\pi_j \cdot 0}
       = 0
\end{align}
which implies that $\hat{\bm{x}}$ and ${\bm{y}}$ are $q$-cycles on $\hat{\mc{K}}$: 
$\hat{\bm{x}} \in  \mc{Z}_{q}(\hat{\mc{K}}; \integer)$
and ${\bm{y}} \in  \mc{Z}_{q}(\hat{\mc{K}}; \integer)$.

\begin{lemma}[1]
\label{lm:homology-on-hatK}
  Given a $q$-cycle  $\bm{z} \in \mc{Z}_{q}(\hat{\mc{K}}; \integer)$ such that $\bm{z}(u_j) = 0$ for all $u_j \in \mc{U}_{q}$,
  and a $(q + 1)$-chain $\bm{\pi} \in \mc{C}_{q + 1}(\hat{\mc{K}}; \integer)\!$ such that $\bm{\pi}(\mu_j) = 0$ for all $\mu_j \in \mc{M}_{q + 1}$,
  let $\hat{\bm{x}} =  \bm{z} - \hat{\bd}_{q + 1} \bm{\pi} \in \mc{C}_{q}(\hat{\mc{K}}; \integer)$ 
  be a $q$-chain homologous to $\bm{z}$ on $\hat{\mc{K}}$, $\hat{\bm{x}} \sim \bm{z}$.
  Let $q$-chains $\bm{x}, \bm{y} \in \mc{C}_{q}(\hat{\mc{K}}; \integer)$ such that 
  $\bm{x}(u_j) = 0$ for all $u_j \in \mc{U}_{q}$ and 
  $\bm{y}(e_i) = 0$ for all $e_i \in \hat{\mc{K}}_{q} \setminus \mc{U}_{q}$,
  be the $\mc{K}_{q}$ and $\mc{U}_{q}$ components of $\hat{\bm{x}}$, respectively:
  $\hat{\bm{x}} = \bm{x} + \bm{y}$.
  
  Then the following statements are true:
  \begin{itemize}
      \item[(1)] $\hat{\bm{x}}$ is a $q$-cycle on $\hat{\mc{K}}$: $\hat{\bm{x}} \in  \mc{Z}_{q}(\hat{\mc{K}}; \integer)$;

      \item[(2)] ${\bm{y}}$ is a null-homologous $q$-cycle on $\hat{\mc{K}}$: 
                 ${\bm{y}} \in \mc{B}_{q}(\hat{\mc{K}}; \integer)$;

      \item[(3)] ${\bm{x}}$ is a $q$-cycle on ${\mc{K}}$ that is homologous to $\bm{z}$: 
                ${\bm{x}} \in  \mc{Z}_{q}({\mc{K}}; \integer)$ and $\bm{x} \sim \bm{z}$ on $\mc{K}$.
  \end{itemize}
\end{lemma}
\begin{proof}
  Let $\hat{\mc{F}}_{q + 1} = \hat{\mc{K}}_{q + 1} \setminus \mc{M}_{q + 1}$
  and $\hat{\mc{E}}_{q} = \hat{\mc{K}}_{q} \setminus \mc{U}_{q}$.
  From the definition of the boundary operator $\hat{\bd}_{q + 1}$ on $\hat{\mc{K}}$ in Eq.~\eqref{eq:up-complex-cobd} and Eq.~\eqref{eq:up-complex-cobd-up}:
  \begin{align*}
      \hat{\bd}_{q + 1} \bm{\pi} 
      &= \sum_{\hat{\mc{F}}_{q + 1}} \bm{\pi}(\tau_j) \cdot  \hat{\bd}_{q + 1} \tau_j
       + \sum_{\mc{M}_{q + 1}}       \bm{\pi}(\mu_j) \cdot  \hat{\bd}_{q + 1} \mu_j
       = \sum_{\hat{\mc{F}}_{q + 1} } \pi_j \cdot  ({\bd}_{q + 1} \tau_j + u_j) 
       + \sum_{\mc{M}_{q + 1}} 0
  \\
      &= \sum_{\hat{\mc{F}}_{q + 1} } \pi_j \cdot  {\bd}_{q + 1} \tau_j
       + \sum_{{\mc{U}}_{q } } \pi_j \cdot  u_j 
       = \bd_{q + 1}  \bm{\pi} + \bm{u}^{\pi}
  \end{align*}
  where we set $\bm{u}^{\pi}\!: \mc{U}_{q } \mapsto \integer^{F} \in \mc{C}_{q }(\hat{\mc{K}}; \integer)$
  such that $\bm{u}^{\pi}(e_i) = 0$ for all $e_i \in \hat{\mc{K}}_{q} \setminus \mc{U}_{q}$.
  Note that $\hat{\bd}_{q + 1} \bm{\pi} \in \mc{B}_{q}(\hat{\mc{K}}; \integer) \subset \mc{Z}_{q}(\hat{\mc{K}}; \integer)$
  and ${\bd}_{q + 1} \bm{\pi} \in \mc{B}_{q}({\mc{K}}; \integer) \subset \mc{Z}_{q}(\hat{\mc{K}}; \integer)$.

  \textbf{(1):} 
  Since $\bm{z} \in \mc{Z}_{q}(\hat{\mc{K}}; \integer)$ such that $\bm{z}(u_j) = 0$ for all $u_j \in \mc{U}_{q}$
  we have 
  \begin{align*}
    \hat{\bd}_{q} \hat{\bm{x}} 
          = \hat{\bd}_{q} \bm{z} - \hat{\bd}_{q} \hat{\bd}_{q + 1} \bm{\pi}
          = 0 - 0
          = 0
    && \iimplies &&
    \hat{\bm{x}} \in \mc{Z}_{q}(\hat{\mc{K}}; \integer)
    &&\text{ and }&&
        \hat{\bm{x}} \sim \bm{z} \text{ on } \hat{\mc{K}}
  \end{align*}
  
  \textbf{(2):} 
  From the above 
  \begin{align*}
    \hat{\bm{x}} = \bm{z} - \hat{\bd}_{q + 1} \bm{\pi}
                 = \bm{z} - \bd_{q + 1} \bm{\pi} - \bm{u}^{\pi}
  \end{align*}
  where 
  $\bm{z} - \bd_{q + 1} \bm{\pi} \!: \hat{\mc{E}}_{q} \mapsto \integer^{E} \in \mc{C}_{q}(\hat{\mc{K}}; \integer)$
  such that  $(\bm{z} - \bd_{q + 1} \bm{\pi})(u_j) = 0$ for all $u_j \in \mc{U}_{q}$,
  and $\bm{u}^{\pi}\!: \mc{U}_{q } \mapsto \integer^{F} \in \mc{C}_{q }(\hat{\mc{K}}; \integer)$ that is zero everywhere on $\hat{\mc{E}}_{q}$.
  
  Let's set $\bm{x} \asgn \bm{z} - \bd_{q + 1} \bm{\pi}$ and $\bm{y} \asgn -\bm{u}^{\pi}$.
  By the definition of  $q$-boundary  on $\hat{\mc{K}}$ in Eq.~\eqref{eq:up-complex-bd-q}:
  \begin{align*}
      \hat{\bd}_{q} \bm{y} 
      = -\hat{\bd}_{q} \bm{u}^{\pi}
         = {\textstyle \sum_{\mc{U}_{q}} -\pi_j \cdot \hat{\bd}_{q} u_j}
         = {\textstyle \sum_{\mc{U}_{q}} -\pi_j \cdot 0}
         = 0 
     && \iimplies && 
     {\bm{y}} \in  \mc{Z}_{q}(\hat{\mc{K}}; \integer)
  \end{align*}
  
  Moreover:
  \begin{align*}
    \bm{y} = \hat{\bm{x}} - \bm{x} 
           = \hat{\bm{x}} - \bm{z} + \bd_{q + 1} \bm{\pi}
           = \hat{\bd}_{q + 1} \bm{\pi} + \bd_{q + 1} \bm{\pi}
     && \iimplies && 
     \bm{y} \sim 0 \text{ on } \hat{\mc{K}}
     && \iimplies && 
     {\bm{y}} \in  \mc{B}_{q}(\hat{\mc{K}}; \integer)
  \end{align*}

  \textbf{(3):} 
  $\bm{x}$ is a $q$-cycle on $\hat{\mc{K}}$:
  \begin{align*}
    0 = \hat{\bd}_{q} \hat{\bm{x}} = \hat{\bd}_{q} {\bm{x}}  + \hat{\bd}_{q} {\bm{y}} = \hat{\bd}_{q} {\bm{x}}  + 0
    &&\iimplies&&
    \hat{\bd}_{q} {\bm{x}} = 0
    &&\iimplies&&
   {\bm{x}} \in  \mc{Z}_{q}(\hat{\mc{K}}; \integer)
  \end{align*}
  and since it is zero on $\mc{U}_{q}$ it is also a $q$-cycle on $\mc{K}$: ${\bm{x}} \in  \mc{Z}_{q}({\mc{K}}; \integer)$.
  Moreover, since there is a $q$-boundary on $\mc{K}$, given by $\bd_{q + 1} \bm{\pi}$, such that 
  $\bm{x} = \bm{z} - \bd_{q + 1} \bm{\pi}$,
  the $q$-cycle $\bm{x}$  is homologous to $\bm{z}$ on $\mc{K}$: $\bm{x} \sim \bm{z}$.
  \begin{align*}
    \bm{x}  = \hat{\bm{x}} - \bm{y} 
    &&\text{ such that }&&
    {\bm{x}} \in  \mc{Z}_{q}({\mc{K}}; \integer)
    &&\text{ and }&&
    \bm{x} \sim \bm{z} \text{ on } \mc{K}
  \end{align*}
\end{proof}

\subsection{Coflux on \texorpdfstring{$\hat{\mc{K}}$}{K-hat}}
\begin{theorem}
\label{th:coflow-preserve-flux}
  Let $\bm{\pi} \in \mc{C}_{q + 1}(\hat{\mc{K}}; \integer)$ such that $\bm{\pi}(\mu_j) = 0$ for all $\mu_j \in \mc{M}_{q + 1}$,
  and $\hat{\bm{x}} = \bm{z} - \hat{\bd}_{q + 1} \bm{\pi} \in \mc{Z}_{q}(\hat{\mc{K}}; \integer)$ be a $q$-cycle homologous to $\bm{z}$ on $\hat{\mc{K}}$.
  Let $\hat{\bm{x}} = \bm{x} + \bm{y}$ such that
  $\bm{x}\!:  {\mc{K}}_{q} \mapsto \integer^{E} \in \mc{Z}_{q}(\hat{\mc{K}}; \integer)$
  and $\bm{y}\!: {\mc{U}}_{q} \mapsto \integer^{F} \in \mc{Z}_{q}(\hat{\mc{K}}; \integer)$.
  
  Then a feasible $q$-coflow $\hat{\bm{\phi}} = \tilde{\bm{f}} + \tilde{\bm{g}}$ preserves the flux value within a homology class,
  in other words:
  \begin{align}
  \label{eq:fixed-coflow-preserve-flux}
       \hat{\bm{\phi}}(\hat{\bm{x}})  = \hat{\bm{\phi}}(\bm{z})  = \tilde{\bm{f}}({\bm{z}}) \leq 0
  \end{align}
  
\end{theorem}
\begin{proof}
  Recall that $\hat{\cobd}_{q + 1} \hat{\bm{\phi}} = \hat{\cobd}_{q + 1} \tilde{\bm{f}} + \hat{\cobd}_{q + 1} \tilde{\bm{g}} = \sum_{\mc{M}} \gamma_j \mu_j$. 
  Then 
  \begin{align*}
      \hat{\cobd}_{q + 1} \hat{\bm{\phi}} (\bm{\pi}) 
      = 0 \cdot \bm{\pi} + \sum_{\mc{M}} 0 \cdot \gamma_j \mu_j
      = 0
  \end{align*}
  
  Then either way we have the lemma's claim:
  \begin{align*}
      \hat{\bm{\phi}}(\hat{\bm{x}})  
      &= \hat{\bm{\phi}}({\bm{z}} - \hat{\bd}_{q + 1} \bm{\pi})  
      = \hat{\bm{\phi}}({\bm{z}}) - \hat{\cobd}_{q + 1} \hat{\bm{\phi}}(\bm{\pi})  
      = \hat{\bm{\phi}}({\bm{z}})
      = \tilde{\bm{f}}({\bm{z}})
  \\
      \hat{\bm{\phi}}(\hat{\bm{x}})  
      &= \hat{\bm{\phi}}({\bm{x}} + \bm{y})  
       = \tilde{\bm{f}}({\bm{x}}) + \tilde{\bm{g}}(\bm{y})  
       = \tilde{\bm{f}}({\bm{z}})  - \tilde{\bm{g}}(\bm{y})   + \tilde{\bm{g}}(\bm{y})  
       = \tilde{\bm{f}}({\bm{z}})  
  \end{align*}
  where we used 
  $\hat{\bm{\phi}}({\bm{z}}) = \tilde{\bm{f}}({\bm{z}}) = \tilde{\bm{f}}({\bm{x}}) - \bm{g}(\bm{\pi}) 
                       = \tilde{\bm{f}}({\bm{x}}) - \hat{\cobd}_{q+1} \tilde{\bm{g}}(\bm{\pi})
                       = \tilde{\bm{f}}({\bm{x}}) +  \tilde{\bm{g}}( \bm{y} ) \leq 0
  $
  from the proof of {the weak duality theorem}~\ref{th:fnorm:weak-duality-SFN},
  which implies $\tilde{\bm{f}}({\bm{x}}) = \tilde{\bm{f}}({\bm{z}}) -  \tilde{\bm{g}}( \bm{y} ) \leq 0$.
  Note that
  \begin{align*}
      \hat{\bm{\phi}}({\bm{y}}) = \tilde{\bm{g}}(\bm{y}) 
                          = \hat{\cobd}_{q+1} \tilde{\bm{g}}(-\bm{\pi})
                          = - {\bm{g}}(\bm{\pi})
                          = {\cobd}_{q+1} \tilde{\bm{f}}(\bm{\pi})
                          =  \tilde{\bm{f}}({\bd}_{q+1} \bm{\pi})
                          =  \tilde{\bm{f}}(\bm{z} - \bm{x})
  \end{align*}
  
\end{proof}

Now we can restate {the fixed max-coflux problem} from Eq.~\eqref{eq:fnorm:flat-norm-lp} and Eq.~\eqref{eq:fnorm:flat-norm-fixed}
and its dual on the extended complex $\hat{\mc{K}}$.
Recall that the fixed $q$-coflow is a $q$-cochain 
$\hat{\bm{\phi}} = \tilde{\bm{f}} + \tilde{\bm{g}} \in \mc{C}^{q}(\hat{\mc{K}}; \integer)\!: \sum_{\mc{S}} \nu_s \to \sum_{\mc{T}} \nu_t$
with a non-empty coboundary on $\hat{\mc{K}}$: $\hat{\cobd}_{q + 1} \bm{\phi} = \sum_{\mc{M}} \gamma_j \mu_j$.
\begin{align}
\label{eq:fnorm:flat-norm-fixed-lp-hatK}
        \begin{array}{lll}\dsp
            \min\limits_{ \hat{\bm{\phi}}  }\ 
            &\dsp 
                    \hat{\bm{\phi}}(\bm{z}) \leq 0
            \\[4pt]
            \text{s.t.} &\dsp  
                  \hat{\cobd} \hat{\bm{\phi}}(\tau_j) =  
                  (\tilde{\bm{f}} + \tilde{\bm{g}})(\hat{\bd} \tau_j) = 0
            \\[2pt]
            &\dsp
                  \hat{\cobd} \hat{\bm{\phi}}(\mu_j) =  
                  (\tilde{\bm{f}} + \tilde{\bm{g}})(\hat{\bd} \mu_j) = \gamma_j
            \\[4pt]
            &\dsp  
                   -\bm{c}_{E} \leq \hat{\bm{\phi}}_{E} = \tilde{\bm{f}} \leq \bm{c}_{E}
            \\[2pt]
            &\dsp  
                    -{\bm{a}}_{{F}} \leq  \hat{\bm{\phi}}_{U} = \tilde{\bm{g}} \leq \bm{a}_{{F}}
            \\[2pt]
            &\dsp  
                  \tilde{\bm{f}} \!: \mc{K}_{q} \mapsto \integer \in \mc{C}^{q}(\hat{\mc{K}})  
            \\[2pt]
            &\dsp  
                  \tilde{\bm{g}} \!: \mc{U}_{q} \mapsto \integer \in \mc{C}^{q}(\hat{\mc{K}})  
            \\[2pt]
            \\[2pt]
        \end{array}
    \overset{dual}{\Longrightarrow}
        \begin{array}{rl}\dsp
            \min\limits_{\hat{\bm{\pi}}, \hat{\bm{x}}}\ 
            &\dsp  
                    \bm{0}_{F} (\bm{\pi}) + 
                    \bm{\gamma}_{F} (\bm{\pi}_{\mc{M}}) + 
                    \abs{{a}(\bm{y})} +
                    \abs{{c}(\bm{x})}
            \\[4pt]
            \text{s.t.} &\dsp 
                    \big[ \hat{\bd} (\bm{\pi} + \bm{\pi}_{\mc{M}}) \big]_{\mc{K}_{q}} + \bm{x} = \bm{z}         
            \\[2pt] 
            &\dsp 
                     \big[ \hat{\bd} (\bm{\pi} + \bm{\pi}_{\mc{M}}) \big]_{\mc{U}_{q}} + \bm{y} = \bm{0}_{F}         
            \\[2pt] 
            &\dsp 
                    \bm{\pi}_{\mc{M}} = \bm{0}_{F} 
            \\[4pt] 
            &\dsp 
                    \bm{\pi}\in \mc{C}_{q + 1}(\hat{\mc{K}}) 
            \\[2pt] 
            &\dsp 
                    \bm{\pi}_{\mc{M}} \in \mc{C}_{q + 1}(\hat{\mc{K}}) 
            \\[2pt] 
            &\dsp 
                    \bm{y}  \in \mc{Z}_{q}(\hat{\mc{K}})  
            \\[2pt]
            &\dsp 
                    \bm{x} \in \mc{Z}_{q}(\hat{\mc{K}})  
        \end{array}
\end{align}
where $\bm{\gamma}_{F} = [\gamma_{1}, \ldots, \gamma_{F}] \in \integer^{F}$ is an integer vector of some flow values,
$\hat{\bm{x}} = \bm{x} + \bm{y}$ is a $q$-cycle homologous to $\bm{z}$ on $\hat{\mc{K}}$,
and $\hat{\bm{\pi}} = \bm{\pi} + \bm{\pi}_{\mc{M}} \equiv \bm{\pi}$ is the bounded $(q + 1)$-chain that has non-zero values only on the original $(q + 1)$-simplices of $\mc{K}$,
and $[ \hat{\bd}_{q + 1} \hat{\bm{\pi}} ]_{\mc{K}_{q}}  =  [ \hat{\bd}_{q + 1} \bm{\pi} ]_{\mc{K}_{q}} = {\bd}_{q + 1} \bm{\pi}$
and $[ \hat{\bd}_{q + 1} \hat{\bm{\pi}} ]_{\mc{U}_{q}}  =  [ \hat{\bd}_{q + 1} \bm{\pi} ]_{\mc{U}_{q}} = \hat{\bd}_{q + 1} \bm{\pi} - {\bd}_{q + 1} \bm{\pi}$.
We can rewrite the dual problem without $\bm{\pi}_{\mc{M}}$ that is always zero:
\begin{align}
\label{eq:fnorm:flat-norm-fixed-dual-lp-hatK}
        \begin{array}{rl}\dsp
            \min\limits_{{\bm{\pi}}, \hat{\bm{x}}}\ 
            &\dsp  
                    \bm{0} \cdot \bm{\pi} + 
                    \abs{{a}(\bm{y})} +
                    \abs{{c}(\bm{x})}
            \\[4pt]
            \text{s.t.} &\dsp 
                     \big[ \hat{\bd}_{q + 1} \bm{\pi} \big]_{\mc{K}_{q}} + \bm{x} = \bm{z}         
            \\[2pt] 
            &\dsp 
                     \big[ \hat{\bd}_{q + 1} \bm{\pi} \big]_{\mc{U}_{q}} + \bm{y} = \bm{0}_{F}         
            \\[2pt] 
            \forall \mu_j \in \mc{M}_{q + 1}: &\dsp 
                    \bm{\pi}(\mu_j) = 0
            \\[4pt] 
            &\dsp 
                    \bm{\pi}\!: \mc{K}_{q + 1} \mapsto \integer^{F} \in \mc{C}_{q + 1}(\hat{\mc{K}}; \integer) 
            \\[2pt] 
            &\dsp 
                    \bm{y} \!: \mc{U}_{q} \mapsto \integer^{F} \in \mc{Z}_{q}(\hat{\mc{K}}; \integer)  
            \\[2pt]
            &\dsp 
                    \bm{x} \!: \mc{K}_{q} \mapsto \integer^{E} \in \mc{Z}_{q}(\hat{\mc{K}}; \integer)  
        \end{array}
    && \iff &&
        \begin{array}{rl}\dsp
            \min\limits_{{\bm{\pi}}, \hat{\bm{x}}}\ 
            &\dsp  
                    \bm{0} \cdot \bm{\pi} + 
                    \abs{{a}(\bm{y})} +
                    \abs{{c}(\bm{x})}
            \\[4pt]
            \text{s.t.} &\dsp 
                     \hat{\bd}_{q + 1} \bm{\pi} + \bm{x} + \bm{y} = \bm{z}         
            \\[2pt] 
            &\dsp 
                    \bm{\pi}(\mu_j) = 0
            \\[2pt] 
            &\dsp 
            \\[4pt] 
            &\dsp 
                    \bm{\pi} \in \mc{C}_{q + 1}(\hat{\mc{K}}; \integer) 
            \\[2pt] 
            &\dsp 
                    \bm{y} \in \mc{Z}_{q}(\hat{\mc{K}}; \integer)  
            \\[2pt]
            &\dsp 
                    \bm{x} \in \mc{Z}_{q}(\hat{\mc{K}}; \integer)  
        \end{array}
\end{align}

\begin{corollary}[of {Lemma \ref{lm:homology-on-hatK}}]
    Let $(\hat{\bm{x}}, \bm{\pi})$ be a feasible solution for the SFN problem on $\hat{\mc{K}}$ in Eq.~\eqref{eq:fnorm:flat-norm-fixed-dual-lp-hatK},
    where $\hat{\bm{x}} = \bm{x} + \bm{y}$ is a $q$-cycle homologous to $\bm{z}$ on $\hat{\mc{K}}$,
    and $\bm{\pi}\in \mc{C}_{q + 1}(\hat{\mc{K}}; \integer)$ such that $\hat{\bd}_{q + 1} \bm{\pi} = \bm{z} - \hat{\bm{x}}$ and $\bm{\pi}(\mu_j) = 0$ for all $\mu_j \in \mc{M}_{q + 1}$
    is the bounded $(q + 1)$-chain.
    
    Then $({\bm{x}}, \bm{\pi})$ is a feasible solution to the SFN problem on ${\mc{K}}$ in Eq.~\eqref{eq:fnorm:flat-norm-fixed},
    where $\bm{x} \sim \bm{z}$ on $\mc{K}$ and $\bm{x} = \bm{z} - \bd_{q + 1} \bm{\pi}$.
\end{corollary}

\section{Weak and strong duality: MSFN}\label{sec:fnorm:duality-theorems}

\subsection{Weak Duality (SFN)}
\label{subsec:fnorm:weak-dual}
Now we can prove the Weak and Strong Duality theorems for the fixed Max-Flux and Flat Norm LPs in Eq.~\eqref{eq:fnorm:flat-norm-fixed}.
\begin{theorem}[Weak Duality (SFN)]
\label{th:fnorm:weak-duality-SFN}
    Let $(\bm{x}, \bm{\pi})$ be a feasible solution for the MSFN problem in Eq.~\eqref{eq:fnorm:flat-norm-lp}
    such that $\bm{x} \sim \bm{z}$ and $\bd_{q + 1} \bm{\pi} = \bm{z} - \bm{x}$ for 
    $\bm{\pi} \in \mc{C}_{q + 1}(\mc{K}; \integer)$,
    and let $(\tilde{\bm{f}}, \bm{g})$ be a feasible solution for the Max-Flux problem in Eq.~\eqref{eq:fnorm:flat-norm-fixed}
    such that $\tilde{\bm{f}}(\bm{z}) \leq 0$.

    
    Then the following holds:
    \begin{align}
    \label{eq:fnorm:weak-duality-SFN}
            0 \leq \tilde{\bm{f}}(-\bm{z})  
              \leq \abs{\bm{c}({\bm{x}})} +  \abs{\bm{a}({\bm{\pi}})} 
    \end{align}
\end{theorem}
\begin{proof}
    $\tilde{\bm{f}}(\bm{z}) \leq 0$
    \begin{align*}
        \tilde{\bm{f}}(\bm{z})  
        &= \tilde{\bm{f}}(\bm{z}) 
            - \cobd_{d + 1} \tilde{\bm{f}}(\bm{\pi}) + \cobd_{d + 1} \tilde{\bm{f}}(\bm{\pi})
        = \tilde{\bm{f}}(\bm{z} - \bd_{d + 1} \bm{\pi}) - \bm{g}(\bm{\pi})
        = \tilde{\bm{f}}(\bm{x}) - \bm{g}(\bm{\pi})
        \leq 0
    \end{align*}

    where 
    \begin{align*}
        \bm{g}(\bm{\pi}) 
        = \bm{g} \left(  \sum_{\tau_j \in \mc{K}_{d + 1}} \pi_j \tau_j \right)
        &= \sum_{\tau_j \in \mc{K}_{d + 1}} 
               \pi_j \left( \bm{g}^{U}(\tau_j) - \bm{g}^{D}(\tau_j)  \right)
   \\[2pt]
        &\leq \sum_{\tau_j \in \mc{K}_{d + 1}} 
               \abs{\pi_j} \left( \bm{g}^{U}(\tau_j) + \bm{g}^{D}(\tau_j)  \right)
        \leq \sum_{\tau_j \in \mc{K}_{d + 1}} 
               \abs{\pi_j} \cdot a_j
       = a (\abs{\bm{\pi}})
    \end{align*}
    
    Since $\tilde{\bm{f}}(\bm{x}) \leq 0$ and $-\bm{c} \leq \tilde{\bm{f}} \leq \bm{c}$,
    we have our claim:
    \begin{align*}
        0 \leq \tilde{\bm{f}}(-\bm{z}) 
        = \bm{g}(\bm{\pi}) - \tilde{\bm{f}}(\bm{x}) 
        = \bm{g}(\bm{\pi}) + \tilde{\bm{f}}(-\bm{x}) 
          \leq  \abs{\bm{a}({\bm{\pi}})}  + \abs{\bm{c}({\bm{x}})}
    \end{align*}
\end{proof}

\subsection{Weak Duality on \texorpdfstring{$\hat{\mc{K}}$}{K-hat} (SFN)}
\label{subsec:fnorm:weak-dual-hatK}
Now we can restate the weak duality theorem~\ref{th:fnorm:weak-duality-SFN} on the extended complex $\hat{\mc{K}}$:
\begin{theorem}[Weak Duality on $\hat{\mc{K}}$]
\label{th:fnorm:weak-duality-SFN-hatK}
    Let $\hat{\bm{x}} = \bm{x} + \bm{y}$ be a $q$-cycle homologous to $\bm{z}$ on $\hat{\mc{K}}$,
    and a $(q + 1)$-chain $\bm{\pi}\in \mc{C}_{q + 1}(\hat{\mc{K}}; \integer)$ 
    such that $\hat{\bd}_{q + 1} \bm{\pi} = \bm{z} - \hat{\bm{x}}$ and $\bm{\pi}(\mu_j) = 0$ for all $\mu_j \in \mc{M}_{q + 1}$.
    So that  $(\hat{\bm{x}}, \bm{\pi})$ is a feasible solution for the SFN problem on $\hat{\mc{K}}$ in Eq.\eqref{eq:fnorm:flat-norm-fixed-dual-lp-hatK},\eqref{eq:fnorm:flat-norm-fixed-lp-hatK}.
    
    
    Let $\hat{\bm{\phi}} = (\tilde{\bm{f}}, \tilde{\bm{g}})$ be a feasible solution for the Max-Flux problem on $\hat{\mc{K}}$ in Eq.~\eqref{eq:fnorm:flat-norm-fixed-lp-hatK}
    such that $\hat{\bm{\phi}}(\bm{z}) = \tilde{\bm{f}}(\bm{z}) \leq 0$.
    
    Then the following holds:
    \begin{align}
    \label{eq:fnorm:weak-duality-SFN-hatK}
            0 \leq \hat{\bm{\phi}}(-\bm{z})  
              \leq \abs{\bm{c}({\bm{x}})} +  \abs{\bm{a}({\bm{\pi}})} 
    \end{align}
\end{theorem}

\subsection{Strong Duality on \texorpdfstring{$\hat{\mc{K}}$}{K-hat} (SFN)}
\label{subsec:fnorm:strong-dual-hatK}

\begin{theorem}[Strong duality on $\hat{\mc{K}}$ (SFN)]
\label{th:fnorm:strong-duality-hatK}
  Let $(\hat{\bm{x}}_{\star} = {\bm{x}}_{\star} + {\bm{y}}_{\star}, \bm{\pi}_{\star})$ be an optimal solution for the SFN problem in Eq.~\eqref{eq:fnorm:flat-norm-fixed-dual-lp-hatK},
  such that $\hat{\bm{x}}_{\star} \sim \bm{z}$ on $\hat{\mc{K}}$ and $\hat{\bd}_{d + 1} \bm{\pi}_{\star} = \bm{z} - \hat{\bm{x}}_{\star}$.
  And, let $\hat{\bm{\phi}}_{\star} = (\tilde{\bm{f}}_{\star}, \tilde{\bm{g}}_{\star})$ be an optimal solution for the Max-Flux/Min-Cost coflow problem in Eq.~\eqref{eq:fnorm:flat-norm-fixed-lp-hatK}
  such that $\hat{\bm{\phi}}_{\star}(\bm{z}) \leq 0$.

  Equality in Eq.~\eqref{eq:fnorm:weak-duality-SFN-hatK} (and hence in Eq.~\eqref{eq:fnorm:weak-duality-SFN})
  is achieved whenever the following Complementary Slackness conditions (orthogonality conditions) hold:
  \begin{align}
    \begin{array}{ccccll}
        \hat{\bm{x}}_{\star}(e_i) > 0 &\iff& {\bm{x}}_{\star}(e_i) > 0     &\iff& \hat{\bm{\phi}}_{\star}(e_i) = - {c_i} \\
        \hat{\bm{x}}_{\star}(e_i) < 0 &\iff& {\bm{x}}_{\star}(e_i) < 0     &\iff& \hat{\bm{\phi}}_{\star}(e_i) = \me{c_i} \\
        \hat{\bm{x}}_{\star}(u_j) > 0 &\iff& \bm{\pi}_{\star}(\tau_j) > 0  &\iff& \hat{\bm{\phi}}_{\star}(u_j) = \me {a_j}  \\
        \hat{\bm{x}}_{\star}(u_j) < 0 &\iff& \bm{\pi}_{\star}(\tau_j) < 0  &\iff& \hat{\bm{\phi}}_{\star}(u_j) =   - {a_j}  \\
    \end{array}
  \end{align}
\end{theorem}
\begin{proof}
  Recall that the cost-capacity, Eq.~\eqref{eq:up-complex-capacity}, of the optimal $q$-cycle $\hat{\bm{x}}_{\star}$ 
      is the minimization objective of SFN in Eq.~\eqref{eq:fnorm:flat-norm-fixed-dual-lp-hatK} and the SFN on the original complex in Eq.~\eqref{eq:fnorm:flat-norm-lp}:
  \begin{align*}
    \hat{c}(\hat{\bm{x}}) = c(\bm{x}) + a(\bm{y}) = c(\bm{x}) + a(-\bm{\pi}) 
    &&\iimplies&&
    \abs{\hat{c}(\hat{\bm{x}})}  = \abs{c(\bm{x})} + \abs{a(\bm{\pi})} \to \min
  \end{align*}
  
  The optimal $q$-cycle $\hat{\bm{x}}_{\star}$ has smallest capacity out of all $q$-cycles homologous to $\bm{z}$ on $\hat{\mc{K}}$, 
  in particular $\abs{\hat{c}(\hat{\bm{x}}_{\star})} \leq \abs{\hat{c}(\bm{z})} = \abs{c(\bm{z})}$.
  Meanwhile, by {Lemma~\ref{th:coflow-preserve-flux}}, the value of any feasible $q$-coflow $\hat{\bm{\phi}}$ on $\bm{z}$ and $\hat{\bm{x}}_{\star}$ is the same: 
  $\hat{\bm{\phi}}(\hat{\bm{x}}_{\star}) = \hat{\bm{\phi}}(\bm{z}) = \tilde{\bm{f}}(\bm{z}) \leq 0$.
  Out of all feasible coflows $\hat{\bm{\phi}}_{\star}$ has the largest absolute flux through $\hat{\bm{x}}_{\star}$,
  which, if equality in Eq.~\eqref{eq:fnorm:weak-duality-SFN-hatK}/\eqref{eq:fnorm:weak-duality-SFN} is reached,
  is equal to the capacity of $\hat{\bm{x}}_{\star}$: 
  \begin{align*}
      \hat{\bm{\phi}}_{\star}(-\hat{\bm{x}}_{\star})  = \abs{\hat{\bm{\phi}}_{\star}(\hat{\bm{x}_{\star}})} = \abs{\hat{c}(\hat{\bm{x}}_{\star})}
  \end{align*}
  It can happen only if $\hat{\bm{\phi}}_{\star}$ saturates all $q$-edges of $-\hat{\bm{x}}_{\star}$ up to their  capacity.
  
  For $e_i \in \mc{K}_{q}$:
  \begin{align*}
      \hat{\bm{\phi}}_{\star}(-\hat{\bm{x}}_{\star}(e_i)) 
      &= \abs{\hat{c}(\hat{\bm{x}}_{\star}(e_i))}
      \iff
  \\[4pt]
      \iff&
        -\hat{\bm{x}}_{\star}(e_i) \cdot \hat{\bm{\phi}}_{\star}(e_i)
      =  \abs{\hat{\bm{x}}_{\star}(e_i)} \cdot \hat{{c}}(e_i)
      &&\iff&&
        -{\bm{x}}_{\star}(e_i) \cdot \tilde{\bm{f}}_{\star}(e_i)
      =  \abs{\bm{x}_{\star}(e_i)} \cdot {{c}}(e_i)
  \\
      \iff&
      \begin{cases}
          \hat{\bm{\phi}}_{\star}(e_i) = - {c}(e_i) ,  & \text{if } \hat{\bm{x}}_{\star}(e_i) > 0 \\
          \hat{\bm{\phi}}_{\star}(e_i) = \me {c}(e_i) ,& \text{if } \hat{\bm{x}}_{\star}(e_i) < 0 \\
      \end{cases}
      &&\iff&&
      \begin{cases}
          \tilde{\bm{f}}_{\star}(e_i) = - {c}(e_i) ,& \text{if } \bm{x}_{\star}(e_i) > 0 \\
          \tilde{\bm{f}}_{\star}(e_i) = \me {c}(e_i) ,& \text{if } \bm{x}_{\star}(e_i) < 0 \\
      \end{cases}
  \end{align*}

  For $u_j \in \mc{U}_{q}$:
  \begin{align*}
      &\hat{\bm{\phi}}_{\star}(-\hat{\bm{x}}_{\star}(u_j)) 
      = \abs{\hat{c}(\hat{\bm{x}}_{\star}(u_j))}
      \iff
  \\[4pt]
      &\iff
        -\hat{\bm{x}}_{\star}(u_j) \cdot \hat{\bm{\phi}}_{\star}(u_j)
      =  \abs{\hat{\bm{x}}_{\star}(u_j)} \cdot \hat{{c}}(u_j)
      &&\iff&&
        -{\bm{y}}_{\star}(u_j) \cdot \tilde{\bm{g}}_{\star}(u_j)
      =  \abs{\bm{y}_{\star}(u_j)} \cdot {{a}}(u_j)
  \\
      &\iff
      \begin{cases}
          \hat{\bm{\phi}}_{\star}(u_j) =   - {a}(u_j) ,  & \text{if } \hat{\bm{x}}_{\star}(u_j) > 0 \\
          \hat{\bm{\phi}}_{\star}(u_j) = \me {a}(u_j) ,  & \text{if } \hat{\bm{x}}_{\star}(u_j) < 0 \\
      \end{cases}
      &&\iff&&
      \begin{cases}
          \tilde{\bm{g}}_{\star}(u_j) =   - {a}(u_j) ,& \text{if } \bm{y}_{\star}(u_j) > 0 \\
          \tilde{\bm{g}}_{\star}(u_j) = \me {a}(u_j) ,& \text{if } \bm{y}_{\star}(u_j) < 0 \\
      \end{cases}
  \\
      &&&\iff&&
      \begin{cases}
          \tilde{\bm{g}}_{\star}(u_j) =   - {a}(u_j) ,& \text{if } \bm{\pi}_{\star}(\tau_j) < 0 \\
          \tilde{\bm{g}}_{\star}(u_j) = \me {a}(u_j) ,& \text{if } \bm{\pi}_{\star}(\tau_j) > 0 \\
      \end{cases}
  \end{align*}

\end{proof}

\subsection{Strong Duality on \texorpdfstring{${\mc{K}}$}{K} (SFN)}
\label{subsec:fnorm:strong-dual}

\begin{theorem}[Strong duality on ${\mc{K}}$ (SFN)]
\label{th:fnorm:strong-duality}
  Let $({\bm{x}}_{\star}, \bm{\pi}_{\star})$ be an optimal solution for the SFN problem in Eq.~\eqref{eq:fnorm:flat-norm-fixed},
  such that ${\bm{x}}_{\star} \sim \bm{z}$ on ${\mc{K}}$ and ${\bd}_{d + 1} \bm{\pi}_{\star} = \bm{z} - {\bm{x}}_{\star}$.
  And, let ${\bm{\phi}}_{\star} = (\tilde{\bm{f}}_{\star}, {\bm{g}}_{\star})$ be an optimal solution for the Max-Flux/Min-Cost coflow problem in Eq.~\eqref{eq:fnorm:flat-norm-fixed}
  such that ${\bm{\phi}}_{\star}(\bm{z}) = \tilde{\bm{f}}_{\star}(\bm{z}) \leq 0$ 
  and ${\bm{\phi}}_{\star}(\bm{x}_{\star}) = \tilde{\bm{f}}_{\star}(\bm{x}_{\star}) + \bm{g}_{\star}(\bm{\pi}_{\star}) \leq 0$.

  Equality in Eq.~\eqref{eq:fnorm:weak-duality-SFN} 
  is achieved whenever the following Complementary Slackness conditions (orthogonality conditions) hold:
  \begin{align}
    \begin{array}{cclccll}
        {\bm{x}}_{\star}(e_i) > 0     &\iff& \tilde{\bm{f}}_{\star}(e_i) =  - {c_i} &\qquad&
        \bm{\pi}_{\star}(\tau_j) > 0  &\iff& {\bm{g}}_{\star}(\tau_j)    = \me{a_j} \\
        {\bm{x}}_{\star}(e_i) < 0     &\iff& \tilde{\bm{f}}_{\star}(e_i) = \me{c_i} &\qquad&
        \bm{\pi}_{\star}(\tau_j) < 0  &\iff& {\bm{g}}_{\star}(\tau_j)    =  - {a_j} \\
    \end{array}
  \end{align}
\end{theorem}
\begin{proof}

  
  From the {Weak Duality theorem~\ref{th:fnorm:weak-duality-SFN}} for any feasible coflow ${\bm{\phi}} = (\tilde{\bm{f}}, {\bm{g}})$
  and any $\bm{x} \overset{\pi}{\sim} \bm{z}$ on $\mc{K}$:
  \begin{align*}
    \bm{\phi}(\bm{z}, \bm{0}_{F}) &= \tilde{\bm{f}}(\bm{z}) + \bm{g}(\bm{0}_{F}) = \tilde{\bm{f}}(\bm{z}) \leq 0
  \\[2pt]
  {\bm{\phi}}(\bm{z}, \bm{0}_{F})  
          &= {\bm{\phi}}\big( \bm{x} + \bd_{d + 1} \bm{\pi}, \bm{0}_{F} \big) 
          = \tilde{\bm{f}}(\bm{x}) + \tilde{\bm{f}}(\bd_{d + 1} \bm{\pi}) 
          = \tilde{\bm{f}}(\bm{x}) - \bm{g}(\bm{\pi})
          = {\bm{\phi}}(\bm{x}, -\bm{\pi})
          \leq 0
    \\[4pt]
        &\iimplies
        {\abs{\bm{a}(\bm{\pi})} \geq } \
        \bm{g}(\bm{\pi}) \geq \tilde{\bm{f}}(\bm{x})
    \\
        &\iimplies
        {\bm{\phi}}(\bm{x}, \bm{0}_{F}) 
        = \tilde{\bm{f}}(\bm{x})  
        = \tilde{\bm{f}}(\bm{z}) + \bm{g}(\bm{\pi}) 
        = {\bm{\phi}}(\bm{z}, \bm{\pi}) 
  \end{align*}
  
  Since our goal is to minimize ${\bm{\phi}}(\bm{z})$, 
  and the co-flux through $\bm{x}$ is bounded from below $\tilde{\bm{f}}(\bm{x}) \geq -\abs{\bm{c}(\bm{x})}$,
  we need to maximize the value $\bm{g}(\bm{\pi})$ as much as possible without violating other constrains:
  \begin{align*}
  {\bm{\phi}}(\bm{z}, \bm{0}_{F}) \leq 0 \to \min 
  &&\iff&& 
    -\abs{\bm{c}(\bm{x})} \leq \tld{\bm{f}}(\bm{x}) \leq 0 \to \min
  \\
   &&&& \tilde{\bm{f}}(\bm{x}) \leq \bm{g}(\bm{\pi}) \leq \abs{\bm{a}(\bm{\pi})} \to \max
  \end{align*}
  
  The optimal pair of chains $({\bm{x}}_{\star}, \bm{\pi}_{\star})$ has the smallest total volume-cost 
  $\abs{c(\bm{x}_{\star})} + \abs{a(\bm{\pi}_{\star})}$ out of all feasible solutions $(\bm{x}, \bm{\pi})$ such that  
  $\bd_{q + 1}\pi = \bm{z} - \bm{x} \iff \bm{x} \sim \bm{z}$ on $\mc{K}$.
  In particular, $\abs{c(\bm{x}_{\star})} + \abs{a(\bm{\pi}_{\star})} \leq \abs{c(\bm{z})} + \abs{a(\bm{0}_{F})} = \abs{ c(\bm{z}) }$.
  
  On the other hand, the flux of any feasible coflow ${\bm{\phi}} = (\tilde{\bm{f}}, {\bm{g}})$ on $\bm{z}$ is same as on $({\bm{x}}_{\star}, -\bm{\pi}_{\star})$:
  $\bm{\phi}(\bm{z}) = \bm{\phi}({\bm{x}}_{\star}, -\bm{\pi}_{\star}) = \tilde{\bm{f}}(\bm{x}) - \bm{g}(\bm{\pi})$.
  Out of all feasible coflows ${\bm{\phi}}_{\star} = (\tilde{\bm{f}}_{\star}, {\bm{g}}_{\star})$ has the largest absolute flux through $\bm{z}$, and thus through $({\bm{x}}_{\star}, \bm{\pi}_{\star})$.
  If equality in Eq.~\eqref{eq:fnorm:weak-duality-SFN} is reached, then this absolute flux should be equal to the total volume-capacity of  $({\bm{x}}_{\star}, \bm{\pi}_{\star})$:
  \begin{align*}
    {\bm{\phi}}_{\star}(-{\bm{x}}_{\star}, \bm{\pi}_{\star})
    = \tilde{\bm{f}}_{\star}(-{\bm{x}}_{\star}) + \bm{g}_{\star} (\bm{\pi}_{\star})
    = \abs{{\bm{\phi}}_{\star}({\bm{x}}_{\star}, -\bm{\pi}_{\star})}
    = \abs{c({\bm{x}}_{\star})} +  \abs{a(\bm{\pi}_{\star})}
  \end{align*}
  
  It can happen only if $\tilde{\bm{f}}_{\star}$ saturates all $q$-edges of $-{\bm{x}}_{\star}$ up to their  capacity
  and ${\bm{g}}_{\star}$ adds maximum allowed amount of flow to the \emph{circulation/curl} of $\tilde{\bm{f}}_{\star}$ 
  on the boundaries of $(q + 1)$-facets of $\bm{\pi}_{\star}$.
  Recall that $\bm{\pi}_{\star}$ is positive on \emph{clockwise} $\tau_j$'s, while $\tilde{\bm{f}}$ is \emph{curling counter-clockwise},
  and hence is negative on the clockwise boundaries of $(q + 1)$-facets. 
  
  For $e_i \in \mc{K}_{q}$:
  \begin{align*}
      \tilde{\bm{f}}_{\star}(-{\bm{x}}_{\star}(e_i)) 
      &= \abs{{c}({\bm{x}}_{\star}(e_i))}
      \iff
  \\[4pt]
      \iff&
        -{\bm{x}}_{\star}(e_i) \cdot \tilde{\bm{f}}_{\star}(e_i)
      =  \abs{\bm{x}_{\star}(e_i)} \cdot {{c}}(e_i)
  \\
      \iff&
      \begin{cases}
          \tilde{\bm{f}}_{\star}(e_i) = - {c}(e_i) ,& \text{if } \bm{x}_{\star}(e_i) > 0 \\
          \tilde{\bm{f}}_{\star}(e_i) = \me {c}(e_i) ,& \text{if } \bm{x}_{\star}(e_i) < 0 \\
      \end{cases}
      &&\iff&&
      \begin{cases}
          {\bm{\phi}}_{\star}(e_i) = - {c}(e_i) ,  & \text{if } {\bm{x}}_{\star}(e_i) > 0 \\
          {\bm{\phi}}_{\star}(e_i) = \me {c}(e_i) ,& \text{if } {\bm{x}}_{\star}(e_i) < 0 \\
      \end{cases}
  \end{align*}
  
  For $\tau_j \in \mc{K}_{q + 1}$:
  \begin{align*}
      {\bm{g}}_{\star}({\bm{\pi}}_{\star}(\tau_j)) 
      &= \abs{{a}({\bm{\pi}}_{\star}(\tau_j))}
      \iff
  \\[4pt]
      \iff&
        {\bm{\pi}}_{\star}(\tau_j) \cdot {\bm{g}}_{\star}(\tau_j)
      =  \abs{\bm{\pi}_{\star}(\tau_j)} \cdot {{a}}(\tau_j)
  \\
      \iff&
      \begin{cases}
          {\bm{g}}_{\star}(\tau_j) = \me {a}(\tau_j) ,& \text{if } \bm{\pi}_{\star}(\tau_j) > 0 \\
          {\bm{g}}_{\star}(\tau_j) =   - {a}(\tau_j) ,& \text{if } \bm{\pi}_{\star}(\tau_j) < 0 \\
      \end{cases}
      &&\iff&&
      \begin{cases}
          {\bm{\phi}}_{\star}(\tau_j) = \me {a}(\tau_j) ,  & \text{if } {\bm{\pi}}_{\star}(\tau_j) > 0 \\
          {\bm{\phi}}_{\star}(\tau_j) =   - {a}(\tau_j) ,  & \text{if } {\bm{\pi}}_{\star}(\tau_j) < 0 \\
      \end{cases}
  \end{align*}  
  
\end{proof}

\section{Directed case (SFN)}
\label{subsec:fnorm:directed}

\subsection{Construction of the dual-graph \texorpdfstring{$\vec{\mc{K}}^{*}$}{K-asterisk}}
We construct the directed \embf{dual graph} 
$\vec{\mc{K}}^{*} = ({\mc{F}}^{*}, \vec{\mc{E}}^{*} = \vec{\mc{E}}^{*}_{+} \cup \vec{\mc{E}}^{*}_{-})$ of ${\mc{K}}$ as follows.

The nodes of ${\mc{K}}^{*}$ are in 1-to-1 corresponds with the facet-simplices:
$\mc{F}^{*} = \vec{\mc{K}}^{*}_{0} \cong \mc{K}_{q + 1}$ are the nodes dual to the original $(q + 1)$-simplices.

\begin{algorithm}[!ht]
    
    \SetKwData{From}{from} \SetKwData{To}{to} \SetKwData{Length}{arc length}
    \SetKwFunction{Capacity}{capacity} \SetKwFunction{Cost}{cost}
    \SetKwFunction{ShortestPathDist}{ShortestPathDist}
    \SetKwInOut{Input}{input}\SetKwInOut{Output}{output}
    
    \Input{ 
            $\bar{\mc{K}}^{*} = (\mc{F}^{*} \cup \mc{V}^{*}, \vec{\mc{E}}^{*} = \vec{\mc{E}}^{*}_{+} \cup \vec{\mc{E}}^{*}_{-} )$, 
            $\bm{z}^{*} \in \mc{Z}^{1}(\mc{K}^{*}; \integer) $,
            $\bm{c} \in \integer^{E}$, $\bm{a} \in \integer^{F}$, $\lambda \in \integerplus$ 
    }
    \Output{ 
            $\hat{\mc{K}}^{*}_{\vvec{\textsc{fn}}} 
                  = (\hat{\mc{F}}^{*}_{\vvec{\textsc{fn}}}, \hat{\mc{E}}^{*}_{\vvec{\textsc{fn}}})$ -- flow network for MSFN,
    }
    \BlankLine
    
    \tcp*[h]{DIST LABELS of VOIDS $\equiv$ HOMOLOGY CLASS of $\bm{z}$}
    
    $\dist^{*}_{\bm{z}} \in \mc{C}^{0}(\mc{K}^{*}; \integer) \leftarrow$ 
    \ShortestPathDist {\From = $\nu^{*}_{0}$, \To = $\bar{\mc{F}}^{*}$, \Length = $\bm{z}^{*}$}\;
    \BlankLine

    \tcp*[h]{VOIDS = SOURCES + TRANSITS + SINKS}
    
    $\mc{S} \leftarrow \big\{ \nu^{*}_{s} \in \mc{V}^{*} \mid \dist^{*}_{\bm{z}}(\nu^{*}_{s}) > 0   \big\}$,  
    $\mc{T} \leftarrow \big\{ \nu^{*}_{t} \in \mc{V}^{*} \mid \dist^{*}_{\bm{z}}(\nu^{*}_{t}) < 0   \big\}$,
    $\mc{Q} \leftarrow \emptyset$\;
    \lIf { $\mc{S} \neq \emptyset$ and $\mc{T} = \emptyset$}{
            $\mc{T} \leftarrow \big\{ \nu^{*}_{t} \in \mc{V}^{*} \mid \dist^{*}_{\bm{z}}(\nu^{*}_{t}) = 0 \big\}$ 
    }
    \lElseIf{ $\mc{S} = \emptyset$ and $\mc{T} \neq \emptyset$}{
            $\mc{S} \leftarrow \big\{ \nu^{*}_{s} \in \mc{V}^{*} \mid \dist^{*}_{\bm{z}}(\nu^{*}_{s}) = 0 \big\}$ 
    }
    \lElse{
            $\mc{Q} \leftarrow \big\{ \nu^{*}_{k} \in \mc{V}^{*} \mid \dist^{*}_{\bm{z}}(\nu^{*}_{k}) = 0   \big\}$
    }
    \BlankLine

    \tcp*[h]{CONSTRUCT FLOW NETWORK}

    $\hat{\mc{F}}^{*}_{\vvec{\textsc{fn}}} \leftarrow \mc{F}^{*} \cup \mc{V}^{*} \cup \{ S^{*}, T^{*}, U^{*}, D^{*} \}$  
    \tcp*[f]{{\small \{source, sink, up, down\}}}
    
    $\vec{\mc{E}}^{*}_{\textsc{st}} \leftarrow 
            \vec{\mc{E}}^{*}
            \cup \{ (S^{*} \to \nu^{*}_k) \}_{\mc{S} \cup \mc{Q}}
            \cup \{ ( \nu^{*}_k \to T^{*}) \}_{\mc{T} \cup \mc{Q}}
            \cup \{ ( T^{*} \to S^{*}) \}
    $
    \tcp*[f]{OHCP network}



    $\vec{\mc{E}}^{*}_{\textsc{u}} \leftarrow 
            \{ (S^{*} \to U^{*}) \} \cup \{ (U^{*} \to \tau^{*}_j) \}_{j = 1}^{F}
    $
    \tcp*[f]{up coboundary}

    $\vec{\mc{E}}^{*}_{\textsc{d}} \leftarrow 
            \{ (D^{*} \to T^{*}) \} \cup \{ (\tau^{*}_j \to D^{*}) \}_{\mc{F}^{*}}
    $
    \tcp*[f]{down coboundary}

    $\hat{\mc{E}}^{*}_{\vvec{\textsc{fn}}} \leftarrow 
            \vec{\mc{E}}^{*}_{\textsc{st}} \cup \vec{\mc{E}}^{*}_{\textsc{u}} \cup \vec{\mc{E}}^{*}_{\textsc{d}}
    $
    \tcp*[f]{FLAT-NORM network}

    \lForEach{$\vec{e}^{*}_{+i} \in \vec{\mc{E}}^{*}_{+}$}{
            \Cost {$\vec{e}^{*}_{+i}$} $\leftarrow  \bm{z}^{*}(\vec{e}^{*}_{+i}) \equiv \phantom{-} \bm{z}({e}_{i})$,
            \Capacity {$\vec{e}^{*}_{+i}$} $\leftarrow  c_i$
    }
    
    \lForEach{$\vec{e}^{*}_{-i} \in \vec{\mc{E}}^{*}_{-}$}{
            \Cost {$\vec{e}^{*}_{-i}$} $\leftarrow  \bm{z}^{*}(\vec{e}^{*}_{-i}) \equiv -\bm{z}({e}_{i})$,
            \Capacity {$\vec{e}^{*}_{-i}$} $\leftarrow  c_i$ 
    }

    \lForEach{$\vec{e}^{*} \in \{ \cobd^{*}_{1} S^{*} \} \cup \{ \cobd^{*}_{1} T^{*} \} $}{
            \Cost {$\vec{e}^{*}$} $\leftarrow  0$,
            \Capacity {$\vec{e}^{*}$} $\leftarrow  \infty$  
    }

    \lForEach{$\vec{u}^{*}_j \in \tilde{\mc{E}}^{*}_{\textsc{u}} \setminus \{ (S^{*} \to U^{*}) \} $}{
            \Cost {$\vec{u}^{*}_{j}$} $\leftarrow  0$,
            \Capacity {$\vec{u}^{*}_j$} $\leftarrow \lambda a_j$
    }

    \lForEach{$\vec{l}^{*}_j \in \tilde{\mc{E}}^{*}_{\textsc{d}} \setminus \{ (D^{*} \to T^{*}) \} $}{
            \Cost {$\vec{l}^{*}_{j}$} $\leftarrow  0$,
            \Capacity {$\vec{l}^{*}_j$} $\leftarrow \lambda a_j$ 
    }

     \label{algo:fnorm:flow-network}
     \caption{SFN Flow Network Setup.}
\end{algorithm}

\begin{figure}[h]
  \centering
  \includegraphics[width=0.45\linewidth]{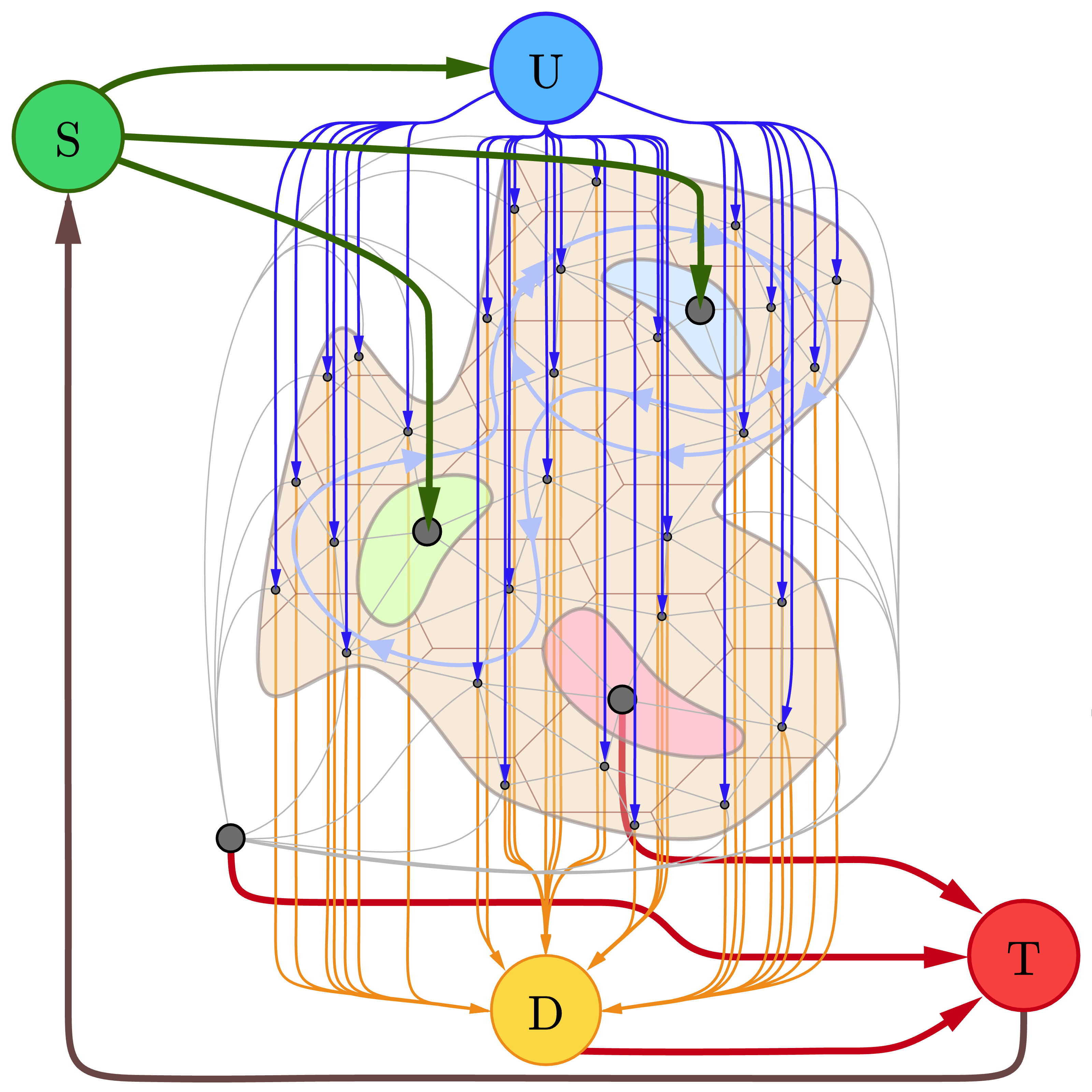}   \  
  \includegraphics[width=0.45\linewidth]{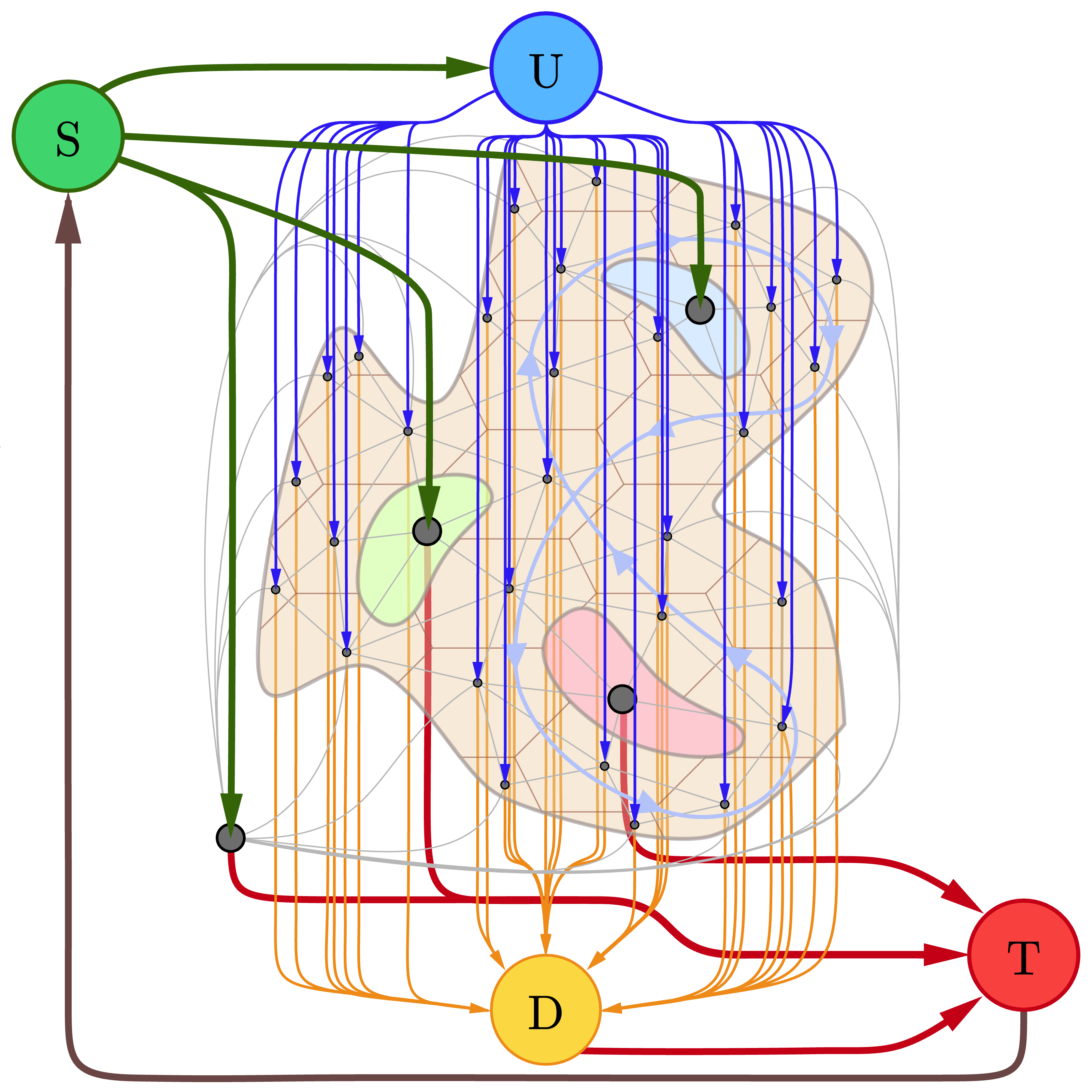}   
  \caption[Flow networks for SFN.]{
        The dual flow networks produced by Algorithm~\ref{algo:fnorm:flow-network} for the two $q$-cycles in Fig.~\ref{fig:types-of-cycles}: the $q$-cycle curling in a single direction is on the \textbf{Left}, and the mixed curling cycle is on the \textbf{Right}.
        The SFN-networks capture the ``area''-weights  by connecting the source $S$ and sink $T$  to the nodes dual to $(q + 1)$-simplices directly or via auxiliary $U$ and $D$ nodes. 
        The capacity of each such transit connection is given by the ``area''-weights of corresponding $(q + 1)$-simplex. 
        These new connections do not depend on the type of $\bm{z}$'s curl and do not affect existing links with the void-nodes. 

        A number of nodes in a SFN-network is $F+ \beta_q + 1 + 4 = F + \beta_q + 5$,
        and the number of di-arcs is at least  $2E + 2F + \beta_q  + 1 + 3 = 2E + 2F + \beta_q  +4 $.        
    }
  \label{fig:fnorm:flow-network}
\end{figure}



\paragraph*{Construction of the flow-network $\bar{\mc{K}}^{*}_{\scalebox{0.5}{ST}}$: }
Finally, we construct the directed \embf{dual graph} 
$\bar{\mc{K}}^{*} = \bar{\mc{K}}^{*}_{\vvec{\textsc{st}}} = (\bar{\mc{F}}^{*} = \mc{F}^{*} \cup \mc{V}^{*}, \vec{\mc{E}}^{*} = \vec{\mc{E}}^{*}_{+} \cup \vec{\mc{E}}^{*}_{-})$ of $\bar{\mc{K}}$ as follows. 
$\bar{\mc{K}}^{*} = \bar{\mc{K}}^{*}_{\vvec{\textsc{st}}} = (\bar{\mc{F}}^{*} = \mc{F}^{*} \cup \mc{V}^{*}, \vec{\mc{E}}^{*} = \vec{\mc{E}}^{*}_{+} \cup \vec{\mc{E}}^{*}_{-})$ of $\bar{\mc{K}}$ as follows. 
The nodes of $\bar{\mc{K}}^{*}$ are in 1-to-1 corresponds with the facet-simplices and voids, including the outer void $\nu_{0}$,
i.e. $\bar{\mc{F}}^{*} = \mc{F}^{*} \cup \mc{V}^{*}  \cong \bar{\mc{K}}_{q + 1} = \mc{K}_{q + 1} \cup \mc{V}$,
and we define the 0-th dual chain group by the isomorphism $\mc{C}_{0}(\mc{K}^{*}) \cong \mc{C}_{q + 1}(\bar{\mc{K}}) \cong \integer^{F + \beta_q}$.
For every oriented edge-simplex $e$ of $\mc{K}$ such that $e = (\eta \dart \tau)$, where $\eta, \tau \in \bar{\mc{K}}_{q + 1}$,
we introduce a directed dual arc $\vec{e}^{*} = (\eta^{*} \to \tau^{*})$ with its weight given by the length-volume of $e$, 
${c}(\vec{e}^{*}_{+}) = {c}(\vec{e}^{*}_{-}) = c(e)$.
We call all such arcs (that are obtained by directly dualizing edge-simplices) the \embf{positive} or \embf{natural} arcs of $\mc{K}^{*}$, and denote the collection of all of them as 
$\vec{\mc{E}}^{*}_{+} = \left\{ \vec{e}^{*}_{+} = (\eta^{*} \to \tau^{*}) \mid (\eta \dart \tau) \in \bar{\mc{K}}_{q} \right\}$.
The set of \embf{negative} or \embf{reversed} arcs, which correspond to the opposite orientation of edge-simplices, is given by
$\vec{\mc{E}}^{*}_{-} = \left\{ \vec{e}^{*}_{-} = (\tau^{*} \to \eta^{*}) \mid (\eta \dart \tau) \in \bar{\mc{K}}_{q} \right\}$.
By default, we include both kinds of arcs to the dual graph $\vec{\mc{E}}^{*} = \vec{\mc{E}}^{*}_{+} \cup \vec{\mc{E}}^{*}_{-}$.
Note that $\mc{K}^{*}$ admits parallel arcs, but not the self-loops.

We dualize the $q$-chains $\bm{x} \in \mc{C}_{q}(\mc{K}; \integer)$ as skew-symmetric functions on arcs: 
$\bm{x}^{*}(\vec{e}^{*}_{+}) = \bm{x}({e})$ and $\bm{x}^{*}(\vec{e}^{*}_{-}) = -\bm{x}({e})$.
Hence $\mc{C}_{1}(\vec{\mc{K}}^{*}) \cong \mc{C}_{q}(\mc{K}) \oplus \mc{C}_{q}(\mc{K}_{-}) \cong \integer^{2E}$, 
where $\mc{C}_{q}(\mc{K}_{-})$ is the $q$-th chain group spanned by the opposite orientations of the edge-simplices.
Note that all $q$-cycles $\bm{z} \in \mc{Z}_{q}(\mc{K})$ are null-homologous on $\bar{\mc{K}}$,
and thus are dualized as $1$-coboundaries, $\bm{z}^{*} \in \mc{B}^{1}(\mc{K}^{*})$.
The dual boundary and coboundary operators are given as 
$\bd^{*}_1\!: \mc{C}_{1}(\mc{K}^{*}) \to \mc{C}_{0}(\mc{K}^{*})  \equiv \cobd_{q + 1}$
and 
$\cobd^{*}_1\!: \mc{C}_{0}(\mc{K}^{*}) \to \mc{C}_{1}(\mc{K}^{*})  \equiv \bd_{q + 1} \oplus -\bd_{q + 1}$.

\paragraph*{Construction of the flow-network $\hat{\mc{K}}^{*}_{\scalebox{0.5}{FN}}$}
The procedure described in Algorithm~\ref{algo:fnorm:flow-network} extends the dual graph ${\mc{K}}^{*}$ to the dual flow network 
$\hat{\mc{K}}^{*}_{\vvec{\textsc{fn}}} = (\hat{\mc{F}}^{*}_{\vvec{\textsc{fn}}}, \hat{\mc{E}}^{*}_{\vvec{\textsc{fn}}})$ 
$\hat{\mc{K}}^{*}_{{\textsc{fn}}} = (\hat{\mc{F}}^{*}_{{\textsc{fn}}}, \hat{\mc{E}}^{*}_{{\textsc{fn}}})$ 
by introducing, among others, the source and sink nodes, $S^{*}$ and $T^{*}$, and correctly connects them to the other nodes.  
The main task of Algorithm~\ref{algo:fnorm:flow-network} is to partition the void-nodes $\nu^{*}_{k} \in \mc{V}^{*}$  
into three mutually exclusive sets (sources, sinks, and transits) 
depending on the direction that $\bm{z}$ winds around the corresponding primal voids on $\bar{\mc{K}}$.
The crucial implication of such partition is that any $q$-coflows or $q$-pseudocoflows 
that originate in source voids and terminate in sink voids (for simplicity assume no transit voids) 
has non-positive flux through $\bm{z}$ or any $q$-cycle homologous to it: 
$\tilde{\bm{f}}(\bm{z}) \leq 0$ and $\tilde{\bm{f}}(\bm{x}) \leq 0$ for any $\bm{x} \sim \bm{z}$ on $\mc{K}$. 
Note that in the case of $q$-coflow $\bm{f}$, such that ${\bm{f}}(\bd_{q + 1} \tau_j) = 0$ for all facet-simplices of $\mc{K}$, 
the flux on $q$-cycles from same homology class is the same: $\bm{f}(\bm{z}) = \bm{f}(\bm{x})$, 
which follows from the Hodge decomposition theorem and can be used to derive Weak and Strong Duality conditions \cite{sullivan1990crystalline}:
$-c(\bm{x}) \leq \bm{f}(\bm{z}) \leq 0$ for any feasible solutions, 
and $-\bm{f}(\bm{z}) = \bm{f}(-\bm{z}) = c(\bm{x})$ for optimal solutions of OHCP and its dual max-flux coflow problem.

\subsection{Directed coflow complexes \texorpdfstring{$\vec{\mc{K}}$}{K-vec}}

\paragraph*{Directed $q$-pseudocoflow on $\vec{\mc{K}}$: }
\label{par:dir-pseudocoflow-hatK}
Let $\pvec{\bm{f}}= [\pvec{\bm{f}}_{+} \  \pvec{\bm{f}}_{-}] \!: \vec{\mc{K}}_{q}  \to \integer_{+}$ be a \embf{non-negative directed} $q$-pseudocoflow, $\pvec{\bm{f}} \in {\mc{C}}^{q}(\vec{\mc{K}}; \integer_{+})$,
where $\vec{\mc{K}}_{q}= \vec{\mc{K}}_{+q} \cup \vec{\mc{K}}_{-q} = {\mc{K}}_{q} \cup -{\mc{K}}_{q}$ are \emph{natural} and \emph{opposite} orientations of original $q$-simplices.
The restored $q$-pseudocoflow $\tld{\bm{f}}$ on the original complex $\tld{\bm{f}} \in \mc{C}^{q}(\mc{K}; \integer)$
is given as $\tld{\bm{f}}(\eta | \tau) = \pvec{\bm{f}}(\eta \dart \tau) - \pvec{\bm{f}}(\tau \dart \eta) \equiv \pvec{\bm{f}}_{+} - \pvec{\bm{f}}_{-}$.

\begin{align}
  \pvec{\bm{f}}(\vec{\bm{z}}) \leq 0 
  &\iff {\cobd}_{q + 1} \pvec{\bm{f}} = 
        \sum_{\mc{F}} \gamma_j \tau_j 
  && \text{on } \vec{\mc{K}}
  \\
  &\iff \bar{\cobd}_{q + 1} \pvec{\bm{f}} = 
        \sum\limits_{\mc{T}} \gamma_t \nu_t - \sum_{\mc{S}} \gamma_s \nu_s 
      + \sum_{\mc{F}} \gamma_j \tau_j 
  && \text{on } \bar{\mc{K}}_{\vvec{\textsc{st}}}
\end{align}
where $-a_j \leq \gamma_j \leq a_j$ for $\tau_j \in K_{q + 1}$, 
$\gamma_{s}, \gamma_{t} > 0$.

A \emph{directed} {$q$-pseudocoflow} $\pvec{\bm{f}}$ is \embf{feasible} if:
\begin{align}
\label{eq:fnorm:feasible-pseudocoflow-dir}
    \me{\bm{0}}_{2E} \leq  \pvec{\bm{f}} & \leq \vec{\bm{c}}_{2E}:&
    \dsp  0 \leq \: &\pvec{\bm{f}}(\vec{e}_i)  \leq {c}_i&
    &\forall \vec{e}_i \in  \vec{\mc{K}}_{q} = \vec{\mc{K}}_{+q} \cup \vec{\mc{K}}_{-q}
\\[4pt]\dsp
    -{\bm{a}}_{{F}} \leq \pvec{\bm{f}} & \leq \bm{a}_{{F}}:&
    \dsp  -{a}_j \leq \: &\pvec{\bm{f}}(\bd_{q + 1} \tau_j)  \leq {a}_j&
    &\forall \tau_j \in {\mc{F}} = {\mc{K}_{q + 1}} 
  \\
    -{\bm{\infty}}_{{B}} \leq \pvec{\bm{f}} & \leq \bm{\infty}_{{B}}:&
    \dsp  -\infty <  \: &\pvec{\bm{f}}(\bd_{q + 1} \nu_k)  < \infty    &
    &\forall \nu_k \in {\mc{V}} 
\end{align}

\paragraph*{Max-coflux on $\vec{\mc{K}}$: }
Recall that $\vec{\bm{z}}(\tau \dart \eta) = - \vec{\bm{z}}(\eta \dart \tau) = - {\bm{z}}(\eta \mid \tau)$,
and $\pvec{\bm{f}}(\bd_{q + 1} \tau) = {\textstyle \sum_{ \vec{\mc{K}}_{q}} \big( \pvec{\bm{f}}(\eta \dart \tau) - \pvec{\bm{f}}(\tau \dart \eta) \big) }$.
Consider the directed max-coflux problem on $\vec{\mc{K}}$: 
\begin{align*}
  \begin{array}{lllll}\dsp
      \max\limits_{\pvec{\bm{f}}}\ 
              &\dsp \pvec{\bm{f}} (-\vec{\bm{z}})    \geq 0       
              &\qquad\iff
              &\dsp \min \pvec{\bm{f}} (\vec{\bm{z}}) \leq 0       
      \\[2pt]
      \subto\ &\dsp -\bm{a} \leq \cobd_{q + 1} \pvec{\bm{f}} \leq \bm{a}             \\[2pt]
              &\dsp  \bm{0}_{2E} \leq  \pvec{\bm{f}} \leq \vec{\bm{c}}_{2E}          
  \end{array}
\end{align*}

or in more details:
\begin{align*}
  \begin{array}{lllrr}\dsp
      \max\limits_{\pvec{\bm{f}}}\ 
          &\dsp - \sum_{(\eta \dart \tau) \in \vec{\mc{K}}_{q}} \pvec{\bm{f}}(\eta \dart \tau) \cdot \vec{\bm{z}}(\eta \dart \tau)   \geq 0
\\[5pt]
      \subto\ 
          &\dsp  -a_j \leq \pvec{\bm{f}}(\bd_{q + 1} \tau_{j}) 
                         = \sum_{(\eta | \tau) \in \mc{K}_{q} } \bigg( \pvec{\bm{f}}(\eta \dart \tau) - \pvec{\bm{f}}(\tau \dart \eta) \bigg) 
                      \leq a_j        
          &&\forall \tau_{j} \in \mc{F} = \mc{K}_{q + 1}    
  \\[3pt]
          &\dsp -\infty \leq \pvec{\bm{f}}(\bd_{q + 1} \nu_{s}) 
                           = \sum_{(\nu_{s} | \tau) \in \mc{K}_q } -\pvec{\bm{f}}(\nu_{s} \dart \tau)   
                           = - \gamma_s            
                        \leq 0
          &&\forall \nu_{s} \in \mc{S} \subset \mc{V}
  \\[3pt]
          &\dsp  \me \: 0 \leq \pvec{\bm{f}}(\bd_{q + 1} \nu_{t}) 
                      = \sum_{(\tau | \nu_{t}) \in \mc{K}_q }  \pvec{\bm{f}}(\tau \dart \nu_{t}) 
                      = + \gamma_t             
                      < \infty
          &&\forall \nu_{t} \in \mc{T} \subset \mc{V}
  \\[5pt]
          &\dsp 0 \leq \pvec{\bm{f}}(\eta \dart \tau) \leq {c}(\eta \dart \tau)
          &&\forall (\eta \dart \tau) \in  \vec{\mc{K}}^{+}_{q} \cup \vec{\mc{K}}^{-}_{q} 
  \end{array}
\end{align*}

The dual of this LP has a variable $-\bm{\pi}(\tau)$ for each $(q + 1)$-face $\tau \in \mc{K}_{q + 1}$ and 
a non-negative variable $\vec{\bm{x}}(\eta \dart \tau) \geq 0$ for each oriented $q$-simplex $(\eta \dart \tau) \in \vec{\mc{K}}_{q}$.
Note that the \emph{positive-clockwise} component $\bm{\pi}^{+} \geq \bm{0}_{F}$ of $\bm{\pi}$ in the decomposition $\bm{\pi} = \bm{\pi}^{+} - \bm{\pi}^{-}$ 
corresponds to the \emph{lower circulation imbalance} constraints $\cobd_{q + 1} \pvec{\bm{f}} \geq -\bm{a}$,
while the \emph{negative-counterclockwise} component $\bm{\pi}^{-} \geq \bm{0}_{F}$
corresponds to the \emph{upper circulation imbalance} constraints $\cobd_{q + 1} \pvec{\bm{f}} \leq \bm{a}$.
\begin{align}
\label{eq:fnorm:dir-prob}
  \begin{array}{rllrrr}\dsp
      \min\limits_{\vec{\bm{x}}, \bm{\pi} }\ 
          &\dsp 
                \sum_{\tau_{j} \in \mc{F}} a_j {\pi}^{+}_{j} + \sum_{\tau_{j} \in \mc{F}} a_j {\pi}^{-}_{j}
              + \sum_{\nu_{t} \in \mc{T}} \gamma_t \pi_t - \sum_{\nu_{s} \in \mc{S}} \gamma_s \pi_s
          &\dsp 
              + \sum_{\vec{e}_{i} \in \vec{\mc{K}}_{q}} c_i \vec{\bm{x}}(\eta \dart \tau) 
  \\[14pt]
      \subto\ 
          &\dsp  \bm{\pi}(\eta) - \bm{\pi}(\tau) + \vec{\bm{x}}(\eta \dart \tau) \geq -\vec{\bm{z}}(\eta \dart \tau) = -   {\bm{z}}(e_i)          
          &\qquad \forall \vec{e}_{+i} = (\eta \dart \tau) \in \vec{\mc{K}}_{+q} 
          \\[4pt]
          &\dsp  \bm{\pi}(\tau) - \bm{\pi}(\eta) + \vec{\bm{x}}(\tau \dart \eta) \geq -\vec{\bm{z}}(\tau \dart \eta) = \me {\bm{z}}(e_i)          
          &\qquad \forall \vec{e}_{-i} = (\tau \dart \eta) \in \vec{\mc{K}}_{-q} 
  \\[8pt]
          &\dsp \vec{\bm{x}}(\vec{e}_{i}) \geq 0
          &\qquad \forall \vec{e}_{i} = (\eta \dart \tau) \in \vec{\mc{K}}_{q} 
          \\
          &\dsp {\bm{\pi}}(\nu_s) = 0; \  {\bm{\pi}}(\nu_t) = 0
          &\qquad \forall \nu_k \in {\mc{V}} 
          \\
          &\dsp \bm{\pi}^{+} \geq \bm{0}_{F}; \  \bm{\pi}^{-} \geq \bm{0}_{F}
          &\qquad \forall \tau_j \in {\mc{K}_{q + 1}} 
  \end{array}
\end{align}
where $\bm{\pi} = \bm{\pi}^{+} - \bm{\pi}^{-}$.

\begin{corollary}[Th.~\ref{th:ohcp:dir-prob-solution} (OHCP Dir-solution)]
\label{th:fnorm:dir-prob-solution}
  Let $\bm{\pi}_{\star}$ and $\vec{\bm{x}}_{\star}$ be the optimal solution for the dual problem above, Eq.~\eqref{eq:fnorm:dir-prob}. 
  Then 
  \begin{itemize}
    \item[(a)]
        $\pi^{\star}_{t} = \bm{\pi}_{\star}(\nu_t) = 0$ and $\pi^{\star}_{s} = \bm{\pi}_{\star}(\nu_s) = 0$.
    \item[(b)]
        A $q$-chain $\bm{x}_{\star} \in \mc{C}_{q}(\mc{K}; \integer)$ defined as 
        $\bm{x}_{\star}(\eta \mid \tau) = \vec{\bm{x}}_{\star}(\tau \dart \eta) - \vec{\bm{x}}_{\star}(\eta \dart \tau) \equiv \bm{x}_{\star}^{-} - \bm{x}_{\star}^{+}$
        is a minimal $q$-cycle homologous to $\bm{z}$ on $\mc{K}$ such that $\bd_{q + 1} \bm{\pi}_{\star} = \bm{z} - \bm{x}_{\star}$. 
  \end{itemize}
\end{corollary}

\subsection{Fixed directed coflow and directed coflow complexes \texorpdfstring{$\hat{\mc{K}}_{FN}$}{K-hat-FN}}

Recall, that on $\hat{\mc{K}}$ we have fixed the pseudocoflow $\tld{\bm{f}}$ in Eq.~\eqref{eq:fnorm:hatK:fixed-coflow}  by introducing 
an auxiliary pseudocoflow $\tld{\bm{g}} \in \mc{C}^{q}(\hat{\mc{K}}; \integer)$ such that  $\tld{\bm{g}}(e_i) = 0 $ for $e_i \in \mc{K}_{q}$ 
and $\tld{\bm{g}}(u_j) = -\gamma_{j}$ for $u_j \in \mc{U}_{q}$.
The coboundary of $\tld{\bm{g}}$ in Eq.~\eqref{eq:fnorm:hatK:auxcoflow:cobd} is given by two disjoint $(q + 1)$-cochains:
$\hat{\cobd}_{q + 1} \tld{\bm{g}}  = \bm{g} - \bm{g}_{\mc{M}}$,  
where $\bm{g} = \bm{g}^{U} - \bm{g}^{D} \!: \mc{K}_{q + 1} \mapsto \integer$ 
is given by $\bm{g} = - {\cobd}_{q + 1} \tld{\bm{f}} =  - \sum_{\mc{F}} \gamma_j \tau_j = - \sum_{\mc{F}} (\gamma_j^{D} - \gamma_j^{U}) \tau_j$,
and $\bm{g}_{\mc{M}} \!: {\mc{M}}_{q + 1} \mapsto \integer$ is defined as $\bm{g}_{\mc{M}}(\mu_j) = -\gamma_j = \bm{g}(\tau_j) $.
The fixed $q$-coflow 
$\hat{\bm{\phi}} \in \mc{C}^{q}(\hat{\mc{K}}; \integer)\!: \sum_{\mc{S}} \nu_s \to \sum_{\mc{T}} \nu_t$
is defined as 
$\hat{\bm{\phi}} = \tld{\bm{f}}  + \tld{\bm{g}}$, Eq.~\eqref{eq:fnorm:hatK:fixed-coflow},
with coboundary laying outside of the original complex:
$\hat{\cobd}_{q + 1} \hat{\bm{\phi}} = \sum_{\mc{M}} \gamma_j \mu_j$.

Before considering the directed version of the fixed coflow $\hvec{\bm{\phi}}$,
let's first construct the directed coflow complex $\hvec{\mc{K}}$
of the extended complex $\hat{\mc{K}} =  \mc{K} \oplus \mc{M}_{q + 1} \oplus \mc{U}_{q}$, defined in Eq.~\eqref{eq:up-complex}.
We already have defined the dir-coflow complex $\vec{\mc{K}}$ of $\mc{K}$ in Eq.~\eqref{eq:def:coflow-network-dir} by duplicating its $q$-skeleton $\vec{\mc{K}}_{q} = \vec{\mc{K}}_{+q} \cup \vec{\mc{K}}_{-q}$,
and a directed feasible $q$-pseudocoflow $\pvec{\bm{f}}\!: \vec{\mc{K}}_{q} \mapsto \integerplus$ was given in Eq.~\eqref{eq:fnorm:feasible-pseudocoflow-dir}.

We want to model a feasible auxiliary pseudo-coflow $\tld{\bm{g}}: \mc{U}_{q} \mapsto \integer$, defined in  Eq.~\eqref{eq:fnorm:hatK:auxcoflow},
as a non-negative function on the  $u$-edges that obeys the area-capacity constraints:
$\tld{\bm{g}}(u_{j}) = \pvec{\bm{g}}(\vec{u}_{+j}) - \pvec{\bm{g}}(\vec{u}_{-j}) = -\gamma_j$
such that $0 \leq \pvec{\bm{g}}(\vec{u}_{j}) \leq a_j$,
where $\vec{u}_{+j} = u_{j}$ and $\vec{u}_{-j} = -u_{j}$ are the \emph{natural and opposite} orientations of  $u_j \in \mc{U}_{q}$,
and $\gamma_{j} = {\cobd}_{q + 1} \tld{\bm{f}}(\tau_j)$. 
We will use the following notation for the non-negative decomposition of $\cobd_{q + 1} \tld{\bm{f}}$'s value on $\tau_j$: 
$\gamma_{j} = \gamma_j^{D} - \gamma_j^{U}$, 
where $\gamma^{D}_{j} = \max(\gamma_{j}, 0) \geq 0$ and $\gamma^{U}_{j} = \max(-\gamma_{j}, 0) \geq 0$.
Let $\vec{\mc{U}}_{q} = \vec{\mc{U}}_{+q} \cup \vec{\mc{U}}_{-q} = \vec{\mc{U}}^{+}_{q} \cup \vec{\mc{U}}^{-}_{q} = \mc{U}_{q} \cup -\mc{U}_{q}$ be the set containing both orientations of $u$-edges, 
where $\vec{u}_{+j} \in \vec{\mc{U}}^{+}_{q}$ and $\vec{u}_{-j} \in \vec{\mc{U}}^{-}_{q}$.
Then
\begin{align}
\label{eq:fnorm:hatK:auxcoflow-to-dir}
    \tld{\bm{g}} = \sum\limits_{\mc{U}_{q}} -\gamma_j u_j 
                  = \sum\limits_{\mc{U}_{q}} \gamma^{U}_{j} u_j           - \sum\limits_{\mc{U}_{q}} \gamma^{D}_{j} u_j 
                  = \sum\limits_{\mc{U}_{q}} \pvec{\bm{g}}(\vec{u}_{+j}) {u}_{j}  - \sum\limits_{\mc{U}_{q}} \pvec{\bm{g}}(\vec{u}_{-j}) {u}_{j} 
\end{align}
where we have defined the directed pseudo-coflow  $\pvec{\bm{g}}\!: \vec{\mc{U}}_{q} \mapsto \integerplus$ as follows:
\begin{align}
\label{eq:fnorm:hatK:dir-auxcoflow}
    \pvec{\bm{g}}(\vec{u}_{+j}) \asgn \gamma^{U}_{j} = 
        \begin{cases}
            \tld{\bm{g}}(u_j), &\text{if } \tld{\bm{g}}(u_j) > 0 \\
            0, &\text{otherwise }
        \end{cases}
    &&\quad&&
    \pvec{\bm{g}}(\vec{u}_{-j}) \asgn \gamma^{D}_{j} = 
        \begin{cases}
            -\tld{\bm{g}}(u_j), &\text{if } \tld{\bm{g}}(u_j) < 0 \\
            0, &\text{otherwise }
        \end{cases}
\end{align}

Let $\pvec{\bm{g}}_{U}\!: \vec{\mc{U}}^{+}_{q} \mapsto \integerplus$ and $\pvec{\bm{g}}_{D}\!: \vec{\mc{U}}^{-}_{q} \mapsto \integerplus$
be the two  components of the directed auxiliary coflow: 
$\pvec{\bm{g}} = [\pvec{\bm{g}}_{U} \  \pvec{\bm{g}}_{D}]   \cong \integer^{2F}_{+} \mapsto \integerplus$
that are given by $\pvec{\bm{g}}_{U}(\vec{u}_{+j}) = \pvec{\bm{g}}(\vec{u}_{+j}) = \gamma^{U}_{j} \geq 0$ and $\pvec{\bm{g}}_{D}(\vec{u}_{-j}) = \pvec{\bm{g}}(\vec{u}_{-j}) = \gamma^{D}_{j} \geq 0$, respectively.
Then Eq.~\eqref{eq:fnorm:hatK:dir-auxcoflow} implies 
$\tld{\bm{g}} = \pvec{\bm{g}}_{U} - \pvec{\bm{g}}_{D}$.

The coboundary of the auxiliary pseudo-coflow in  Eq.~\eqref{eq:fnorm:hatK:auxcoflow:cobd} is given by two disjoint $(q + 1)$-cochains:
$\hat{\cobd}_{q + 1} \tld{\bm{g}}  = \bm{g} - \bm{g}_{\mc{M}}$,  
where $\bm{g}\!: \mc{K}_{q + 1} \mapsto \integer$ and $\bm{g}_{\mc{M}} \!: {\mc{M}}_{q + 1} \mapsto \integer$ 
are defined as $\bm{g}_{\mc{M}}(\mu_j) = \bm{g}(\tau_j)= -\gamma_j =  \gamma_j^{U} - \gamma_j^{D}$.
Let ${\bm{g}}_{U}\!: {\mc{K}}_{q + 1} \mapsto \integerplus$ and ${\bm{g}}_{D}\!: {\mc{K}}_{q + 1} \mapsto \integerplus$
such that $\bm{g} = \bm{g}_{U} - \bm{g}_{D}$.
It easy to see that $\bm{g}_{U}(\tau_j) = \hat{\cobd}_{q + 1} \pvec{\bm{g}}_{U}(\tau_j)$
and $\bm{g}_{D}(\tau_j) = \hat{\cobd}_{q + 1} \pvec{\bm{g}}_{D}(\tau_j)$.
In order to properly handle the coboundary of $\tld{\bm{g}}$ outside of the original complex,
let's consider the set $\vec{\mc{M}}_{q + 1}$ of the \emph{natural $\vec{\mu}_{+j} = \mu_{j}$} and \emph{opposite $\vec{\mu}_{-j} = -\mu_{j}$} orientations of  $\mu$-facets:
$\vec{\mc{M}}_{q + 1} = \vec{\mc{M}}^{+}_{q + 1} \cup \vec{\mc{M}}^{-}_{q + 1} = \mc{M}_{q + 1} \cup -\mc{M}_{q + 1}$,
where $\vec{\mu}_{+j} \in \vec{\mc{M}}^{+}_{q + 1}$ and $\vec{\mu}_{-j}  \in \vec{\mc{M}}^{-}_{q + 1}$.
In the similar fashion to Eq.~\eqref{eq:fnorm:hatK:auxcoflow-to-dir},
we define a non-negative $(q + 1)$-cochain $\bm{g}_{\vec{\mc{M}}}\!: \vec{\mc{M}}_{q + 1} \mapsto \integerplus$:
\begin{align}
\label{eq:fnorm:hat:dir-auxcoflow-cobd-M}
    \bm{g}_{\mc{M}} &= \sum\limits_{\mc{M}_{q + 1}} -\gamma_j \mu_j 
                     = \sum\limits_{\mc{M}_{q + 1}} \gamma^{U}_{j} \mu_j           - \sum\limits_{\mc{M}_{q + 1}} \gamma^{D}_{j} \mu_j 
                     = \sum\limits_{\mc{M}_{q + 1}} {\bm{g}}_{\vec{\mc{M}}}(\vec{\mu}_{+j}) {\mu}_{j}  - \sum\limits_{\mc{M}_{q + 1}} {\bm{g}}_{\vec{\mc{M}}}(\vec{\mu}_{-j}) {\mu}_{j} 
\end{align}
where 
$\bm{g}_{\vec{\mc{M}}}(\vec{\mu}_{+j}) \asgn \bm{g}_{\mc{M}}(\mu_j) = \gamma^{U}_{j} \geq 0$ if $\bm{g}_{\mc{M}}(\mu_j) > 0$, 
and 
$\bm{g}_{\vec{\mc{M}}}(\vec{\mu}_{-j}) \asgn - \bm{g}_{\mc{M}}(\mu_j) = \gamma^{D}_{j} \geq 0$ if $\bm{g}_{\mc{M}}(\mu_j) < 0$,  respectively.
Let's denote the two components of $\bm{g}_{\vec{\mc{M}}}$ defined on $\vec{\mc{M}}^{+}_{q + 1}$ and $\vec{\mc{M}}^{+}_{q + 1}$
as $\bm{g}^{+}_{\vec{\mc{M}}}$ and $\bm{g}^{-}_{\vec{\mc{M}}}$, respectively\footnote{
    $\bm{g}_{\vec{\mc{M}}} = [\bm{g}^{+}_{\vec{\mc{M}}} \  \bm{g}^{-}_{\vec{\mc{M}}}] \cong \integer^{2F} \mapsto \integerplus$
    for $\bm{g}^{+}_{\vec{\mc{M}}}\!: \vec{\mc{M}}^{+}_{q + 1} \mapsto \integerplus$ 
    and $\bm{g}^{-}_{\vec{\mc{M}}}\!: \vec{\mc{M}}^{-}_{q + 1} \mapsto \integerplus$.
}.
Then, since one of $\gamma^{U}_{j} \geq 0$ or $\gamma^{D}_{j} \geq 0$ is always zero, 
Eq.~\eqref{eq:fnorm:hat:dir-auxcoflow-cobd-M} implies $\bm{g}_{\mc{M}} = \bm{g}^{+}_{\vec{\mc{M}}} - \bm{g}^{-}_{\vec{\mc{M}}}$:
$\bm{g}_{\mc{M}}({\mu}_{j}) = \gamma^{U}_{j} - \gamma^{D}_{j} = \bm{g}^{+}_{\vec{\mc{M}}}(\vec{\mu}_{+j}) - \bm{g}^{-}_{\vec{\mc{M}}}(\vec{\mu}_{-j})$.

Let's define the coboundaries of the oriented $u$-edges 
$\vec{u}_{+j} \in \vec{\mc{U}}^{+}_{q}$ and $\vec{u}_{-j} \in \vec{\mc{U}}^{-}_{q}$
as follows:
\begin{align}
\label{eq:dir-up-complex:cobd-up}
    \hat{\cobd}_{q + 1} \vec{u}_{+j} = \tau_{j} - \vec{\mu}_{+j}
    && 
    \hat{\cobd}_{q + 1} \vec{u}_{-j} = \vec{\mu}_{-j} - \tau_{j} 
\end{align}

Then 
\begin{align*}
    \hat{\cobd}_{q + 1} \pvec{\bm{g}}_{U} 
    &= \sum_{\vec{\mc{U}}^{+}_{q}} \gamma^{U}_{j} \cdot \hat{\cobd}_{q + 1} \vec{u}_{+j}  
     = \sum_{\mc{K}_{q + 1}} \gamma^{U}_{j} \tau_j - \sum_{\vec{\mc{M}}^{+}_{q + 1}} \gamma^{U}_{j} \vec{\mu}_{+j}
\\[2pt]
    \hat{\cobd}_{q + 1} \pvec{\bm{g}}_{D} 
    &= \sum_{\vec{\mc{U}}^{-}_{q}} \gamma^{D}_{j} \cdot \hat{\cobd}_{q + 1} \vec{u}_{-j}  
     = \sum_{\vec{\mc{M}}^{-}_{q + 1}} \gamma^{D}_{j} \vec{\mu}_{-j} - \sum_{\mc{K}_{q + 1}} \gamma^{D}_{j} \tau_j 
\\[4pt]
    \hat{\cobd}_{q + 1} \pvec{\bm{g}} = \hat{\cobd}_{q + 1} \pvec{\bm{g}}_{U} + \hat{\cobd}_{q + 1} \pvec{\bm{g}}_{D} 
    &= \sum_{\mc{K}_{q + 1}} \gamma^{U}_{j} \tau_j - \sum_{\vec{\mc{M}}^{+}_{q + 1}} \gamma^{U}_{j} \vec{\mu}_{+j}
      + \sum_{\vec{\mc{M}}^{-}_{q + 1}} \gamma^{D}_{j} \vec{\mu}_{-j} - \sum_{\mc{K}_{q + 1}} \gamma^{D}_{j} \tau_j 
    \\[2pt]  
    &= \sum_{\mc{K}_{q + 1}} (\gamma^{U}_{j} - \gamma^{D}_{j}) \tau_j 
    - \sum_{\vec{\mc{M}}^{+}_{q + 1}} \gamma^{U}_{j} \vec{\mu}_{+j}
    + \sum_{\vec{\mc{M}}^{-}_{q + 1}} \gamma^{D}_{j} \vec{\mu}_{-j}
    \\[2pt]
    &= \bm{g} - \bm{g}^{+}_{\vec{\mc{M}}} + \bm{g}^{-}_{\vec{\mc{M}}}
    = \bm{g} - \bm{g}_{{\mc{M}}} 
\end{align*}

This proves the following lemma.

\begin{lemma}
  Any (feasible) auxiliary pseudo-coflow $\tld{\bm{g}} \in \mc{C}^{q}(\hat{\mc{K}}; \integer)\!: \mc{U}_{q} \mapsto \integer$
  and its coboundary on and outside of $\mc{K}$
  can be computed by a corresponding non-negative directed auxiliary pseudo-coflow 
  $\pvec{\bm{g}} = [\pvec{\bm{g}}_{U} \  \pvec{\bm{g}}_{D}]   \!: \vec{\mc{U}}^{+}_{q} \cup \vec{\mc{U}}^{-}_{q}   \mapsto \integerplus$
  by assigning $\tld{\bm{g}} = \pvec{\bm{g}}_{U} - \pvec{\bm{g}}_{D}$.
\end{lemma}

Finally, let's update the boundary operators $\hat{\bd}$ in accordance with Eq.~\eqref{eq:dir-up-complex:cobd-up}:
\begin{align}
\label{eq:dir-up-complex:cobd}
    &\hat{\bd}_{q + 1} \tau_j \asgn \bd_{q + 1} \tau_j + \vec{u}_{+j} - \vec{u}_{-j}
    &&
    \hat{\bd}_{q + 1} \vec{\mu}_{+j} \asgn - \vec{u}_{+j} 
    &&
    \hat{\bd}_{q + 1} \vec{\mu}_{-j} \asgn  \vec{u}_{-j} 
  \\
\label{eq:dir-up-complex:bd-q}
    &&&\hat{\bd}_{q} \vec{u}_{+j} \asgn 0&&
    \hat{\bd}_{q} \vec{u}_{-j} \asgn 0&&
\end{align}

Then we define dir-coflow complex as $\hvec{\mc{K}} = \vec{\mc{K}} \oplus \vec{\mc{U}}_{q} \oplus \vec{\mc{M}}_{q + 1}$,
where $\vec{\mc{U}}_{q} = \vec{\mc{U}}^{+}_{q} \cup \vec{\mc{U}}^{-}_{q}$ and $\vec{\mc{M}}_{q + 1}= \vec{\mc{M}}^{+}_{q + 1} \cup \vec{\mc{M}}^{-}_{q + 1}$,
and the fixed directed coflow as $\vec{\bm{\phi}} = \pvec{\bm{f}} + \pvec{\bm{g}}$:
$\hat{\cobd}_{q + 1} \vec{\bm{\phi}} (\tau_j) 
      = \hat{\cobd}_{q + 1} \pvec{\bm{f}}(\tau_j) +  \hat{\cobd}_{q + 1} \pvec{\bm{g}}(\tau_j)
      = \hat{\cobd}_{q + 1} \pvec{\bm{f}}(\tau_j) +  \hat{\cobd}_{q + 1} \pvec{\bm{g}}_{U}(\tau_j) - \hat{\cobd}_{q + 1} \pvec{\bm{g}}_{D}(\tau_j)
$.

\section{Conclusions}
We proved the duality theorems~\ref{th:fnorm:weak-duality-SFN-hatK} and \ref{th:fnorm:strong-duality} for mSFN  by constructing the "hat"-complex $\hat{\mc{K}}$.
This construction preserves the original homology class of the input $q$-cycle $\bm{z}$ as show by Lemma~\ref{lm:homology-on-hatK}, 
and by Theorem~\ref{lm:homology-on-hatK} the fixed coflux preserves the flux through $\bm{z}$.
The ``hat''-complex  can be extended to a direct hat-complex.
The dual complex of a direct hat-complex is a dual flow network for a mSFN problem, whose construction is given in The Algorithm~\ref{algo:fnorm:flow-network}.

%
%
%

  \chapter{Stability of Flat Norm distance}

\section{Multiscale Flat Norm  Distance }\label{sec:prelim}

\begin{definition}[Geometric graph]
A graph $\mathscr{G}\left(\mathscr{V},\mathscr{E}\right)$ with node set $\mathscr{V}$ and edge set $\mathscr{E}$ is said to be a geometric graph of $\,\mathbb{R}^d$ if the set of nodes $\mathscr{V}\subset\mathbb{R}^d$ and the edges are Euclidean straight line segments $\left\{\overline{uv}~|~e:=\left(u,v\right)\in\mathscr{E}\right\}$ which intersect (possibly) at their endpoints.
\end{definition}
\begin{definition}[Structurally similar geometric graphs]
Two geometric graphs $\mathscr{G}_0\left(\mathscr{V}_0,\mathscr{E}_0\right)$ and $\mathscr{G}_1\left(\mathscr{V}_1,\mathscr{E}_1\right)$ are said to be \emph{structurally similar} at the level of $\gamma \geq 0$, termed $\gamma$-similar, whenever $\dist\left(\mathscr{G}_0,\mathscr{G}_1\right)\leq\gamma$ for some distance function $\dist$ between the two graphs.
\end{definition}
We could consider a given network as a set of edge geometries.
Hence we could consider the problem of comparing geometric graphs $\mathscr{G}_0$ and $\mathscr{G}_1$ as that of comparing the set of edge geometries $\mathscr{E}_0$ and $\mathscr{E}_1$. In this paper, we propose a suitable distance that allows us to compare between a pair of geometric graphs or a pair of geometries.
We use the multiscale flat norm, which has been well explored in the field of geometric measure theory, to define such a distance between the geometries. 

\subsection{Multiscale Flat Norm}
\label{subsec:multiscale-flat-norm}
We use the multiscale simplicial flat norm proposed by Ibrahim et al.~\cite{ibrahim2011simplicial} to compute the distance between two networks.
We now introduce some background for this computation. 
A $d$-dimensional \emph{current} $T$ (referred to as a $d$-current) is a generalized $d$-dimensional geometric object with orientations (or direction) and multiplicities (or magnitude).
An example of a $2$-current is a surface with finite area (multiplicity) and a specific orientation (clockwise or counterclockwise).
We use $\mathcal{C}_d$ to denote the set of all $d$-currents,
and $\mathcal{C}_d(\real^{p})$ to denote the set of $d$-currents embedded in $\real^{p}$.
$\mathbf{V}_d(T)$ or $\left|{T}\right|$ denotes the $d$-dimensional \emph{volume} of $T$, e.g., length in 1D or area in 2D.
The boundary of $T$, denoted by $\partial T$, is a $(d-1)$-current.
The \emph{multiscale flat norm} of a $d$-current $T \in \mathcal{C}_d$, at scale $\lambda \geq 0$ is defined as
\begin{equation}
    \mathbb{F}_\lambda\left(T\right) = \min_{S \in \mathcal{C}_{d + 1}}\left\{\mathbf{V}_d\left(T-\partial S\right)+\lambda \mathbf{V}_{d+1}\left(S\right)\right\},
\label{eq:flatnorm-def}
\end{equation}
where the minimum is taken over all $(d+1)$-currents $S$.
Computing the flat norm of a 1-current (curve) $T$ identifies the optimal 2-current (area patches) $S$ that minimizes the sum of the length of current $T-\partial S$ and the area of patch(es) $S$.
Fig.~\ref{fig:demo-flatnorm-basic} shows the flat norm computation for a generic 1D current $T$ (blue).
The 2D area patches $S$ (magenta) are computed such that the expression in Eq.~(\ref{eq:flatnorm-def}) is minimized for the chosen value of $\lambda$ that ends up using most of the patch under the sharper spike on the left but only a small portion of the patch under the wider bump to the right.
\begin{figure}[ht!]
    \centering
    \includegraphics[width=0.98\textwidth]{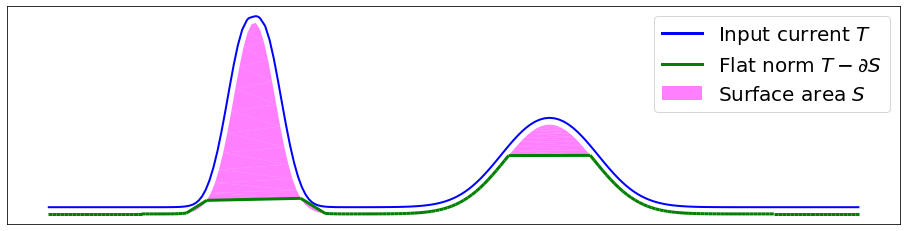}
    \caption[Multiscale flat norm of a 1D current.]
    {Multiscale flat norm of a 1D current $T$ (blue).
      The flat norm is the sum of length of the resulting 1D current $T-\partial S$ (green) and the area of 2D patches $S$ (magenta).
      We show $T-\partial S$ slightly separated for easy visualization.}
    \label{fig:demo-flatnorm-basic}
\end{figure}

The scale parameter $\lambda$ can be intuitively understood as follows.
Rolling a ball of radius $1/\lambda$ on the $1$-current $T$ traces the output current $T-\partial S$ and the untraced regions constitute the patches $S$.
Hence we observe that for a large $\lambda$, the radius of the ball is very small and hence it traces major features while smoothing out (i.e., missing) only minor features (wiggles) of the input current.
But for a small $\lambda$, the ball with a large radius smoothes out larger scale features (bumps) in the current.
Note that for smaller $\lambda$, the cost of area patches is smaller in the minimization function and hence more patches are used for computing the flat norm. We can use the flat norm to define a natural distance between a pair of 1-currents $T_1$ and $T_2$ as follows~\cite{ibrahim2011simplicial}.
\begin{equation}
    \mathbb{F}_\lambda\left(T_1,T_2\right) = \mathbb{F}_\lambda\left(T_1-T_2\right)\label{eq:flatnorm-def2}
\end{equation}

We compute the flat norm distance between a pair of input geometries (synthetic and actual) as the flat norm of the current $T=T_1-T_2$ where $T_1$ and $T_2$ are the currents corresponding to individual geometries.
Let $\Sigma$ denote the set of all line segments in the input current $T$.
We perform a constrained triangulation of $\Sigma$ to obtain a $2$-dimensional finite oriented simplicial complex $K$.
A constrained triangulation ensures that each line segment $\sigma_i\in\Sigma$ is an edge in $K$,
and that $T$ is an oriented $1$-dimensional subcomplex of $K$.

Let $m$ and $n$ denote the numbers of edges and triangles in $K$.
We can denote the input current $T$ as a $1$-chain $\sum_{i=1}^{m}t_i\sigma_i$ where $\sigma_i$ denotes an edge in $K$ and $t_i$ is the corresponding multiplicity.
Note that $t_i=-1$ indicates that orientation of $\sigma_i$ and $T$ are opposite, $t_i=0$ denotes that $\sigma_i$ is not contained in $T$, and $t_i=1$ implies that $\sigma_i$ is oriented the same way as $T$.
Similarly, we define the set $S$ to be the $2$-chain of $K$ and denote it by $\sum_{i=1}^{m}s_i\omega_i$ where $\omega_i$ denotes a $2$-simplex in $K$ and $s_i$ is the corresponding multiplicity.

The boundary matrix $\left[\partial\right]\in\mathbb{Z}^{m\times n}$ captures the intersection of the $1$ and $2$-simplices of $K$.
The entries of the boundary matrix $\left[\partial\right]_{ij}\in\{-1,0,1\}$.
If edge $\sigma_i$ is a face of triangle $\omega_j$, then $\left[\partial\right]_{ij}$ is nonzero and it is zero otherwise.
The entry is $-1$ if the orientations of $\sigma_i$ and $\omega_j$ are opposite and it is $+1$ if the orientations agree. 

We can respectively stack the $t_i$'s and $s_i$'s in $m$ and $n$-length vectors $\mathbf{t}\in\mathbb{Z}^m$ and $\mathbf{s}\in\mathbb{Z}^n$. The $1$-chain representing $T-\partial S$ is denoted by $\mathbf{x}\in\mathbb{Z}^m$ and is given as $\mathbf{x} = \mathbf{t} - \left[\partial\right]\mathbf{s}$.
The multiscale flat norm defined in Eq.~(\ref{eq:flatnorm-def}) can be computed by solving the following optimization problem:
\begin{equation}
\begin{aligned}
    \mathbb{F}_\lambda\left(T\right) = \min_{\mathbf{s}\in\mathbb{Z}^n} & \sum_{i=1}^{m}w_i\left|x_i\right|+\lambda \left(\sum_{j=1}^{n}v_j\left|s_j\right|\right)\\
    \textrm{s.t.} & \quad \mathbf{x} = \mathbf{t} - \left[\partial\right]\mathbf{s}, 
    \quad \mathbf{x} \in \mathbb{Z}^m,
\end{aligned}
\label{eq:opt-flatnorm-def}
\end{equation}
where $\mathbf{V}_d\left(\tau\right)$ in Eq.~(\ref{eq:flatnorm-def}) denotes the volume of the $d$-dimensional simplex $\tau$.
We denote volume of the edge $\sigma_i$ as $\mathbf{V}_1(\sigma_i)=w_i$ and set it to be the Euclidean length, and volume of a triangle $\tau_j$ as $\mathbf{V}_2(\tau_j)=v_j$ and set it to be the area of the triangle.

In this work, we consider geometric graphs embedded on the geographic plane and are associated with longitude and latitude coordinates.
We compute the Euclidean length of edge $\sigma_i$ as $w_i=R\Delta\phi_i$ where $\Delta\phi_i$ is the Euclidean normed distance between the geographic coordinates of the terminals of $\sigma_i$ and $R$ is the radius of the earth.
Similarly, the area of triangle $\tau_j$ is computed as $v_j=R^2\Delta\Omega_j$ where $\Delta\Omega_j$ is the solid angle subtended by the geographic coordinates of the vertices of $\tau_j$.

Using the fact that the objective function is piecewise linear in $\mathbf{x}$ and $\mathbf{s}$, the minimization problem can be reformulated as an integer linear program (ILP) as follows:
\begin{subequations}
\begin{align}
    \mathbb{F}_\lambda\left(T\right) = \min & \sum_{i=1}^{m}w_i\left(x_i^{+}+x_i^{-}\right)+\lambda \left(\sum_{j=1}^{n}v_j\left(s_j^{+}+s_j^{-}\right)\right)\\
    \textrm{s.t.} & \quad \mathbf{x}^{+}-\mathbf{x}^{-} = \mathbf{t} - \left[\partial\right]\left(\mathbf{s}^{+}-\mathbf{s}^{-}\right)\\
    & \quad \mathbf{x}^{+},\mathbf{x}^{-} \geq 0,\quad \mathbf{s}^{+},\mathbf{s}^{-} \geq 0\\
    & \quad \mathbf{x}^{+},\mathbf{x}^{-} \in \mathbb{Z}^m,\quad \mathbf{s}^{+},\mathbf{s}^{-} \in \mathbb{Z}^n \label{seq:intger-constraints}
\end{align}
\label{eq:opt-flatnorm}
\end{subequations}
\noindent The linear programming relaxation of the ILP in Eq.~(\ref{eq:opt-flatnorm}) is obtained by ignoring the integer constraints Eq.~(\ref{seq:intger-constraints}).
We refer to this relaxed linear program (LP) as the \emph{\bfseries flat norm LP}.
Ibrahim et al.~\cite{ibrahim2011simplicial} showed that the boundary matrix $[\partial]$ is totally unimodular for our application setting.
Hence the flat norm LP will solve the ILP, and hence the flat norm can be computed in polynomial time.

Algorithm~\ref{alg:distance} describes how we compute the distance between a pair of geometries with the associated embedding on a metric space $\mathcal{M}$.
We assume that the geometries (networks) $\mathscr{G}_1\left(\mathscr{V}_1,\mathscr{E}_1\right)$ and $\mathscr{G}_2\left(\mathscr{V}_2,\mathscr{E}_2\right)$ with respective node sets $\mathscr{V}_1,\mathscr{V}_2$ and edge sets $\mathscr{E}_1,\mathscr{E}_2$ have no one-to-one correspondence between the $\mathscr{V}_i$'s or $\mathscr{E}_i$'s.
Note that each vertex $v\in\mathscr{V}_1,\mathscr{V}_2$ is a point and each edge $e\in\mathscr{E}_1,\mathscr{E}_2$ is a straight line segment in $\mathcal{M}$. We consider the collection of edges $\mathscr{E}_1,\mathscr{E}_2$ as input to our algorithm.
First, we orient the edge geometries in a particular direction (left to right in our case) to define the currents $T_1$ and $T_2$, which have both magnitude and direction.
Next, we consider the bounding rectangle $\mathscr{E}_{\textrm{bound}}$ for the edge geometries and define the set $\Sigma$ to be triangulated as the set of all edges in either geometry and the bounding rectangle.
We perform a constrained Delaunay triangulation \cite{Si2010} on the set $\Sigma$ to construct the $2$-dimensional simplicial complex $K$.
The constrained triangulation ensures that the set of edges in $\Sigma$ is included in the simplicial complex $K$.
Then we define the currents $T_1$ and $T_2$ corresponding to the respective edge geometries $\mathscr{E}_1$ and $\mathscr{E}_2$ as $1$-chains in $K$.
Finally, the flat norm LP is solved to compute the simplicial flat norm.


\begin{algorithm}[ht!]
\caption{Distance between a pair of geometries}
\label{alg:distance}
    \textbf{Input}: Geometries $\mathscr{E}_1,\mathscr{E}_2$\\
    \textbf{Parameter}: Scale $\lambda$
    \begin{algorithmic}[1]
            \STATE Orient each edge from left to right: $\Tilde{\mathscr{E}}_1 \asgn \mathsf{Orient}\left(\mathcal{E}_1\right);~~\Tilde{\mathcal{E}}_2:=\mathsf{Orient}\left(\mathcal{E}_2\right)$.
            \STATE Find bounding rectangle for the pair of geometries: \\$\mathcal{E}_{\textrm{bound}}=\mathsf{rect}\left(\Tilde{\mathcal{E}}_1,\Tilde{\mathcal{E}}_2\right)$.
            \STATE Define the set of line segments to be triangulated: \\$\Sigma=\Tilde{\mathcal{E}}_1\cup\Tilde{\mathcal{E}}_2\cup\mathcal{E}_{\textrm{bound}}$.
            \STATE Perform constrained triangulation on set $\Sigma$ to construct $2$-complex $K$.
            \STATE Define the currents $T_1,T_2$ as $1$-chains of oriented edges $\Tilde{\mathscr{E}}_1$ and $\Tilde{\mathscr{E}}_2$ in $K$.
            \STATE Solve the flat norm LP to compute flat norm $\mathbb{F}_\lambda\left(T_1-T_2\right)$.
    \end{algorithmic}
    \textbf{Output}: Flat norm distance $\mathbb{F}_\lambda\left(T_1-T_2\right)$.
\end{algorithm}

\subsection{Normalized Flat Norm}
\label{subsec:normalized-fn}
Recall that in our context of synthetic power distribution networks, the primary goal of comparing a synthetic network to its actual counterpart is to infer the quality of the replica or the \emph{digital duplicate} synthesized by the framework.
The proposed approach using the flat norm for structural comparison of a pair of geometries provides us a method to perform global as well as local comparison.
While we can produce a global comparison by computing the flat norm distance between the two networks, it may not provide us with complete information on the quality of the synthetic replicate.
On the other hand, a local comparison can provide us details about the framework generating the synthetic networks.
For example, a synthetic network generation framework might produce higher quality digital replicates of actual power distribution networks for urban regions as compared to rural areas.
A local comparison highlights this attribute and identifies potential use case scenarios of a given synthetic network generation framework.

Furthermore, availability of actual power distribution network data is sparse due to its proprietary nature.
We may not be able to produce a global comparison between two networks due to unavailability of network data from one of the sources.
Hence, we want to restrict our comparison to only the portions in the region where data from either network is available, which also necessitates a local comparison between the networks.

For a local comparison, we consider uniform sized regions and compute the flat norm distance between the pair of geometries within the region.
However, the computed flat norm is dependent on the length of edges present within the region from either network.
Hence we define the \emph{normalized} multiscale flat norm, denoted by $\widetilde{\mathbb{F}}_{\lambda}$, for a given region as
\begin{equation}
\label{eq:flat-norm-normalized-def}
    \widetilde{\mathbb{F}}_{\lambda}\left(T_1-T_2\right) = \frac{\mathbb{F}_{\lambda}\left(T_1-T_2\right)}{|T_1|+|T_2|}\,.
\end{equation}

For a given parameter $\epsilon$, a local region is defined as a square of size $2\epsilon\times 2\epsilon$ steradians.
Let $T_{1,\epsilon}$ and $T_{2,\epsilon}$ denote the currents representing the input geometries inside the local region characterized by $\epsilon$. Note that the ``amount'' or the total length of network geometries within a square region varies depending on the location of the local region. In this case, the lengths of the network geometries are respectively $|T_{1,\epsilon}|$ and $|T_{2,\epsilon}|$. Therefore, we use the ratio of the total length of network geometries inside a square region to the parameter $\epsilon$ to characterize this ``amount'' and denote it by $|T|/\epsilon$ where 
\begin{equation}
    |T|/\epsilon = \frac{|T_{1,\epsilon}|+|T_{2,\epsilon}|}{\epsilon}\,.
\end{equation}
Note that while performing a comparison between a pair of network geometries in a local region using the multiscale flat norm, we need to ensure that comparison is performed for similar length of the networks inside similar regions.
Therefore, the ratio $|T|/\epsilon$, which indicates the length of networks inside a region scaled to the size of the region, becomes an important aspect of characterization while performing the flat norm based comparison.




\section{Notion of stability for the flat norm distance}
\label{sec:fn-stability}

We now investigate two approaches to define a notion of stability for the flat norm distance.
For any measure of discrepancy between objects, the notion of \emph{stability} is not only desirable from a theoretical standpoint but is also necessary for practical applications. 
The comparison metric is said to be \emph{stable} if \emph{small changes in the input geometries lead to only small changes in the measured discrepancy}.
But such a formulation introduces a ``chicken and an egg'' problem---in order to evaluate the stability of a proposed metric we need an alternative baseline metric to measure the small change in the input. 
Of course, the baseline metric should be stable as well. 
This constitutes the first approach. 
The alternative approach is to derive directly an upper bound on the proposed metric under well-defined controlled perturbations of input geometries.

The Hausdorff distance metric $\mathbb{D}_H$, which has been extensively used in the literature for comparing geometrically embedded networks, is stable in this sense and is hence a natural choice for use as the baseline metric.
At the same time, the Hausdorff distance is not sensitive enough to adequately measure small changes in the input geometries. 
Let us consider a $\e$-ball around each node in the network for a chosen perturbation radius $\e > 0$.
We then uniformly sample a point in each circular region and use them as the perturbed embeddings of the nodes.
The Hausdorff distance will change if the perturbation is either \emph{significantly large} to overshadow the current value by moving some node far enough,
or it is \emph{very specific} and affects the maximizer nodes of $\mathbb{D}_H$.
As will be shown in the next subsection (Sec.~\ref{subsec:FN-VS-HDF}) on a few simple counter-examples and the real-world networks introduced in the previous section (Sec.~\ref{subsec:implementation:fn-power-grids}),
knowing the value of the Hausdorff distance or how it changed is not enough to infer any useful information about the flat norm distance, and vice versa.
Even though our examples are 1-dimensional, the behavior can be observed in any dimension.

In the subsequent subsection (Sec.~\ref{subsec:FN-BOUND}),
under some mild assumptions about the scale and the radius of perturbation, 
we derive an upper bound on the flat norm distance for the case of simple piecewise linear curves in $\real^2$.
We formalize these curves as a special class of integral 1-currents, namely the piecewise linear currents,
and study a class of (positive) $\e$-perturbations of their nodes. 
It allows us to track the change of the components of the flat norm distance while perturbing each node one by one,
which in turn allows us to construct a non-trivial upper bound on $\mathbb{F}_{\lambda}$ between the original 1-current and its final perturbed version.

\subsection{Comparing the Hausdorff and the flat norm distance}
\label{subsec:FN-VS-HDF}
We refer the reader to standard textbooks on geometric measure theory \cite{Federer1969,Morgan2016} for the formal definition of integral currents and other related concepts.
For our purposes, it is sufficient to consider an integral current as a collection of oriented manifolds with or without boundary, and with integer multiplicities as well as orientations for each submanifold.
Recall from Sec.~\ref{subsec:multiscale-flat-norm} that $\mc{C}_d(\real^{d + 1})$ denotes the set of all oriented $d$-dimensional integral currents ($d$-current) embedded in $\real^{d + 1}$,
and $\supp(T) \subset \real^{d+1}$ is the $d$-dimensional support for $T \in \mc{C}_d(\real^{d + 1})$.
Let $X \in \mc{C}_d(\real^{d + 1})$ with $(d-1)$-boundary $\bd X \in \mc{C}_{d - 1}(\real^{d + 1})$,
then the set of all $d$-currents embedded in $\real^{d+1}$ spanned by the boundary of $X$ 
is denoted as $\mc{C}_{d}[\bd X; \real^{d + 1}] \subset \mc{C}_{d}(\real^{d + 1})$,
or simply $\mc{C}_{d}[\bd X]$ if the embedding space is clear from the context.

Let $T_0, T_1 \in \mc{C}_d(\real^{d + 1})$ be two integral $d$-currents in $\real^{d + 1}$, 
and $\norm{v - u}_d$  be the Euclidean distance between $u,v \in \real^d$.
The Hausdorff distance between currents $T_1$ and $T_0$ is given by
\begin{align*}
    \mathbb{D}_H(T_1, T_0)
    &=   \max \left\lbrace 
                \sup\limits_{v \in \supp{T_0}} \mathbb{D}(v, T_1),
                \sup\limits_{v \in \supp{T_1}} \mathbb{D}(v, T_0),
    \right\rbrace
\end{align*}
where $\mathbb{D}(v, T)$ is  the distance from a point to a current given as
\begin{align*}
    \mathbb{D}(v, T) = \inf\limits_{u \in \supp{T}} \norm{v - u}_d\,.
\end{align*}

Let $X = T_1 - T_0 - \bd S$ be the $d$-component of the flat norm decomposition in Eq.~\eqref{eq:flatnorm-def}.
Note that $\bd X = \bd T_1 - \bd T_0 - \bd \bd S = \bd T_1 - \bd T_0$,
i.e., $X \in C_d[\bd T_1 - \bd T_0]$,
and it can be rendered to zero only if $\bd T_1 - \bd T_0 \equiv 0$. 
Hence, the volume of the minimal $d$-current spanned by $\bd T_1 - \bd T_0$ provides a lower bound on $\mathbb{F}_{\lambda}(T_1 - T_0)$.
On the other hand the Hausdorff distance between boundaries  $\mathbb{D}_H(\bd T_1, \bd T_0)$ doesn't provide any meaningful insights about the actual value of  $\mathbb{D}_H(T_1, T_0)$.
This prompts us to suspect that the Hausdorff distance between two $d$-currents does not have any meaningful relations with the value of the flat norm distance.
In fact, we expect $\mathbb{F}_{\lambda}(T_1 - T_0)$ to be more sensitive to perturbations of input geometries.
Moreover, as can be seen from the examples below and the discussion in the following Section \ref{subsec:FN-BOUND},
given that the scale $\lambda > 0$ is small enough,
when $T_1$ is perturbed within a $\e$-tube
the range of incurred changes of the flat norm distance depends, mainly, on the size of perturbation $\e > 0$ and the input volume $\mathbf{V}_d(T_1)$, and not on $\mathbb{D}_H(T_1, T_0)$.
Although the examples in this Section are given for 1-currents in $\real^2$ for illustrative purposes, this conclusion holds in the general case of $d$-currents as well.

\begin{example}[Fixing $\mathbb{D}_H(\tld{T}_1, T_0)$] \label{eg:Hdf-example-1}
Let $T_1$ and $T_0$ be two 1-currents in $\real^2$
with common boundaries, $\bd T_1 = - \bd T_0$,
and let $H = \mathbb{D}_H(T_1, T_0)$ be the Hausdorff distance between them.
For some small $\e > 0$, consider $\tld{T}_1$---a perturbation of  $T_1$  within  an $\e$-tube---such that the boundaries and the Hausdorff distance do not change:
\begin{align*}
\begin{array}{ccccc}\dsp
    \bd \tld{T}_1 = \bd T_1 
    &\text{ and }&\dsp
    \mathbb{D}_H(\tld{T}_1, T_0) = H.
\end{array}
\end{align*}
Note that $\mathbb{D}_H(\bd T_1, \bd T_0) = \mathbb{D}_H(\bd T_1, \bd \tld{T}_1) = 0$.
See Fig.~\ref{fig:currents:example-currents-and-neighborhoods} for  examples of perturbations $\tld{T}_1$.

\begin{figure}[ht!]
    \begin{minipage}{.50\textwidth}
                \centering
                \includegraphics[width=0.95\textwidth]{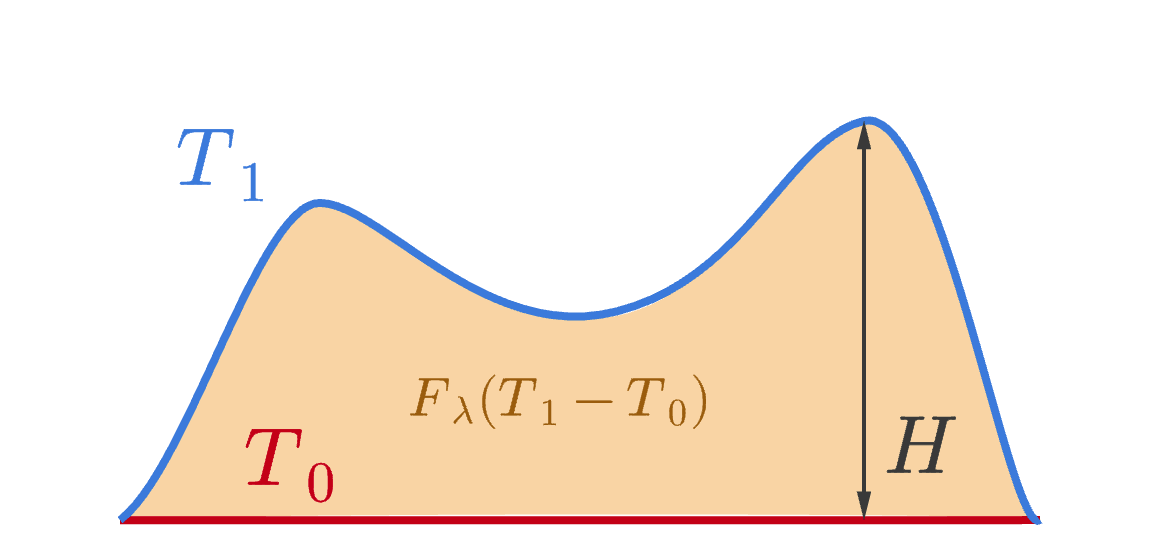} \\
                 \includegraphics[width=0.9\textwidth]{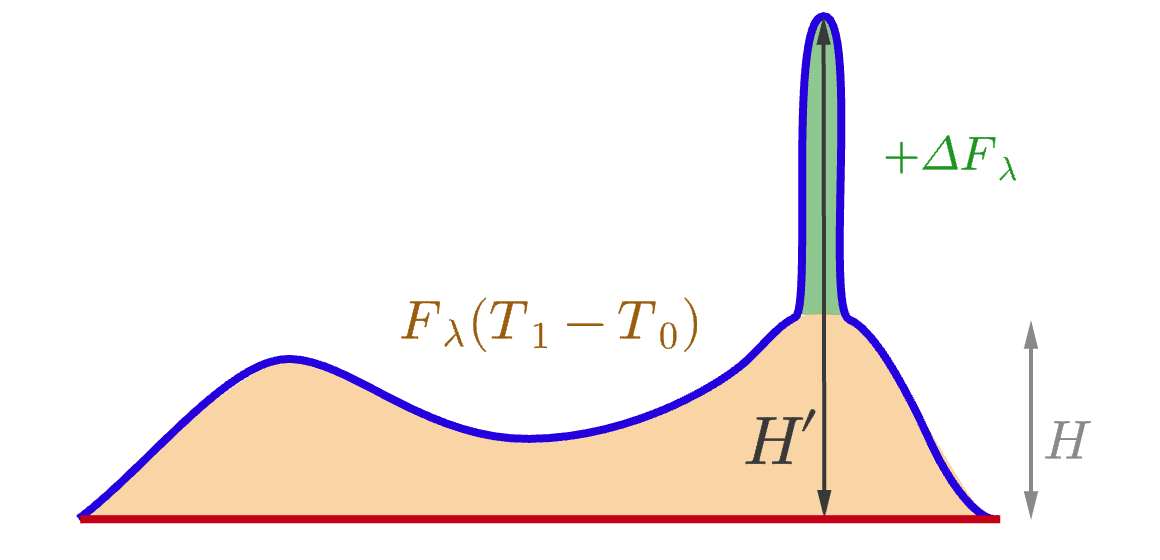}
    \end{minipage}%
    \begin{minipage}{.50\textwidth}
                \centering
                \includegraphics[width=0.95\textwidth]{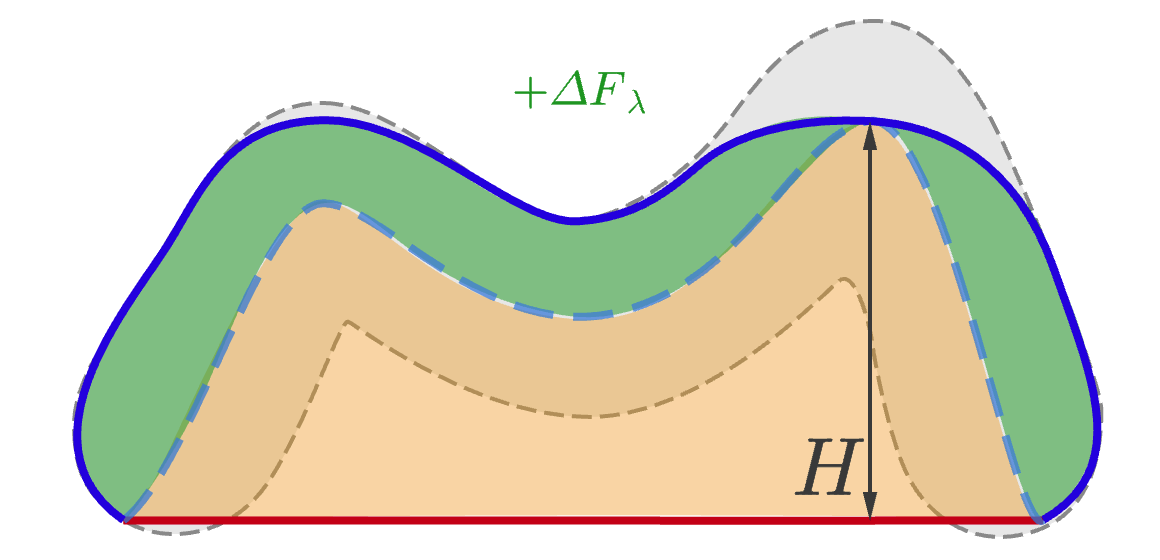} \\
                \includegraphics[width=0.95\textwidth]{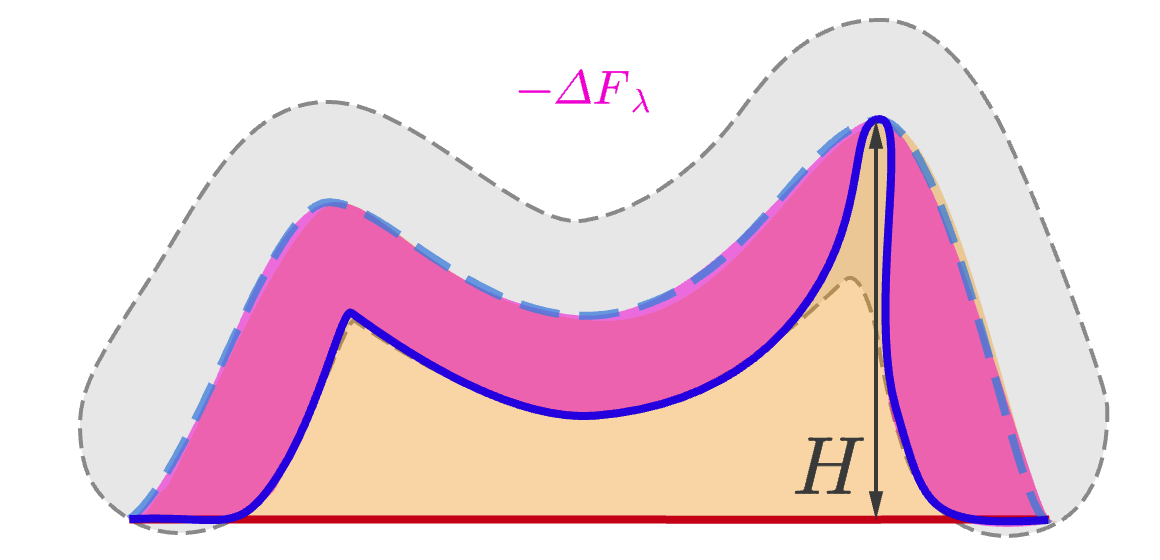}
    \end{minipage}\\ 
    \caption[Comparing the Hausdorff distance and flat norm distance.]{
        \textbf{Top Left:}
            The input curves \textcolor{AzureBlue}{$T_1$} and \textcolor{BrightRed}{$T_0$} with shared endpoints
            and Hausdorff distance $\mathbb{D}_H(T_1, T_0) = H$.
            At a small enough scale $\lambda > 0$, the flat norm distance \textcolor{DarkOrange}{$\mathbb{F}_{\lambda}(T_1 - T_0)$} corresponds to the orange patch in between them.
        \textbf{Right: Example 1.}
            The example perturbations \textcolor{FullBlue}{$\tld{T}_1$} (solid  blue)  that lie within a $\e$-neighborhood (gray) of \textcolor{AzureBlue}{$T_1$}(dashed blue).
            The Hausdorff distance between \textcolor{FullBlue}{$\tld{T}_1$} and \textcolor{BrightRed}{$T_0$} remains same, i.e., $H$.
            The green patch (Right, Top) captures the  \textcolor{ForestGreen}{increment $\Delta \mathbb{F}$},
            and beneath it (Right, Bottom) the pink area corresponds to the \textcolor{DeepPink}{decrement $\Delta \mathbb{F}$}.
        \textbf{Bottom Left: Example 2.}
            The example perturbation of $T_1$ that moves \emph{only} the maximizer of $\mathbb{D}_H(T_1, T_0)$
            further away from $T_0$ so that $\mathbb{D}_H(\tld{T}_1, T_0) = H^{\prime} >> H$.
            The flat norm distance increases by  \textcolor{ForestGreen}{$\Delta \mathbb{F}$} that corresponds to the area of the created spike, which can be arbitrary small.
    }
\label{fig:currents:example-currents-and-neighborhoods}
\end{figure}

We could have cases where $\tld{T}_1$ lies mostly at the upper envelope of this $\e$-tube,
causing the flat norm distance to increase by $\Delta \mathbb{F}_{\lambda} = \Abs{\mathbb{F}_{\lambda}(T_1 - T_0) - \mathbb{F}_{\lambda}(\tld{T}_1 - T_0)}$ (highlighted in green),
or mostly at the lower envelope causing a decrease in the flat norm distance, respectively (highlighted in pink).
In both cases, one would expect the ideal measure of discrepancy between $\tld{T}_1$ and $T_0$ to change significantly as well (compared to the one between $T_1$ and $T_0$).
The flat norm distance accurately captures all such changes (to keep the example simple, we consider the default flat norm distance and not the normalized version).
At the same time, both such variations could have the same Hausdorff distance $H$ from $T_0$ as $T_1$, which completely misses all the changes applied to $T_1$ in either case.
\end{example}

\begin{example}[Fixing $\mathbb{F}_{\lambda}(\tld{T}_1 - T_0)$] \label{eg:Hdf-example-2}
A modification of this example can illustrate the other extreme case---when Hausdorff distance changes by a lot but the flat norm distance does not change much at all, see bottom row in Fig.~\ref{fig:currents:example-currents-and-neighborhoods}.
Consider moving \emph{only} the highest point on $T_1$ further away from $T_0$ so that Hausdorff distance becomes $H^{\prime} >> H$,
as shown on the bottom left figure of Fig.~\ref{fig:currents:example-currents-and-neighborhoods}.
We keep $T_1$ a connected curve, thus creating a sharp spike in it.
While the Hausdorff distance between the curves has increased dramatically, the flat norm distance sees only a minute increase as measured by the area under the spike.
Moreover, the increment $\Delta \mathbb{F}_{\lambda}$ can be decreased to almost zero by narrowing the spike.
Once again, the flat norm distance accurately captures the intuition that the curves have \emph{not} changed much when just a single point moves away while the rest of the curve stays the same.
Hence the flat norm provides a more robust metric that better captures significant changes while maintaining stability to small perturbations (also see Section \ref{subsec:FN-BOUND} for theoretical bounds).
\end{example}

\subsection{Empirical study} \label{ssec:Hdf-example-empirical}
We observe similar behavior to those illustrated by the theoretical example (Fig.~\ref{fig:currents:example-currents-and-neighborhoods}) in our computational experiments.
Fig.~\ref{fig:flatnorm-stability-empirical} shows scatter plots denoting empirical distribution of percentage deviation of the two metrics from the original values $\left(\%\Delta\mathbb{D}_{H},\%\Delta\widetilde{\mathbb{F}}_{\lambda}\right)$ for a local region. The perturbations are considered for three different radii shown in separate plots. We note that the percentage deviations in the two metrics are comparable in most cases. In other words, neither metric behaves abnormally for a small perturbation in one of the networks.
\begin{figure}[htb!]
    \centering
    \includegraphics[width=0.95\textwidth]{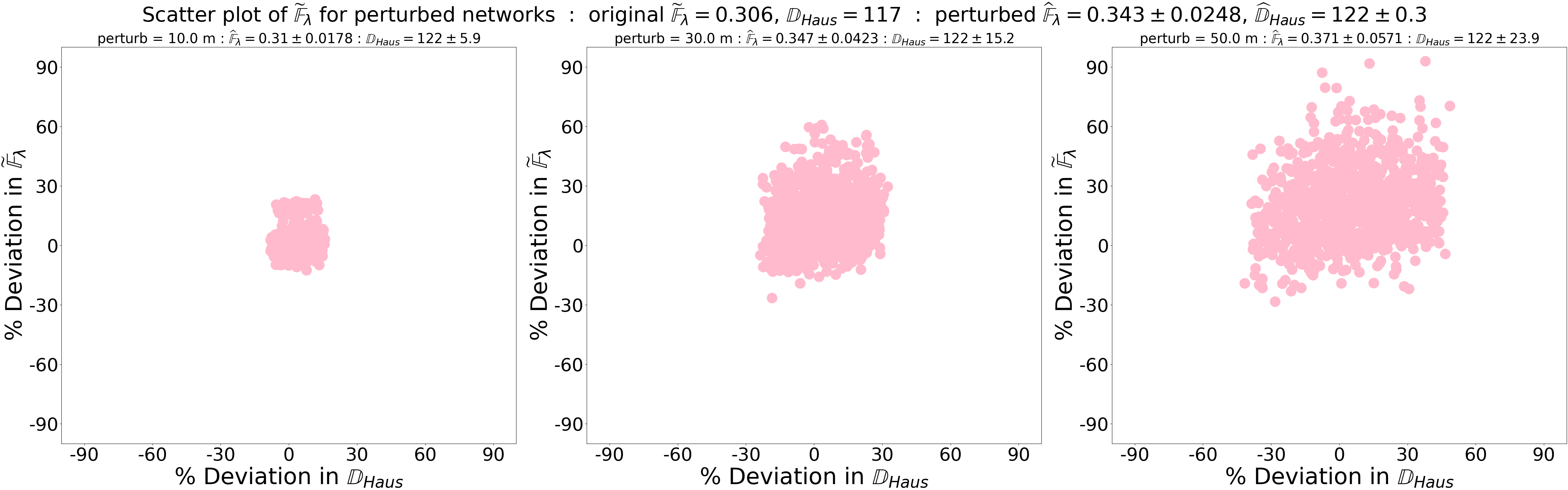}
    \caption[Scatter plots showing the effect of network perturbation on the normalized flat norm and Hausdorff distances for a local region.]
    {Scatter plots showing the effect of network perturbation on the normalized flat norm and Hausdorff distances for a local region.
      The percentage deviation in the metrics for the perturbations is shown along each axis.
      We do not observe significantly large deviations in any one metric for a given perturbation.}
    \label{fig:flatnorm-stability-empirical}
\end{figure}

Next, we compare the sensitivity of the two metrics to outliers.
Here, we consider a single random node in one of the networks and perturb it.
Fig.~\ref{fig:flatnorm-outlier-empirical} shows the sensitivity of the metrics to these outliers.
The original normalized flat norm and Hausdorff distance metrics are shown by the horizontal and vertical dashed lines respectively.
The points along the horizontal dashed line denote the cases where the Hausdorff distance metric is more sensitive to the outliers, while the normalized flat norm metric remains the same.
These cases occur when the perturbed random node determines the Hausdorff distance, similar to the second Example where Hausdorff distance increased from $H$ to $H^{\prime}$.
On the flip side, the points along the vertical dashed line denote the Hausdorff distance remaining unchanged while the normalized flat norm metric shows variation.
Just as in the theoretical example (Fig.~\ref{fig:currents:example-currents-and-neighborhoods}), such variation in the normalized flat norm metric implies a variation in the network structure.
However, such variation is not captured by the Hausdorff distance metric.
Hence, our proposed metric is capable of identifying structural differences due to perturbations while remaining stable when widely separated nodes (which are involved in Hausdorff distance computation) are perturbed.
The other points which are neither on the horizontal nor the vertical dashed lines indicate that either metric can identify the structural variation due to the perturbation.

\begin{figure}[htb!]
    \centering
    \includegraphics[width=0.95\textwidth]{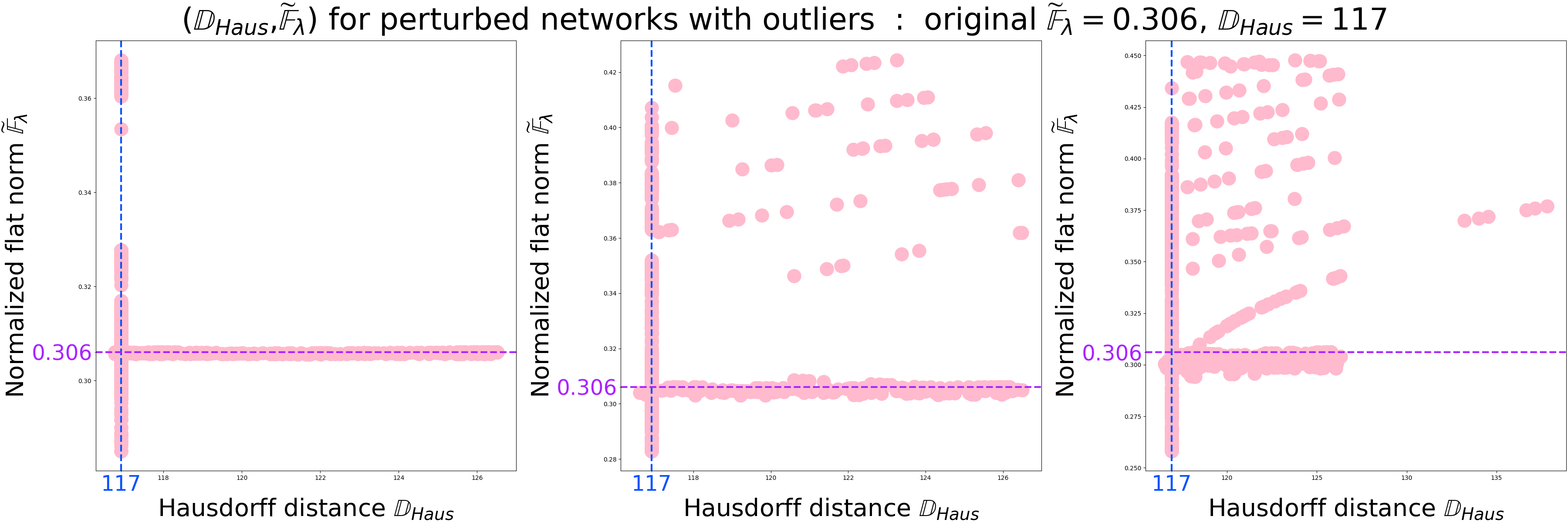}
    \caption[Scatter plots showing the effect of a few outliers on normalized flat norm and Hausdorff distance for a local region.]
    {Scatter plots showing the effect of a few outliers on normalized flat norm and Hausdorff distance for a local region.
    The original normalized flat norm and Hausdorff distance are highlighted by the dashed horizontal and vertical lines.
    We observe multiple cases where the Hausdorff distance is more sensitive to outliers compared to $\tilde{\mathbb{F}}_{\lambda}$.}
    \label{fig:flatnorm-outlier-empirical}
\end{figure}

\section{Stability of the Flat Norm}
\label{subsec:FN-BOUND}

The results of the previous subsection are applicable in all dimensions, i.e., one cannot expect to bound changes in the flat norm by small functions of the changes in the Hausdorff distance.
In this subsection, we adopt a more direct approach to the investigation of the stability of our discrepancy measure between geometric objects. 
Our goal is to construct an upper bound on $\mathbb{F}_{\lambda}$  from the bottom up based only on the input geometries and an appropriately defined radius of perturbation. 
To this end, we consider simple piecewise linear (PWL) curves spanned by a pair of points with no self-intersections embedded in $\real^2$.
Here, \emph{simple} means that there is no self intersection or branching in the curve that connects two points.
Despite its apparent simplicity,  this class of curves is of particular interest, since they can potentially approximate any continuous non-intersecting curve in $\real^2$.
More directly, power grid networks that form the main motivated for our work can be seen as collections of such simple curves.
Although, the results of this subsection are proven only for a pair of simple PWL currents, the empirical findings on the real-world networks, presented in the next section (Sec.~\ref{sec:stats}) comply surprisingly well with the upper bounds established for simple curves (see Fig.~\ref{fig:flatnorm-hists-lambdas}).

We conceptualize a simple piecewise linear curve $\mc{T} \subset \real^2$ between points $s$ and $t$ as the 1-current $T$ embedded in $\real^2$ that is equipped with an edge set $\mathbf{E}(T) = \left\lbrace e_{1}, e_{2} , \ldots, e_{n-1}, e_{n} \right\rbrace$ given by the linear segments of $\mc{T}$,
and a node set $\mathbf{N}(T) = \left\lbrace v_0, v_1, \ldots, v_{n - 1}, v_{n} \right\rbrace$, where $v_0 = s$ and $v_n = t$, and $v_i = e_i \cap e_{i+1}$ for $i=1,\dots,n-1$.
Such currents are defined as a formal sum of their edges $T = e_1 + \ldots + e_n$, and can be thought of as discretized linear approximations of simple continuous curves in $\real^2$.
We refer to this type of currents as \emph{PWL currents},
and denote the set of all PWL currents with $n$ edges spanned by $s$ and $t$ (i.e., starting at $s$ and ending and $t$) as $\mc{L}_n[s, t]$ or $\mc{L}_n[s, t; \real^2]$.
A \emph{subcurrent} of $T \in \mc{L}_{n}[s, t]$ spanned by nodes $v_i$ and $v_j$, where $i < j$, is denoted $T[v_i, v_j] \subseteq T$, and $\mathbf{E}\big( T[v_i, v_j] \big) = \left\lbrace e_{i + 1} , \ldots, e_{j} \right\rbrace$,
which implies $T[v_i, v_j] \in \mc{L}_{j - i}[v_i, v_j]$.
The length of $T$ is given by $\abs{T} = \sum_{j = 1}^{n} \abs{e_j}$.

We start with a PWL current $T_0$ in $\real^2$ and its copy $T_1$.
Next, we consider a sequence of perturbations of $T_1$'s nodes within some $\e$-neighborhood to obtain $\tld{T}_1$,
while tracking the components of the flat norm distance between the original and the perturbed copy.
Recall that the components of the flat norm distance between generic inputs $T_0$ and $T_1$ are the perimeter of the unfilled void $\abs{T_1 - T_0 - \bd S}$ and the area of a 2-current $S$ given as $\mc{A} \left(S\right) = \mathbf{V}_2\left(S\right)$, see Fig.~\ref{fig:demo-flatnorm-basic}.
We refer to them as the \emph{length} (or \emph{perimeter}) and the \emph{area} components of flat norm distance, respectively:
\begin{align}
\label{eq:flat-norm-1-currents}   
    \mathbb{F}_\lambda\left(T_1 - T_0\right) 
    &= 
        \min\limits_{S \in \mc{C}_2(\real^2) }
        \big\{ \Abs{T_1 - T_0 -\bd S}  + \lambda \mc{A} \left(S\right) \big\}.
\end{align}

Since $T_0$ and $T_1$ are identical to start with, the flat norm distance between them is zero (for $S = \emptyset$).
Now let us take a look at the components of the flat norm distance $\mathbb{F}_\lambda(\tld{T}_1 - T_0)$ between the original current and the perturbed copy $\tld{T}_1$.
Let $X \subset \real^2$ be the 2-dimensional void with the boundary given by $\tld{T}_1 - T_0$,
and $S \in C_2(\real^2)$ be a 2-current that fills in, possibly partially, the void $X$.
The area component in the optimal decomposition of $\mathbb{F}_\lambda(\tld{T}_1 - T_0)$ is  bounded by the area of $X$, $\mc{A}(S) \leq \mc{A}(X)$,
and is maximized when the void is fully filled, i.e., $S \equiv X$.
On the other hand, the length component can be, potentially, arbitrarily large due to the complexity of $\bd S$:
\begin{align}
\label{eq:pwl-fn-length-bound}    
  0 \leq \abs{\tld{T}_1 - T_0 -\bd S} 
  \leq \abs{\tld{T}_1 - T_0} + \abs{\bd S}.
\end{align}

Here we make an important assumption about the values of parameters $\lambda > 0$ and $\e > 0$,
formalized below, that allows us to circumvent the potential unboundedness of the perimeter component. 
We mention that the problem of identifying the ranges of values of parameters that fit the assumption is out of the scope of this paper, and will be the focus of future research.

\begin{assumption}[Filled voids]
\label{ref:main-assumption}
  For any original current $T_0$ embedded in $\real^2$,  
  the scale parameter $\lambda = \lambda(T_0) > 0$ is small enough 
  such that for any size of the perturbation $\e = \e(\lambda, T_0) > 0$ taken within some range $0 < \e < M_{\lambda}(T_0)$,
  \textbf{the optimal decomposition of the flat norm distance between $T_0$ and its consecutive perturbations always fills in all the voids that appear as a result of the perturbations}.
\end{assumption}
We note that this assumption does not introduce any additional challenges in implementing the flat norm distance, 
since we always can find a small enough $\lambda$ by scaling it by one-half until all gaps between the input geometries are filled.

Given parameters $\lambda > 0$ and $\e > 0$ under Assumption \ref{ref:main-assumption},
the minimization objective of $\mathbb{F}_{\lambda}(\tld{T}_1 - T_0)$ in Eq.~\eqref{eq:flat-norm-1-currents} is achieved by a 2-current $S \equiv X$, where $X$ is a 2D-void in between $\tld{T}_1$ and $T_0$. 
Hence, $\bd S = \bd X = \tld{T}_1 - T_0$, i.e., $S \in \mc{C}_2[\bd X ] = \mc{C}_2[\tld{T}_1 - T_0]$.
This implies that the length component in Eq.~\eqref{eq:pwl-fn-length-bound} renders to 0---its minimum value.
Conversely, if we would leave the void unfilled, i.e., $S \equiv 0$,
then the area component as well as its boundary become zero, $\mc{A}(S) = 0$ and $\abs{\bd S} = 0$,
while the length component is equal to the void's perimeter: $\abs{\tld{T}_1 - T_0 - \bd S} = \abs{\tld{T}_1 - T_0} = \abs{\bd X}$.
Hence, under Assumption \ref{ref:main-assumption}, the optimization objective of $\mathbb{F}_{\lambda}(\tld{T}_1 - T_0)$ reduces to the scaled area of $S$ for $S \in \mc{C}_2[\tld{T}_1 - T_0]$: 
\begin{align}
\label{eq:pwl-fn-objective}     
  \mathbb{F}_{\lambda}(\tld{T}_1 - T_0) 
   &= \min\limits_{S \in \mc{C}_2[\bd X] }
           \left\{ \abs{\tld{T}_1 - T_0 -\bd S}  + \lambda \mc{A} \left(S\right) \right\}
 = \lambda \mc{A}(S).
\end{align}

This collapsed minimization objective implies that for the scale and the perturbation radius given by our assumption, 
the void produced by perturbing the copy of $T_0$ 
is filled in by a 2-current $S$ spanned by $\tld{T}_1 - T_0$,
such that the scaled area of $S$ is upper bounded by its (non-scaled) perimeter:
\begin{align}
\label{eq:pwl-fn-upper-bound}   
  \mathbb{F}_{\lambda}(\tld{T}_1 - T_0) = 
  \lambda \mc{A}(S) < \abs{\bd S} = \abs{\tld{T}_1 - T_0} .
\end{align}

\subsection{\texorpdfstring{$\e$}{Delta}-perturbations of PWL currents}
\label{subsec:PWL-PERTURB}

We consider a PWL current $T_0 \in \mc{L}_{n}[s, t]$ spanned by $s, t \in \real^2$,
called the \emph{original current},
and its copy $T_1 = T_0$, e.g., see Fig.~\ref{fig:pwl-perturbation-00}.
Obviously, $\mathbb{F}_\lambda\left(T_1 - T_0\right) = 0$ at any scale $\lambda > 0$. 

\begin{figure}[ht!]
   \centering
   \includegraphics[width=0.60\textwidth]{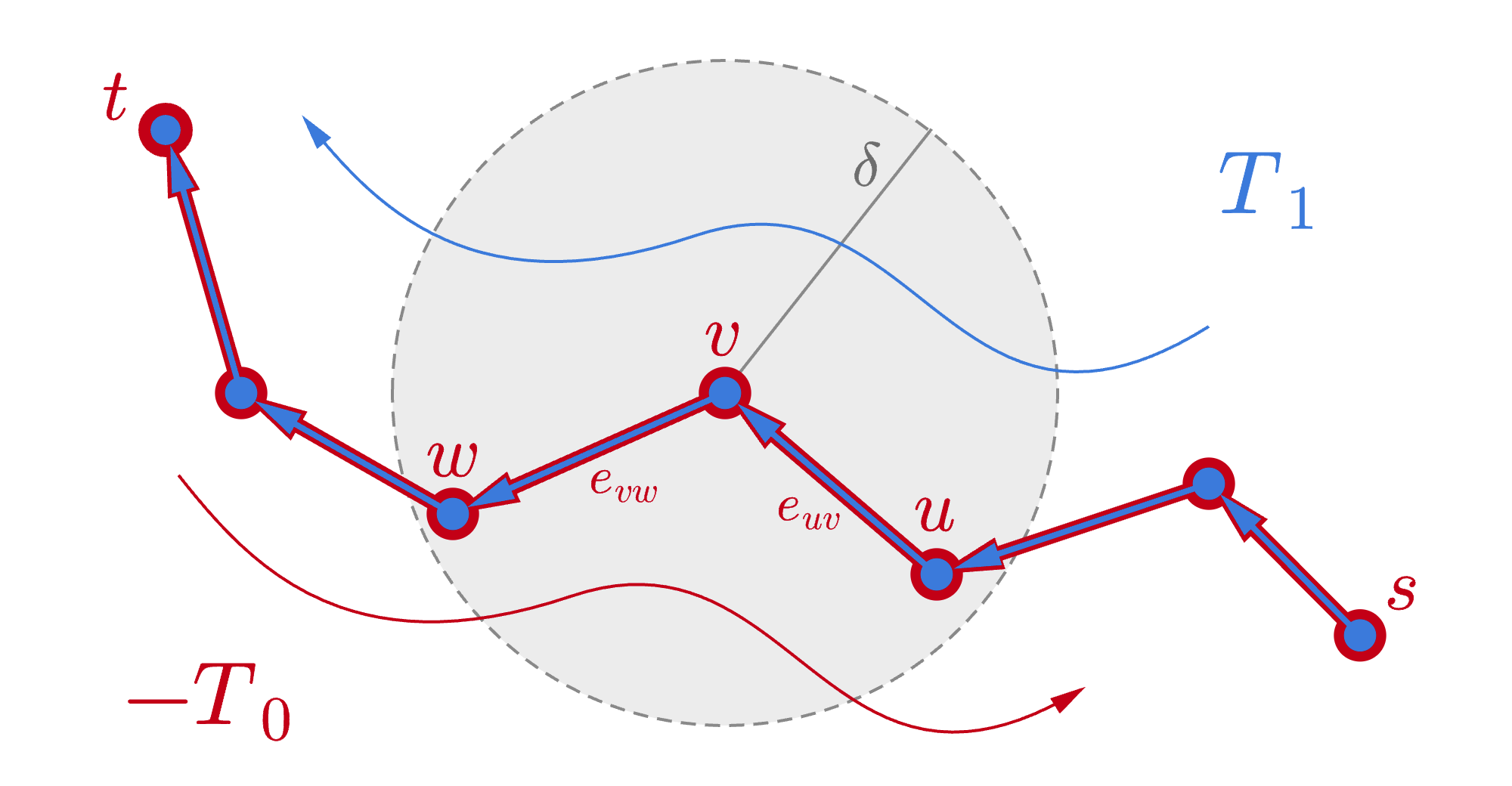}
   \caption[A PWL current \textcolor{BrightRed}{${T_0}$} and its copy \textcolor{AzureBlue}{$T_1$}.]
       {The original PWL current \textcolor{BrightRed}{${T_0}$}$\in \mc{L}_n[s, t]$ and its copy \textcolor{AzureBlue}{$T_1$}.
        The arrows show the orientations in $T_1 - T_0$,
        and the gray disk is the $\e$-ball \textcolor{gray}{$\mc{B}_{\e}(v)$} centered at an interior vertex $v$.
   }
\label{fig:pwl-perturbation-00}
\end{figure}

Let $v = v_i \in T_1$  be an interior  vertex of $T_1$, $1 \leq i \leq n - 1$.
Given a \emph{radius of perturbation} $\e > 0$, 
consider a mapping $v \to \tld{v}$ such that $\tld{v}$ stays within an open $\e$-ball centered at $v$, 
i.e., $\tld{v} \in \mc{B}_{\e}(v) = \left\{ x \in \real^2 \mid \norm{v - x}_2 < \e \right\}$.
Let $u = v_{i-1}$ and $w = v_{i + 1}$ be the adjacent nodes of $v$,
and let the corresponding incident edges be $e_{uv} = (u, v) = (v_{i-1}, v_{i}) = e_{i}$ and $e_{vw} = (v, w) = (v_{i}, v_{i + 1}) = e_{i + 1}$.
Then, let $\tld{e}_{uv} = (u, \tld{v})$ and $\tld{e}_{vw} = (\tld{v},  w)$
be the edge-like currents connecting the neighbors of $v$ to its perturbation $\tld{v}$, see Fig.~\ref{fig:pwl-perturbation-QQ}. 
We define a \textbf{$\boldsymbol{\e}$-perturbation} of $v \in T_1$ as the collection of mappings of $v$ along with its two connected edges under the assumption that the new edges $\tld{e}_{uv}$ and $\tld{e}_{vw}$ \emph{\textbf{do not cross any of the original edges}}.
\begin{align*}
  v \perturbe \tld{v} = \big\{
          v \to \tld{v}, e_{uv} \to \tld{e}_{u v}, e_{vw} \to \tld{e}_{v w} 
    \big\}
    ~\text{ subject to }~ \tld{v} \in \mc{R}_{\e}(v)
\end{align*}
where $\mc{R}_{\e}(v) \subseteq \mc{B}_{\e}(v)$ is the \emph{allowed region} of $\e$-perturbation
defined as a subregion of the $\e$-ball centered at $v$ that is in the direct line of sight of the vertices adjacent to $v$ (see Fig.~\ref{fig:pwl-perturbation-AA}):
\begin{align}
\label{eq:pwl-e-perturb-region}   
  \mc{R}_{\e}(v) = \mc{R}_{\e}(v; T_1) &= 
  \big\{ 
          \tld{v} \in \mc{B}_{\e}(v) \mid \tld{e}_{uv} \cap T_1 = \emptyset,\, \tld{e}_{vw} \cap T_1 = \emptyset 
  \big\}.
\end{align}

\begin{figure}[ht!]
   \centering
   \includegraphics[width=0.55\textwidth]{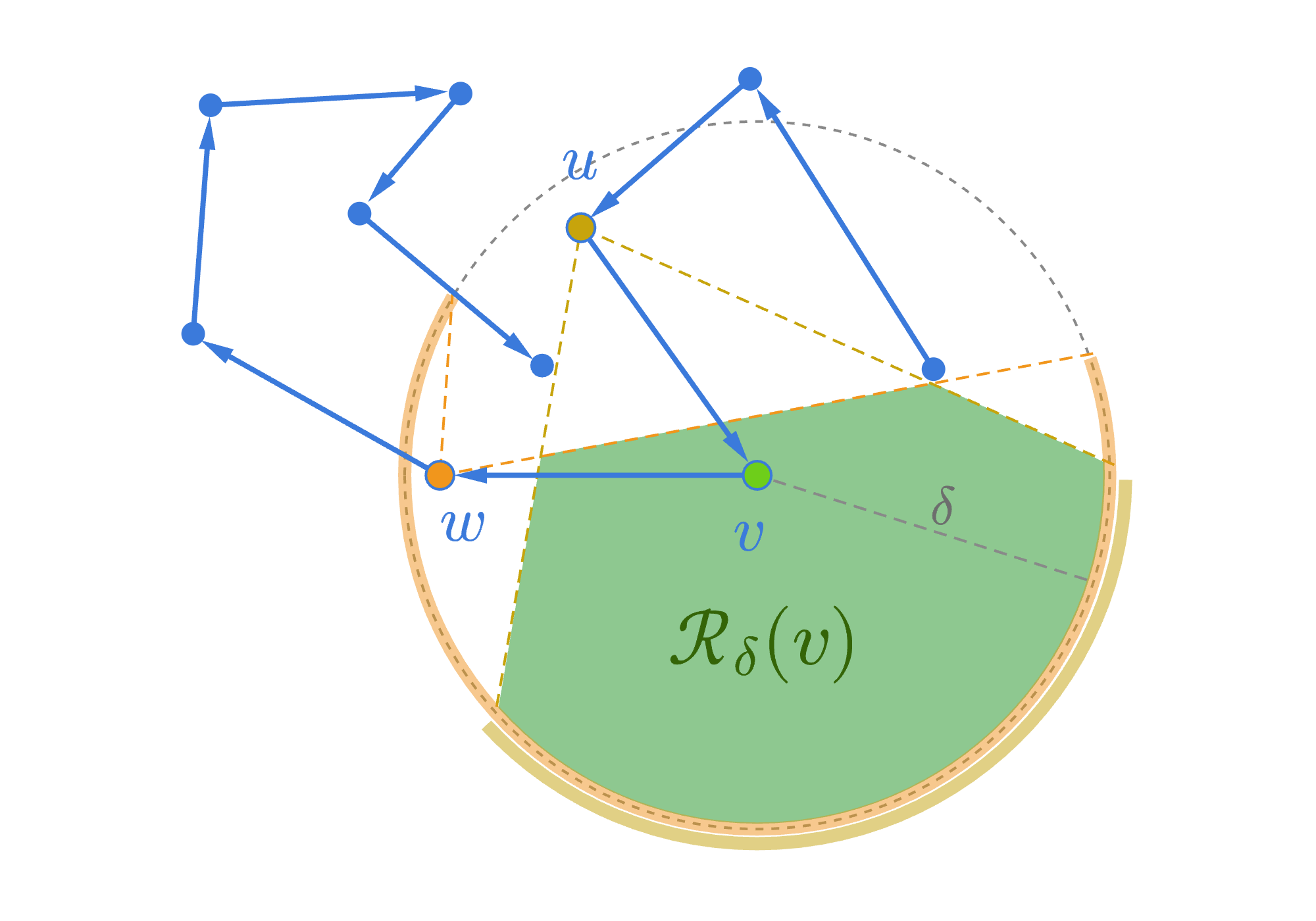}
   \caption[The allowed region \textcolor{ForestGreen}{$\mc{R}_{\e}(v)$} of a $\e$-perturbation  {$v \perturbE \tilde{v}$}.]
       {The \textcolor{DarkYellow}{yellow} and \textcolor{DarkOrange}{orange} circular arcs indicate the subregions of the $\e$-ball that are in the line of sight of nodes $u$ and $w$.
     The allowed region \textcolor{ForestGreen}{$\mc{R}_{\e}(v)$} of a $\e$-perturbation $v \perturbe \tilde{v}$ is shown in green,
     in which it is guaranteed that the perturbed edges will not cross any of the original edges (see Eq.~\eqref{eq:pwl-e-perturb-region}).
   }
   \label{fig:pwl-perturbation-AA}
\end{figure}

A {$\e$-perturbation} of a vertex $v \perturbe \tld{v}$ maps $T_1$ into $\tld{T}_1$ by acting on its node and edge sets as defined below,
and defines a $\e$-perturbation  $T_1 \perturbe \tld{T}_1$.
We say that the $\e$-perturbation of a current $T_1 \perturbe \tld{T}_1$ is \emph{induced} by $v \perturbe \tld{v}$.
Note that the non-overlapping conditions  in Eq.~\eqref{eq:pwl-e-perturb-region},
implies that $\tld{e}_{uv} \cap \tld{T}_1 = \tld{e}_{vw} \cap \tld{T}_1= \emptyset$,
which means that $\tld{T}_1$ is injective (has no self intersections). 
\begin{align*}
  \mathbf{N}(\tld{T}_1)
  &= v \perturbe \tld{v} \big( \mathbf{N}(T_1) \big)
   = \left\{ s,  v_1, \ldots, u, \tld{v}, w,  \ldots, v_{n-1}, t \right\} ~~~\text{ and }
   \\
  \mathbf{E}(\tld{T}_1)
  &= v \perturbe \tld{v} \big( \mathbf{E}(T_1) \big)
   = \left\{ e_s,  e_2, \ldots, \tld{e}_{u v}, \tld{e}_{v w}, \ldots, e_{n-1}, e_t \right\}.
\end{align*}

\noindent The $\e$-perturbations of boundaries are given by the following maps:
\begin{align*}
    s \perturbe \tld{s} &= 
    \big\{
        s \to \tld{s}, e_{s} \to \tld{e}_{s}
    \big\}
    \text{\hspace*{0.6in} and }
    &
    t \perturbe \tld{t} &= 
    \big\{
        t \to \tld{t}, e_{t} \to \tld{e}_{t}
    \big\},
\end{align*} 
where $\tld{s} \in \mc{R}_{\e}(s) = \mc{B}_{\e}(s) \cap \{ \tld{s} \mid \tld{e}_{s} \cap T_1 = \emptyset \}$ 
and $\tld{t} \in \mc{R}_{\e}(t) = \mc{B}_{\e}(t) \cap \{ \tld{t} \mid \tld{e}_{t} \cap T_1 = \emptyset \}$.

\subsubsection{Flat Norm of \texorpdfstring{$\e$}{delta}-perturbations}
\label{subsec:FN-PWL-PERTURB}

\begin{figure}[b!]
    \centering
    \includegraphics[width=0.50\textwidth]{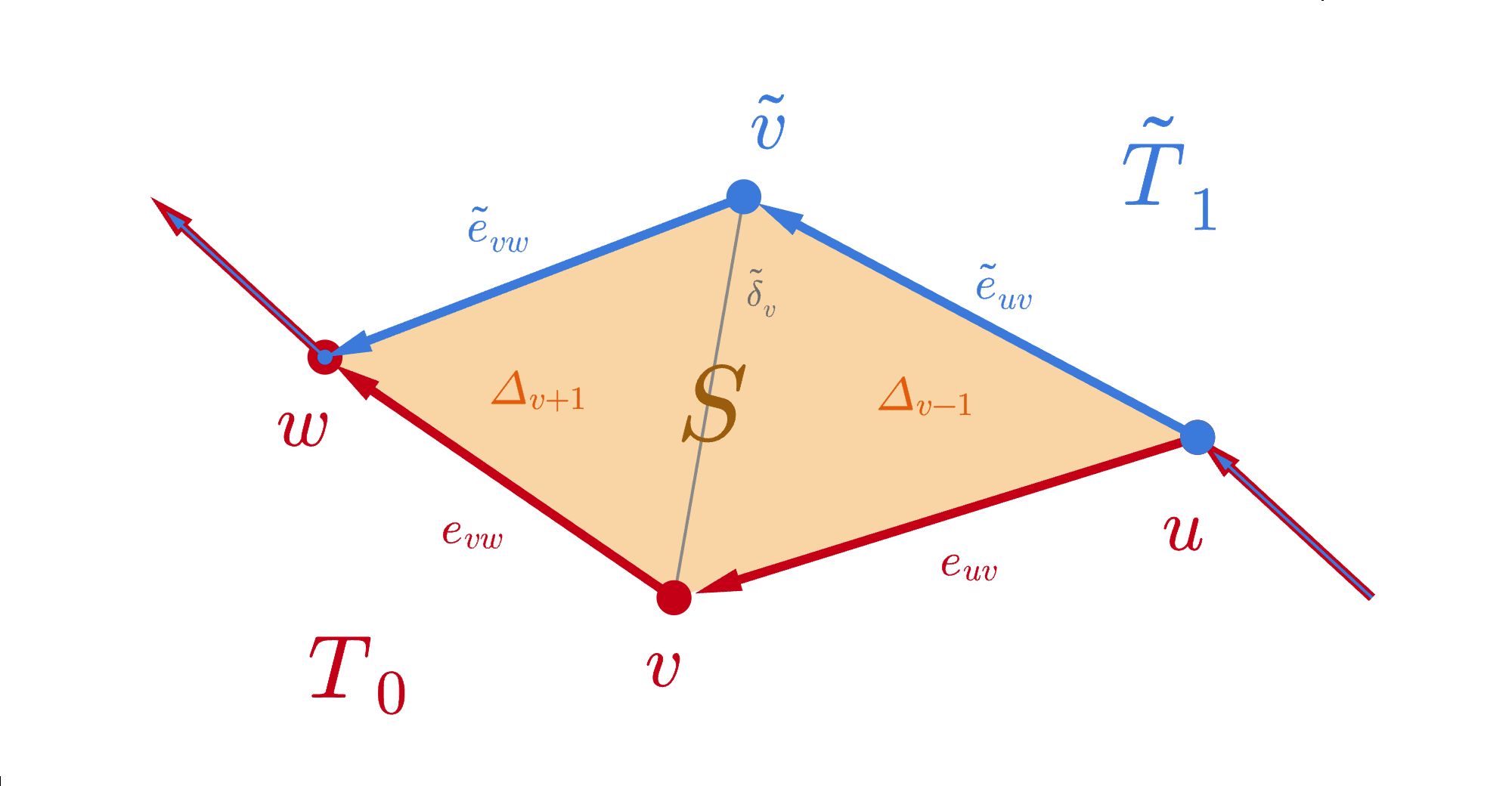}
    \caption[The quadrilateral $S$ that appears as result of a $\e$-perturbation $T_1 \perturbE \tilde{T}_1$.]
        {The quadrilateral $S = [u, v, w, \tilde{v}]  = \Delta_{v + 1} + \Delta_{v - 1}$
       appears as result of a $\e$-perturbation $T_1 \perturbe \tilde{T}_1$ induced by a perturbation of a non-boundary vertex $v \perturbe \tilde{v}$.
   }
    \label{fig:pwl-perturbation-QQ}
\end{figure}

Having specified the necessary definitions and procedures,
we are now ready to prove our first result regarding a single $\e$-perturbation of an interior node.

\begin{theorem}
\label{thm:fn-bound:single-perturbation}
  Let $T_0 \in \mc{L}_n[s,t; \real^2]$ for $n \geq 2$ and $T_1 = T_0$ be its copy.
  Given a small enough scale $\lambda > 0$ and an appropriate radius of perturbation $\e > 0$,
  consider a $\e$-perturbation $T_1 \perturbe \tld{T}_1$ induced by perturbation of an interior node $v \perturbe \tld{v}$.
  Then the flat norm distance between $T_0$ and $\tld{T}_1$ 
  is upper bounded as follows:
  \begin{align*}
    \mathbb{F}_{\lambda}(\tld{T}_1 - T_0) \leq \frac{\lambda \e}{2} \aBs{ T_0[u, w] } 
  \end{align*}
  where $u$ and $w$ are vertices of $T_0$ adjacent to $v$.
\end{theorem}
\begin{proof}
    Recall that by the Assumption~\ref{ref:main-assumption},
    the optimal 2-current $S$ is spanned by $\tld{T}_1 - T_0$,
    and by the construction of $\mc{R}_{\e}(v)$ the new edges $\tld{e}_{uv}$ and $\tld{e}_{vw}$ do not overlap with the original current. 
    Then $\bd S =  \tld{T}_1 - T_0 =  e_{uv} +  e_{vw} - \tld{e}_{vw} - \tld{e}_{uv}$,
    which is the boundary of a quadrilateral spanned by the vertices $[u, v, w, \tld{v}]$, see Fig.~\ref{fig:pwl-perturbation-QQ}.
    To keep the notation simple, we just write $S = [u, v, w, \tld{v}] \in \mc{C}_2[\tld{T}_1 - T_0 ]$.
    Consider the diagonal $\tld{\e}_{v} = (v, \tld{v}) \in \mc{L}_1[v, \tld{v}]$ that connects $v$ to its perturbation.
    Its length is bounded by the radius of perturbation $\abs{\tld{\e}_{v}} \leq \e$ and
    it partitions $S$ into a pair of triangles $\Delta_{v + 1}$ and $\Delta_{v - 1}$:
    \begin{align}         
    \label{eq:pwl-perturb-quad-S}
      S = [u, v, w, \tld{v}] = [v, w, \tld{v}] + [\tld{v}, u, {v}] = \Delta_{v + 1} + \Delta_{v - 1}
    \end{align}
    with boundaries given by $\bd \Delta_{v + 1} = e_{vw} - \tld{e}_{vw} - \tld{\e}_{v}$
    and $\bd \Delta_{v - 1} = e_{uv}  + \tld{\e}_{v} - \tld{e}_{uv}$.
    The corresponding areas are 
    $\mc{A}(\Delta_{v + 1}) = \Half \abs{\tld{\e}_{v}} \abs{e_{v w}} \sin \theta_{v + 1}$ and 
    $\mc{A}(\Delta_{v - 1}) = \Half \abs{\tld{\e}_{v}} \abs{e_{u v}} \sin \theta_{v - 1}$,
    where $\theta_{i}$ is the angle between the diagonal $\tld{\e}$ and the original edge in the corresponding triangle, (see Fig.~\ref{fig:pwl-perturbation-QQ}).
    Note that both $\Delta_{v + 1}$ and $\Delta_{v - 1}$, and consequently $S$,  
    attain the largest possible area
    when the perturbed vertex $\tld{v}$ lands on the boundary of the $\e$-ball $\mc{B}_{\e}(v)$
    and $\tld{\e}_{v}$ is perpendicular to the original edges, i.e., $\theta_{v \pm 1} = \sfrac{\pi}{2}$.
    Let $\Delta_v$ be one of the triangles and $e_v = e_{vw}$ or $e_v = e_{uv}$ be its original edge,
    then the following are the upper bounds on the area of $\Delta_v$ and $S$:
    \begin{align}
    \label{eq:pwl-perturb-area-upper-bound}         
      \mc{A}(\Delta_{v}) = \Half \abs{\tld{\e}_{v}} \abs{e_{v}} \sin \theta
                      &\leq \frac{\e}{2} \abs{e_{v}}
      \\
    \label{eq:pwl-perturb-quad-area-upper-bound}    
      \mc{A}(S) = \mc{A}(\Delta_{v + 1}) + \mc{A}(\Delta_{v - 1})
              &\leq \frac{\e}{2} \big( \abs{e_{vw}}  + \abs{e_{uv}} \big)
    \end{align}
    
    Note that $e_{uv} + e_{vw}$ defines a subcurrent of $T_0$ spanned by $u$ and $w$,
    namely $T_0[u, w]$.
    Finally, recall that under our main Assumption~\ref{ref:main-assumption},
    $\mathbb{F}_{\lambda}(\tld{T}_1 - T_0)$ is given by $\lambda \mc{A}(S)$ (see Eq.~\eqref{eq:pwl-fn-objective}), 
    which together with the upper bound on $\mc{A}(S)$ in Eqn.~\ref{eq:pwl-perturb-quad-area-upper-bound} implies the main statement of the Theorem.
\end{proof}

Furthermore, observe that by the triangle inequality (see Fig.~\ref{fig:pwl-perturbation-QQ})
the length of the perturbed edges in the boundary of $S$ is bounded as follows:
\begin{align}
\label{eq:pwl-perturb-edges-upper-bound}
  \abs{e_v} - \e \leq \abs{e_v} - \abs{\tld{\e}} \leq
  \abs{\tld{e}_v} 
  \leq \abs{e_v} + \abs{\tld{\e}} \leq \abs{e_v} + \e
\end{align}
which implies that $\aBs{\bd \Delta_v} = \abs{e_v} + \abs{ \tld{e}_v} + \abs{\tld{\e}} \geq 2 \abs{{e}_v} - \abs{\tld{\e}} + \abs{\tld{\e}} = 2 \abs{e_v}$,
and thus the following bounds hold: 
\begin{align*}
  2 \abs{e_v}  \leq & ~\aBs{\bd \Delta_v} \leq \dsp 2 \abs{e_v} + 2 \e \quad \text{ and }  \\
  2\big( \abs{e_{v w}} + \abs{e_{u v}} \big) - 2\e  \leq & ~~ \aBs{\bd S} \,\leq \dsp
                        2 \big( \abs{e_{v w }} + \abs{e_{u v}} \big) + 2\e.  \\
\end{align*}

\subsubsection{Sequential $\e$-perturbations} \label{sssec:FN-PWL-PERTURB:SEQUENTIAL}

We now want to derive results similar to Theorem \ref{thm:fn-bound:single-perturbation}
for the case where perturbations are applied to a subcurrent of $T_1$ given by a subset of its adjacent nodes.
To this end, we consider a sequence of $\e$-perturbations of the copy current
$T_1 \perturbe \tld{T}_1 \perturbe \ldots \perturbe \tld{T}^{n - 1}_1$ 
induced by sequential perturbations of the interior points 
$v_1 \perturbe \tld{v}_1, \ldots, v_{n - 1} \perturbe \tld{v}_{n - 1}$,
which we denote as $[v_1, \ldots, v_{n - 1}] \perturbe [\tld{v}_1, \ldots, \tld{v}_{n - 1}]$.

The procedure of $\e$-perturbations described above does not guarantee ``out of the box''
additivity of the area components $\mc{A}(S_i)$ of the corresponding flat norm distances 
$\mathbb{F}_\lambda(\tld{T}^{i}_1 - \tld{T}^{i - 1}_1)$.
The problem arises when a $\e$-perturbation $\tld{v}_i$ lands within a region $S_j$ produced by $v_j \perturbe \tld{v}_j$ for some $j < i$, where $\bd S_j = \tld{T}^{j}_{1} - \tld{T}^{j - 1}_{1}$.
It means that $S_i \cap S_j \neq \emptyset$ and $\mc{A}(S_i + S_j) \neq \mc{A}(S_i) + \mc{A}(S_j)$. 
But since $S_i$ and $S_j$ are embedded in $\real^2$ 
the area of a formal sum $S_i + S_j$ is not larger than the area of their union: 
$\mc{A}\big( S_i + S_j \big) \leq \mc{A} \left(S_i \cup S_j \right) \leq \mc{A}(S_i) + \mc{A}(S_j)$.
Therefore, to find an upper bound for $\mathbb{F}_\lambda(\tld{T}^{n-1}_1 - T_0)$
we can consider only perturbation sequences $T_1 \perturbe \ldots \perturbe \tld{T}^{n - 1}_1$ that produce non-overlapping regions $S_i$,
and hence, additive area components $\mc{A}(S_i)$.

One way to achieve this is to force $\tld{v}$ for any choice of $v$ to land on the same ``side'' of $T_1$, and hence of $T_0$,
by restricting the $\e$-ball $\mc{B}_{\e}(v)$
to a \emph{positive cone} $\mc{B}^{\splus}_{\e}(v)$ given by the edges incident to $v$ in $T_1$. 
As an example in Fig.~\ref{fig:pwl-perturbation-AA-d}, the light green circular arc indicates the segment of the $\e$-ball that corresponds to $\mc{B}^{\splus}_{\e}(v)$.
Let $v = v_i \in \mathbf{N} (T_1)$ be an interior vertex of $T_1$, i.e., $1 \leq i \leq n -1$,
and $e_{uv} = e_{i} = (v_{i - 1}, v_{i})$ and $e_{vw} = e_{i + 1} = (v_{i}, v_{i + 1})$
are the incident edges of $v$ in $T_1$.
Then we get that
\begin{align}
\label{eq:pwl-eplus-perturb-cone}       
  \mc{B}^{\splus}_{\e}(v) = \mc{B}^{\splus}_{\e}(v; T_1)
  &= \bigg\{
      x \in \mc{B}_{\e}(v) 
      \,\bigg\vert 
      \det\big[x - v,\, e_{uv} \big]^{\tr} > 0 \text{ and } \det\big[x - v,\, e_{vw} \big]^{\tr} > 0
  \bigg\}
\end{align}
where $\det [ x, y ]^{\tr} = x_1 y_2 - x_2 y_1$ for $x = (x_1, x_2) \in \real^2$  and $y = (y_1, y_2) \in \real^2$.
In the case of boundary vertices $s$ or $t$, 
the positive cone reduces to the half-disk bounded by a line that contains $e_s$ or $e_t$, respectively.
As previously specified in Eq.~\eqref{eq:pwl-e-perturb-region}, the allowed region of $\tld{v}$ is specified by the non-overlapping conditions:
\begin{align}
\label{eq:pwl-eplus-perturb-region}     
  \mc{R}^{\splus}_{\e}(v) = \mc{R}^{\splus}_{\e}(v; T_1) 
  &= 
    \bigg\{ 
            \tld{v} \in \mc{B}^{\splus}_{\e}(v) 
            \,\bigg\vert \,
            \tld{e}_{uv} \cap T_1 = \tld{e}_{vw} \cap T_1 = \emptyset 
    \bigg\}
  = \mc{B}^{\splus}_{\e}(v; T_1) \cap \mc{R}_{\e}(v; T_1).
\end{align}

\begin{figure}[ht!]
   \centering
   \includegraphics[width=0.55\textwidth]{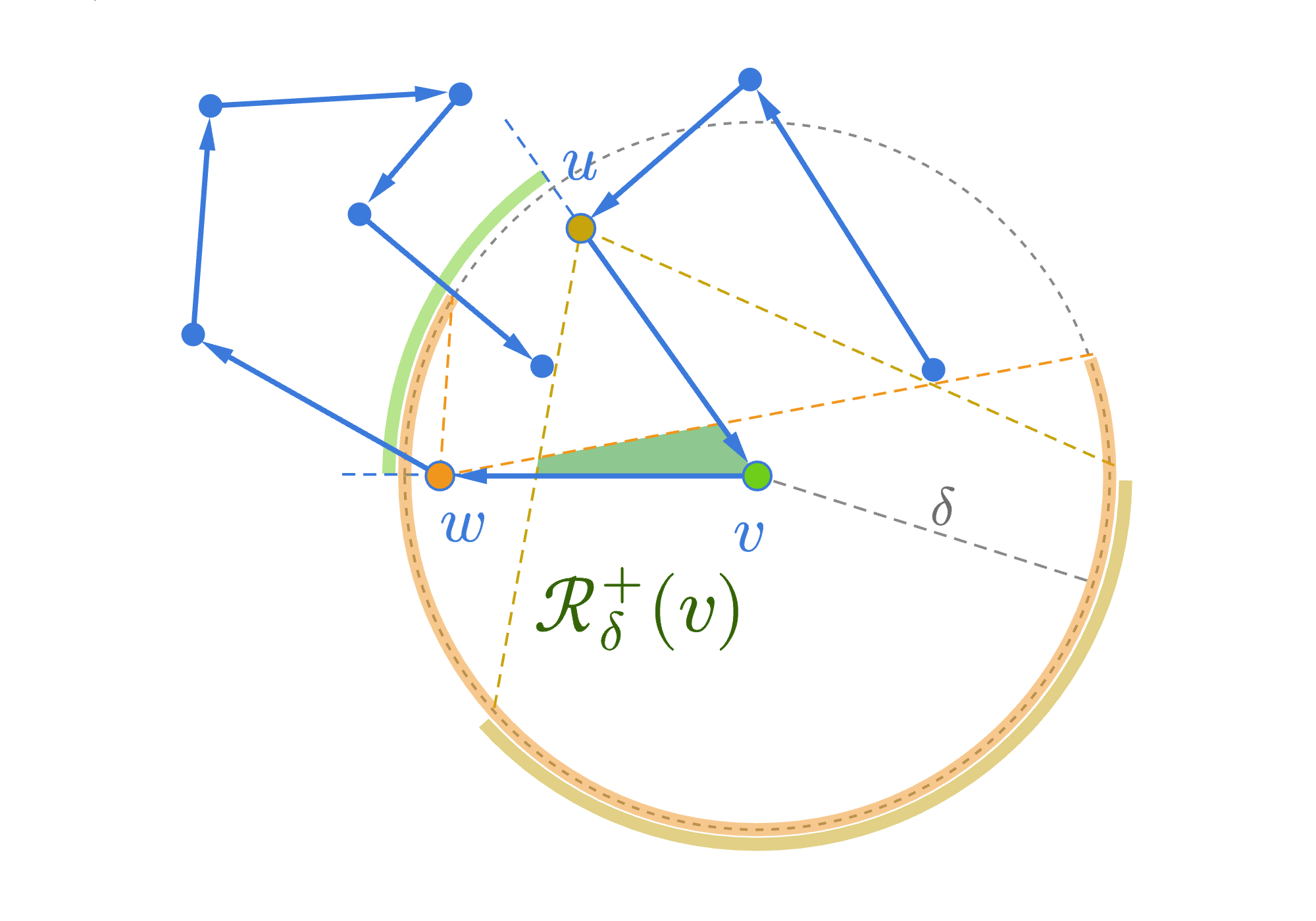}
   \caption[The allowed region \textcolor{ForestGreen}{$\mc{R}^{\splus}_{\e}(v)$}  of a positive $\e$-perturbation $v \perturbEplus \tilde{v}$.]{
     The \textcolor{GrassGreen}{light green} arc indicates the positive cone $\mc{B}^{\splus}_{\e}(v)$ (see Eq.~\eqref{eq:pwl-eplus-perturb-cone}).
     The allowed region \textcolor{ForestGreen}{$\mc{R}^{\splus}_{\e}(v)$} (see Eq.~\eqref{eq:pwl-eplus-perturb-region}) of a positive $\e$-perturbation $v \perturbEplus \tilde{v}$ is shown in green.
     It ensures that \emph{the sequential perturbations produce non-overlapping regions $S_i$} (Proposition~\ref{thm:eplus-non-overlapping}).
     Compare this figure to the allowed region shown in Fig.~\ref{fig:pwl-perturbation-AA}.
   } 
\label{fig:pwl-perturbation-AA-d}
\end{figure}

We call a perturbation $v_i \perturbe \tld{v}_i$
a \textit{\textbf{positive $\boldsymbol{\e}$-perturbation}} of $v_i \in \tld{T}^{i - 1}_1$
if $\tld{v}_i \in \mc{R}^{\splus}_{\e}(v_i; \tld{T}^{i - 1}_1)$.
We denote it as $v_i \perturbeplus \tld{v}_i$, and it induces $\tld{T}^{i - 1}_1 \perturbeplus \tld{T}^{i}_1$.
It easy to see that a sequence of $k$ positive $\e$-perturbations $T_1 \perturbeplus \ldots \perturbeplus \tld{T}^{k}_1$
produces non-overlapping regions $S_1, \ldots, S_k$.

\begin{proposition}
\label{thm:eplus-non-overlapping}
  Let $T_1 \in \mc{L}_n[s,t; \real^2]$ for $n \geq 2$.
  Consider a sequence  of perturbations 
  $T_1 \perturbeplus \tld{T}^{k_0}_1 \perturbeplus \ldots \perturbeplus \tld{T}^{k}_1$ 
  induced by positive $\e$-perturbations of adjacent nodes $[v_{k_0}, \ldots, v_{k}]$
  for some $0 \leq k_0 < k \leq n$.
  Then the regions $S_{k_0}, \ldots, S_k$ produced by the corresponding perturbations in the sequence do not overlap.
\end{proposition}

\begin{proof}
  
  First, when two adjacent nodes are perturbed, $v_{i - 1} \perturbeplus \tld{v}_{i - 1}$ and $v_{i} \perturbeplus \tld{v}_{i}$,
  the edge $e_{i} = (v_{i - 1}, v_{i})$ is perturbed twice:
  $e_i \to \tld{e}_i \to \tld{\tld{e}}_i$,
  where $\tld{e}_i \in \mc{L}_{1}[\tld{v}_{i-1}, v_i]$
  and $\tld{\tld{e}}_i \in \mc{L}_{1}[\tld{v}_{i-1}, \tld{v}_i]$.
  Therefore, the  boundaries of corresponding regions $S_{i - 1}$ and $S_i$ are:
  \begin{align}
    \bd S_{i - 1} = & \, \tld{T}^{i - 1}_1 - \tld{T}^{i - 2}_1 = \tld{e}_{i - 1} + e_{i} - \tld{\tld{e}}_{i - 1} -\tld{e}_{i} ~\text{ and }
    \\
    \bd S_{i} = & ~~ \tld{T}^{i}_1 - \tld{T}^{i - 1}_1 ~= \tld{e}_{i} + e_{i + 1} - \tld{\tld{e}}_{i} -\tld{e}_{i + 1}. 
  \end{align}
  
  It is worth pointing out that when only interior nodes are perturbed, i.e., $1 \leq k_0$ and $k \leq n$,
  the edges $e_{k_0} = (v_{k_0 - 1}, v_{k_0}) \in S_{k_0}$ and $e_{k + 1} = (v_{k}, v_{k + 1}) \in S_k$ are perturbed only once.
  Thus, the edges incident to the boundaries of $T_1$, namely $e_s$ and $e_t$, 
  are perturbed to $\tld{\tld{e}}_s$ or $\tld{\tld{e}}_t$ if and only if the boundary nodes $s$ and $t$ were perturbed
  together with all the nodes in between. 
  In this case the corresponding area components $S_0$ and $S_n$
  are given by  triangles $\Delta_s = [{s}, \tld{v}_1, \tld{s}]$ 
  and $\Delta_t = [\tld{t}, \tld{v}_{n-1}, t]$, respectively.

  Second, observe that any point $x$ within $S_i$ is in $\mc{B}^{\splus}_{\e}(v_i)$ as well, 
  which means that $\det\big[x - v_i,\, \tld{e}_{i} \big]^{\tr} > 0 $ and  $\det\big[x - v_i,\, e_{i + 1} \big]^{\tr} > 0$.
  This point also belongs to $\mc{B}^{-}_{\e}(\tld{v}_i) = \big\{ x \in \mc{B}_{\e}(\tld{v}_i)  \mid \det\big[x - \tld{v}_i,\, \tld{\tld{e}}_{i} \big]^{\tr} < 0 \text{ and } \det\big[x - \tld{v}_i,\, \tld{e}_{i + 1} \big]^{\tr} < 0 \big\}$,
  since $x$ is enclosed by $\bd S_i$.
  
  Given a sequence $T_1 \perturbeplus \tld{T}^{k_0}_1 \perturbeplus \ldots \perturbeplus \tld{T}^{k}_1$, 
  let us assume that $\tld{v}_k \in S_j$ for some $k_0 \leq j < k$,
  while $\tld{v}_i \in \mc{R}^{\splus}_{\e}(v_i; \tld{T}^{i - 1}_1)$ for all $k_0 \leq i \leq k$. 
  If $k - j > 1$ then $\bd S_k$ and $\bd S_j$ do not share any edges,
  and hence placing $\tld{v}_k$ inside $S_j$ requires an intersection of edges, which contradicts $\tld{v}_k \in \mc{R}^{\splus}_{\e}(v_k; \tld{T}^{k-1}_1)$.
  If $j = k - 1$ then $\tld{v}_k \in S_k \iff \det\big[\tld{v}_k - v_k,\, \tld{e}_{k} \big]^{\tr} > 0 $
  and $\tld{v}_k \in S_{k - 1} \iff \det\big[\tld{v}_k - \tld{v}_{k - 1},\, \tld{e}_{k} \big]^{\tr} < 0 $,
  where $\tld{e}_k = (\tld{v}_{k-1}, v_k)$,
  which is also a contradiction.
  Therefore $S_k$ does not overlap with any of the previously produced regions $S_{k_0}, \ldots, S_{k -1}$.
\end{proof}

\begin{corollary}[Additivity of $\mathbb{F}_{\lambda}$]
\label{thm:corollary:fn-additivity}
Let $T_0 \in \mc{L}_n[s,t; \real^2]$ for $n \geq 2$, and $T_1 = T_0$ is its copy.
  Given a small enough scale $\lambda > 0$ and an appropriate radius of perturbation $\e > 0$,
  consider a sequence of perturbations $T_1 \perturbeplus \tld{T}^{k_0}_1 \perturbeplus \ldots \perturbeplus \tld{T}^{k}_1$ 
  induced by the positive $\e$-perturbations  of adjacent nodes $[v_{k_0}, \ldots, v_k]$ in $T_1$
  for some $0 \leq k_0 < k \leq n$.
  Then the flat norm distance between the adjacent perturbations is additive
  and sums up to $ \mathbb{F}_{\lambda}(\tld{T}^{k}_1 - T_0)$:
  \begin{align*}
    \mathbb{F}_{\lambda}(\tld{T}^{k}_1 - {T}_0) 
    &= \sum_{i = k_0}^{k} \mathbb{F}_{\lambda}(\tld{T}^{i}_1 - \tld{T}^{i - 1}_1)  
  \end{align*}
  where we set $\tld{T}^{k_0 - 1} = T_1$.
\end{corollary}

\begin{proof}
    Let $S = S_{k_0} + \ldots + S_k$ be the 2-current that combines all the regions produced by $T_1 \perturbeplus \tld{T}^{k_0}_1 \perturbeplus \ldots \perturbeplus \tld{T}^{k}_1$. 
    Since $S_i \cap S_j = \emptyset$ for $i \neq j$  and $\bd S_i = \tld{T}^{i}_1 - \tld{T}^{i - 1}_1$, the boundary of $S$ is given by
    \begin{align*}
      \bd S &= \bd S_{k_0} + \bd S_{k_0 + 1} + \ldots + \bd S_{k - 1} + \bd S_{k} 
    \\
            &= (\tld{T}^{k_0}_1 - {T}_1) + (\tld{T}^{k_0 + 1}_1 - \tld{T}^{k_0}_1) + \ldots + (\tld{T}^{k - 1}_1 - \tld{T}^{k - 2}_1) + (\tld{T}^{k}_1 - \tld{T}^{k - 1}_1)
            = \tld{T}^{k}_1 - T_1. 
    \end{align*}
    
    This means that $S$ is spanned by $\tld{T}^{k}_1 - T_1 = \tld{T}^{k}_1 - T_0$, 
    which under the Assumption~\ref{ref:main-assumption} implies that
    \begin{align*}
      \mathbb{F}_{\lambda}(\tld{T}^{k}_1 - T_0) 
       = \lambda  \mc{A}(S)
       = \lambda \sum_{i = k_0}^{k} \mc{A}(S_i)
      &= \sum_{i = k_0}^{k} \mathbb{F}_{\lambda}(\tld{T}^{i}_1 - \tld{T}^{i - 1}_1).
    \end{align*}
\end{proof}

\begin{figure}[ht!]
   \centering
   \includegraphics[width=0.70\textwidth]{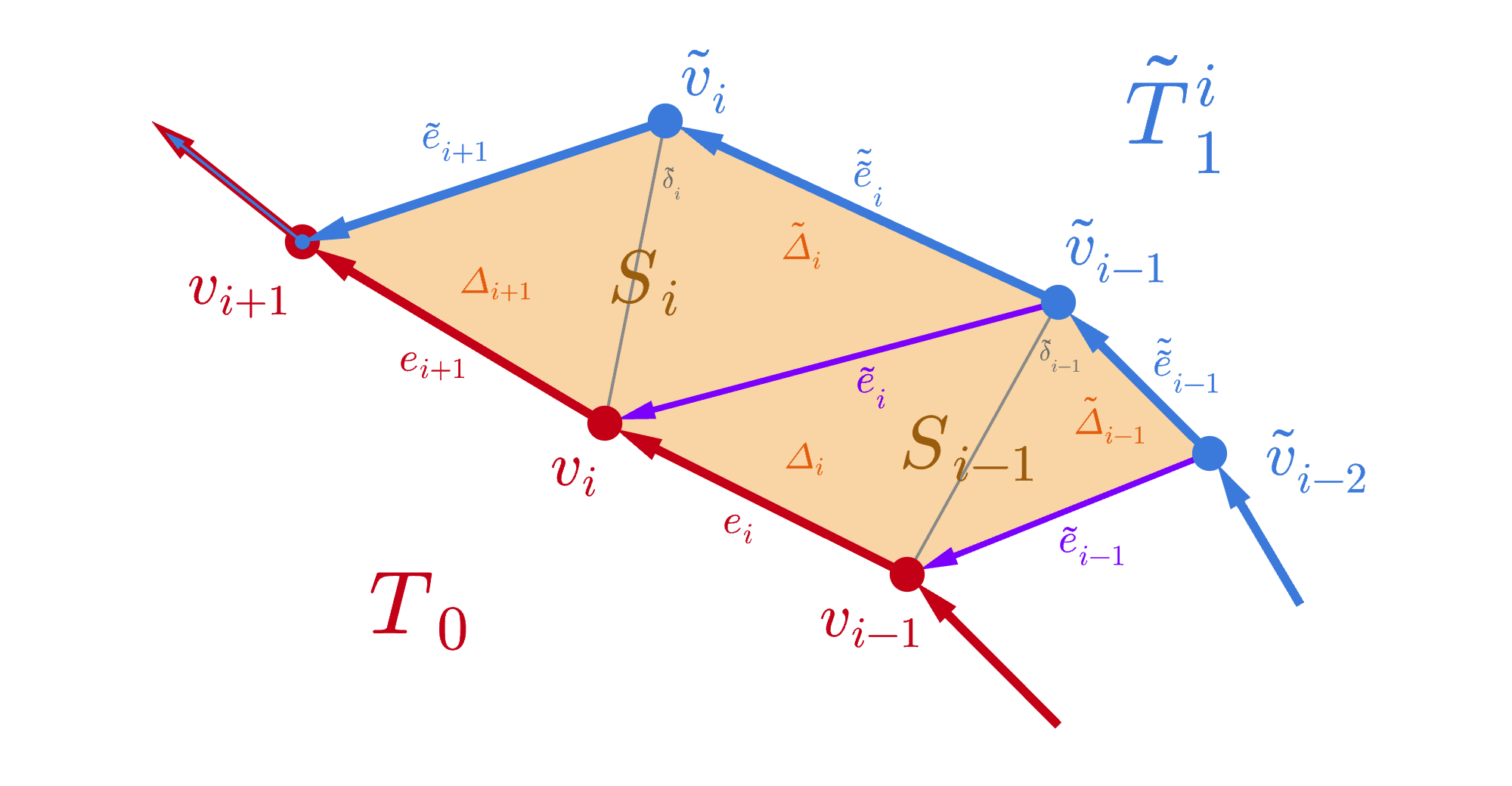}
   \caption[The $i$-th positive $\e$-perturbation.]{
        The $i$-th positive $\e$-perturbation.
        The purple edges show history of the previously perturbed edges \textcolor{DodgerPurple}{$\tld{e}_v$}
        that have been mapped to \textcolor{AzureBlue}{$\tld{\tld{e}}_v$} in $\tld{T}^{i}_1$.
   } 
\label{fig:pwl-perturbation-01-v}
\end{figure}

We now present in Theorem \ref{thm:fn-bound:k-perturbations} an upper bound on the flat norm distance between the input current $T_0$ and its perturbed version resulting from a sequence of perturbations of its \emph{internal} nodes, i.e., nodes in $\mathbf{N} (T_0) \setminus \{s,t\}$.
We subsequently extend the result to include perturbations of all nodes including the boundary nodes in Corollary \ref{thm:corollary:fn-bound-all}.

\begin{theorem}
\label{thm:fn-bound:k-perturbations}
  Let $T_0 \in \mc{L}_n[s,t; \real^2]$ for $n \geq 3$, and $T_1 = T_0$ is its copy.
  Given a small enough scale $\lambda > 0$ and an appropriate radius of perturbation $\e > 0$,
  consider a sequence of perturbations $T_1 \perturbeplus \tld{T}^{k_0}_1 \perturbeplus \ldots \perturbeplus \tld{T}^{k}_1$ 
  induced by the positive $\e$-perturbations  of adjacent \emph{non-boundary} nodes $[v_{k_0}, \ldots, v_k]$ in $T_1$
  for some $1 \leq k_0 <  k \leq n - 1$.
  Then the flat norm distance between $T_0$ and $\tld{T}^{k}_1$ is upper bounded as follows:
  \begin{align*}
    \mathbb{F}_{\lambda}(\tld{T}^{k}_1 - {T}_0) 
    &\leq 
          \frac{\lambda \e}{2} \bigg( 
                    \aBs{T_0[v_{k_0 - 1}, v_{k + 1}]} + \aBs{T_0[v_{k_0}, v_k]} +  (k - k_0)\e 
               \bigg).
  \end{align*}
  
\end{theorem}
\begin{proof}
    As previously was observed in Eq.~\eqref{eq:pwl-perturb-quad-S},
    $S_i$ can be partitioned by $\tld{\e}_{i} = (v_i, \tld{v}_i)$ into a pair of triangles
    $\Delta_{v_{i + 1}} = [v_i, v_{i+1}, \tld{v}_i]$ and $\tld{\Delta}_{v_i} = [\tld{v}_i, \tld{v}_{i - 1}, v_i]$,
    with the respective boundaries $\bd \Delta_{v_{i + 1}} = e_{i + 1} - \tld{e}_{i + 1} - \tld{\e}_{i}$ and 
    $\bd \tld{\Delta}_{v_{i}} = \tld{e}_{i} - \tld{\tld{e}}_{i} + \tld{\e}_{i}$.
    The area of these triangles is bounded as in Eq.~\eqref{eq:pwl-perturb-area-upper-bound}, 
    and the upper bound on the length of the perturbed edges $\tld{e}_{v}$ is given by corresponding triangle inequalities in Eq.~\ref{eq:pwl-perturb-edges-upper-bound}.
    Hence, we get the following upper bound for $\mc{A}(S_i)$:
    \begin{align}
    \label{eq:pwl-perturb-seq-area-bound}     
      \mc{A}(S_i) = \mc{A}(\Delta_{v_{i + 1}}) + \mc{A}(\tld{\Delta}_{v_{i}})
      &\leq \frac{\e}{2} \big( \abs{e_{i + 1}}   + \abs{\tld{e}_{i}} \big)
       \leq \frac{\e}{2} \big( \abs{e_{i + 1}}   + \abs{{e}_{i}} + \e \big)
    \end{align}

    Note that the upper bound on the flat norm distance for the first perturbation in the sequence $T_1 \perturbeplus \tld{T}^{k_0}$
    is given by Theorem~\ref{thm:fn-bound:single-perturbation} instead of Eq.~\eqref{eq:pwl-perturb-seq-area-bound},
    namely $\mathbb{F}_{\lambda}(\tld{T}^{k_0}_1 - T_1) = \lambda \mc{A}(S_{k_0}) \leq \frac{\lambda \e}{2} \big( \abs{e_{k_0}}  + \abs{e_{k_0 + 1}} \big)$.
    As was shown above in the Corollary~\ref{thm:corollary:fn-additivity} 
    under the Assumption~\ref{ref:main-assumption},
    we get additivity of the flat norm distances between consecutive perturbations, and hence the additivity of the upper bounds.
    Then we get the Theorem's claim after doing some algebra:
    \begin{align*}
      \mathbb{F}_{\lambda}(\tld{T}^{k}_1 - T_0) 
      &= \sum_{i = k_0}^{k} \mathbb{F}_{\lambda}(\tld{T}^{i}_1 - \tld{T}^{i - 1}_1)  
       = \lambda \sum_{i = k_0 + 1}^{k} \mc{A}(S_i) + \lambda \mc{A}(S_{k_0})
    \notag
    \\
      &\leq \frac{\lambda \e}{2} \left( 
                  \sum_{i = k_0 + 1}^{k}(\abs{\tld{e}_i} + \abs{e_{i + 1}}) + (\abs{e_{k_0}} + \abs{e_{k_0 + 1}}) 
            \right)
    \notag
    \\
      &\leq \frac{\lambda \e}{2} \left( 
                  \sum_{i = k_0}^{k + 1} \abs{e_{i}} + \sum_{i = k_0 + 1}^{k}\abs{{e}_i} +  (k - k_0)\e 
            \right)
    \notag
    \\
      &= \frac{\lambda \e}{2} \bigg( 
              \aBs{T_0[v_{k_0 - 1}, v_{k + 1}]} + \aBs{T_0[v_{k_0}, v_k]} +  (k - k_0)\e 
         \bigg).
    \notag
    \end{align*}

\end{proof}

We immediately get the flat norm bounds on the positive $\e$-perturbations of all interior nodes,
as well as for perturbations of all nodes as corollaries.

\begin{corollary}[Complete perturbation of interior]
\label{thm:corollary:fn-bound-interior}
  Given the setup of Theorem~\ref{thm:fn-bound:k-perturbations},
  consider a sequence of perturbations $T_1 \perturbeplus \ldots \perturbeplus \tld{T}^{n-1}_1$ 
  induced by the positive $\e$-perturbations of all interior nodes $[v_{1}, \ldots, v_{n - 1}]$.
  Then an upper bound on $\mathbb{F}_{\lambda}\big(\tld{T}^{n-1}_1 - T_0 \big)$ 
  is given by 
  \begin{align*}
    \mathbb{F}_{\lambda}\big(\tld{T}^{n-1}_1 - T_0 \big) 
    &\leq    
          \frac{\lambda \e}{2} \bigg( 
              \aBs{T_0} +  \aBs{ T_0[v_1, v_{n - 1}] } + (n - 2) \e 
          \bigg)
    \\
    &=    
          \lambda \e \bigg( 
              \aBs{T_0 } - \left( \frac{\abs{e_s} + \abs{e_t}}{2} \right) + (n - 2) \frac{\e}{2} 
          \bigg).
  \end{align*}
\end{corollary}

\begin{corollary}[Complete perturbation of $T_1$]
\label{thm:corollary:fn-bound-all}
  Given the setup of Theorem~\ref{thm:fn-bound:k-perturbations} with $n \geq 2$,
  consider a sequence of perturbations $T_1  \perturbeplus \ldots \perturbeplus \tld{T}^{n+1}_1$ 
  induced by the positive $\e$-perturbations of each node of $T_1$. 
  Then an upper bound on $\mathbb{F}_{\lambda}\big(\tld{T}^{n + 1}_1 - T_0 \big)$ 
  is given by 
  \begin{align}
  \label{eq:pwl-perturb-all-fn-bound}     
    \mathbb{F}_{\lambda}(\tld{T}^{n + 1}_1 - T_0)
    &\leq    \frac{\lambda \e}{2} \big( 2 \abs{T_0} + n \e \big)
    =        \lambda \e \left( \abs{T_0} + \frac{n \e}{2} \right).
  \end{align}
\end{corollary}

\begin{proof}
    
    Due to the additivity of the flat norm distance  between  adjacent $\e^{\scalebox{0.4}{+}}$-perturbations, 
    we can perturb the interior nodes $[v_{1}, \ldots, v_{n - 1}]$ first,
    and the boundary nodes after that, since now all edges will be perturbed twice and the order will not affect the upper bound. 
    The perturbation of the interior nodes induces a sequence $T_1 \perturbeplus \ldots \perturbeplus \tld{T}^{n-1}_1$ from Corollary~\ref{thm:corollary:fn-bound-interior},
    while $[s, t] \perturbeplus [\tld{s}, \tld{t}]$ induces 
    $\tld{T}^{n - 1}_1 \perturbeplus \tld{T}^{n}_1 \perturbeplus \tld{T}^{n + 1}_1$.

    The area components produced by $s \perturbeplus \tld{s}$ and $t \perturbeplus \tld{t}$ are spanned by the differences of corresponding perturbations
    are given by triangles $S_0 = \Delta_s = [{s}, \tld{v}_1, \tld{s}] \in \mc{C}_2[\tld{T}^{n}_1 - \tld{T}^{n-1}_1] $ 
    and $S_n = \Delta_t = [\tld{t}, \tld{v}_{n-1}, t]  \in \mc{C}_2[\tld{T}^{n+1}_1 - \tld{T}^{n}_1] $.
    Their boundaries are:
    \begin{align*}
      \bd S_0 = \bd \Delta_{s} = \tld{e}_s - \tld{\tld{e}}_s - \tld{\e}_{s}
      &&\text{ and }&&
      \bd S_n = \bd \Delta_{t} = \tld{e}_t - \tld{\tld{e}}_t + \tld{\e}_{t}
    \end{align*}
    where $\tld{\e}_{v} \in \mc{L}_1(v, \tld{v})$ such that $\abs{\tld{\e}_{v}} \leq \e$, see Fig.~\ref{fig:pwl-perturbation-BB}.
    Then the upper bound on the flat norm distance becomes
    \begin{align}
      \mathbb{F}_{\lambda}(\tld{T}^{n + 1}_1 - T_0) 
      &=     \mathbb{F}_{\lambda}(\tld{T}^{n + 1}_1 - \tld{T}^{n}_1) 
             + \mathbb{F}_{\lambda}(\tld{T}^{n}_1 - \tld{T}^{n - 1}_1) 
             + \mathbb{F}_{\lambda}(\tld{T}^{n-1}_1 - T_0) 
    \notag
    \\
      &\leq    
             \frac{\lambda \e}{2} \left( 
                  \sum_{i = 2}^{n-1} \abs{\tld{e}_{i}} 
                  + \sum_{i = 1}^{n} \abs{{e}_{i}} 
                  + \abs{\tld{e_s}} + \abs{\tld{e_t}}
            \right) 
    \notag
    \\
      &=    
           \frac{\lambda \e}{2} \left( 
                 \sum_{i = i}^{n} \abs{\tld{e}_{i}} 
                 + \sum_{i = 1}^{n} \abs{{e}_{i}}  
            \right)
    \label{eq:fn-bound:pwl-perturb-all-intermedite}     
    \\
      &\leq   \frac{\lambda \e}{2} \big( 
                    2 \abs{T_0} + n \e 
              \big)
      =    \lambda \e \left( 
              \abs{T_0} + \frac{n \e}{2} 
            \right).
    \notag
    \end{align}

\end{proof}

\begin{figure}[ht!]
   \centering
   \includegraphics[width=0.70\textwidth]{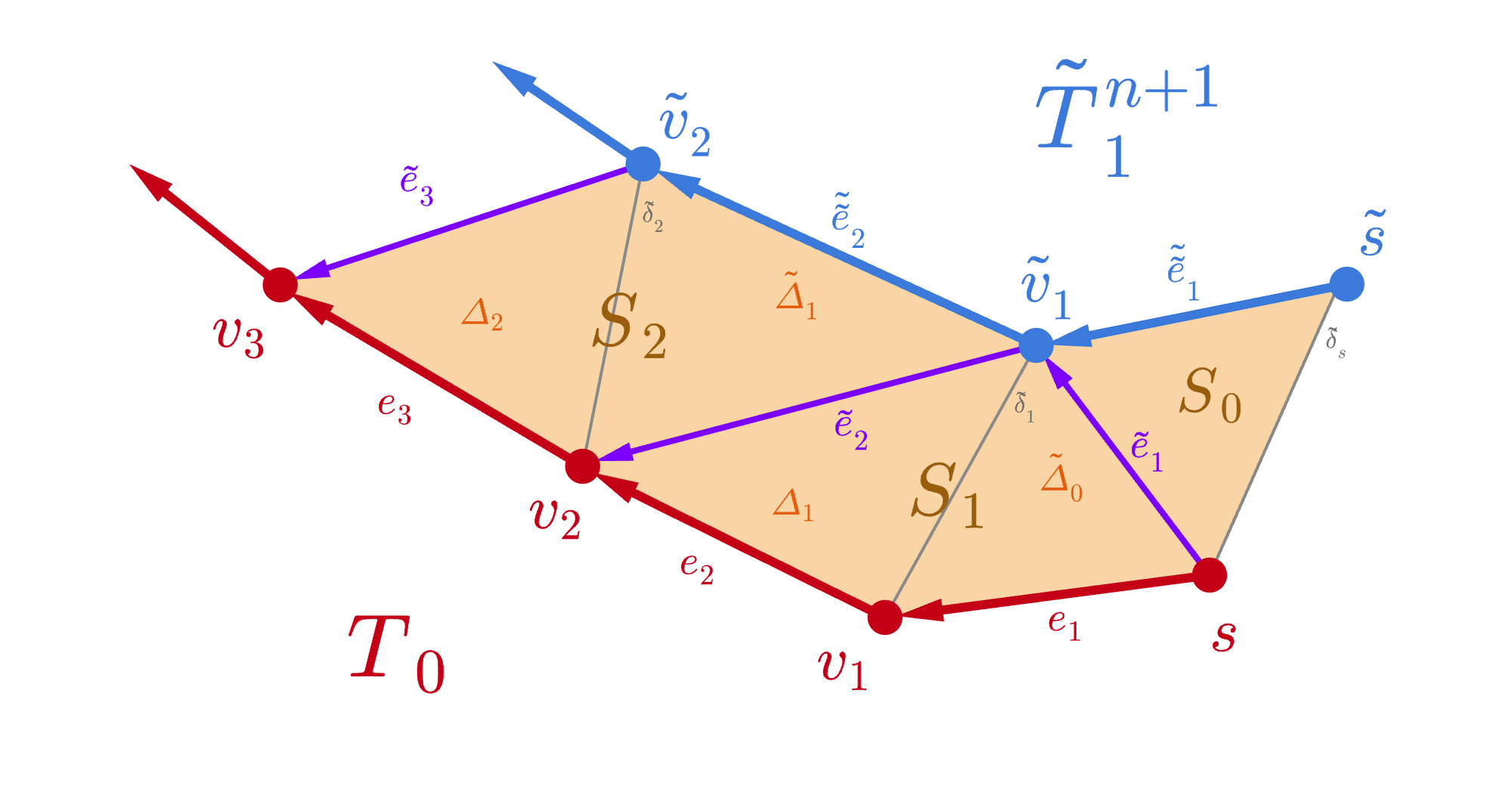}
   \caption[Completely perturbed $T_1$.]{
        Completely perturbed $T_1$.
        The positive $\e$-perturbations $\tilde{T}^{n-1}_1 \perturbEplus \tilde{T}^{n}_1 \perturbEplus \tilde{T}^{n+1}_1$
        induced by perturbations of the boundary vertices $[s, t] \perturb [\tld{s}, \tld{t}]$.
        The area components $S_0$ and $S_n$ that were produced in the result are given by triangles instead of quadrilaterals.
   }
\label{fig:pwl-perturbation-BB}  
\end{figure}

\subsection{Normalized flat norm of \texorpdfstring{$\e$}{Delta}-perturbations}
\label{sssec:FN-PWL-PERTURB:NORMALIZED}

Recall from Section~\ref{subsec:normalized-fn} (Eq.~\eqref{eq:flat-norm-normalized-def}) that the normalized flat norm distance is given as
\begin{align*}
    \tld{\mathbb{F}}_{\lambda}\big(\tld{T}_1 - T_0 \big) 
    &=
        \frac{\mathbb{F}_{\lambda}\big(\tld{T}_1 - T_0 \big) }{\aBs{T_0} + \aBs{\tld{T}_1}}.
\end{align*}
Note that the normalized flat norm distance has a natural upper bound given by $\sfrac{\mathbb{F}_{\lambda}\big(\tld{T}^{n-1}_1 - T_0 \big) }{\aBs{T_0}}$
that may be of interest when measurements of one of the input geometries are not available or are not reliable. 
The following corollary of Theorem~\ref{thm:fn-bound:k-perturbations} follows from Eq.~\eqref{eq:pwl-perturb-all-fn-bound}
by dividing both sides by $\Abs{T_0}$.

\begin{corollary}
\label{thm:norm-fn-bound:loose}
  Given the setup of Theorem~\ref{thm:fn-bound:k-perturbations} for $n \geq 2$,
  consider a sequence of perturbations $T_1  \perturbeplus \ldots \perturbeplus \tld{T}^{n+1}_1$ 
  induced by the positive $\e$-perturbations of each node of $T_1$.
  Then an upper bound on the normalized flat norm distance  $\tld{\mathbb{F}}_{\lambda}\big(\tld{T}^{n + 1}_1 - T_0 \big)$ 
  is given by 
  \begin{align*}
    \tld{\mathbb{F}}_{\lambda}\big(\tld{T}^{n+1}_1 - T_0 \big) 
    &\leq \lambda \e \left( 1 + \frac{n}{2 \aBs{T_0}} \e \right)
     = \lambda \e \left( 1 + \frac{\e}{2 \hat{e}} \right)
  \end{align*}
  where $\hat{e} = \frac{1}{n} \abs{T_0}$ is the average edge length of $T_0$.
\end{corollary}


\begin{theorem}
\label{thm:norm-fn-bound:tight}
  Let $T_0 \in \mc{L}_n[s,t; \real^2]$ for $n \geq 2$, and $T_1 = T_0$ be its copy.
  Given a small enough scale $\lambda > 0$ and an appropriate radius of perturbation $\e > 0$,
  consider a sequence of perturbations $T_1  \perturbeplus \ldots \perturbeplus \tld{T}^{n+1}_1$ 
  induced by the positive $\e$-perturbations of each node of $T_1$. 
  Then an upper bound on the normalized flat norm distance 
  $\mathbb{F}_{\lambda}\big(\tld{T}^{n + 1}_1 - T_0 \big)$ 
  is given by 
  \begin{align}
  \label{eq:norm-fn-bound:tight}      
    \tld{\mathbb{F}}_{\lambda}\big(\tld{T}^{n + 1}_1 - T_0 \big) 
    &\leq
        \frac{\lambda \e}{2} \left( 
                \frac{1}{2} + \frac{\hat{e} + \e}{\hat{e} + \hat{e}_{\tld{n}} }
        \right)
       = \lambda \e \left( 
                \frac{1}{4} + \frac{\hat{e} + \e}{ 2(\hat{e} + \hat{e}_{\tld{n}}) }
        \right)
  \end{align}
  where $\hat{e}_{\tld{n}} = \frac{1}{n} \aBs{\tld{T}^{n + 1}_1}$ is the average length of perturbed edges.
\end{theorem}
\begin{proof}

    To derive an upper bound for the normalized flat norm distance  
    observe that by the triangle inequality in $\Delta_i$ and $\tld{\Delta}_{i}$ we get the following bounds:
    \begin{align*}
    &\begin{array}{ccccc}
      \Delta_i: &&\dsp e_i - \e \leq &\tld{e}_i& \leq e_i + \e 
      \\[1em]
      \tld{\Delta}_i: &&\dsp \tld{\tld{e}}_i - \e \leq &\tld{e}_i& \leq \tld{\tld{e}} + \e 
      \\[1em]
      \iimplies &&\dsp \frac{e_i + \tld{\tld{e}}_i}{2} - \e \leq &\tld{e}_i&\dsp \leq \frac{e_i + \tld{\tld{e}}_i}{2} + \e .
    \end{array}
    \end{align*}

    Let $\bbar{T}_1 = \tld{e}_1 + \ldots + \tld{e}_n$.
    Note that $T_0 = e_1 + \ldots + e_n$ and $\tld{T}^{n + 1}_1 = \tld{\tld{e}}_1 + \ldots + \tld{\tld{e}}_n$.
    We then continue the derivation from the intermediate result in Eq.~\eqref{eq:fn-bound:pwl-perturb-all-intermedite}
    obtained during the proof of Corollary~\ref{thm:corollary:fn-bound-all} of Theorem~\ref{thm:fn-bound:k-perturbations}:
    \begin{align*}
      \mathbb{F}_{\lambda}(\tld{T}^{n + 1}_1 - T_0)  
      &\leq \frac{\lambda \e}{2} \left( 
                   \sum_{i = 1}^{n} \abs{\tld{e}_{i}} 
                   + \sum_{i = 1}^{n} \abs{{e}_{i}}  
              \right)
        = \frac{\lambda \e}{2} \bigg( \aBs{\bbar{T}_1 } + \aBs{T_0} \bigg)
      \\
      &\leq \frac{\lambda \e}{2} \left(
          \frac{\aBs{T_0} + \aBs{ \tld{T}^{n + 1}_1 } }{2}
          + \abs{T_0} + n \e
      \right).
    \end{align*}
    
    Recall that $\hat{e}_{\tld{n}} = \sfrac{\aBs{\tld{T}^{n + 1}_1}}{n}$.
    Then,
    \begin{align*}
        \frac{\mathbb{F}_{\lambda}(\tld{T}^{n + 1}_1 - T_0)}{\aBs{T_0} + \aBs{ \tld{T}^{n + 1}_1}}
      &\leq
        \frac{\lambda \e}{2} \left( 
              \frac{1}{2} + \frac{\aBs{T_0}}{\aBs{T_0} + \aBs{ \tld{T}^{n + 1}_1}}
              + \frac{n}{\abs{T_0} + \abs{ \tld{T}^{n + 1}_1}} \e
        \right)
      \\
      &=
        \frac{\lambda \e}{2} \left( 
              \frac{1}{2} + \frac{\hat{e}}{\hat{e} + \hat{e}_{\tld{n}} }
              + \frac{\e}{\hat{e} + \hat{e}_{\tld{n}} } 
        \right).
    \end{align*}

\end{proof}

Combining the generic upper bound in Eq.~\eqref{eq:pwl-fn-upper-bound} that holds for small enough $\lambda > 0$ and an appropriate $\e > 0$,
we can rewrite Eq.~\eqref{eq:norm-fn-bound:tight} as
\begin{align}
\label{eq:norm-fn-bound:tight-truncated}
  \tld{\mathbb{F}}_{\lambda}(\tld{T}^{n + 1}_1 - T_0) 
  &\leq \min \left\{ 
          \lambda \e \left( \frac{1}{4} + \frac{\hat{e} + \e}{ 2(\hat{e} + \hat{e}_{\tld{n}}) }\right),
          \,1
        \right\}.
\end{align}

\begin{corollary}
  In the case of a non-shrinking perturbation sequence $T_1 \perturbeplus \ldots \perturbeplus \tld{T}^{n + 1}_1$ such that $\aBs{\tld{T}^{n + 1}_1} \geq \aBs{T_1} = \aBs{T_0}$,
  the upper bound on the normalized flat norm distance is given as
  \begin{align*}
    \tld{\mathbb{F}}_{\lambda}(\tld{T}^{n + 1}_1 - T_0) 
    &\leq 
      \lambda \e \left( 
              \frac{3}{4} 
              + \frac{\e}{4 \hat{e} } 
      \right).
  \end{align*}

\end{corollary}





\chapter{Applications: Structural Validation Of Synthetic Power Distribution Networks}


\section{Implementation of flat norm.}
\label{sec:implementation}

The power grid is the most vital infrastructure that provides crucial support for the delivery of basic services to most segments of society. 
Once considered a passive entity in power grid planning and operation, the power distribution system poses significant challenges in the present day. 
The increased adoption of rooftop solar photovoltaics (PVs) and electric vehicles (EVs) augmented with residential charging units has altered the energy consumption profile of an average consumer.
Access to extensive datasets pertaining to power distribution networks and residential consumer demand is vital for public policy researchers and power system engineers alike. 
However, the proprietary nature of power distribution system data hinders their public availability.
This has led researchers to develop frameworks that synthesize realistic datasets pertaining to the power distribution system~\cite{highres_net,overbye_2020,nrel_net,rounak2020,anna_naps,schweitzer}. 
These frameworks create digital replicates similar to the actual power distribution networks in terms of their structure and function. 
Hence the created networks can be used as \emph{digital duplicates} in simulation studies of policies and methods before implementation in real systems. 

The algorithms associated with these frameworks vary widely---ranging from first principles based approaches~\cite{rounak2020,anna_naps} to learning statistical distributions of network attributes~\cite{schweitzer} to using deep learning models such as generative adversarial neural networks~\cite{feeder_gan}. 
Validating the synthetic power distribution networks with respect to their physical counterpart is vital for assessing the suitability of their use as effective digital duplicates. 
Since the underlying assumptions and algorithms of each framework are distinct from each other, some of them may excel compared to others in reproducing digital replicates with better precision for selective regions. 
To this end, we require well-defined metrics to rank the frameworks and judge their strengths and weaknesses in generating digital duplicates of power distribution networks for a particular geographic region.

The literature pertaining to frameworks for synthetic distribution network creation include certain validation results that compare the generated networks to the actual counterpart~\cite{highres_net,validate2020,schweitzer}. But the validation results are mostly limited to comparing the statistical network attributes such as degree and hop distributions and power engineering operational attributes such as node voltages and edge power flows. Since power distribution networks represent real physical systems, the created digital replicates have associated geographic embedding. Therefore, a structural comparison of synthetic network graphs to their actual counterpart becomes pertinent for power distribution networks with geographic embedding.
Consider an example where a digital twin is used to analyze impact of a weather event~\cite{samiul2021}.
Severe weather events such as hurricanes, earthquakes and wild fires occur in specific geographic trajectories, affecting only portions of societal infrastructures. 
In order to correctly identify them during simulations, the digital twin should structurally resemble the actual infrastructure.

\subsection{Geometry comparison using flat norm.}
\label{subsec:implementation:geom}
We show a simple example depicting the use of flat norm to compute the distance between a pair of geometries that are two line segments of equal length meeting at their midpoints in Fig.~\ref{fig:demo-flatnorm}.
As the angle between the two line segments decreases from $90$ to $15$ degrees, the computed flat norm also decreases.
\begin{figure}[ht!]
    \centering
    \includegraphics[width=0.47\textwidth]{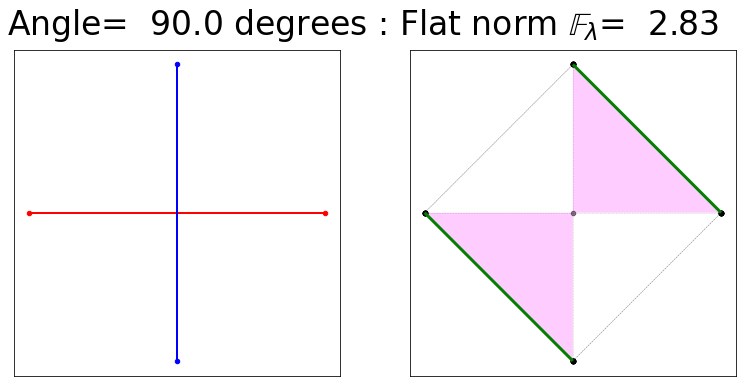}
    \includegraphics[width=0.47\textwidth]{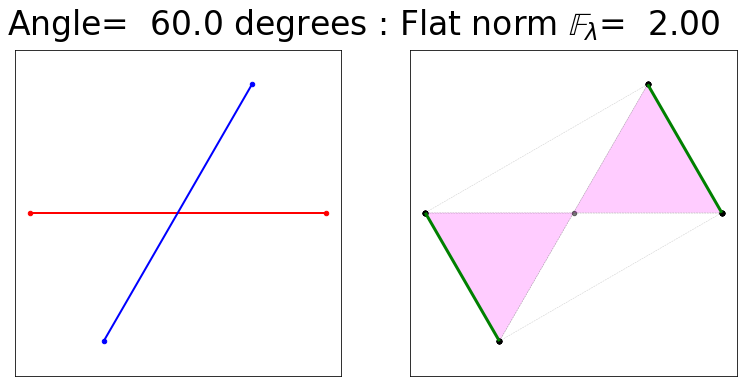}
    \includegraphics[width=0.47\textwidth]{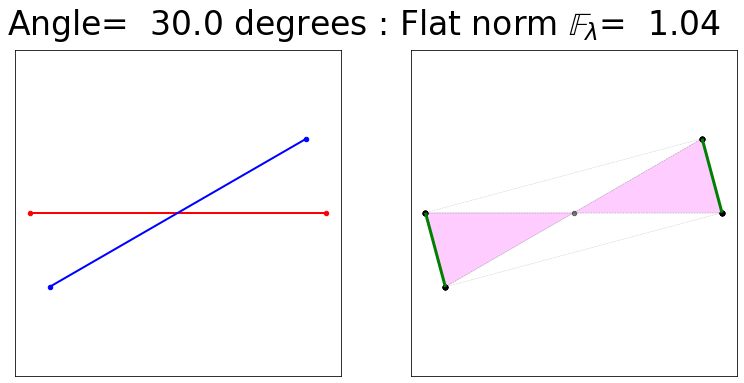}
    \includegraphics[width=0.47\textwidth]{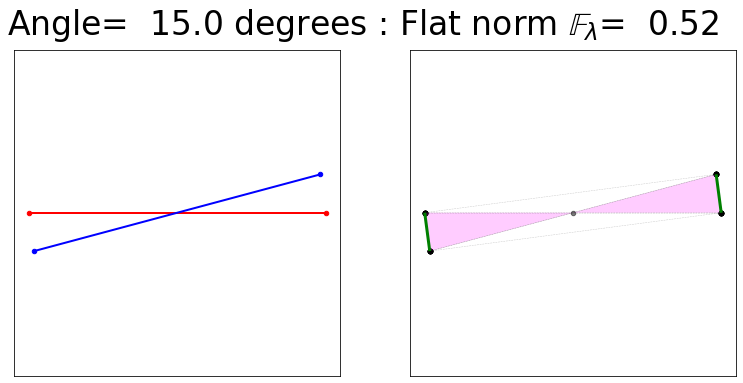}
    \caption[Variation in flat norm for a pair of line segments.] 
    {Variation in flat norm for pairs of geometries as the angle between them decreases.
      When the geometries are perpendicular to each other, flat norm distance is the maximum and it decreases as the angle decreases.}
    \label{fig:demo-flatnorm}
\end{figure}

\subsection{Flat norm computation for a pair of geometries.}
\label{subsec:implementation:fn-geom}
~We demonstrate the steps involved in computing the flat norm for a pair of input geometries in Fig.~\ref{fig:demo-toy}.
The input geometries are a collection of line segments shown in blue and red (top left). We construct the set $\Sigma$ by combining all the edges of either geometry along with the bounding rectangle (top right). Thereafter, we perform a constrained triangulation to construct the $2$-dimensional simplicial complex $K$ (bottom left). Finally, we compute the multiscale simplicial flat norm with $\lambda=1$ (bottom right).
Note that this computation captures the length deviation (shown by green edges) and the lateral displacement (shown by the magenta patches).
\begin{figure}[ht!]
    \centering
    \includegraphics[width=0.98\textwidth]{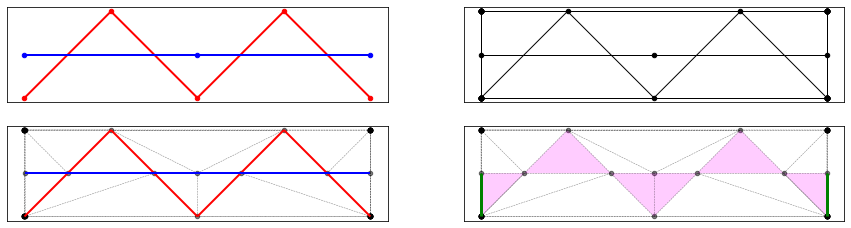}
    \caption{Steps in computing the flat norm for a pair of input geometries.}
    \label{fig:demo-toy}
\end{figure}

\subsection{Flat norm computation for a pair of networks.}
\label{subsec:implementation:fn-power-grids}
The steps involved in computing the multiscale flat norm distance between a pair of geometries are shown in Fig.~\ref{fig:example-1}.
They include the actual power distribution network (red) for a region in a county from USA and the synthetic network (blue) constructed for the same region~\cite{rounak2020}.
First, we orient each edge in either network from left to right.
Thereafter, we find the enclosing rectangular boundary around the pair of networks.
We perform a constrained Delaunay triangulation which ensures that the edges in the geometries and the convex boundary are selected as edges of the triangles.
Finally the flat norm LP (relaxation of the ILP in Eq.~(\ref{eq:opt-flatnorm})) is solved to compute the flat norm distance between the networks.
\begin{figure}[p!]
    \centering
    \includegraphics[height=0.80\textheight]{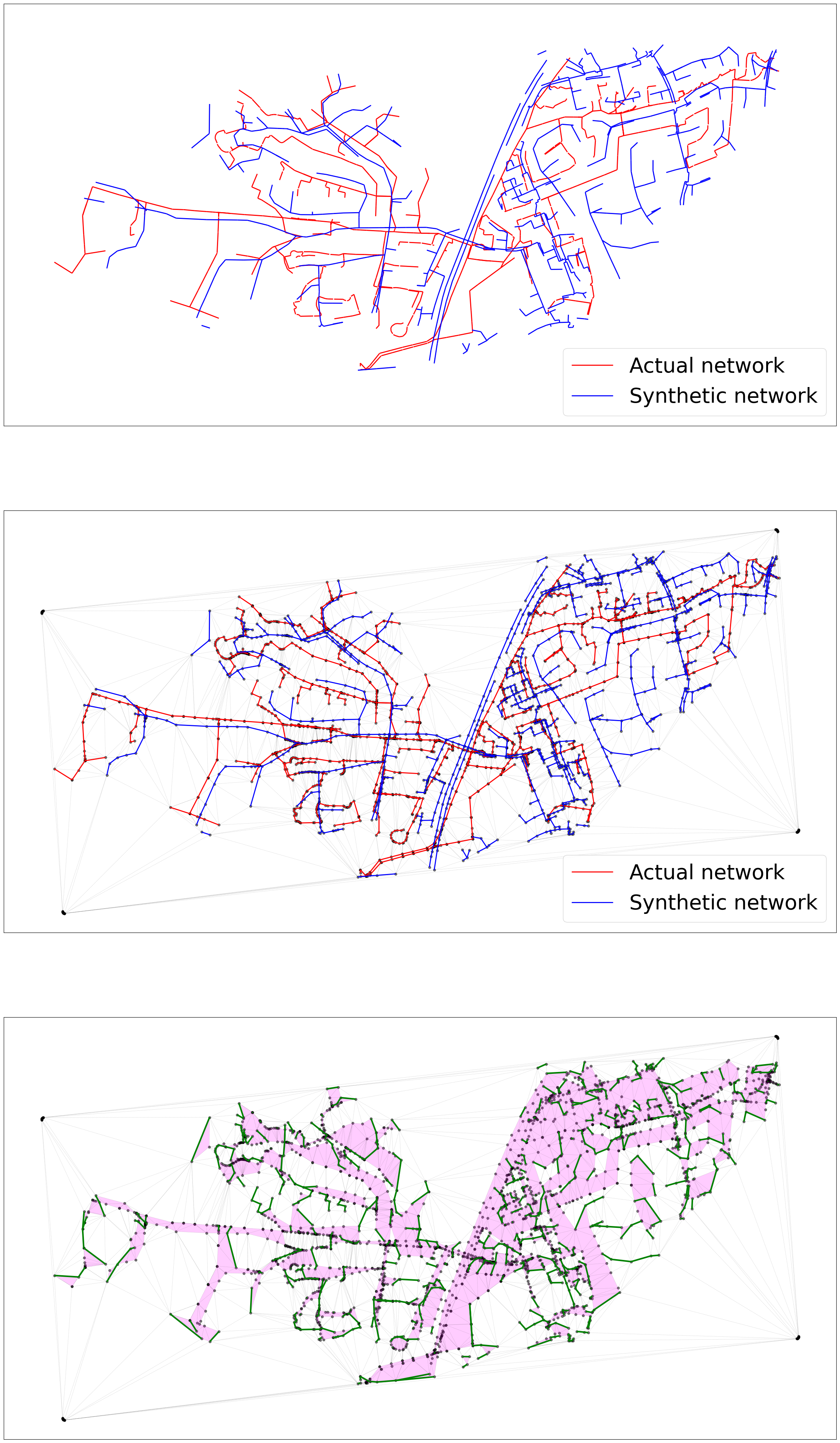}
    \caption[Steps in computing the flat norm between two networks.]
    {Steps showing the flat norm distance computation between two networks (shown in blue and red in the top two plots).
      First, the convex rectangular boundary around the pair of networks is identified.
      A constrained triangulation is computed such that the edges in the networks and convex boundary are edges of triangles (middle).
      The flat norm LP is solved to compute the simplicial flat norm, which includes the sum of areas of the magenta triangles and lengths of green edges (bottom).
    }
    \label{fig:example-1}
\end{figure}

\begin{figure}[p!]
    \centering
    \includegraphics[width=0.6\textwidth]{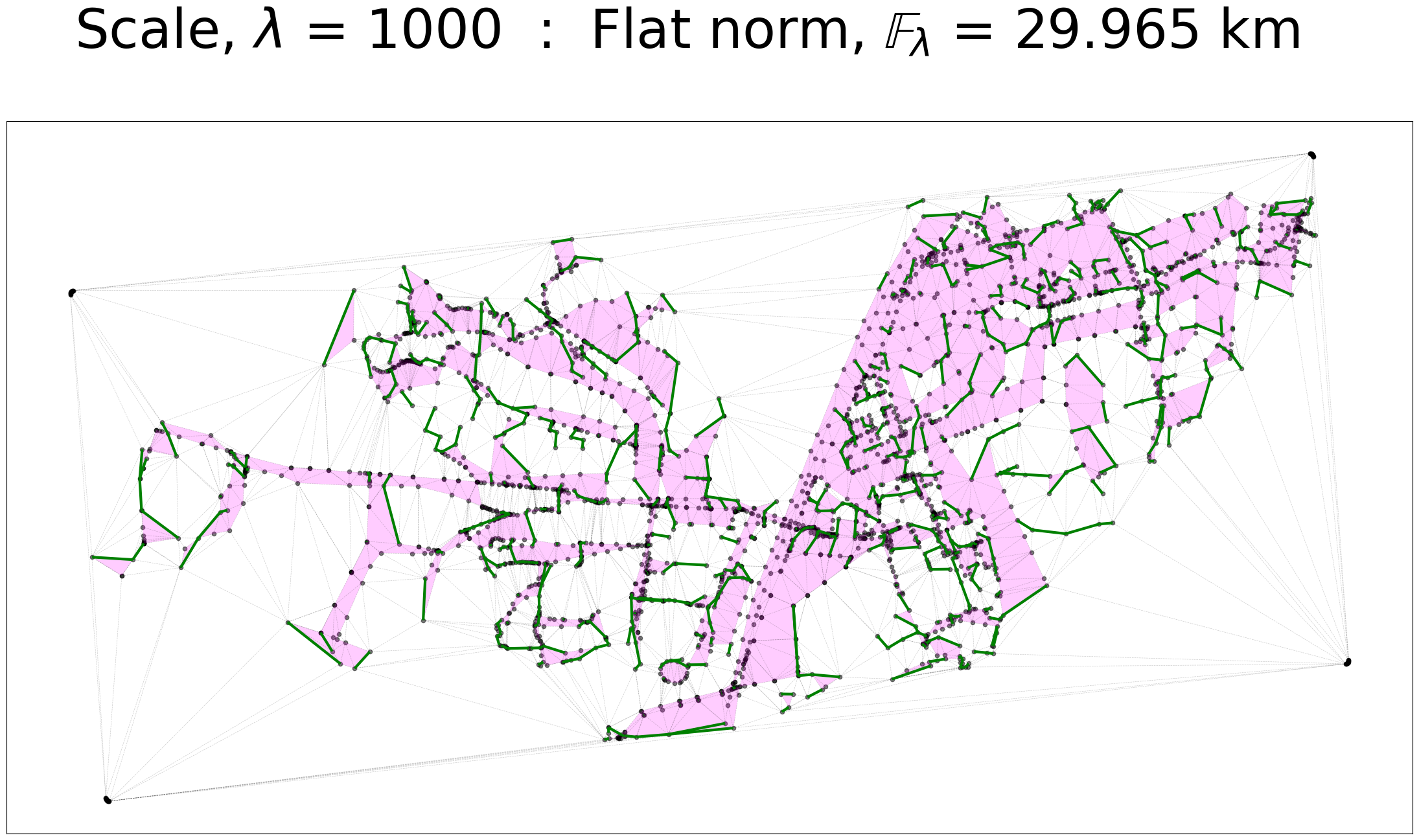} \\
    \vspace*{0.2in}
    \includegraphics[width=0.6\textwidth]{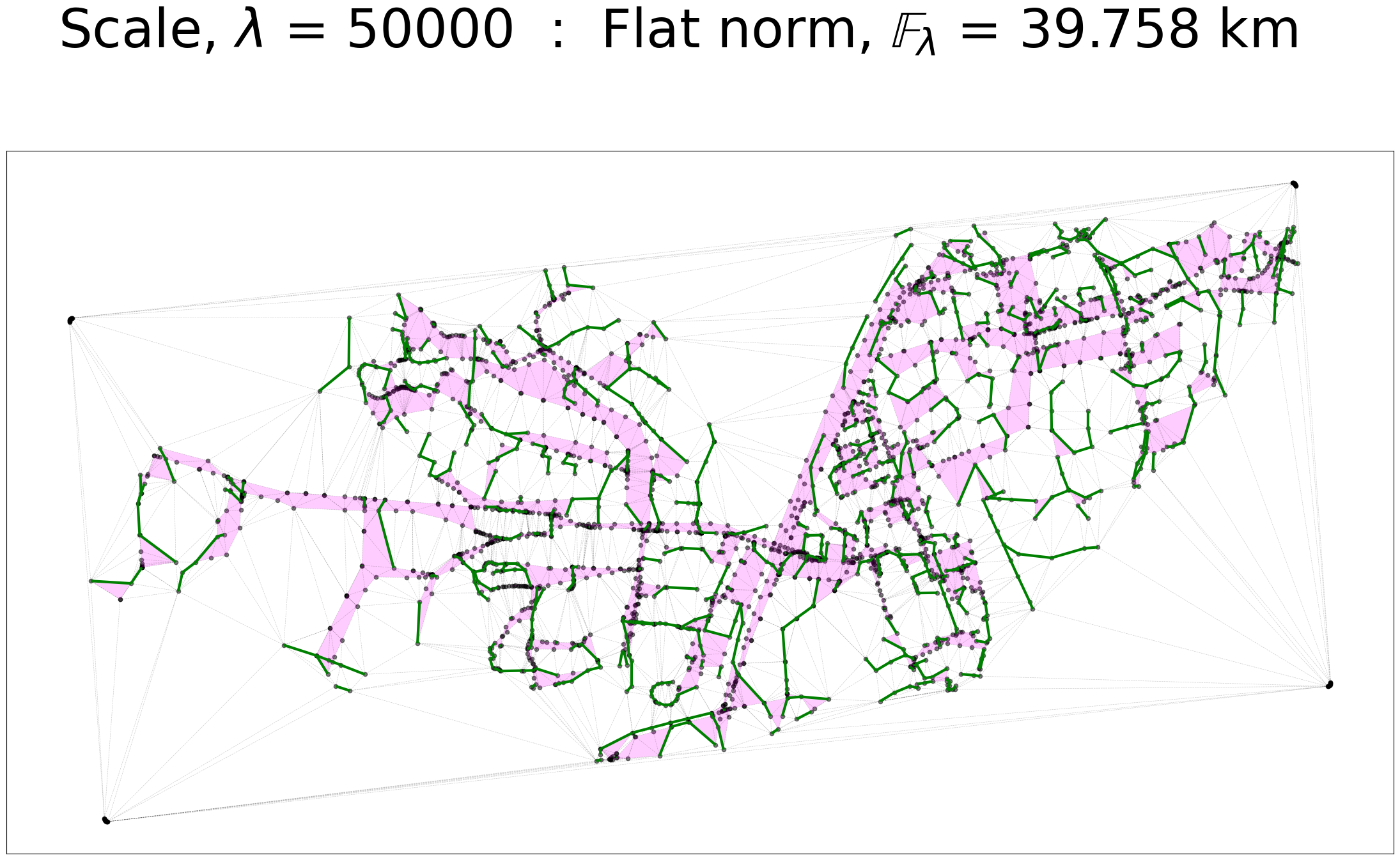} \\
    \vspace*{0.2in}
    \includegraphics[width=0.6\textwidth]{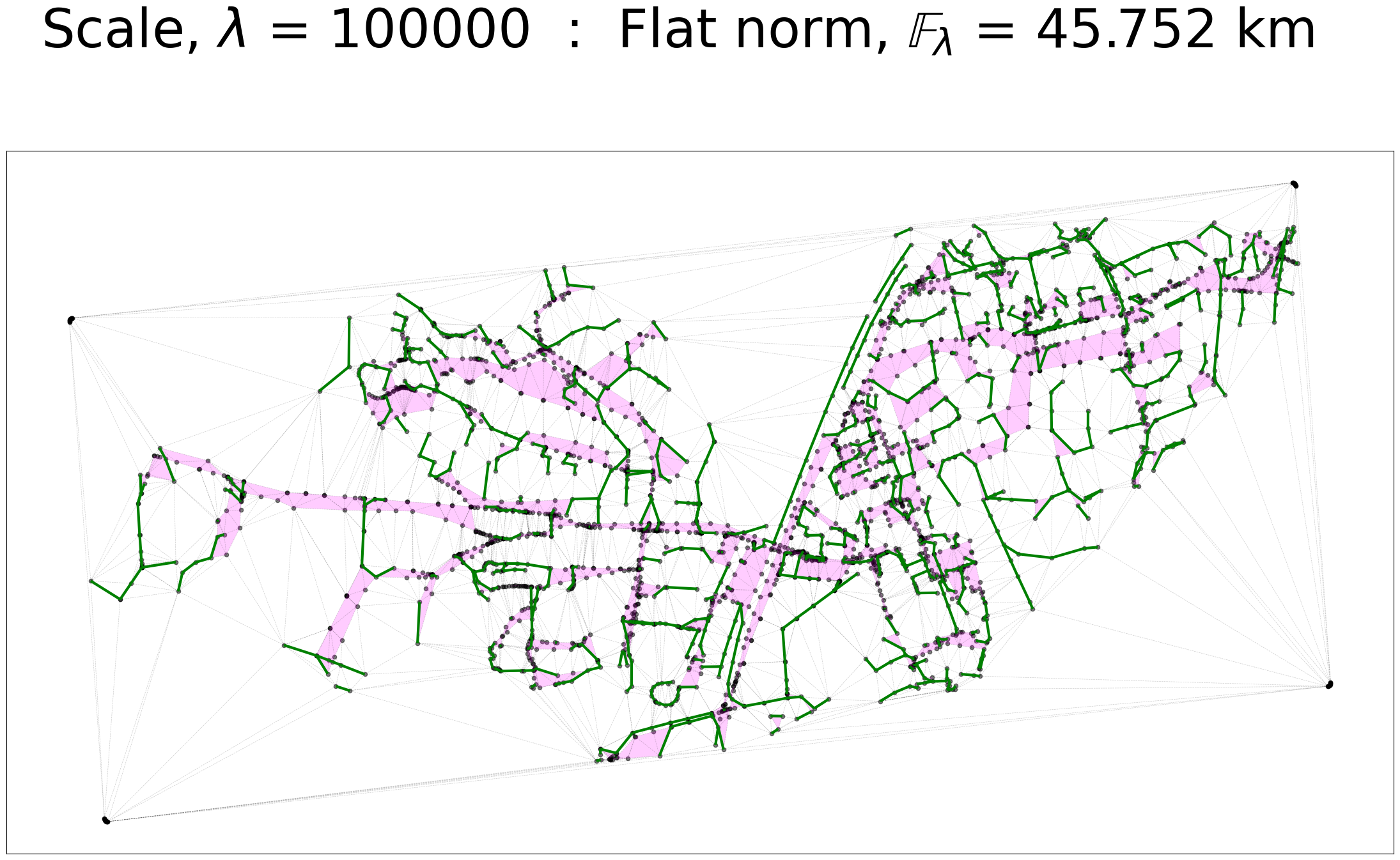}
    \caption[Variation in flat norm decomposition  for a pair of network geometries depending on the scale parameter $\lambda$.]%
    {The flat norm computed between the pair of network geometries for three values of the scale parameter $\lambda$ ranging between $\lambda=1000$ to $\lambda=10000$.}
    \label{fig:lambda-var}
\end{figure}

The multiscale flat norm produces different distance values for different values of the scale parameter $\lambda$.
Fig.~\ref{fig:lambda-var} shows the flat norm distance between the actual and synthetic power network for the same region for multiple values of the scale parameter $\lambda$.
We observe that as $\lambda$ becomes larger, the 2D patches used in computing the flat norm become smaller as it becomes more expensive to use the area term in the flat norm LP minimization problem.

The variation of the computed flat norm for different values of the scale parameter is summarized in Fig.~\ref{fig:scale-variation}.
As the scale parameter is increased, fewer area patches are considered in the simplicial flat norm computation.
This is captured by the blue decreasing curve in the plot.
The computed flat norm increases for larger values of the scale parameter $\lambda$ as more and more individual currents contribute their unscaled length toward the flat norm value instead of becoming a boundary of some area component, which, if there are any, now also contribute more because of the increased scale  $\lambda$.
We show the plot with two different vertical scales: the left scale indicates the deviation in length (measured in km) and the right scale shows the deviation expressed through the area patches (measured in sq.km).
\begin{figure}[ht!]
    \centering
    \includegraphics[width=0.90\textwidth]{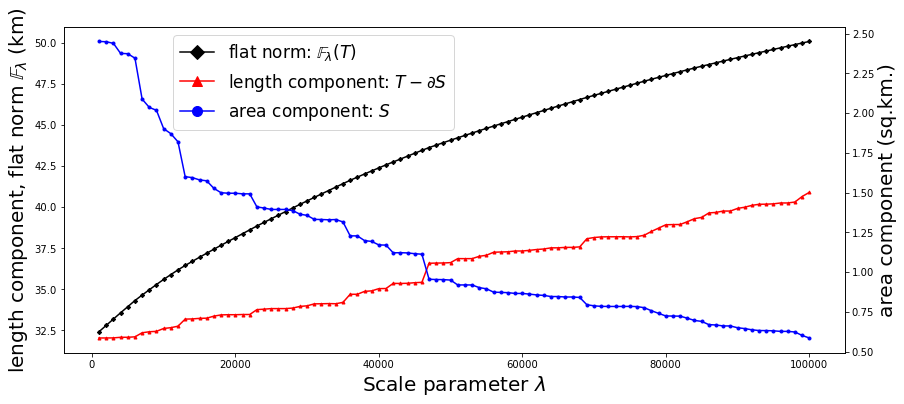}
    \caption[Variation in the flat norm value for a pair of network geometries depending on the scale parameter $\lambda$.]
        {Effect of varying the scale parameter $\lambda$ in the flat norm computation.
      The flat norm for a $1$-dimensional current consists of two parts: a length component and a scaled surface area component.
      The variations in the length component and the unscaled surface area component (right vertical scale) are also shown.}
    \label{fig:scale-variation}
\end{figure}

\subsection{Comparing Network Geometries}
The primary goal of computing the flat norm is to compare the pair of input geometries.
As mentioned earlier, the flat norm provides an accurate measure of the difference between the geometries by considering both the length deviation and area patches in between the geometries.
Further, we normalize the computed flat norm to the total length of the geometries.
In this section, we show examples where we computed the normalized flat norm for the pair of network geometries (actual and synthetic) for a few regions.

\begin{figure}[ht!] 
    \centering
    \includegraphics[width=0.46\textwidth]{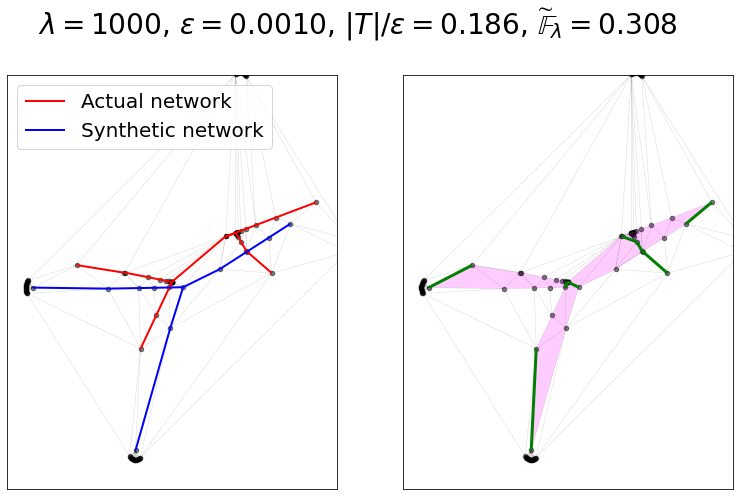}
    \includegraphics[width=0.46\textwidth]{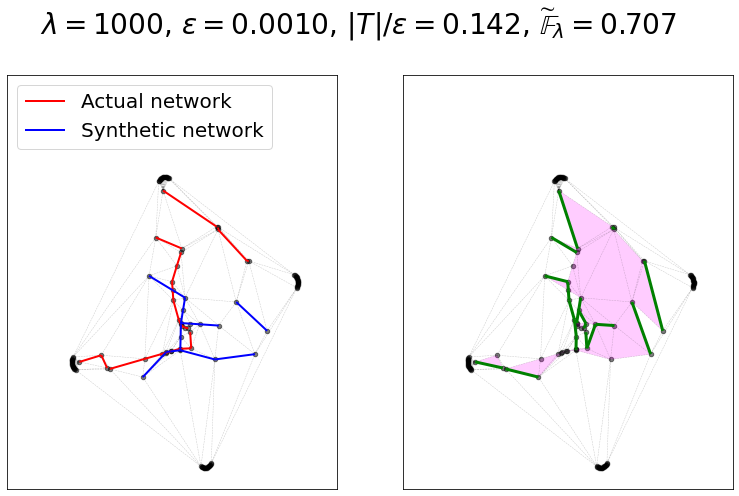}
    \includegraphics[width=0.46\textwidth]{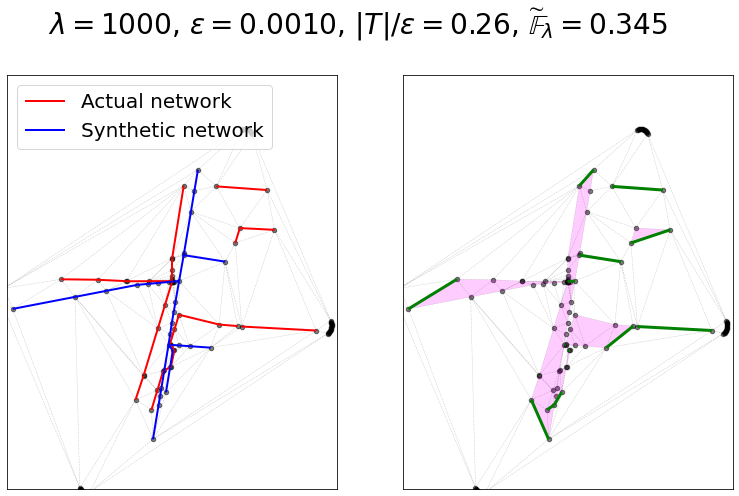}
    \includegraphics[width=0.46\textwidth]{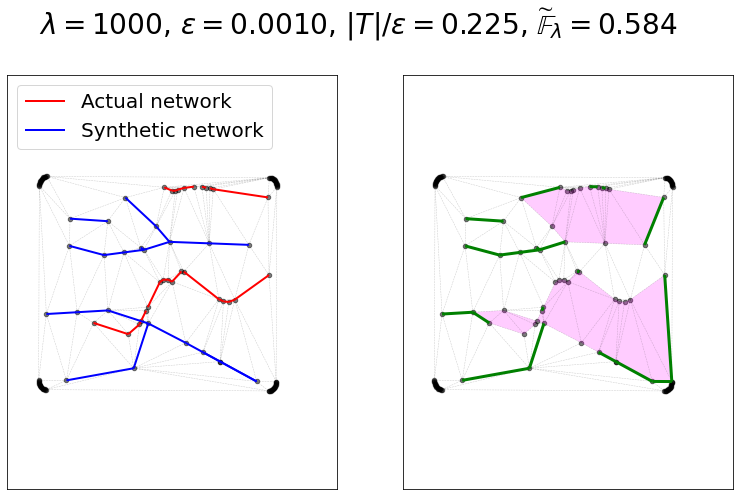}
    \caption[Normalized flat norm distances for pairs of similar regions in the network.]
    {Normalized flat norm (with scale $\lambda=1000$) distances for pairs of regions in the network of same size ($\epsilon=0.001$) with similar $|T|/\epsilon$ ratios (two pairs each in the top and bottom rows).
      The pairs of geometries for the first plot (on left) are quite similar, which is reflected in the low flat norm distances between them.
      The network geometries on the right plots are more dissimilar and hence the flat norm distances are high.}
    \label{fig:comparing}
\end{figure}
The top two plots in Fig.~\ref{fig:comparing} show two regions characterized by $\epsilon=0.001$ and almost similar $|T|/\epsilon$ ratios. This indicates that the length of network scaled to the region size is almost equal for the two regions. From a mere visual perspective, we can conclude that the first pair of network geometries resemble each other where as the second pair are fairly different. This is further validated from the results of the flat norm distance between the network geometries computed with the scale $\lambda=1000$, since the first case produces a smaller flat norm distance compared to the latter. The bottom two plots show another example of two regions with almost similar $|T|/\epsilon$ ratios and enable us to infer similar conclusions. The results strengthens our case of using flat norm as an appropriate measure to perform a local comparison of network geometries.
We can choose a suitable $\gamma > 0$ which differentiates between these example cases and use the proposed flat norm distance metric to identify structurally similar network geometries.
However, the choice of $\gamma$ has to be made empirically.
This necessitates a statistical study of randomly chosen local regions in different sections of the networks, which is performed in the Section~\ref{sec:stats}.

\section{Statistical analysis of normalized flat norm}
\label{sec:stats}

We use the proposed multiscale flat norm to compare a pair of network geometries
from power distribution networks for a region in a county in USA.
The two networks considered are the actual power distribution network for the region and the synthetic network generated using the methodology proposed by Meyur et al.~\cite{rounak2020}.
We provide a brief overview of these networks.

\medskip
\noindent\textbf{Actual network.}~The actual power distribution network was obtained from the power company serving the location.
Due to its proprietary nature, node and edge labels were redacted from the shared data.
Further, the networks were shared as a set of handmade drawings, many of which had not been drawn to a well-defined scale.
We digitized the drawings by overlaying them on OpenStreetMaps~\cite{osm} and georeferencing to particular points of interest~\cite{arcgis}.
Geometries corresponding to the actual network edges are obtained as shape files.

\medskip
\noindent\textbf{Synthetic network.}~The synthetic power distribution network is generated using a framework with the underlying assumption that the network follows the road network infrastructure to a significant extent~\cite{rounak2020}.
To this end, the residences are connected to local pole top transformers located along the road network to construct the low voltage (LV) secondary distribution network.
The local transformers are then connected to the power substation following the road network leading to the medium voltage (MV) primary distribution network.
That is, the primary network edges are chosen from the underlying road infrastructure network such that the structural and power engineering constraints are satisfied. 

\smallskip
In this section we study the empirical distribution of the normalized flat norm $\widetilde{\mathbb{F}}_{\lambda}$ for different local regions and argue that it indeed captures the similarity between input geometries.
We use Algorithm~(\ref{alg:sample}) to sample random square shaped regions of size $2\epsilon\times 2\epsilon$ steradians from a given geographic location.
\begin{algorithm}[tbhp]
\caption{Sample square regions from location}
\label{alg:sample}
    \textbf{Input}: Geometries $\mathscr{E}_1,\mathscr{E}_2$, number of regions $N$\\
    \textbf{Parameter}: Size of region $\epsilon$
    \begin{algorithmic}[1]
    \STATE Find bounding rectangle for the pair of geometries: $\mathscr{E}_{\textrm{bound}}=\mathsf{rect}\left(\mathscr{E}_1,\mathscr{E}_2\right)$.
    \STATE Initialize set of regions: $\mathscr{R} \leftarrow \{\}$.
    \WHILE{$\left|\mathscr{R}\right| \leq N$}
      \STATE Sample a point $\left(x,y\right)$ uniformly from region bounded by $\mathscr{E}_{\textrm{bound}}$.
      \STATE Define the square region $r\left(x,y\right)$ formed by the corner points $\left\{ \left(x - \epsilon, y - \epsilon\right), \left(x + \epsilon, y + \epsilon\right) \right\}$.
      
      \IF{$r\left(x,y\right) \cap \mathscr{E}_1 \cap \mathscr{E}_2 \neq \emptyset$}
          \STATE Add region $r\left(x,y\right)$ to the set of sampled regions: $\mathscr{R} \leftarrow \mathscr{R} \cup \left\{r\left(x,y\right)\right\}$.
      \ENDIF
    \ENDWHILE
\end{algorithmic}
\textbf{Output}: Set of sampled regions: $\mathscr{R}$.
\end{algorithm}
We perform our empirical studies for two urban locations of a county in USA. These locations have been identified as `Location A' and `Location B' for the remainder of this paper. We consider local regions of sizes characterized by $\epsilon\in\{0.0005, 0.001, 0.0015, 0.002\}$. For each location, we randomly sample $N=50$ local regions for each value of $\epsilon$ using Algorithm~(\ref{alg:sample}) and hence we consider $50\times4=200$ regions. For every sampled region, we use Algorithm~(\ref{alg:distance}) to compute the multiscale flat norm between the network geometries contained within the region with scale parameter $\lambda\in\{10^3,25\times10^3,50\times10^3,75\times10^3,10^5\}$. Thereafter, we normalize the computed flat norm using Eq.~(\ref{eq:flat-norm-normalized-def}). Additionally, we compute the global normalized flat norm for the entire location and indicate it by $\widetilde{\mathbb{F}}_{\lambda}^{G}$. The corresponding square box bounding the entire location is characterized by $\epsilon_{G}$. We also denote the total length of networks in each location scaled by the size of the location by the ratio $|T_G|/\epsilon_{G}$. The detailed statistical results for the experiments are included in the Appendix.

\subsection{Empirical distribution of \texorpdfstring{$\widetilde{\mathbb{F}}_{\lambda}$}{normalized flat norm}}

First, we show the histogram of normalized flat norms for Location A and Location B with the five different values for the scale parameter $\lambda$, Fig.~\ref{fig:flatnorm-hists-lambdas}.
Each histogram shows the empirical distribution of normalized flat norm values $\widetilde{\mathbb{F}}_{\lambda}$ for $200$ uniformly sampled local regions ($50$ regions for each $\epsilon$).
We also record the global normalized flat norm between the network geometries of the location $\widetilde{\mathbb{F}}_{\lambda}^{G}$ and denote it by the solid blue line in each histogram.
We show the mean normalized flat norm $\widehat{\mathbb{F}}_{\lambda}$ using the solid green line, with dashed green lines indicating the standard deviation of the distribution.
As we can see from the above histograms, the distribution is skewed toward the right for {high values of the scale parameter $\lambda$}.
This follows from our previous discussion of the dependence of the flat norm on the scale parameter:
for a large $\lambda$, the area patches are weighed higher in the objective function of the flat norm LP (Eq.~(\ref{eq:opt-flatnorm})).
Therefore, the contribution of lengths of the input currents $T_1$ and $T_2$ towards the flat norm distance becomes more dominant 
at \emph{the high values of the scale parameter $\lambda$},
so that the flat norm  $\mathbb{F}_\lambda\left(T_1-T_2\right)$ is slowly approaching the total network length $|T_1| + |T_2|$.
Hence, the normalized flat norm is approaching $1$.
For the remainder of the paper, we will continue our discussion with scale parameter $\lambda=1000$ since the empirical distributions of normalized flat norm corresponding to $\lambda=1000$ indicate almost Gaussian distribution.

\begin{figure}[ht!]
    \centering
    \includegraphics[width=\textwidth]{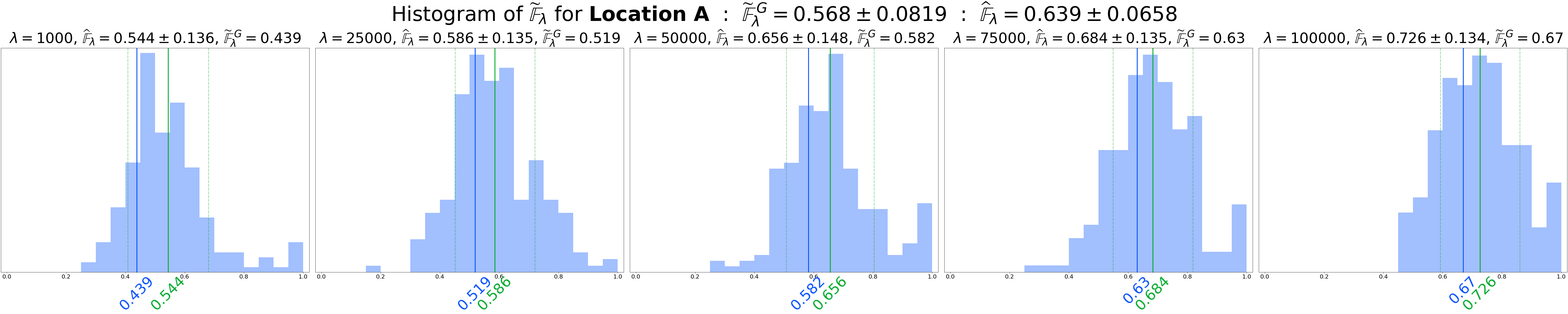}\\
    \smallskip
    \includegraphics[width=\textwidth]{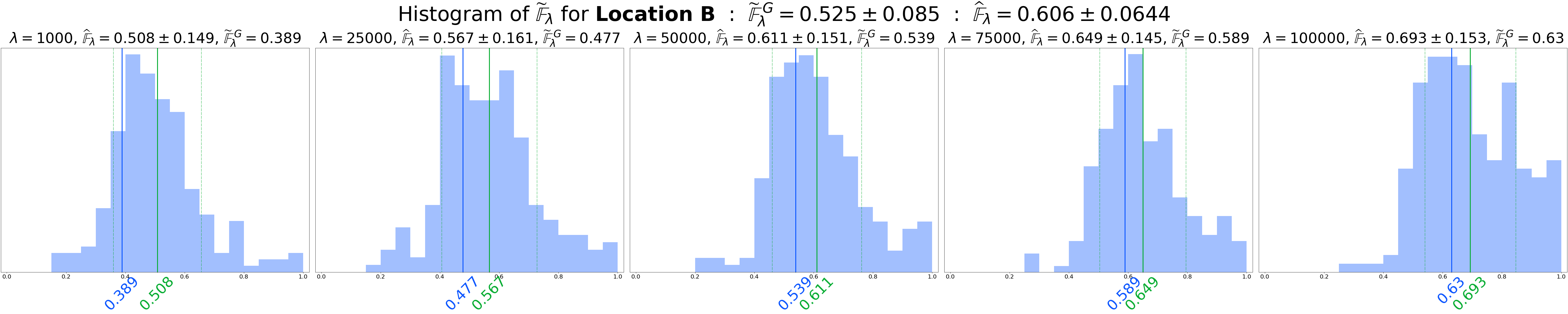}
    \caption[Distribution of normalized flat norm depending on $\lambda$ computed on uniformly sampled local regions.]
    {Distribution of normalized flat norm computed using five different values of $\lambda$ for $200$ uniformly sampled local regions in Location A (top) and Location B (bottom).
    The blue line in each histogram denotes the global normalized flat norm $\widetilde{\mathbb{F}}_{\lambda}^{G}$ computed for the location with the corresponding scale $\lambda$. The solid green line denotes the mean normalized flat norm $\widehat{\mathbb{F}}_{\lambda}$ for the uniformly sampled local regions computed with scale $\lambda$. The dashed green lines show the spread of the distribution.}
    \label{fig:flatnorm-hists-lambdas}
\end{figure}

Next, we consider the empirical distribution of normalized flat norm computed with scale parameter $\lambda=1000$ for uniformly sampled local regions in Location A and Location B, Fig.~\ref{fig:flatnorm-hists-epsilons}.
We show separate histograms for four different-sized local regions (different values of $\epsilon$).
Note that for \emph{small-sized local regions} (low $\epsilon$), the distribution is skewed toward the right. 
This is because when we consider small regions, we often capture very isolated network geometries and the flat norm computation is close to the total network length $\mathbb{F}_\lambda\left(T_1-T_2\right) \rightarrow |T_1|+|T_2|$, which again leads the normalized flat norm to be close to $1$.
Such occurrences are avoided in larger local regions, and therefore we do not observe skewed distributions.

\begin{figure*}[ht!]
    \centering
    \includegraphics[width=0.95\textwidth]{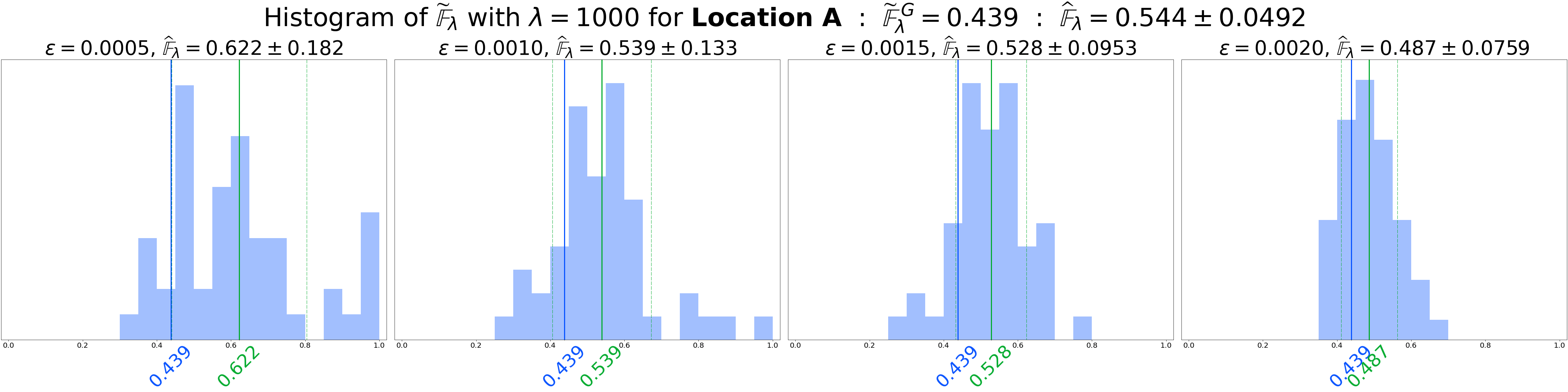}\\
    \smallskip
    \includegraphics[width=0.95\textwidth]{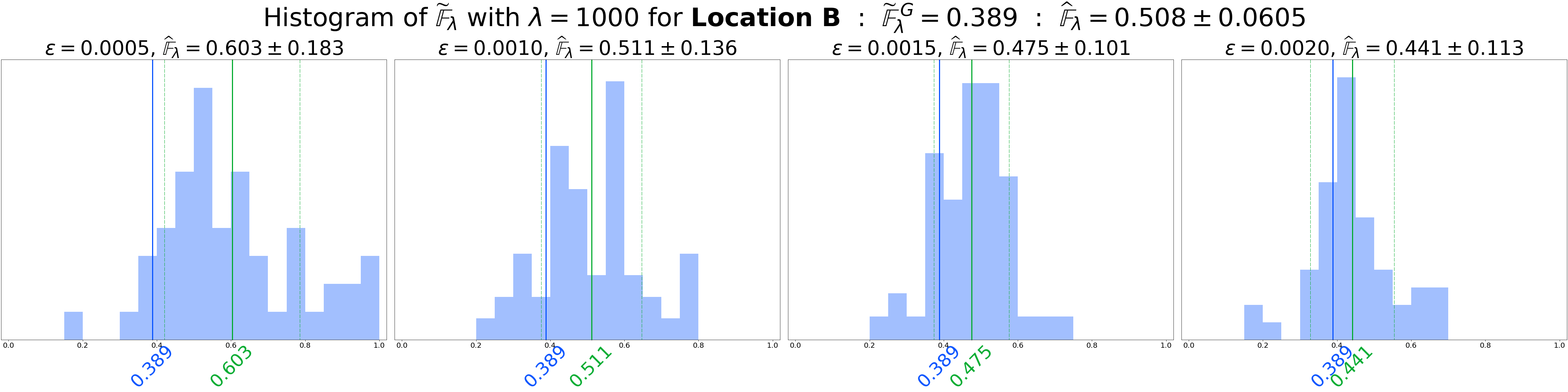}
    \caption[Distribution of normalized flat norm depending on $\epsilon$ computed on uniformly sampled local regions at the fixed scale  $\lambda=1000$.]
    {Distribution of normalized flat norm computed using $\lambda=1000$ for $50$ uniformly sampled local regions with four different sizes $\epsilon$ in Location A (top) and Location B (bottom). The blue line in each histogram denotes the global normalized flat norm $\widetilde{\mathbb{F}}_{\lambda}^{G}$. The solid green line denotes the mean normalized flat norm $\widehat{\mathbb{F}}_{\lambda}$ for the uniformly sampled local regions. The dashed green lines show the spread of the distribution.}
    \label{fig:flatnorm-hists-epsilons}
\end{figure*}

\subsection{Distribution of \texorpdfstring{$\widetilde{\mathbb{F}}_{\lambda}$}{normalized flat norm distance} across local regions}
\begin{figure*}
    \centering
    \includegraphics[width=0.90\textwidth]{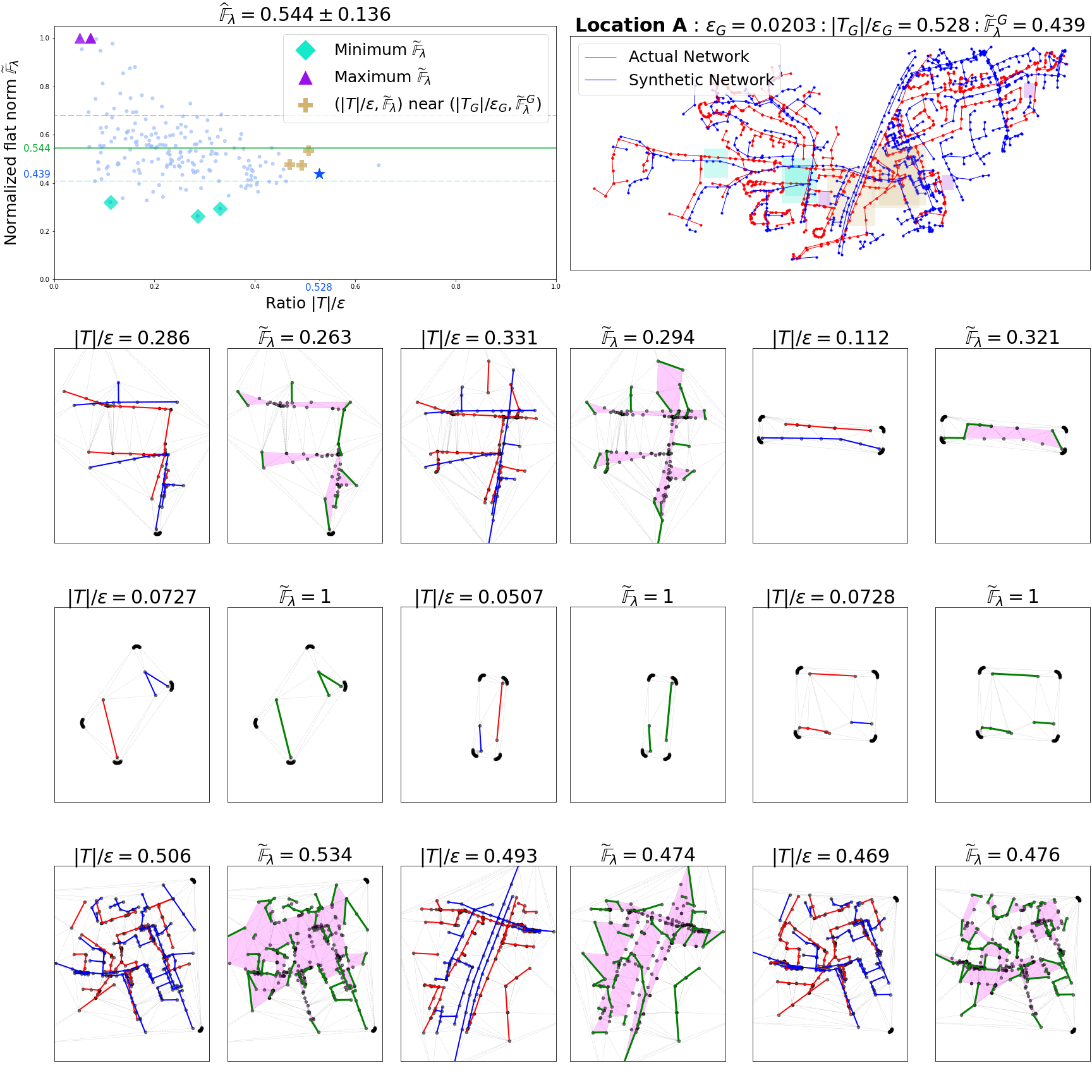}
    \caption[Normalized flat norm computed for entire Location A and few local regions within it.]%
    {Plots showing normalized flat norm computed for entire Location A and few local regions within it. The scatter plot (top left plot) shows the empirical distribution of $\left(|T|/\epsilon,\widetilde{\mathbb{F}}_{\lambda}\right)$ values with the global normalized flat norm $\left(|T_G|/\epsilon_G,\widetilde{\mathbb{F}}_{\lambda}^{G}\right)$ for the region (blue star). Nine local regions (three with small $\widetilde{\mathbb{F}}_{\lambda}$, three with large $\widetilde{\mathbb{F}}_{\lambda}$ and three with $\left(|T|/\epsilon,\widetilde{\mathbb{F}}_{\lambda}\right)$ values close to the global value $\left(|T_G|/\epsilon_G,\widetilde{\mathbb{F}}_{\lambda}^{G}\right)$) are additionally highlighted. The local regions are highlighted along with the pair of network geometries (top right plot). The normalized flat norm computation (with scale $\lambda=1000$) for the local regions are shown in bottom plots.}
    \label{fig:regionA}
\end{figure*}

  The scatter plot in the top left of Fig.~\ref{fig:regionA} shows the empirical distribution of $\left(|T|/\epsilon,\widetilde{\mathbb{F}}_{\lambda}\right)$ values.
  The scatter plot highlights as a blue star the global value $\left(|T_G|/\epsilon_G,\widetilde{\mathbb{F}}_{\lambda}^{G}\right)$ of Location A, which indicates the normalized flat norm computed for the entire location.
  The global normalized flat norm (with a scale parameter $\lambda=1000$) for Location A is $\widetilde{\mathbb{F}}_{\lambda}^{G}=0.439$ and the ratio $|T_G|/\epsilon_G=0.528$.
  Further, nine additional points are highlighted in the scatter plot denoting nine local regions within Location A.
  The solid green line denotes the mean of the normalized flat norm values and the dashed green lines indicate the spread of the values around the mean.

\begin{figure*}
    \centering
    \includegraphics[width=0.90\textwidth]{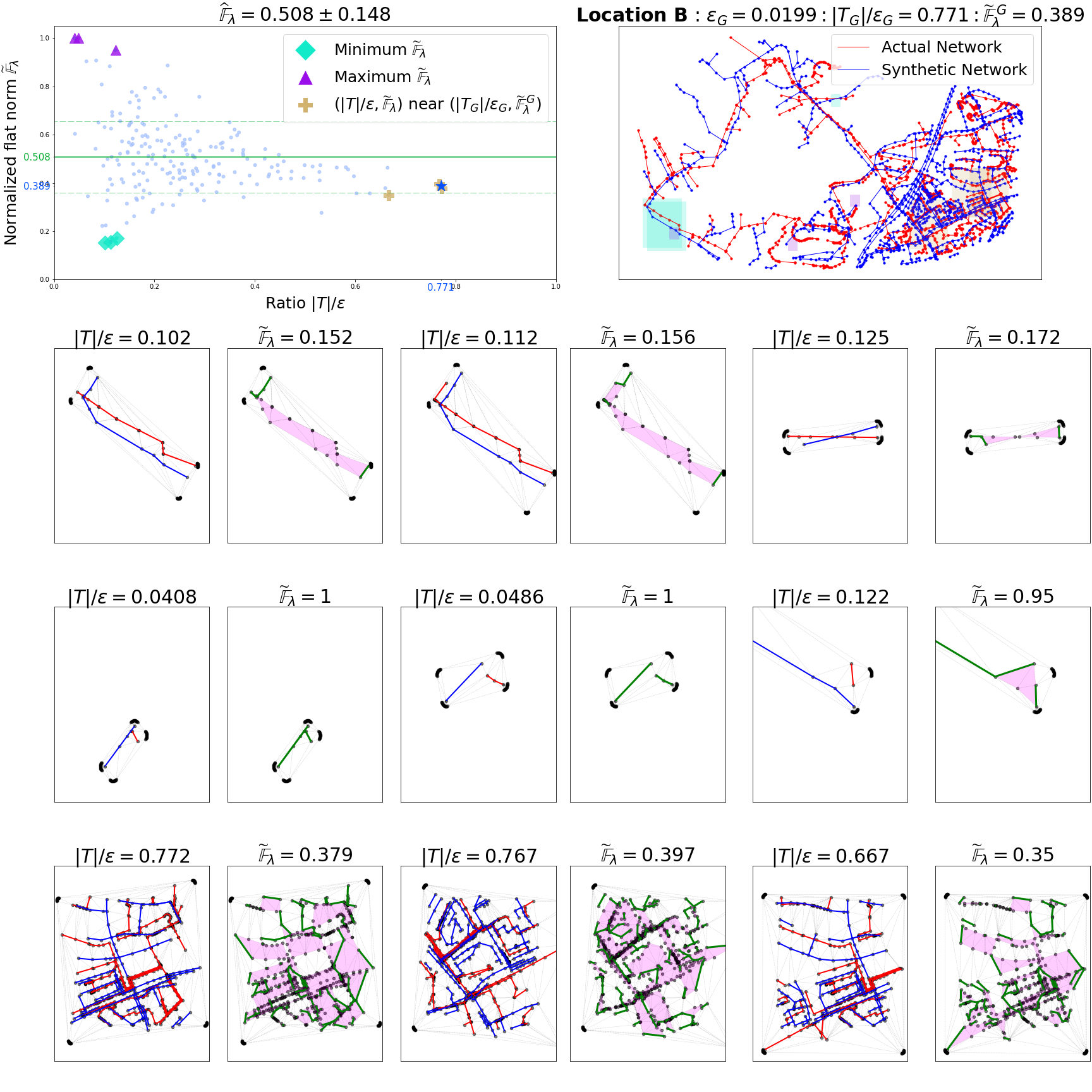}
    \caption[Normalized flat norm computed for entire Location B and few local regions within it.]
    {Plots showing normalized flat norm computed for entire Location B and few local regions within it. The scatter plot (top left plot) shows the empirical distribution of $\left(|T|/\epsilon,\widetilde{\mathbb{F}}_{\lambda}\right)$ values with the global normalized flat norm $\left(|T_G|/\epsilon_G,\widetilde{\mathbb{F}}_{\lambda}^{G}\right)$ for the region (blue star). 
    Nine local regions (three with small $\widetilde{\mathbb{F}}_{\lambda}$, three with large $\widetilde{\mathbb{F}}_{\lambda}$ and three with $\left(|T|/\epsilon,\widetilde{\mathbb{F}}_{\lambda}\right)$ values close to the global value $\left(|T_G|/\epsilon_G,\widetilde{\mathbb{F}}_{\lambda}^{G}\right)$) are additionally highlighted. The local regions are highlighted along with the pair of network geometries (top right plot). The normalized flat norm computation (with scale $\lambda=1000$) for the local regions are shown in bottom plots.}
    \label{fig:regionB}
\end{figure*}

The nine local regions are selected such that three of them have the minimum $\widetilde{\mathbb{F}}_{\lambda}$ in the location (highlighted by cyan colored diamonds), three of them have the maximum $\widetilde{\mathbb{F}}_{\lambda}$ in the location (highlighted by purple triangles), and the remaining three local regions have the $\left(|T|/\epsilon,\widetilde{\mathbb{F}}_{\lambda}\right)$ values close to the global value $\left(|T_G|/\epsilon_G,\widetilde{\mathbb{F}}_{\lambda}^{G}\right)$ for the location (highlighted by tan plus symbols).
The network geometries within each region and the flat norm computation with scale $\lambda=1000$ are shown in the bottom plots.
The computed flat norm $\widetilde{\mathbb{F}}_{\lambda}$ and ratio $|T|/\epsilon$ values are shown above each plot.
The local regions are also highlighted (cyan, purple, and tan colored boxes, respectively) in the top right plot where the actual and synthetic network geometries within the entire location are overlaid.
Fig.~\ref{fig:regionB} shows similar local regions from Location B.

From a mere visual inspection of both Figs.~\ref{fig:regionA} and~\ref{fig:regionB}, we notice that the network geometries in each local region shown in the first row of the bottom plots resemble and almost overlap each other.
The computed normalized flat norm $\widetilde{\mathbb{F}}_{\lambda}$ values for these local regions agree with this observation.
Similarly, the large value of the normalized flat norm justifies the observation that network geometries for local regions depicted in the second row of the bottom plots do not resemble each other.
These observations validate our choice of using normalized flat norm as a suitable measure to compare network geometries for local regions.

\section{Conclusions}

We have proposed a fairly general metric to compare a pair of network geometries embedded on the same plane.
Unlike standard approaches that map the geometries to points in a possibly simpler space and then measuring distance between those points~\cite{KeBaCaLe2009}, or comparing ``signatures'' for the geometries,  our metric works directly in the input space and hence allows us to capture all details in the input.
The metric uses the multiscale flat norm from geometric measure theory, and can be used in more general settings as long as we can triangulate the region containing the two geometries.
It is impossible to derive \emph{standard} stability results for this distance measure that imply only small changes in the flat norm metric when the inputs change by a small amount---there is no alternative metric to measure the \emph{small change in the input}.
For instance, our theoretical example (in Fig.~\ref{fig:currents:example-currents-and-neighborhoods}) shows that the commonly used Hausdorff metric cannot be used for this purpose.
Instead, we have derived upper bounds on the flat norm distance between a piecewise linear 1-current and its perturbed version as a function of the radius of perturbation under certain assumptions provided the perturbations are performed carefully (see Section \ref{subsec:FN-BOUND}).
On the other hand, we do get natural stability results for our distance following the properties of the flat norm---small changes in the input geometries lead to only small changes in the flat norm distance between them~\cite{Federer1969,Morgan2016}.

We use the proposed metric to compare a pair of power distribution networks: (i) actual power distribution networks of two locations in a county of USA obtained from a power company and (ii) synthetically generated digital duplicate of the network created for the same geographic location.
The proposed comparison metric is able to perform global as well as local comparison of network geometries for the two locations.
We discuss the effect of different parameters used in the metric on the comparison.
Further, we validate the suitability of using the flat norm metric for such comparisons using computation as well as theoretical examples.

\end{mainchapters}


\makebibliography{main.bib}

\begin{thebibliography}{10}

\bibitem{roads2014}
Mahmuda Ahmed, Brittany~Terese Fasy, and Carola Wenk.
\newblock Local persistent homology based distance between maps.
\newblock In {\em Proceedings of the 22nd ACM SIGSPATIAL International Conference on Advances in Geographic Information Systems}, SIGSPATIAL '14, page 43–52, New York, NY, USA, 2014. Association for Computing Machinery.

\bibitem{BaDiBiChSuWa2019}
Yunsheng Bai, Hao Ding, Song Bian, Ting Chen, Yizhou Sun, and Wei Wang.
\newblock {SimGNN: A Neural Network Approach to Fast Graph Similarity Computation}.
\newblock In {\em Proceedings of the Twelfth ACM International Conference on Web Search and Data Mining}, WSDM '19, page 384–392, New York, NY, USA, 2019. Association for Computing Machinery.

\bibitem{highres_net}
Antoine Bidel, Tom Schelo, and Thomas Hamacher.
\newblock Synthetic distribution grid generation based on high resolution spatial data.
\newblock In {\em 2021 IEEE International Conference on Environment and Electrical Engineering and 2021 IEEE Industrial and Commercial Power Systems Europe (EEEIC / ICPS Europe)}, pages 1--6. IEEE, 2021.

\bibitem{roads_haus}
Maria~Antonia Brovelli, Marco Minghini, Monia Molinari, and Peter Mooney.
\newblock Towards an automated comparison of openstreetmap with authoritative road datasets.
\newblock {\em Transactions in GIS}, 21(2):191--206, 2017.

\bibitem{streets2006}
Alessio Cardillo, Salvatore Scellato, Vito Latora, and Sergio Porta.
\newblock Structural properties of planar graphs of urban street patterns.
\newblock {\em Phys. Rev. E}, 73:066107, 2006.

\bibitem{chambers2012homology}
Erin~W Chambers, Jeff Erickson, and Amir Nayyeri.
\newblock Homology flows, cohomology cuts.
\newblock {\em SIAM Journal on Computing}, 41(6):1605--1634, 2012.

\bibitem{chen2011hardness}
Chao Chen and Daniel Freedman.
\newblock Hardness results for homology localization.
\newblock {\em Discrete \& Computational Geometry}, 45(3):425--448, 2011.

\bibitem{chen2022maximum}
Li~Chen, Rasmus Kyng, Yang~P Liu, Richard Peng, Maximilian~Probst Gutenberg, and Sushant Sachdeva.
\newblock Maximum flow and minimum-cost flow in almost-linear time.
\newblock In {\em 2022 IEEE 63rd Annual Symposium on Foundations of Computer Science (FOCS)}, pages 612--623. IEEE, 2022.

\bibitem{embeddabilityR3}
Arnaud de~Mesmay, Yo'av Rieck, Eric Sedgwick, and Martin Tancer.
\newblock Embeddability in $\mathbb{R}^{3}$ is np-hard.
\newblock In {\em Proceedings of the Twenty-Ninth Annual ACM-SIAM Symposium on Discrete Algorithms}, pages 1316--1329. SIAM, 2018.

\bibitem{ohcp2011}
Tamal~K Dey, Anil~N Hirani, and Bala Krishnamoorthy.
\newblock Optimal homologous cycles, total unimodularity, and linear programming.
\newblock {\em SIAM Journal on Computing}, 40(4):1026--1044, 2011.

\bibitem{dey2020computing}
Tamal~K Dey, Tao Hou, and Sayan Mandal.
\newblock Computing minimal persistent cycles: Polynomial and hard cases.
\newblock In {\em Proceedings of the Fourteenth Annual ACM-SIAM Symposium on Discrete Algorithms}, pages 2587--2606. SIAM, 2020.

\bibitem{dunfield2011least}
Nathan~M Dunfield and Anil~N Hirani.
\newblock The least spanning area of a knot and the optimal bounding chain problem.
\newblock In {\em Proceedings of the twenty-seventh annual symposium on Computational geometry}, pages 135--144, 2011.

\bibitem{duval2015cuts}
Art~M Duval, Caroline~J Klivans, and Jeremy~L Martin.
\newblock Cuts and flows of cell complexes.
\newblock {\em Journal of Algebraic Combinatorics}, 41:969--999, 2015.

\bibitem{eppstein_1995}
David Eppstein.
\newblock Subgraph isomorphism in planar graphs and related problems.
\newblock In {\em Proceedings of the Sixth Annual ACM-SIAM Symposium on Discrete Algorithms}, SODA '95, page 632–640, USA, 1995. Society for Industrial and Applied Mathematics.

\bibitem{escolar2016optimal}
Emerson~G Escolar and Yasuaki Hiraoka.
\newblock Optimal cycles for persistent homology via linear programming.
\newblock In {\em Optimization in the Real World}, pages 79--96. Springer, 2016.

\bibitem{arcgis}
ESRI.
\newblock Georeferencing a raster to a vector.

\bibitem{Federer1969}
Herbert Federer.
\newblock {\em Geometric Measure Theory}.
\newblock Die Grundlehren der mathematischen Wissenschaften, Band 153. Springer-Verlag, 1969.

\bibitem{glaunes2005transport}
Joan Glaunes.
\newblock {\em Transport par diff{\'e}omorphismes de points, de mesures et de courants pour la comparaison de formes et l’anatomie num{\'e}rique}.
\newblock PhD thesis, Universit{\'e} Paris, 2005.

\bibitem{goldberg2002combinatorial}
Timothy~E Goldberg.
\newblock {\em Combinatorial Laplacians of simplicial complexes}.
\newblock PhD thesis, Bard College, 2002.

\bibitem{hagberg2008exploring}
Aric Hagberg, Pieter Swart, and Daniel S~Chult.
\newblock Exploring network structure, dynamics, and function using networkx.
\newblock Technical report, Los Alamos National Lab.(LANL), Los Alamos, NM (United States), 2008.

\bibitem{ibrahim2011simplicial}
Sharif Ibrahim, Bala Krishnamoorthy, and Kevin~R. Vixie.
\newblock Simplicial flat norm with scale.
\newblock {\em Journal of Computational Geometry}, 4(1):133--159, 2013.

\bibitem{KeBaCaLe2009}
David~G. Kendall, Dennis~M. Barden, {T.~Keith} Carne, and Huiling Le.
\newblock {\em Shape and Shape Theory}.
\newblock Wiley Series in Probability and Statistics. Wiley, 2009.

\bibitem{validate2020}
Venkat Krishnan, Bruce Bugbee, Tarek Elgindy, Carlos Mateo, Pablo Duenas, Fernando Postigo, Jean-S\'{e}ebastien Lacroix, Tom\'{a}s G\'{o}mez~San Roman, and Bryan Palmintier.
\newblock Validation of synthetic u.s. electric power distribution system data sets.
\newblock {\em IEEE Transactions on Smart Grid}, 11(5):4477--4489, 2020.

\bibitem{overbye_2020}
Hanyue Li, Jessica~L. Wert, Adam~Barlow Birchfield, Thomas~J. Overbye, Tomas Gomez~San Roman, Carlos~Mateo Domingo, Fernando Emilio~Postigo Marcos, Pablo~Duenas Martinez, Tarek Elgindy, and Bryan Palmintier.
\newblock Building highly detailed synthetic electric grid data sets for combined transmission and distribution systems.
\newblock {\em IEEE Open Access Journal of Power and Energy}, 7:478--488, 2020.

\bibitem{minRepresentatives2021}
Lu~Li, Connor Thompson, Gregory Henselman-Petrusek, Chad Giusti, and Lori Ziegelmeier.
\newblock {Minimal Cycle Representatives in Persistent Homology Using Linear Programming: An Empirical Study With User’s Guide}.
\newblock {\em Frontiers in Artificial Intelligence}, 4(681117), 2021.

\bibitem{feeder_gan}
Ming Liang, Yao Meng, Jiyu Wang, David~L. Lubkeman, and Ning Lu.
\newblock Feedergan: Synthetic feeder generation via deep graph adversarial nets.
\newblock {\em IEEE Transactions on Smart Grid}, 12(2):1163--1173, 2021.

\bibitem{lim2020hodge}
Lek-Heng Lim.
\newblock Hodge laplacians on graphs.
\newblock {\em Siam Review}, 62(3):685--715, 2020.

\bibitem{mac1936combinatorial}
Saunders Mac~Lane.
\newblock {\em A combinatorial condition for planar graphs}.
\newblock Seminarium Matemat., 1936.

\bibitem{MaWe2024}
Sushovan Majhi and Carola Wenk.
\newblock Distance measures for geometric graphs.
\newblock {\em Computational Geometry}, 118:102056, 2024.

\bibitem{nrel_net}
Carlos Mateo, Fernando Postigo, Fernando de~Cuadra, Tom\'{a}s G\'{o}mez~San Roman, Tarek Elgindy, Pablo Due\~{n}as, Bri-Mathias Hodge, Venkat Krishnan, and Bryan Palmintier.
\newblock Building large-scale u.s. synthetic electric distribution system models.
\newblock {\em IEEE Transactions on Smart Grid}, 11(6):5301--5313, 2020.

\bibitem{embeddabilityRd}
Jir{\'\i} Matousek, Martin Tancer, and Uli Wagner.
\newblock Hardness of embedding simplicial complexes in $\mathbb{R}^{d}$.
\newblock In {\em Society for Industrial and Applied Mathematics and Association for Computing Machinery. Proceeding of the ACM-SIAM Symposium on Discrete Algorithms}, page 855. Society for Industrial and Applied Mathematics, 2009.

\bibitem{maxwell2021algorithmic}
William Maxwell.
\newblock {\em Algorithmic Problems on Simplicial Complexes}.
\newblock PhD thesis, Oregon State University, 2021.

\bibitem{generalizedMaxflow2021}
William Maxwell and Amir Nayyeri.
\newblock Generalized max flows and min cuts in simplicial complexes.
\newblock {\em URL: http://arxiv. org/abs/2106.14116}, 2021.

\bibitem{mendez2023directed}
David M{\'e}ndez and Rub{\'e}n~J S{\'a}nchez-Garc{\'\i}a.
\newblock A directed persistent homology theory for dissimilarity functions.
\newblock {\em Journal of Applied and Computational Topology}, 7(4):771--813, 2023.

\bibitem{meyur2023structural}
Rounak Meyur, Kostiantyn Lyman, Bala Krishnamoorthy, and Mahantesh Halappanavar.
\newblock Structural validation of synthetic power distribution networks using the multiscale flat norm.
\newblock In {\em International Conference on Computational Science}, pages 55--69. Springer, 2023.

\bibitem{rounak2020}
Rounak Meyur, Madhav Marathe, Anil Vullikanti, Henning Mortveit, Samarth Swarup, Virgilio Centeno, and Arun Phadke.
\newblock Creating realistic power distribution networks using interdependent road infrastructure.
\newblock In {\em 2020 IEEE International Conference on Big Data (Big Data)}, pages 1226--1235. IEEE, 2020.

\bibitem{rounak_pnas}
Rounak Meyur, Anil Vullikanti, Samarth Swarup, Henning Mortveit, Virgilio Centeno, Arun Phadke, Vince Poor, and Madhav Marathe.
\newblock Ensembles of realistic power distribution networks.
\newblock {\em Proceedings of the National Academy of Sciences}, 119(26):e2123355119, 2022.

\bibitem{efficiency2020}
Ignacio Morer, Alessio Cardillo, Albert D\'{\i}az-Guilera, Luce Prignano, and Sergi Lozano.
\newblock Comparing spatial networks: A one-size-fits-all efficiency-driven approach.
\newblock {\em Phys. Rev. E}, 101:042301, 2020.

\bibitem{Morgan2016}
Frank Morgan.
\newblock {\em Geometric Measure Theory: A Beginner's Guide}.
\newblock Academic Press, 5th edition, 2016.

\bibitem{MoVi2007}
Simon~P. Morgan and Kevin~R. Vixie.
\newblock {$L^1\text{TV}$} computes the flat norm for boundaries.
\newblock {\em Abstract and Applied Analysis}, 2007:14, 2007.

\bibitem{graphsim-2020}
Seongmin Ok.
\newblock A graph similarity for deep learning.
\newblock In H.~Larochelle, M.~Ranzato, R.~Hadsell, M.F. Balcan, and H.~Lin, editors, {\em Advances in Neural Information Processing Systems}, volume~33, pages 1--12. Curran Associates, Inc., 2020.

\bibitem{osm}
{Open Street Map Foundation}.
\newblock Open street maps.

\bibitem{PaGaMiHa2018}
Benjamin Paa{\ss}en, Claudio Gallicchio, Alessio Micheli, and Barbara Hammer.
\newblock Tree edit distance learning via adaptive symbol embeddings.
\newblock In Jennifer Dy and Andreas Krause, editors, {\em Proceedings of the 35th International Conference on Machine Learning}, volume~80 of {\em Proceedings of Machine Learning Research}, pages 3976--3985. PMLR, 2018.

\bibitem{patchkoria1977cohomology}
Alex Patchkoria.
\newblock Cohomology of monoids with coefficients in semimodules.
\newblock {\em Bull. Georgian Acad. Sci}, 86(3):545--548, 1977.

\bibitem{riba2020}
Pau Riba, Andreas Fischer, Josep Llad\'{o}s, and Alicia Forn\'{e}s.
\newblock Learning graph edit distance by graph neural networks.
\newblock {\em Pattern Recognition}, 120:108132, 2021.

\bibitem{samiul2021}
Kamol~Chandra Roy, Samiul Hasan, Aron Culotta, and Naveen Eluru.
\newblock Predicting traffic demand during hurricane evacuation using real-time data from transportation systems and social media.
\newblock {\em Transportation Research Part C: Emerging Technologies}, 131:103339, 2021.

\bibitem{anna_naps}
Shammya~Shananda Saha, Eran Schweitzer, Anna Scaglione, and Nathan~G. Johnson.
\newblock A framework for generating synthetic distribution feeders using openstreetmap.
\newblock In {\em 2019 North American Power Symposium (NAPS)}, pages 1--6. IEEE, 2019.

\bibitem{schweitzer}
Eran Schweitzer, Anna Scaglione, Antonello Monti, and Giuliano~Andrea Pagani.
\newblock Automated generation algorithm for synthetic medium voltage radial distribution systems.
\newblock {\em IEEE Journal on Emerging and Selected Topics in Circuits and Systems}, 7(2):271--284, 2017.

\bibitem{Si2010}
Hang Si.
\newblock Constrained {D}elaunay tetrahedral mesh generation and refinement.
\newblock {\em Finite Elements in Analysis and Design}, 46:33--46, 2010.

\bibitem{sullivan1990crystalline}
John~M. Sullivan.
\newblock {\em A Crystalline Approximation Theorem for Hypersurfaces}.
\newblock PhD thesis, Princeton University, 1990.

\bibitem{Tantardini2019}
Mattia Tantardini, Francesca Ieva, Lucia Tajoli, and Carlo Piccardi.
\newblock Comparing methods for comparing networks.
\newblock {\em Scientific Reports}, 9(1):17557, 2019.

\bibitem{van2020deterministic}
Jan van~den Brand.
\newblock {\em A Deterministic Linear Program Solver in Current Matrix Multiplication Time}, pages 259--278.
\newblock SIAM, 2020.

\bibitem{veinott1968integral}
Arthur~F Veinott, Jr and George~B Dantzig.
\newblock Integral extreme points.
\newblock {\em SIAM review}, 10(3):371--372, 1968.

\bibitem{vixie2010multiscale}
Kevin~R Vixie, Keith Clawson, Thomas~J Asaki, Gary Sandine, Simon~P Morgan, and Brandon Price.
\newblock Multiscale flat norm signatures for shapes and images.
\newblock {\em Applied Mathematical Sciences}, 4(14):667--680, 2010.

\bibitem{xu2015}
Hangjun Xu.
\newblock An algorithm for comparing similarity between two trees.
\newblock \url{https://arxiv.org/abs/1508.03381}.
\newblock Last accessed 26 Sept 2022.

\end{thebibliography}

\begin{appendices}

\chapter{MSFN and OHCP: Example}
\begin{figure}[h]
    \centering
    \includegraphics[width=0.8\linewidth]{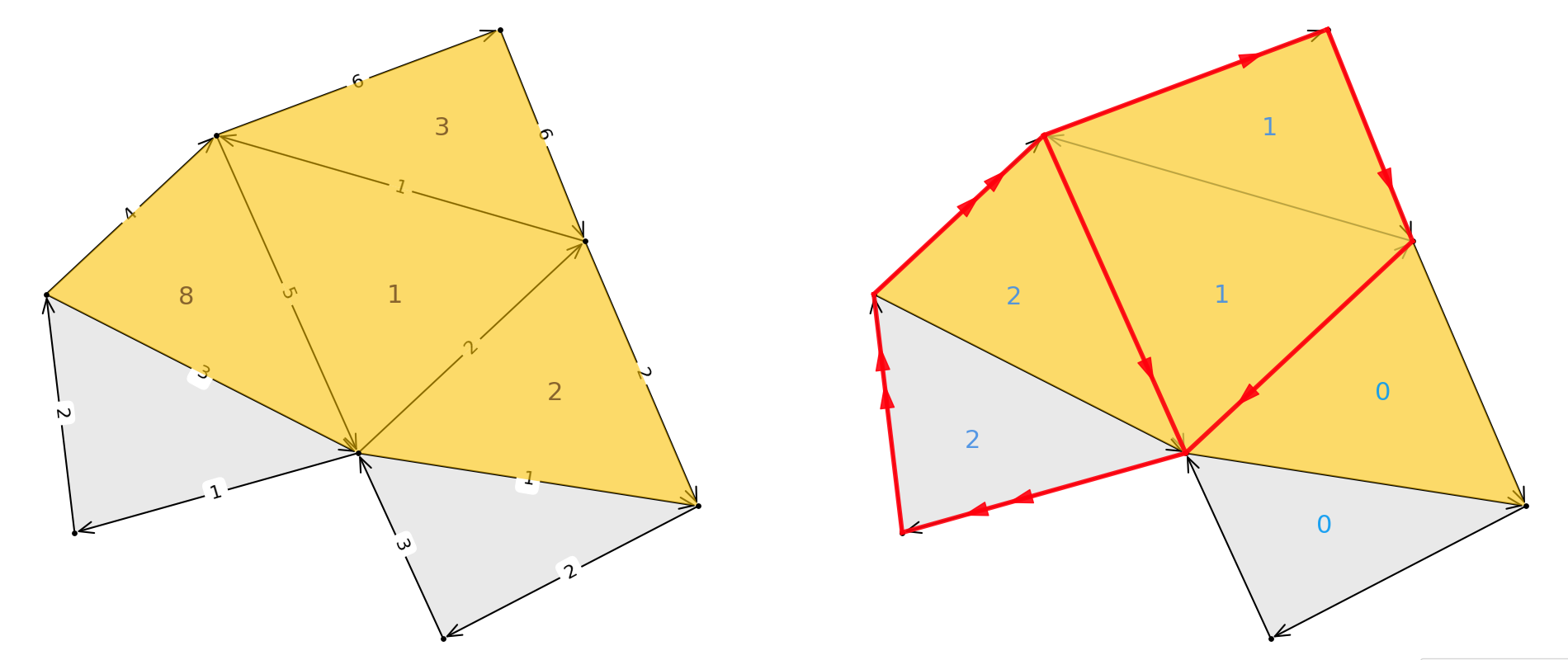}
    \caption[An example of a 2-complex $\mc{K}$ with 2 inner voids and a non-trivial $1$-cycle $\bm{z}$.]
        {LEFT: An example of a 2-complex $\mc{K}$ with the length and area weights displayed. Gray triangles are voids $\nu_1$ and $\nu_2$.
            RIGHT: Shortest path distances from $\nu_{0}$ relative to \textcolor{BrightRed}{$\bm{z}$}
            (shown in red, the number of arrows corresponds to $\abs{\bm{z}(e_i) }$).
        }
    \label{fig:ort-and-dual}
\end{figure}

\begin{figure}[!h]
    \centering
    \includegraphics[width=0.3\linewidth]{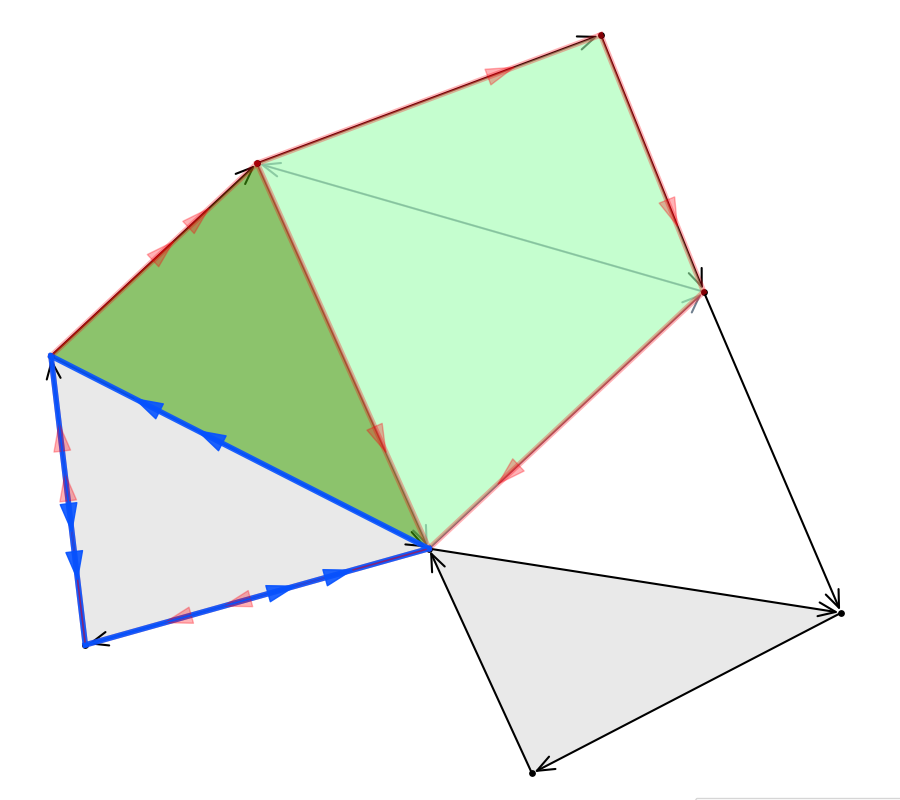} \hspace{10pt}
    \includegraphics[width=0.3\linewidth]{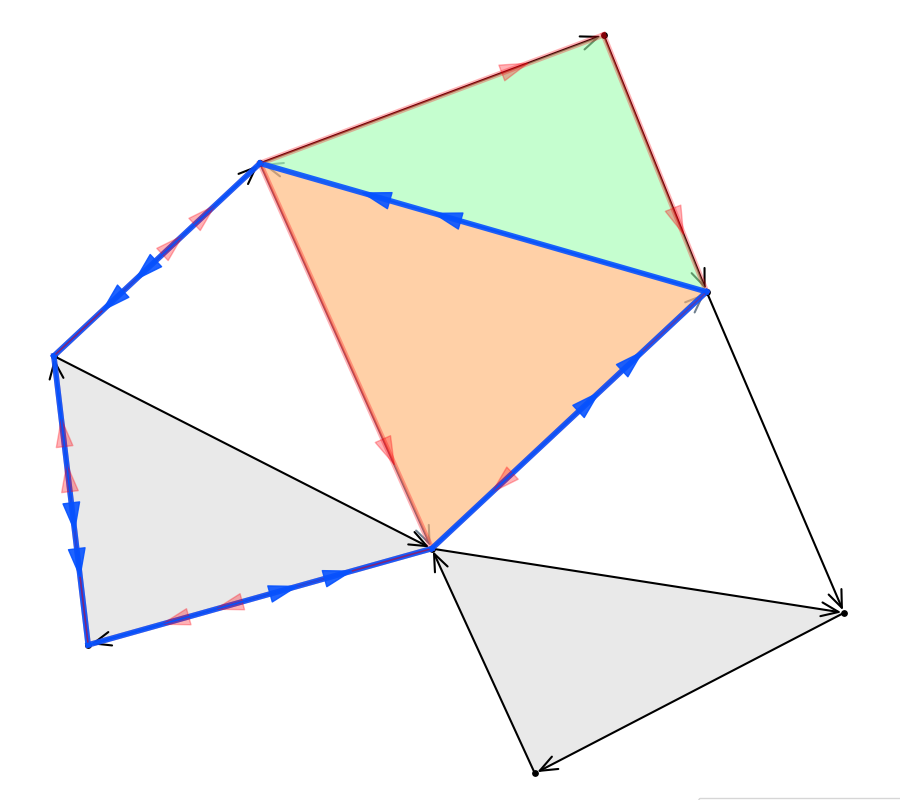}
    \includegraphics[width=0.3\linewidth]{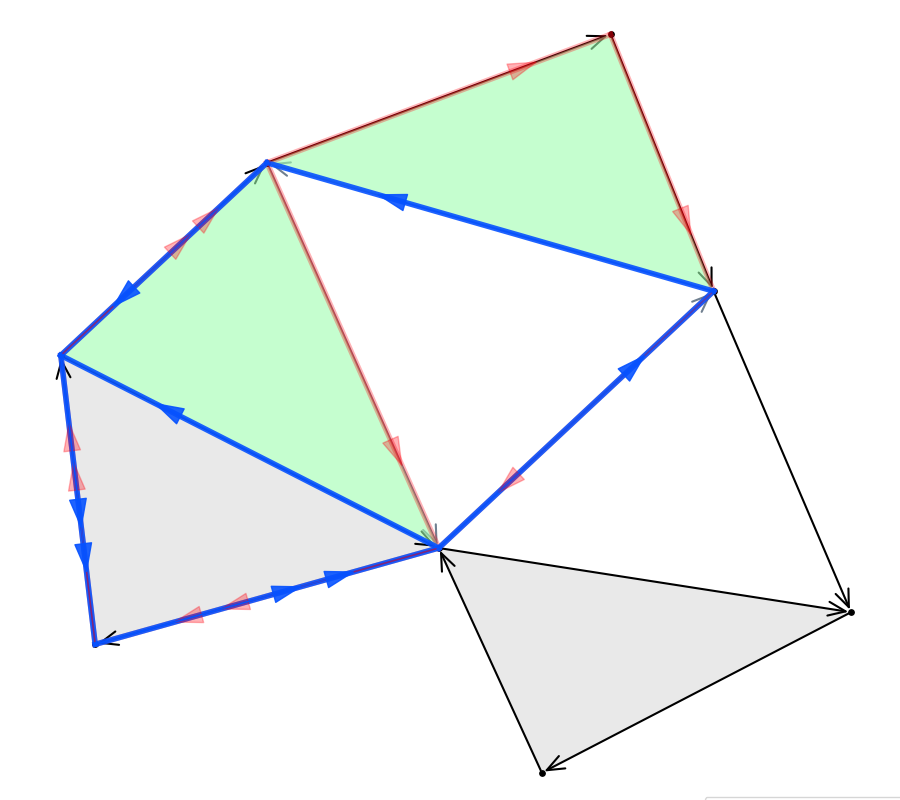}

    \caption[OHCP and mSFN optimal solutions for $\mc{K}$  and $\bm{z}$.]
        {LEFT: OHCP solution.
        MID: SFN solution.
        RIGHT: mSFN solution for $\lambda = \sfrac{1}{2}$ (edge-lengths doubled).
        The optimal $1$-cycle $\bm{x} \sim \bm{z}$ is shown as blue counter clockwise cycle \textcolor{FullBlue}{$\bm{-x}$}.
        The number of arrows corresponds to $\abs{\bm{x}(e_i) }$.
        Highlighted triangles correspond to $\bm{\pi}\!: \bd \bm{\pi}  = \textcolor{BrightRed}{\bm{z}} \textcolor{FullBlue}{\bm{-x}}$:
       \textcolor{LimeGreen}{$\bm{\pi}(\tau_j) = +1$} -- green,
       \textcolor{ForestGreen}{$\bm{\pi}(\tau_j) = +2$} -- dark green,
       \textcolor{RedOrange}{$\bm{\pi}(\tau_j) = -1$} -- red/orange.
    }
\end{figure}



\end{appendices}

\end{document}